\documentclass[preprint, aps, prd]{revtex4-1}

\def\mc{\mathcal}
\def\ul{\underline}

\usepackage[dvips]{graphicx}
\usepackage{amssymb}
\usepackage{amssymb,amsmath}
\usepackage{graphicx}
\usepackage{dsfont}
\usepackage{caption}
\usepackage{subcaption}
\usepackage{mathtools}
\usepackage{verbatim}
\usepackage{graphicx}
\usepackage{multirow}
\usepackage[outline]{contour}
\usepackage{xcolor,colortbl}
\makeatother
\begin{document}
\title{Supersymmetric quantum mechanics from wrapped D4-branes}

\author{Patharadanai Nuchino} \email[REVTeX Support:
]{danai.nuchino@hotmail.com and parinya.ka@hotmail.com} 
\affiliation{Quantum and Gravity Theory Research Group, Department of Physics, Faculty of Science, Ramkhamhaeng University, 282 Ramkhamhaeng Road, Bang Kapi, Bangkok 10240, Thailand}
 \author{Parinya Karndumri}
\affiliation{String Theory and
Supergravity Group, Department of Physics, Faculty of Science,
Chulalongkorn University, 254 Phayathai Road, Pathumwan, Bangkok
10330, Thailand}

\date{\today}
\begin{abstract}
We find a large class of holographic solutions describing D4-branes wrapped on four-manifolds $\mathcal{M}_4$ with constant curvature leading to gravity duals of supersymmetric quantum mechanics in the IR via twisted compactifications. The manifolds $\mathcal{M}_4$ considered here are four-dimensional spheres and hyperbolic spaces, products of two Riemann surfaces, and Kahler four-cycles. The solutions are obtained from the maximal gauged supergravity in six dimensions with $CSO(p,q,5-p-q)$ and $CSO(p,q,4-p-q)\ltimes \mathbb{R}^4$ gauge groups. These gauged supergravities can be embedded in type IIA theory via consistent truncations on $H^{p,q}\times \mathbb{R}^{5-p-q}$ and $H^{p,q}\times\mathbb{R}^{4-p-q}\times S^1$, respectively. The solutions take the form of $t\times \mathcal{M}_4$-sliced domain walls interpolating between locally flat domain walls and singular geometries in the IR. Upon uplifted to type IIA theory, many solutions admit physical IR singularities and could holographically describe supersymmetric quantum mechanics arising from twisted compactifications of D4-branes on $\mathcal{M}_4$.
\end{abstract}
\maketitle

\tableofcontents
%%%%%%%%%%%%%%%%%%%%%%%%%%%%%%%%%%%%%%%%%%%%%%%%%%%%%%%%%%%%%%%%%%%%%%%%%%%%%%%%%%%%%%%%%%%%%%%%%%%%%%%%
\section{Introduction}
For almost three decades, the AdS/CFT correspondence \cite{maldacena,Gubser_AdS_CFT,Witten_AdS_CFT} and the generalization to non-conformal field theories called DW/QFT correspondence \cite{DW_QFT1,DW_QFT2,DW_QFT3} have attracted much attention in the holographic study of strongly-coupled field theories using gravity duals. Both AdS$_{d+1}$/CFT$_{d}$ and DW$_{d+1}$/QFT$_d$ dualities have been used to study various aspects of field theories in different dimensions with $d\geq 2$. Unlike higher dimensional counterparts, AdS$_{2}$/CFT$_{1}$ and DW$_{2}$/QFT$_1$ correspondences have been far less explored to date. The dual $(0+1)$-dimensional field theories in this case are superconformal quantum mechanics and supersymmetric quantum mechanics. Some examples of the latter arise naturally in string/M-theory in the form of matrix models \cite{BFSS,BMN}. These models are expected to play an important role in string theory, D-brane physics, and non-commutative geometries, see \cite{SQM_review1,SQM_review2} for recent reviews.
\\
\indent Most of the important results in AdS$_{d+1}$/CFT$_{d}$ or DW$_{d+1}$/QFT$_d$ dualities are generally obtained by studying various types of solutions to gauged supergravity in $d+1$ dimensions. Since the construction of two-dimensional gauged supergravity has appeared only recently in \cite{2D_SUGRA1}, see also \cite{2D_hidden} for earlier results, the study of DW$_{2}$/QFT$_1$ correspondence within the framework of gauged supergravity has only been carried out in a handful of previous works \cite{2D_D0,2D_MQM_holo,deformed_Matrix_henning}, see also \cite{Matrix_shperical_brane} for another approach using spherical branes. These provide gravity duals of matrix quantum mechanics on the world-volume of D0-branes. Alternatively, QFT$_1$ or supersymmetric quantum mechanics can also arise from Dp-branes wrapped on a $p$-dimensional compact manifold. These configurations can preserve some amount of the original supersymmetry by a topological twist \cite{Witten_twist,Maldacena_nogo}.  
\\
\indent In the present paper, we are interested in the case of D4-branes wrapped on a four-manifold $\mc{M}_4$ with constant curvature, see \cite{Gauntlett1,Gauntlett2,D6_wrapped_M4,Calos_D6_M4,2D_Bobev,Wraped_M5,BH_microstate_6D2,Minwoo_6D_BH2} for examples of other D-branes and M-branes wrapped on a four-manifold and, in particular, \cite{Rafael_SQM_D6} for a solution describing supersymmetric quantum mechanics from D6-branes wrapped on six-cycles. The $\mc{M}_4$ manifolds under consideration here are given by a Riemannian four-manifold of constant curvature, a product of two Riemann surfaces, and a Kahler four-cycle. In the DW/QFT correspondence, the dynamics of the maximal five-dimensional gauge theory on the world-volume of D4-branes in type IIA theory can be explored by studying the effective maximal gauged supergravity in six dimensions \cite{6D_Max_Gauging}. Accordingly, we will study solutions of D4-branes wrapped on $\mc{M}_4$ by performing topological twists on the (flat) domain wall solutions dual to five-dimensional super Yang-Mills (SYM) theory along $\mc{M}_4$. A large number of flat domain walls from maximal gauged supergravity in six dimensions has been found in \cite{6D_DW_I,6D_DW_II,6D_DW_III}, and solutions describing D4-branes wrapped on two- and three-manifolds have also studied recently in \cite{6D_twist_I}. 
\\
\indent We will consider six-dimensional gauged supergravity with $CSO(p,q,5-p-q)$ and $CSO(p,q,4-p-q)\ltimes \mathbb{R}^4$ gauged groups. Both of these gauged supergravities can be obtained from an $S^1$-reduction of the maximal gauged supergravity in seven dimensions with $CSO(p,q,5-p-q)$ and $CSO(p,q,4-p-q)$ gauge groups \cite{7D_Max_Gauging}. It has been shown in \cite{Malek_IIA_IIB} that seven-dimensional $CSO(p,q,4-p-q)$ gauged supergravity can be obtained from type IIA theory via a consistent truncation on $H^{p,q}\times \mathbb{R}^{4-p-q}$. Accordingly, further truncation on $S^1$ leads to six-dimensional gauged supergravity with $CSO(p,q,4-p-q)\ltimes \mathbb{R}^4$, or equivalently, a truncation of type IIA theory on $H^{p,q}\times \mathbb{R}^{4-p-q}\times S^1$ leads to $CSO(p,q,4-p-q)\ltimes \mathbb{R}^4$ gauged supergravity in six dimensions. On the other hand, the maximal $CSO(p,q,5-p-q)$ gauged supergravity in seven dimensions can also be obtained as a truncation of eleven-dimensional supergravity on $H^{p,q}\times \mathbb{R}^{5-p-q}$ \cite{Henning_KK}. Another dimensional reduction on $S^1$ gives $CSO(p,q,5-p-q)$ gauged supergravity in six dimensions as a consistent truncation of eleven-dimensional supergravity on $H^{p,q}\times \mathbb{R}^{5-p-q}\times S^1$. Together with the fact that type IIA theory can be obtained from an $S^1$-reduction of M-theory, the six-dimensional $CSO(p,q,5-p-q)$ gauged supergravity can be considered as a truncation of type IIA theory on $H^{p,q}\times \mathbb{R}^{5-p-q}$ after a truncation of M-theory on $S^1$. The full truncation ansatz for the case of $S^4\sim H^{5,0}$ has been constructed in \cite{Pope_typeII_S3_S4}. However, complete truncation ansatze for both type IIA theory on $H^{p,q}\times \mathbb{R}^{4-p-q}$ or eleven-dimensional supergravity on $H^{p,q}\times \mathbb{R}^{5-p-q}$ have not been worked to date although the relevant twist matrices and the truncation ansatze for the metric and internal components of the form fields have already been given in \cite{Malek_IIA_IIB,Henning_KK}. Therefore, the complete truncations to six dimensions are currently not available either. In any case, all solutions of these six-dimensional gauged supergravities can in principle be embedded in type IIA theory as described above. 
\\
\indent The solutions of interest in this paper take the form of $\mathbb{R}\times \mc{M}_4$-sliced domain walls with the $\mathbb{R}$ factor being the time coordinate. We will look for solution interpolating between asymptotically locally flat domain walls dual to maximal SYM in five dimensions and curved domain wall with $t\times \mc{M}_4$-sliced world-volume. Holographically, these solutions would describe RG flows across dimensions from maximal five-dimensional SYM to one-dimensional SYM or supersymmetric quantum mechanics in the IR. In general, due to the presence of topological twists, the latter is obtained from twisted compactifications of the former on $\mc{M}_4$. We also point out that, in holographic studies involving non-conformal field theories, supergravity solutions generically contain singular geometries in the IR. 
\\
\indent To determine whether the resulting solutions are physically acceptable, we employ the condition introduced in \cite{Maldacena_nogo}. According to this condition, the $(00)$-component of type IIA metric $\hat{g}_{00}$ should not diverge near the IR singularities since fixed proper energy excitations in supergravity should correspond to lower and lower energy excitations in the dual field thoery. Upon uplifted to type IIA theory, the solutions with physical IR singularities would describe different configurations involving D4-branes wrapped on $\mc{M}_4$. It should also be noted that although the complete ansatze for the aforementioned truncations are not available, we can determine the ten-dimensional metric component $\hat{g}_{00}$ using partial results given in \cite{Malek_IIA_IIB,Henning_KK} and check whether a given singularity is physical or not. For convenience, we also give the relevant expressions for $\hat{g}_{00}$ for all of the solutions in appendix \ref{g00}. In addition, it should be pointed out that  there is another criterion to determine whether a given singularity is physically acceptable or not directly within the framework of lower-dimensional gauged supergravity theories. This is the Gubser's criterion given in \cite{Gubser_Sing}. According to this criterion, a singularity in gauged supergravity is physical if the scalar potential is bounded above when the singularity is approached. In order to apply this criterion, we need to construct the corresponding two-dimensional Lagrangian by a compactification of the Lagrangian of the six-dimensional gauged supergravity on $\mc{M}_4$ and identify the effective scalar potential in two dimensions, see for example the appendix of \cite{Maldacena_nogo}. However, due to some complications in transforming the effective two-dimensional Lagrangian to Einstein frame, it is not clear how to implement this criterion in the present case. Accordingly, we will not consider Gubser's criterion in this paper.         
\\
\indent The paper is organized as follows. Section  \ref{6DN=(2,2)gSUGRA} briefly reviews maximal $N=(2,2)$ gauged supergravity in six dimensions. Then, the details of gaugings under $GL(5)\subset SO(5, 5)$ are given together with the embedding of $CSO(p,q,5-p-q)$ and $CSO(p,q,4-p-q)\ltimes \mathbb{R}^4$ gauge groups in the global symmetry group $SO(5,5)$. The solutions describing D4-branes wrapped on $\mc{M}_4$ from the resulting gauged supergravities are respectively given in sections \ref{15_Sec} and \ref{40_Sec}. Conclusions and discussions are given in section \ref{conclusion_sec}. For convenience, we collect some useful formulae and relations in appendices \ref{formula} and \ref{g00} together with a number of numerical solutions in appendices \ref{24_Numer_App} and \ref{40_Numer_App}.
%%%%%%%%%%%%%%%%%%%%%%%%%%%%%%%%%%%%%%%%%%%%%%%%%%%%%%%%%%%%%%%%%%%%%%%%%%%%%
\section{Six-dimensional $N=(2,2)$ gauged supergravity with $CSO(p,q,5-p-q)$ and $CSO(p,q,4-p-q)\ltimes \mathbb{R}^4$ gauge groups}\label{6DN=(2,2)gSUGRA}
We first review the general structure of six-dimensional $N=(2,2)$ gauged supergravity in the embedding tensor formalism constructed in \cite{6D_Max_Gauging}. We will mainly setup the notations and collect relevant formulae for finding supersymmetric solutions. More detailed construction can be found in \cite{6D_Max_Gauging}. 

\subsection{Maximal gauged supergravity in six dimensions}
Maximal supersymmetry only allows for the supergravity multiplet. In six dimensions, the bosonic field content consists of the vielbein $e^{\hat{\mu}}_\mu$, two-form potentials $B_{\mu\nu, m}$ and ${B_{\mu\nu}}^m$ ($m=1,\ldots,5$), three-forms $C_{\mu\nu\rho, A}$ ($A=1,\ldots,16$), vector fields ${A_\mu}^A$, and scalars ${V_A}^{\alpha\dot{\alpha}}$ ($\alpha=1,\ldots,4$ and $\dot{\alpha}=\dot{1},\ldots,\dot{4}$) that parametrize the coset space $SO(5,5)/SO(5)\times SO(5)$. Curved and flat spacetime indices are denoted by $\mu$, $\nu$ and $\hat{\mu}$, $\hat{\nu}$, respectively. Lower and upper $m$ indices respectively label fundamental and anti-fundamental representations of $GL(5)\subset SO(5,5)$. The index $A$ describes Majorana-Weyl spinors of the $SO(5,5)$ duality symmetry while the indices $\alpha$ and $\dot{\alpha}$ are respectively two sets of $SO(5)$ spinor indices in the $SO(5)\times SO(5)$ local symmetry. 
\\
\indent Fermionic field content contains the gravitini $\psi_{+\mu\alpha}$ and $\psi_{-\mu\dot{\alpha}}$, and the spin-$\frac{1}{2}$ fields $\chi_{+a\dot{\alpha}}$ and $\chi_{-\dot{a}\alpha}$ where $a=1,\ldots,5$ and $\dot{a}=\dot{1},\ldots,\dot{5}$ respectively denote two sets of $SO(5)$ vector indices in $SO(5)\times SO(5)$. We use $\pm$ to indicate spacetime chiralities of the spinors. In addition, all spinors are symplectic-Majorana-Weyl. With all these, all component fields can be written as
\begin{equation}\label{6DSUGRAmultiplet}
\left(e^{\hat{\mu}}_\mu, B_{\mu\nu, M}, C_{\mu\nu\rho, A}, {A_\mu}^A, {V_A}^{\alpha\dot{\alpha}}, \psi_{+\mu\alpha}, \psi_{-\mu\dot{\alpha}}, \chi_{+a\dot{\alpha}}, \chi_{-\dot{a}\alpha}\right).
\end{equation}
We also note that due to the self-duality of the two-form fields, the Lagrangian of the ungauged supergravity is only invariant under $GL(5)$ subgroup of the full duality symmetry $SO(5,5)$. In this case, only the electric two-forms $B_{\mu\nu, m}$ appear in the Lagrangian while the magnetic duals ${B_{\mu\nu}}^m$ can be introduced on-shell.  
\\
\indent In the embedding tensor formalism, gaugings can be described by introducing the minimal coupling of various fields via the covariant derivative
\begin{equation}\label{gauge_covariant_derivative}
D_\mu=\partial_\mu-g{A_\mu}^A\ {\Theta_A}^{MN}\boldsymbol{t}_{MN}
\end{equation}
where $g$ is a gauge coupling constant, and  ${\Theta_A}^{MN}$ is the embedding tensor. The corresponding gauge generators are given by $X_A={\Theta_A}^{MN}\boldsymbol{t}_{MN}$. In general, these are particular linear combinations of the $SO(5,5)$ generators $\boldsymbol{t}_{MN}$ and form a closed Lie algebra of the gauge group $G_0\subset SO(5,5)$. The latter implies that the embedding tensor needs to satisfy the quadratic constraint of the form
\begin{equation}
\left[X_A,X_B\right]\ = \ -{(X_A)_B}^C\,X_C
\end{equation}
with ${(X_A)_B}^C$ denoting gauge generators in chiral spinor representation of $SO(5,5)$ given below.
\\  
\indent In addition, consistency with supersymmetry requires the embedding tensor to transform as $\mathbf{144}_c$ representation of $SO(5,5)$. Accordingly, ${\Theta_A}^{MN}$ can be parametrized in terms of a vector-spinor $\theta^{AM}$ of $SO(5,5)$ as 
\begin{equation}
{\Theta_A}^{MN}\ =\ -\theta^{B[M}(\Gamma^{N]})_{BA}\ \equiv \ \left(\Gamma^{[M}\theta^{N]}\right)_A
\end{equation}
with $\theta^{AM}$ subject to the constraint
\begin{equation}\label{MainLC}
(\Gamma_M)_{AB}\,\theta^{BM}\ =\ 0
\end{equation}
where $(\Gamma_M)_{AB}$ is an $SO(5,5)$ gamma matrix. In terms of $\theta^{AM}$, the quadratic constraint reduces to the following two conditions
\begin{equation}\label{QC}
\theta^{AM}\theta^{BN}\eta_{MN}\ =\ 0\qquad \textrm{and} \qquad  \theta^{AM}\theta^{B[N}(\Gamma^{P]})_{AB}=0\, .
\end{equation} 
$\eta_{MN}$ is the off-diagonal $SO(5,5)$ invariant tensor given by
\begin{equation}\label{off-diag-eta}
\eta_{MN}\ =\ \eta^{MN}\ =\ \begin{pmatrix} 	0 & \mathds{1}_5 \\
							\mathds{1}_5 & 0    \end{pmatrix}\, .
\end{equation}
With $SO(5,5)$ generators in vector and spinor representations given by
\begin{equation}
{(\boldsymbol{t}_{MN})_P}^Q=4\eta_{P[M}\delta^Q_{N]}\qquad\text{and}\qquad {(\boldsymbol{t}_{MN})_A}^B\ =\ {(\Gamma_{MN})_A}^B,
\end{equation}
the corresponding gauge generators take the forms
\begin{equation}\label{DefGaugeGen}
{(X_A)_M}^N=2\left(\Gamma_{M}\theta^{N}\right)_A+2\left(\Gamma^{N}\theta_{M}\right)_A\qquad\text{and}\qquad {(X_A)_B}^C= \left(\Gamma^{M}\theta^{N}\right)_A{(\Gamma_{MN})_B}^C\, .
\end{equation}
\indent Unlike the ungauged supergravity, the most general gauged supergravity can involve a symmetry that lies outside the off-shell $GL(5)$ symmetry. Under $GL(5)$, $\theta^{AM}$ can be decomposed as $\theta^{AM}=(\theta^{Am},{\theta^{A}}_m)$. Gaugings triggered by $\theta^{Am}$ are called electric in the sense that only electric two-forms participate in the gauged theory while magnetic gaugings obtained from ${\theta^A}_m$ also involve magnetic two-forms together with additional three-form tensor fields $C_{\mu\nu\rho, A}$. In this case, the magnetic two-forms ${B_{\mu\nu}}^m$ can appear in the gauged Lagrangian by coupling to the three-form potentials via the topological term. The electric and magnetic two-forms are combined into a vector representation of the full global symmetry group $SO(5,5)$ denoted by $B_{\mu\nu, M}=(B_{\mu\nu, m}, {B_{\mu\nu}}^m)$. 
\\
\indent The gauge covariance of the field strength tensors for all these fields requires the presence of higher-degree form fields leading to hierarchies of non-abelian vector and tensor fields of various ranks. In particular, the covariant field strengths of vector and two-form fields are defined by
\begin{eqnarray}
{\mathcal{H}_{\mu\nu}}^{A}&=&{F_{\mu\nu}}^{A}-\sqrt{2}g\theta^{AM}B_{\mu\nu, M},\label{ModTensor1}\\
\mathcal{H}_{\mu\nu\rho, M}&=&3D_{[\mu}B_{\nu\rho],M}+3\sqrt{2}(\Gamma_M)_{AB}{A_{[\mu}}^{A}(\partial_\nu {A_{\rho]}}^{B}+\frac{1}{3}g {X_{[CD]}}^B{A_\nu}^{C} {A_{\rho]}}^{D})\nonumber\\&&-\sqrt{2}g\eta_{MN}\theta^{AN}C_{\mu\nu\rho, A}\label{ModTensor2}
\end{eqnarray}
with the usual non-abelian gauge field strengths
\begin{equation}\label{Ful2Form}
{F_{\mu\nu}}^{A}=2\partial_{[\mu}{A_{\nu]}}^{A}+g{X_{[BC]}}^{A}{A_{\mu}}^{B}{A_{\nu}}^{C}\, .
\end{equation}
These field strengths satisfy the following Bianchi identities
\begin{eqnarray}
D_{[\mu}{\mathcal{H}_{\nu\rho]}}^{A}&=&-\frac{\sqrt{2}}{3}g\theta^{AM}\mathcal{H}_{\mu\nu\rho, M}, \label{DefBianchi1}\\
D_{[\mu}\mathcal{H}_{\nu\rho\lambda], M}&=&\frac{3}{2\sqrt{2}}(\Gamma_M)_{AB}{\mathcal{H}_{[\mu\nu}}^{A}{\mathcal{H}_{\rho\lambda]}}^{B}-\frac{1}{2\sqrt{2}}g\eta_{MN}\theta^{AN}\mathcal{H}_{\mu\nu\rho\lambda, A}\label{DefBianchi2}
\end{eqnarray}
with the four-form field strengths defined by
\begin{eqnarray}\label{4-form}
\mathcal{H}_{\mu\nu\rho\lambda, A}&=&4D_{[\mu}C_{\nu\rho\lambda],A}-(\Gamma^M)_{AB}\left[6\sqrt{2}B_{\mu\nu, M}{\mathcal{H}_{\rho\lambda}}^{B}+6g\theta^{BN}B_{[\mu\nu, M}B_{\rho\lambda], N}\right.\nonumber \\
&&\left.\phantom{\sqrt{2}}+8(\Gamma_M)_{CD}{A_{[\mu}}^{B}{A_\nu}^{C}\partial_\rho {A_{\lambda]}}^{D}+2(\Gamma_M)_{CF}{X_{DE}}^F{A_{[\mu}}^{B}{A_\nu}^{C} {A_\rho}^{D} {A_{\lambda]}}^{E}\right].\nonumber \\
\end{eqnarray}
In addition, all of the aforementioned field strength tensors are subject to the following duality relations
\begin{eqnarray}
0&=&g{\theta^A}_m({\mathcal{H}_{\mu\nu\rho}}^m-\frac{1}{6}K^{mn}\epsilon_{\mu\nu\rho\lambda\sigma\tau}{\mathcal{H}^{\lambda\sigma\tau}}_n),\label{3formSD}\\
0&=&g{\theta^A}_m(\mathcal{H}_{\mu\nu\rho\lambda, A}+\frac{1}{2}M_{AB}\epsilon_{\mu\nu\rho\lambda\sigma\tau}\mathcal{H}^{\sigma\tau, B})\label{2_4SD}
\end{eqnarray}
where $K^{mn}$ and $M_{AB}$ are scalar matrices written in terms of the scalar coset representative.
\\
\indent The matrix $M_{AB}$ is defined by
\begin{equation}
M_{AB}=\Omega_{\alpha\beta}\Omega_{\dot{\alpha}\dot{\beta}}{V_A}^{\alpha\dot{\alpha}}{V_B}^{\beta\dot{\beta}} 
\end{equation}
with $\Omega_{\alpha\beta}$ and $\Omega_{\dot{\alpha}\dot{\beta}}$ being the two $USp(4)\sim SO(5)$ symplectic forms used for lowering indices $\alpha$ and $\dot{\alpha}$, respectively. Raising of indices $\alpha$ and $\dot{\alpha}$ is implemented by using $\Omega^{\alpha\beta}=(\Omega_{\alpha\beta})^*$ and $\Omega^{\dot{\alpha}\dot{\beta}}=(\Omega_{\dot{\alpha}\dot{\beta}})^*$. On the other hand, the matrix $K^{mn}$ is defined in terms of the coset representative in vector representation of $SO(5,5)$ of the form
\begin{equation}
{\mathcal{V}_M}^{\underline{A}}=\begin{pmatrix} {\mathcal{V}_m}^{a}& {\mathcal{V}_m}^{\dot{a}}\\{\mathcal{V}^m}^{a}& {\mathcal{V}^m}^{\dot{a}}\end{pmatrix}
\end{equation}
with $\underline{A}=(a,\dot{a})$. This is related to the coset representative in the chiral spinor representation ${V_A}^{\alpha\dot{\alpha}}$ by the following relations
\begin{eqnarray}
{\mathcal{V}_M}^a&=&\frac{1}{16}V^{A\alpha\dot{\alpha}}(\Gamma_M)_{AB}{(\gamma^a)_{\alpha\dot{\alpha}}}^{\beta\dot{\beta}}{V^B}_{\beta\dot{\beta}},\label{VVrel1}\\{\mathcal{V}_M}^{\dot{a}}&=&-\frac{1}{16}V^{A\alpha\dot{\alpha}}(\Gamma_M)_{AB}{(\gamma^{\dot{a}})_{\alpha\dot{\alpha}}}^{\beta\dot{\beta}}{V^B}_{\beta\dot{\beta}}\, \label{VVrel2}
\end{eqnarray}
in which $V^{A\alpha\dot{\alpha}}=\Omega^{\alpha\beta}\Omega^{\dot{\alpha}\dot{\beta}}{V^A}_{\beta\dot{\beta}}$ and ${V^A}_{\alpha\dot{\alpha}}$ are inverse elements of the coset representative ${V_A}^{\alpha\dot{\alpha}}$.
In these relations, $(\Gamma_M)_{AB}$ and ${(\Gamma_{\ul{A}})_{\alpha\dot{\alpha}}}^{\beta\dot{\beta}}=({(\gamma_a)_{\alpha\dot{\alpha}}}^{\beta\dot{\beta}},{(\gamma_{\dot{a}})_{\alpha\dot{\alpha}}}^{\beta\dot{\beta}})$ are respectively $SO(5,5)$ gamma matrices in non-diagonal $\eta_{MN}$ and diagonal $\eta_{\ul{A}\ul{B}}$ bases, see Appendix A of \cite{6D_DW_I} for more detail.  
\\  
\indent With all these, the matrix $K^{mn}$ is defined by
\begin{equation}
K^{mn}=\mc{V}^{ma}({\mc{V}_n}^a)^{-1}P_+-\mc{V}^{m\dot{a}}({\mc{V}_n}^{\dot{a}})^{-1}P_-
\end{equation}
in which $P_\pm$ are projections on the self- and anti-self-dual parts given by 
\begin{equation}
P_\pm=\frac{1}{2}(1\pm j).
\end{equation}
The operation $j$ implements the duality transformation as 
\begin{equation}
jH_{\mu\nu\rho}=\tilde{H}_{\mu\nu\rho}=\frac{1}{6}e\epsilon_{\mu\nu\rho\lambda\sigma\kappa}H^{\lambda\sigma\kappa}.
\end{equation}
\indent The bosonic Lagrangian of the general six-dimensional $N=(2,2)$ gauged supergravity takes the form
\begin{eqnarray}
e^{-1}\mathcal{L}&=&\frac{1}{4}R-\frac{1}{16}{P_{\mu}}^{a\dot{a}}{P^\mu}_{a\dot{a}}-\frac{1}{4}M_{AB}{\mathcal{H}_{\mu\nu}}^{A}\mathcal{H}^{\mu\nu, B}-\frac{1}{12}K^{mn}\mc{H}_{\mu\nu\rho, m}{\mc{H}^{\mu\nu\rho}}_n\nonumber \\
& &-\mathbf{V}+e^{-1}\mathcal{L}_{\text{top}}\label{BosLag}
\end{eqnarray}
where $\mathcal{L}_{\text{top}}$ is the topological term given in \cite{6D_Max_Gauging}. The kinetic term for the scalar fields is written in terms of the scalar vielbein
\begin{equation}
{P_{\mu}}^{a\dot{a}}=\frac{1}{4}{(\gamma^a)}_{\alpha\beta}{(\gamma^{\dot{a}})}_{\dot{\alpha}\dot{\beta}}V^{A\alpha\dot{\alpha}}D_\mu {V_{A}}^{\beta\dot{\beta}}\label{PDef}
\end{equation}
where ${(\gamma^a)}_{\alpha\beta}=\frac{1}{4}\Omega_{\beta\gamma}{(\gamma^a)_{\alpha\dot{\alpha}}}^{\gamma\dot{\alpha}}$ and ${(\gamma^{\dot{a}})}_{\dot{\alpha}\dot{\beta}}=\frac{1}{4}\Omega_{\dot{\beta}\dot{\gamma}}{(\gamma^{\dot{a}})_{\alpha\dot{\alpha}}}^{\alpha\dot{\gamma}}$. The scalar potential is given in terms of the embedding tensor and the coset representatives by
\begin{equation}\label{scalarPot1}
\mathbf{V}=\frac{g^2}{2}\theta^{AM}\theta^{BN}{\mathcal{V}_M}^a{\mathcal{V}_N}^b\left[{V_A}^{\alpha\dot{\alpha}}{(\gamma_a)_\alpha}^\beta{(\gamma_b)_\beta}^\gamma V_{B\gamma\dot{\alpha}}\right].
\end{equation}
\indent For all solutions given in the present paper, the three-form field strengths identically vanish. For simplicity, we will accordingly set $\mc{H}_{\mu\nu\rho, M}=0$ from now on. It is straightforward to verify that this is indeed a consistent truncation in all cases considered here. 
\\
\indent With this, supersymmetry transformations of the fermionic fields read
\begin{eqnarray}
\delta\psi_{+\mu\alpha}&=& D_\mu\epsilon_{+\alpha}+\frac{g}{4}\hat{\gamma}_\mu {T_\alpha}^{\dot{\beta}}\epsilon_{-\dot{\beta}}+\frac{1}{8}({\hat{\gamma}_\mu}^{\ \nu\rho}-6\delta^\nu_\mu\hat{\gamma}^\rho){\mathcal{H}_{\nu\rho}}^A\Omega_{\alpha\beta}{V_A}^{\beta\dot{\beta}}\epsilon_{-\dot{\beta}},\label{1stSUSY}\\
\delta\psi_{-\mu\dot{\alpha}}&=& D_\mu\epsilon_{-\dot{\alpha}}-\frac{g}{4}\hat{\gamma}_\mu {T^{\beta}}_{\dot{\alpha}}\epsilon_{+\beta}+\frac{1}{8}({\hat{\gamma}_\mu}^{\ \nu\rho}-6\delta^\nu_\mu\hat{\gamma}^\rho){\mathcal{H}_{\nu\rho}}^A\Omega_{\dot{\alpha}\dot{\beta}}{V_A}^{\beta\dot{\beta}}\epsilon_{+\beta},\label{2ndSUSY}\\
\delta\chi_{+a\dot{\alpha}}&=&\frac{1}{4}{P^\mu}_{a\dot{a}}\hat{\gamma}_\mu{(\gamma^{\dot{a}})_{\dot{\alpha}}}^{\dot{\beta}}\epsilon_{-\dot{\beta}}+2g{(T_{a})^\beta}_{\dot{\alpha}}\epsilon_{+\beta}-\frac{g}{2}{T^{\alpha}}_{\dot{\alpha}}{(\gamma_a)_\alpha}^\beta\epsilon_{+\beta}\nonumber\\&&+\frac{1}{4}{\mathcal{H}_{\mu\nu}}^A\hat{\gamma}^{\mu\nu}\Omega_{\dot{\alpha}\dot{\beta}}{V_A}^{\alpha\dot{\beta}}{(\gamma_a)_\alpha}^\beta\epsilon_{+\beta},\label{3rdSUSY}\\
\delta\chi_{-\dot{a}\alpha}&=&\frac{1}{4}{P^\mu}_{a\dot{a}}\hat{\gamma}_\mu{(\gamma^a)_\alpha}^\beta\epsilon_{+\beta}+2g{(T_{\dot{a}})_{\alpha}}^{\dot{\beta}}\epsilon_{-\dot{\beta}}+\frac{g}{2}{T_{\alpha}}^{\dot{\alpha}}{(\gamma_{\dot{a}})_{\dot{\alpha}}}^{\dot{\beta}}\epsilon_{-\dot{\beta}}\nonumber\\&&+\frac{1}{4}{\mathcal{H}_{\mu\nu}}^A\hat{\gamma}^{\mu\nu}\Omega_{\alpha\beta}{V_A}^{\beta\dot{\alpha}}{(\gamma_{\dot{a}})_{\dot{\alpha}}}^{\dot{\beta}}\epsilon_{-\dot{\beta}}.\label{4thtSUSY}
\end{eqnarray}
The covariant derivatives of supersymmetry parameters, $\epsilon_{+\alpha}$ and $\epsilon_{-\dot{\alpha}}$, are
\begin{eqnarray}
D_\mu\epsilon_{+\alpha}&=& \partial_\mu\epsilon_{+\alpha}+\frac{1}{4}{\omega_\mu}^{\nu\rho}\hat{\gamma}_{\nu\rho}\epsilon_{+\alpha}+\frac{1}{4}{Q_\mu}^{ab}{(\gamma_{ab})_\alpha}^\beta\epsilon_{+\beta},\label{CoDivEp+}\\
D_\mu\epsilon_{-\dot{\alpha}}&=& \partial_\mu\epsilon_{-\dot{\alpha}}+\frac{1}{4}{\omega_\mu}^{\nu\rho}\hat{\gamma}_{\nu\rho}\epsilon_{-\dot{\alpha}}+\frac{1}{4}{Q_\mu}^{\dot{a}\dot{b}}{(\gamma_{\dot{a}\dot{b}})_{\dot{\alpha}}}^{\dot{\beta}}\epsilon_{-\dot{\beta}}\label{CoDivEp-}
\end{eqnarray}
with $\hat{\gamma}_\mu=e_\mu^{\hat{\mu}}\hat{\gamma}_{\hat{\mu}}$. Matrices $\hat{\gamma}_{\hat{\mu}}$ are spacetime gamma matrices, see the convention in Appendix B of \cite{6D_DW_I}. For simplicity, we will suppress all spacetime spinor indices.
The $SO(5)\times SO(5)$ composite connections ${Q_\mu}^{ab}$ and ${Q_\mu}^{\dot{a}\dot{b}}$ are given by
\begin{eqnarray}
{Q_{\mu}}^{ab}&=&\frac{1}{8}{(\gamma^{ab})}_{\alpha\beta}\Omega_{\dot{\alpha}\dot{\beta}} V^{A\alpha\dot{\alpha}}D_\mu {V_A}^{\beta\dot{\beta}},\label{QuDef}\\
{Q_{\mu}}^{\dot{a}\dot{b}}&=&\frac{1}{8}\Omega_{\alpha\beta}{(\gamma^{\dot{a}\dot{b}})}_{\dot{\alpha}\dot{\beta}}V^{A\alpha\dot{\alpha}}D_\mu {V_A}^{\beta\dot{\beta}}.\label{QdDef}
\end{eqnarray}
The T-tensors in the fermionic supersymmetry transformations are defined as
\begin{equation}\label{TTenDef}
(T^a)^{\alpha\dot{\alpha}}={\mathcal{V}_M}^a\theta^{AM}{V_A}^{\alpha\dot{\alpha}},\qquad (T^{\dot{a}})^{\alpha\dot{\alpha}}=-{\mathcal{V}_M}^{\dot{a}}\theta^{AM}{V_A}^{\alpha\dot{\alpha}}
\end{equation}
with
\begin{equation}
T^{\alpha\dot{\alpha}}\equiv (T^a)^{\beta\dot{\alpha}}{(\gamma_a)_\beta}^\alpha=-(T^{\dot{a}})^{\alpha\dot{\beta}}{(\gamma_{\dot{a}})_{\dot{\beta}}}^{\dot{\alpha}}.
\end{equation}
In terms of the T-tensors, the scalar potential is given by
\begin{equation}\label{scalarPot2}
\mathbf{V}=g^2\left[(T^a)^{\alpha\dot{\alpha}}(T_a)_{\alpha\dot{\alpha}}-\frac{1}{2}T^{\alpha\dot{\alpha}}T_{\alpha\dot{\alpha}}\right].
\end{equation}

\subsection{Gauging $CSO(p,q,5-p-q)$ and $CSO(p,q,4-p-q)\ltimes \mathbb{R}^4$ groups}
In this paper, we are only interested in $CSO(p,q,5-p-q)$ and $CSO(p,q,4-p-q)\ltimes \mathbb{R}^4$ gauge groups that can be embedded in $GL(5)\subset SO(5, 5)$. The embedding tensor $\theta^{AM}$ in $\mathbf{144}_c$ representation of $SO(5,5)$ decomposes under $GL(5)$ as
\begin{equation}\label{mainthetaDec}
\mathbf{144}_c\ \rightarrow\ \overline{\mathbf{5}}^{+3}\,\oplus\,\mathbf{5}^{+7}\,\oplus\,\mathbf{10}^{-1}\,\oplus\,\mathbf{15}^{-1}\,\oplus\,\mathbf{24}^{-5}\,\oplus\,\overline{\mathbf{40}}^{-1}\,\oplus\,\overline{\mathbf{45}}^{+3}\, .
\end{equation}
$CSO(p,q,5-p-q)$ gauge group with $p+q\leq 5$ corresponds to the embedding tensor in $\mathbf{15}^{-1}$ representation while $CSO(p,q,4-p-q)\ltimes\mathbb{R}^4$ gauge group with $p+q\leq 4$ is obtained from the embedding tensor in $\overline{\mathbf{40}}^{-1}$ representation. Both of these gauge groups arise from a dimensional reduction on $S^1$ of the maximal gauged supergravity in seven dimensions with $CSO(p,q,5-p-q)$ and $CSO(p,q,4-p-q)$ gauge groups constructed in \cite{7D_Max_Gauging}. Moreover, as previously mentioned, the six-dimesional gauged supergravity with these two gauge groups can also be embedded in type IIA theory via consistent truncations on $H^{p,q}\times \mathbb{R}^{5-p-q}$ and $H^{p,q}\times\mathbb{R}^{4-p-q}\times S^1$, respectively.
\\
\indent Decomposing the $SO(5,5)$ spinor representation under $GL(5)$ as $\mathbf{16}_s\ \rightarrow\ \overline{\mathbf{5}}^{+3}\,\oplus\,\mathbf{10}^{-1}\,\oplus\,\mathbf{1}^{-5}$, we find that the gauge generators can be written as
\begin{equation}\label{GaugeGenSplit}
X_A\ =\ \mathbb{T}_{Am}X^m+\mathbb{T}_{A}^{mn}X_{mn}+\mathbb{T}_{A\ast}X_\ast\, .
\end{equation}
Generators $X^m$, $X_{mn}$, and $X_\ast$ correspond respectively to $\overline{\mathbf{5}}^{+3}$, $\mathbf{10}^{-1}$, and $\mathbf{1}^{-5}$ representations of $GL(5)$. The decomposition matrices $\mathbb{T}_{Am}$, $\mathbb{T}_{A}^{mn}$, and $\mathbb{T}_{A\ast}$ are given in appendix \ref{formula}. Similarly, the vector fields in $\mathbf{16}_c$ representation decompose as
\begin{equation}
{A_\mu}^A\ =\ \mathbb{T}^{Am}A_{\mu,m}+\mathbb{T}^{A}_{mn}{A_\mu}^{mn}+\mathbb{T}^{A}_{\ast}A_\mu^{\ast}
\end{equation}
with matrices $\mathbb{T}^{Am}$, $\mathbb{T}^{A}_{mn}$, and $\mathbb{T}^{A}_{\ast}$ being inverses of $\mathbb{T}_{Am}$, $\mathbb{T}_{A}^{mn}$, and $\mathbb{T}_{A\ast}$ given by complex conjugations, $\mathbb{T}^A=(\mathbb{T}_A)^{-1}=(\mathbb{T}_A)^*$. Accordingly, the gauge covariant derivative can be written as
\begin{equation}
D_\mu=\nabla_\mu-gA_{\mu,m} X^m-g{A_\mu}^{mn} X_{mn}-gA_\mu^{\ast} X_\ast\, .
\end{equation}  
\indent To identify relevant components of the embedding tensor, we recall the decomposition of $SO(5,5)$ vector representation $\mathbf{10}\ \rightarrow\ \mathbf{5}^{+2}\,\oplus\,\overline{\mathbf{5}}^{-2}$ under $GL(5)$. We can then write the embedding tensor as $\theta^{AM}=(\theta^{Am},{\theta^{A}}_m)$ with the two components $\theta^{Am}$ and ${\theta^{A}}_m$ containing the following irreducible $GL(5)$ representations
\begin{eqnarray}
\theta^{Am}&:&\qquad \overline{\mathbf{5}}^{+3}\,\oplus\,\mathbf{10}^{-1}\,\oplus\,\mathbf{24}^{-5}\,\oplus\,\overline{\mathbf{40}}^{-1},\label{splitthetaDec1}\\ 
{\theta^{A}}_m&:&\qquad \overline{\mathbf{5}}^{+3}\,\oplus\,\mathbf{5}^{+7}\,\oplus\,\mathbf{10}^{-1}\,\oplus\,\mathbf{15}^{-1}\,\oplus\,\overline{\mathbf{45}}^{+3}\, .\label{splitthetaDec2}
\end{eqnarray}
Gaugings in $\mathbf{15}^{-1}$ representation can be parametrized by a symmetric tensor $Y_{mn}=Y_{(mn)}$ leading to the non-vanishing components of the embedding tensor of the form     
\begin{eqnarray}\label{15_rep_theta}
\theta^{Am}=0\qquad \textrm{and}\qquad {\theta^A}_m=\mathbb{T}^{An}Y_{mn}\, .
\end{eqnarray}
On the other hand, gaugings from $\overline{\mathbf{40}}^{-1}$ representation are characterized by a tensor $U^{np,m}=U^{[np],m}$ satisfying $U^{[np,m]}=0$. This leads to the embedding tensor of the form
\begin{eqnarray}
\theta^{Am}=\mathbb{T}^{A}_{np}U^{np,m}\qquad \textrm{and}\qquad {\theta^A}_m=0\, .
\end{eqnarray} 
We also point out that these gaugings are respectively purely magnetic and purely electric, see \cite{6D_DW_I} for more detail.
\\
\indent To construct an explicit form of the scalar coset representative, we first consider the decomposition of $25$ non-compact generators of $SO(5,5)$ under $GL(5)$ as
\begin{equation}\label{15repscalarDEC}
25\ \rightarrow\ \underbrace{\mathbf{1}+\mathbf{14}}_{\hat{\boldsymbol{t}}_{a\dot{b}}}\,+\underbrace{\mathbf{10}}_{\boldsymbol{s}_{mn}}\, .
\end{equation}
$\hat{\boldsymbol{t}}_{a\dot{b}}$ denote non-compact generators of $GL(5)$ defined by
\begin{equation}
\hat{\boldsymbol{t}}_{a\dot{b}}=\frac{1}{2}\left({\mathbb{M}_{a}}^M{\mathbb{M}_{\dot{b}}}^N+{\mathbb{M}_{\dot{b}}}^M{\mathbb{M}_{a}}^N\right)\boldsymbol{t}_{MN}
\end{equation}
where ${\mathbb{M}_{\underline{A}}}^M=({\mathbb{M}_{a}}^M,{\mathbb{M}_{\dot{a}}}^M)$ is a transformation matrix whose explicit form is given in \eqref{offDiagTrans}. 
\\
\indent In accordance with \eqref{15repscalarDEC}, we denote all the $25$ scalar fields as 
\begin{equation}
\Phi^I=\{\varphi,\phi_1,\ldots,\phi_{14},\varsigma_1,\ldots,\varsigma_{10}\}\label{6D_scalar}
\end{equation}
with $I=1,\ldots,25$. The scalar $\varphi$ is the dilaton corresponding to the $SO(1,1)$ factor in $GL(5)\sim SL(5)\times SO(1,1)$ generated by
\begin{equation}\label{SO(1,1)Gen}
\boldsymbol{d}=\hat{\boldsymbol{t}}_{1\dot{1}}+\hat{\boldsymbol{t}}_{2\dot{2}}+\hat{\boldsymbol{t}}_{3\dot{3}}
+\hat{\boldsymbol{t}}_{4\dot{4}}+\hat{\boldsymbol{t}}_{5\dot{5}}\, .
\end{equation}
On the other hand, fourteen scalars $\{\phi_1,...,\phi_{14}\}$ corresponding to generators $\hat{\boldsymbol{t}}_{a\dot{b}}$ parametrize the $SL(5)/SO(5)$ submanifold of $SO(5,5)/SO(5)\times SO(5)$. Finally, the remaining ten scalars $\{\varsigma_1,...,\varsigma_{10}\}$ correspond to the shift generators $\boldsymbol{s}_{mn}=\boldsymbol{t}_{mn}$ and will be called shift scalars. From a higher-dimensional point of view, these scalars arise from the $S^1$ reduction of the vector fields in the maximal seven-dimensional gauged supergravity.

%%%%%%%%%%%%%%%%%%%%%%%%%%%%%%%%%%%%%%%%%%%%%%%%%%%%%%%%%%%%%%%%%%%%%%%
\section{Supersymmetric quantum mechanics from $CSO(p,q,5-p-q)$ gauged supergravity}\label{15_Sec}
We begin the study of supersymmetric solutions describing D4-branes wrapped on a four-manifold $\mc{M}_4$ from the maximal gauged supergravity with $CSO(p,q,5-p-q)$ gauge group. From the embedding tensor given in \eqref{15_rep_theta}, we can choose a basis in which the symmetric tensor $Y_{mn}$ takes the form
\begin{equation}\label{diagYmn}
Y_{mn}\ =\ \text{diag}(\underbrace{1,..,1}_p,\underbrace{-1,..,-1}_q,\underbrace{0,..,0}_r)
\end{equation}
for $p+q+r=5$. In this case, the only non-vanishing gauge generators are given by 
\begin{equation}
X_{mn}=\mathbb{T}^{A}_{mn}X_A=2Y_{p[m}{\boldsymbol{t}^p}_{n]}
\end{equation}
with ${\boldsymbol{t}^m}_{n}$ being $GL(5)$ generators. Therefore, only ten gauge fields ${A_\mu}^{mn}={A_\mu}^{[mn]}$ appear in the gauge covariant derivatives 
\begin{equation}
D_\mu=\nabla_\mu-g{A_\mu}^{mn} X_{mn}\, .
\end{equation}
The gauge generators satisfy the relations
\begin{equation}
[X_{mn},X_{pq}]={(X_{mn})_{pq}}^{rs}X_{rs}
\end{equation}
with ${(X_{mn})_{pq}}^{rs}=2{(X_{mn})_{[p}}^{[r}\delta_{q]}^{s]}$. Therefore, the corresponding gauge group is determined to be
\begin{equation}
G_0\ =\ CSO(p,q,r)\ =\ SO(p,q) \ltimes \mathbb{R}^{(p+q)\cdot r}.
\end{equation}
\indent Recall that gaugings in $\mathbf{15} ^{-1}$ representation are magnetic, we find that, in general, the two- and three-from fields ${B_{\mu\nu}}^m$ and $C_{\mu\nu\rho, A}$ can appear in the Lagrangian. With $\theta^{Am}=0$, the two-form field strengths take the form
\begin{equation}
{\mathcal{H}_{\mu\nu}}^{A}={F_{\mu\nu}}^{A}-\sqrt{2}g\mathbb{T}^{An}Y_{nm}{B_{\mu\nu}}^m
\end{equation}
which can be decomposed as
\begin{eqnarray}
\mathcal{H}_{\mu\nu,m}&=&\mathbb{T}_{Am}{\mathcal{H}_{\mu\nu}}^{A}=-\sqrt{2}gY_{mn}{B_{\mu\nu}}^n,\nonumber\\
{\mathcal{H}_{\mu\nu}}^{mn}&=&\mathbb{T}_{A}^{mn}{\mathcal{H}_{\mu\nu}}^{A}={F_{\mu\nu}}^{mn}\, .
\end{eqnarray}
${F_{\mu\nu}}^{mn}$ denote the usual non-abelian gauge field strengths given by 
\begin{equation}
{F_{\mu\nu}}^{mn}=\mathbb{T}_{A}^{mn}{F_{\mu\nu}}^{A}=2\partial_{[\mu}{A_{\nu]}}^{mn}+g{(X_{pq})_{rs}}^{mn}{A_{\mu}}^{pq}{A_{\nu}}^{rs}.\label{10_rep_F_def}
\end{equation} 

%%%%%%%%%%%%%%%%%%%%%%%%%%%%%%%%%%%%%%%%%%%%%%%%%%%%%%%%%%%%%%%%%%%%%%%%%%%%%%%%%%%%%%%%%%%%%%%%%%%%%%%%%%%%%%%%%%%%%%%%%%%%%%%%%%%%%%%%%
\subsection{D4-branes wrapped on a Riemannian four-manifold}\label{15_Sig4_Sec}
\begin{figure}[h!]
  \centering
    \includegraphics[width=\linewidth]{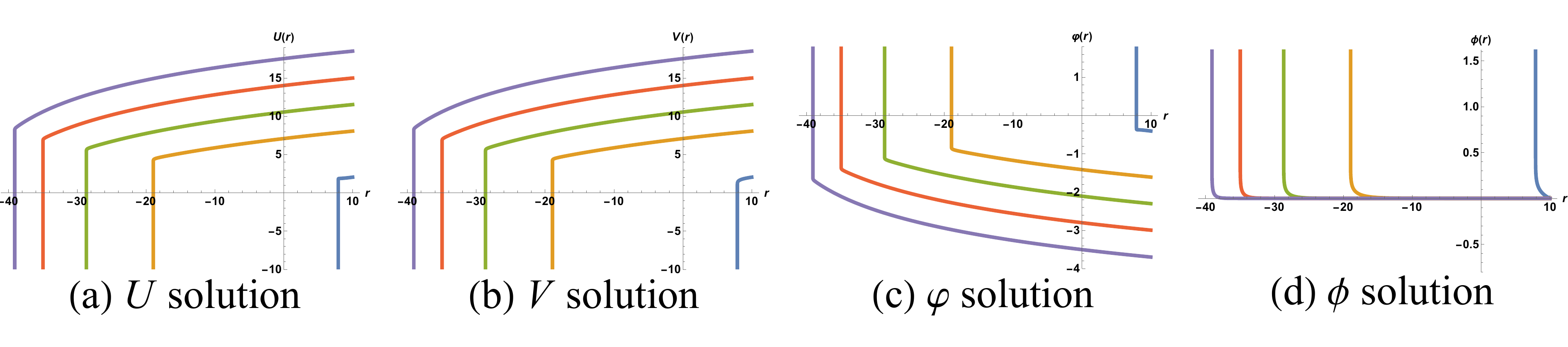}
\caption{Interpolating solutions between the locally $SO(5)$ flat domain wall as $r\rightarrow+\infty$ and $t\times S^4$-sliced curved domain walls for $SO(4)$ twist in $SO(5)$ gauge group. The blue, orange, green, red, and purple curves refer to $g=0.15, 0.50, 1, 2$, and $4.04$, respectively.}
\label{15_S4_SO(4)_special_SO(5)gg_flows}
\end{figure}

\begin{figure}[h!]
  \centering
    \includegraphics[width=\linewidth]{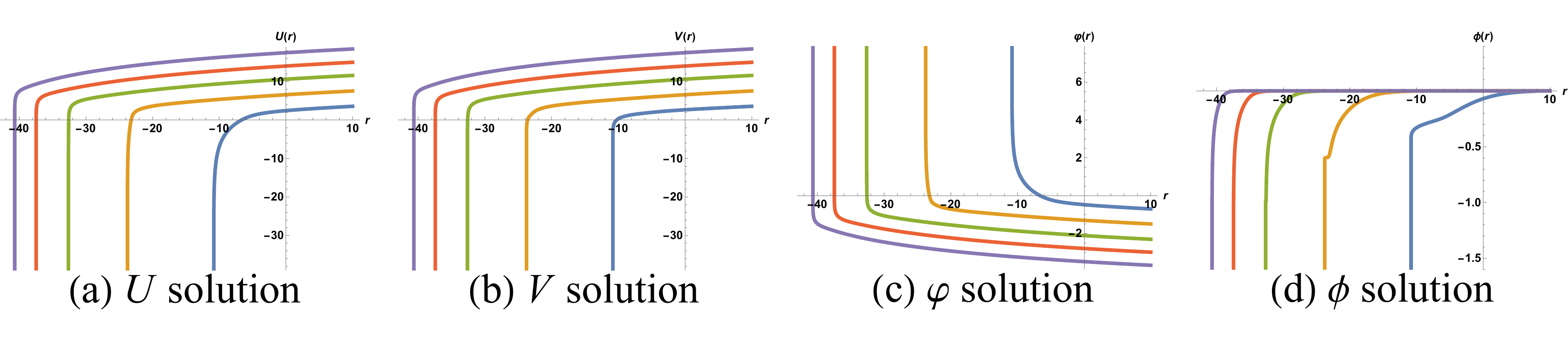}
\caption{Interpolating solutions between the locally $SO(5)$ flat domain wall as $r\rightarrow+\infty$ and $t\times H^4$-sliced curved domain walls for $SO(4)$ twist in $SO(5)$ gauge group. The blue, orange, green, red, and purple curves refer to $g=0.20, 0.45, 1, 2$, and $4$, respectively.}
\label{15_H4_SO(4)_special_SO(5)gg_flows}
\end{figure}

We first consider the case of supersymmetric solutions describing D4-branes wrapped on a Riemannian four-manifold $\Sigma^4_k$ with constant curvature. The curvature of $\Sigma^4_k$ is characterized by a constant $k$ with values $k=1,0,-1$ corresponding to a four-dimensional sphere $S^4$, a flat space $\mathbb{R}^4$, or a hyperbolic space $H^4$, respectively. Although there are many possibilities to perform topological twists in this case, we restrict to the case of $\Sigma^4_k$ being a constant curvature manifold for simplicity and also for the sake of comparison with a similar work in seven dimensions studied in \cite{7D_max_twist}. 
\\
\indent The metric ansatz for the six-dimensional spacetime is given by
\begin{equation}\label{SO(4)6Dmetric}
ds_6^2=-e^{2U(r)}dt^2+dr^2+e^{2V(r)}ds^2_{\Sigma^{4}_{k}}\, .
\end{equation}
$ds^2_{\Sigma^{4}_{k}}$ is the metric on $\Sigma^4_k$ with the explicit form written as
\begin{equation}\label{Sigma4metric}
ds^2_{\Sigma^4_k}=d\chi^2+f_k(\chi)^2\left[d\psi^2+\sin^2{\psi}(d\theta^2+\sin^2{\theta}d\zeta^2)\right]
\end{equation}
for $\chi, \psi, \theta \in[0,\frac{\pi}{2}]$, $\zeta\in[0,2\pi]$. The function $f_k(\chi)$ is defined by
\begin{equation}\label{fFn}
f_k(\chi)=\begin{cases}
                        	\sin{\chi}, \ \  \quad\text{ for } k=+1 \\
                       	\chi, \ \ \ \ \ \ \quad\text{ for } k=0\\
			\sinh{\chi}, \quad\text{ for } k=-1
                    \end{cases}.
\end{equation}
\indent With the following choice of vielbein
\begin{eqnarray}
e^{\hat{t}}&=&e^{U}dt, \qquad e^{\hat{r}}=dr,  \qquad e^{\hat{\chi}}=e^{V}d\chi, \qquad  e^{\hat{\psi}}= e^{V}f_{k}(\chi)d\psi,\nonumber \\
e^{\hat{\theta}}&=&e^{V}f_{k}(\chi)\sin{\psi}d\theta, \qquad e^{\hat{\zeta}}=e^{V}f_{k}(\chi)\sin{\psi}\sin{\theta}d\zeta,\label{Sigma4bein}
\end{eqnarray}
we obtain the non-vanishing components of the spin connection given by
\begin{eqnarray}
{\omega_{\hat{t}}}^{\hat{t}\hat{r}}&=&U', \qquad {\omega_{\hat{i}}}^{\hat{j}\hat{r}}= V'\delta^{\hat{j}}_{\hat{i}}, \qquad
{\omega_{\hat{\psi}}}^{\hat{\psi}\hat{\chi}}={\omega_{\hat{\theta}}}^{\hat{\theta}\hat{\chi}}={\omega_{\hat{\zeta}}}^{\hat{\zeta}\hat{\chi}}=e^{-V}\frac{f'_{k}(\chi)}{f_{k}(\chi)},\nonumber\\ {\omega_{\hat{\theta}}}^{\hat{\theta}\hat{\psi}}&=&{\omega_{\hat{\zeta}}}^{\hat{\zeta}\hat{\psi}}=e^{-V}\frac{\cot\psi}{f_{k}(\chi)}, \qquad\quad {\omega_{\hat{\zeta}}}^{\hat{\zeta}\hat{\theta}}=e^{-V}\frac{\cot\theta}{f_{k}(\chi)\sin{\psi}}\, .
\label{AdS3xSigma4SpinCon}
\end{eqnarray}
We have split the six-dimensional coordinates as $x^\mu=(t,r,x^i)$, $i=\chi,\psi,\theta,\zeta$, with indices $\hat{i},\hat{j},\ldots=\hat{\chi}, \hat{\psi}, \hat{\theta}, \hat{\zeta}$ being flat indices on $\Sigma^4_k$. The $r$-derivatives will generally be denoted by $'$ with an exception that a $'$ on any function with an explicit argument refers to the derivative of the function with respect to that argument as in $f'_k(\chi)=\frac{df_k(\chi)}{d\chi}$. 
\\
\indent To preserve some amount of supersymmetry, we perform a topological twist by turning on some gauge fields to cancel the spin connection ${\omega_{\hat{i}}}^{\hat{j}\hat{k}}$ on $\Sigma^4_k$. In the present case, we achieve this by turning on $SO(4)$ gauge fields as follows
\begin{eqnarray}
{A_{\hat{\psi}}}^{\hat{\psi}\hat{\chi}}&=&{A_{\hat{\theta}}}^{\hat{\theta}\hat{\chi}}={A_{\hat{\zeta}}}^{\hat{\zeta}\hat{\chi}}=e^{-V}\frac{p}{4k}\frac{f'_{k}(\chi)}{f_{k}(\chi)},\nonumber\\ {A_{\hat{\theta}}}^{\hat{\theta}\hat{\psi}}&=&{A_{\hat{\zeta}}}^{\hat{\zeta}\hat{\psi}}=e^{-V}\frac{p}{4k}\frac{\cot\psi}{f_{k}(\chi)}, \qquad {A_{\hat{\zeta}}}^{\hat{\zeta}\hat{\theta}}=e^{-V}\frac{p}{4k}\frac{\cot\theta}{f_{k}(\chi)\sin{\psi}}
\end{eqnarray}
where $p$ is a constant magnetic charge. In order to perform this topological twist, we consider gauged supergravity with the gauge group containing an $SO(4)$ subgroup. These gauge groups are given by $SO(5)$, $SO(4,1)$, and $CSO(4,0,1)$ which can be characterized by a tensor $Y_{mn}$ in the embedding tensor of the form
\begin{equation}\label{SO(4)Ytensor}
Y_{mn}=\text{diag}(1,1,1,1,\lambda)
\end{equation}
for $\lambda=+1,-1,0$. Among the $25$ scalar fields, there are two $SO(4)$ singlet scalars parametrized by the coset representative
\begin{equation}\label{15_SO(4)_coset}
V=e^{\varphi\boldsymbol{d}+\phi\mathcal{Y}}.
\end{equation}
The first scalar is the dilaton $\varphi(r)$ associated with the $SO(1,1)$ generator $\boldsymbol{d}$, and the other one $\phi(r)$ corresponds to the non-compact generator
\begin{equation}\label{15_SO(4)_non_com}
\mathcal{Y}=\hat{\boldsymbol{t}}_{1\dot{1}}+\hat{\boldsymbol{t}}_{2\dot{2}}+\hat{\boldsymbol{t}}_{3\dot{3}}+\hat{\boldsymbol{t}}_{4\dot{4}}-4\,\hat{\boldsymbol{t}}_{5\dot{5}}.
\end{equation}
\indent The topological twist can be achieved by imposing the twist condition 
\begin{equation}\label{simple_twist_con}
gp=k
\end{equation}
together with the following projection conditions
\begin{eqnarray}
\hat{\gamma}_{\hat{\psi}\hat{\chi}}\epsilon_{+\alpha}&=&{(\gamma_{12})_\alpha}^\beta\epsilon_{+\beta},\qquad \hat{\gamma}_{\hat{\theta}\hat{\psi}}\epsilon_{+\alpha}={(\gamma_{23})_\alpha}^\beta\epsilon_{+\beta},\qquad\hat{\gamma}_{\hat{\zeta}\hat{\theta}}\epsilon_{+\alpha}={(\gamma_{34})_\alpha}^\beta\epsilon_{+\beta},\label{Sigma4ProjCon}\\
\text{and}\qquad\hat{\gamma}_{\hat{\psi}\hat{\chi}}\epsilon_{-\dot{\alpha}}&=&{(\gamma_{\dot{1}\dot{2}})_{\dot{\alpha}}}^{\dot{\beta}}\epsilon_{-\dot{\beta}},\qquad \hat{\gamma}_{\hat{\theta}\hat{\psi}}\epsilon_{-\dot{\alpha}}={(\gamma_{\dot{2}\dot{3}})_{\dot{\alpha}}}^{\dot{\beta}}\epsilon_{-\dot{\beta}},\qquad\hat{\gamma}_{\hat{\zeta}\hat{\theta}}\epsilon_{-\dot{\alpha}}={(\gamma_{\dot{3}\dot{4}})_{\dot{\alpha}}}^{\dot{\beta}}
\epsilon_{-\dot{\beta}}\, \label{Sigma4ProjCon-}.
\end{eqnarray}
\indent To obtain the corresponding BPS equations, we need to further impose an additional projector of the form
\begin{equation}\label{15_DW_Proj}
\hat{\gamma}_r\epsilon_\pm=\epsilon_\mp
\end{equation}
due to the radial dependence of scalars. We will also take the supersymmetry parameters to depend only on $r$. With all these projectors, the unbroken supersymmetry has two supercharges. Each chirality of $\epsilon_\pm$ consists of two unbroken supercharges due to the projectors \eqref{Sigma4ProjCon} and \eqref{Sigma4ProjCon-}, and the two chiralities are related by the $\hat{\gamma}_r$-projector \eqref{15_DW_Proj}. 
\\
\indent The non-vanishing components of the $SO(4)$ gauge field strengths are given by
\begin{equation}\label{SO(4)2form}
{\mathcal{H}_{\hat{i}\hat{j}}}^{\tilde{m}\tilde{n}}={F_{\hat{i}\hat{j}}}^{\tilde{m}\tilde{n}}=-\frac{1}{2}e^{-2V}p\,\delta^{\tilde{m}}_{[\hat{i}}\delta^{\tilde{n}}_{\hat{j}]}
\end{equation}
in which we have split the $GL(5)$ index as $m=(\tilde{m},5)$ with $\tilde{m},\tilde{n}=1,2,3,4$ and $\hat{i},\hat{j}=\{1,2,3,4\}=\{\hat{\chi}, \hat{\psi}, \hat{\theta}, \hat{\zeta}\}$. These identically satisfy the Bianchi identity \eqref{DefBianchi1} and the two-/four-form duality \eqref{2_4SD}. However, the above gauge field strengths also lead to a non-vanishing term in the Bianchi identity given in \eqref{DefBianchi2}. This term takes the form
\begin{equation}\label{Sig4_gamFF}
(\Gamma_m)_{AB}\mathbb{T}^{A}_{np}\mathbb{T}^{B}_{qr}{\mathcal{H}_{[\hat{i}\hat{j}}}^{np}{\mathcal{H}_{\hat{k}\hat{l}]}}^{qr}=\frac{p^2}{4\sqrt{2}}e^{-4 V}\delta_m^5\varepsilon_{\hat{i}\hat{j}\hat{k}\hat{l}}\, .
\end{equation}
Therefore, to satisfy the Bianchi identity \eqref{DefBianchi2}, we turn on a magnetic two-form potential of the form
\begin{equation}
{B_{\hat{t}\hat{r}}}^5=-\frac{3p^2}{8 g^2}  e^{-2 (2V+3\varphi+8\phi)}
\end{equation}
giving rise to the following two-form field strength
\begin{equation}\label{another_SO(4)2form}
\mathcal{H}_{\hat{t}\hat{r},5}=\frac{3\lambda p^2}{4\sqrt{2}g} e^{-2 (2V+3\varphi+8\phi)}\, .
\end{equation}
\indent However, with this new contribution in $\mathcal{H}_{\mu\nu,5}$, the two-/four-form duality \eqref{2_4SD} is no longer satisfied. To remedy this, we turn on the four-form field strength of the form
\begin{equation}\label{Sig4_H4}
{\theta^A}_m\mathcal{H}_{\hat{i}\hat{j}\hat{k}\hat{l}, A}=-{\theta^A}_mM_{AB}\mathbb{T}^{Bn}\varepsilon_{\hat{i}\hat{j}\hat{k}\hat{l}}{\mathcal{H}^{\hat{t}\hat{r}}}_n=\frac{3\lambda^2p^2}{4\sqrt{2}g}e^{-4 V}\delta_m^5\varepsilon_{\hat{i}\hat{j}\hat{k}\hat{l}}\, .
\end{equation}
With all these, all the Bianchi identities as well as the field equations for vector and tensor fields are satisfied. 
\\
\indent It should be pointed out that for $CSO(4,0,1)$ gauge group with $\lambda=0$, the four-form field strength in \eqref{Sig4_H4} vanishes identically. In this case, the problematic term in \eqref{Sig4_gamFF} cannot be canceled and must vanish implying $p=0$. This leads to untwisted solutions with all the gauge fields vanishing and corresponds to the flat domain wall solutions studied in \cite{6D_DW_I}. Accordingly, we will not consider the non-semisimple gauge group $CSO(4,0,1)$ in the following analysis. 
\\
\indent Substituting the above ansatz into the conditions $\delta \psi_{+\mu\alpha}=0$, $\delta \psi_{-\mu\dot{\alpha}}=0$, $\delta\chi_{+a\dot{\alpha}}=0$, and $\delta\chi_{-\dot{a}\alpha}=0$, and imposing the projectors \eqref{Sigma4ProjCon}, \eqref{Sigma4ProjCon-}, and \eqref{15_DW_Proj}, we find the following BPS equations
\begin{eqnarray}
U'&=&\frac{g}{4\sqrt{2}}e^{\varphi-4\phi}(4+\lambda e^{20\phi})-\frac{3p}{4\sqrt{2}}e^{-2V-\varphi+4\phi}+\frac{9\lambda p^2}{16\sqrt{2}g}e^{-4V-3\varphi-8\phi},\label{15_M4_BPSeq1}\\
V'&=&\frac{g}{4\sqrt{2}}e^{\varphi-4\phi}(4+\lambda e^{20\phi})+\frac{3p}{4\sqrt{2}}e^{-2V-\varphi+4\phi}-\frac{3\lambda p^2}{16\sqrt{2}g}e^{-4V-3\varphi-8\phi},\label{15_M4_BPSeq2}\\
\varphi'&=&-\frac{g}{20\sqrt{2}}e^{\varphi-4\phi}(4+\lambda e^{20\phi})+\frac{3p}{20\sqrt{2}}e^{-2V-\varphi+4\phi}-\frac{9\lambda p^2}{80\sqrt{2}g}e^{-4V-3\varphi-8\phi},\label{15_M4_BPSeq3}\qquad\,\ \\
\phi'&=&\frac{g}{5\sqrt{2}}e^{\varphi-4\phi}(1-\lambda e^{20\phi})-\frac{3p}{20\sqrt{2}}e^{-2V-\varphi+4\phi}-\frac{3\lambda p^2}{40\sqrt{2}g}e^{-4V-3\varphi-8\phi}\label{15_M4_BPSeq4}
\end{eqnarray}
together with a simple form of the Killing spinors
\begin{equation}
\epsilon_\pm=e^{\frac{U(r)}{2}}\epsilon_{\pm}^{0}\, .\label{DW_Killing_spinor}	
\end{equation}
$\epsilon_\pm^{0}$ are constant symplectic-Majorana-Weyl spinors satisfying the projectors \eqref{Sigma4ProjCon}, \eqref{Sigma4ProjCon-}, and \eqref{15_DW_Proj}. It can also be verified that these equations satisfy all bosonic field equations derived from the Lagrangian \eqref{BosLag}. We also note that these equations do not admit a fixed point solution of the form $AdS_2\times\Sigma^4_k$ at which $\varphi'=\phi'=V'=0$ and $U'=\textrm{const}$. 
\\
\indent The resulting BPS equations cannot be solved analytically, so we will look for numerical solutions. We are interested in solutions interpolating between asymptotically locally flat domain walls studied in \cite{6D_DW_I} and another domain wall with the world-volume given by $t\times \Sigma^4_k$. According to the DW/QFT correspondence, we expect these solutions to be holographically dual to supersymmetric quantum mechanics preserving two supercharges in the IR. This quantum mechanics is obtained from twisted compactification on a Riemannian four-manifold of the maximal SYM theory in five dimensions dual to the flat domain wall solutions of \cite{6D_DW_I}. 
\\
\indent The first class of solutions is given by solutions that are asymptotically $SO(5)$ symmetric domain wall of the form
\begin{equation}\label{SO(4)_SO(5)_flat_DW_asym}
U\sim V\sim 5\ln gr,\qquad\varphi\sim-\ln gr,\qquad\phi\sim \frac{1}{g^{16} r^{16}}\, .
\end{equation}
We also note that the $SO(4)$ singlet scalar $\phi\rightarrow 0$ as $r\rightarrow \infty$ while the dilaton $\varphi$ is invariant under $SO(5)$. This asymptotic geometry can only exist in $SO(5)$ gauge group. Examples of numerical solutions interpolating between this locally $SO(5)$ flat domain wall and $t\times \Sigma^4_k$-sliced curved domain walls are given in figures \ref{15_S4_SO(4)_special_SO(5)gg_flows} and \ref{15_H4_SO(4)_special_SO(5)gg_flows} for $k=1$ and $k=-1$ with various values of the gauge coupling constant $g$. For the solutions with $t\times \mathbb{R}^4$ slices and $k=0$, the domain walls are entirely flat since the twist condition implies $p=0$. Accordingly, we have omitted these solutions.   

\begin{figure}[h!]
  \centering
    \includegraphics[width=0.8\linewidth]{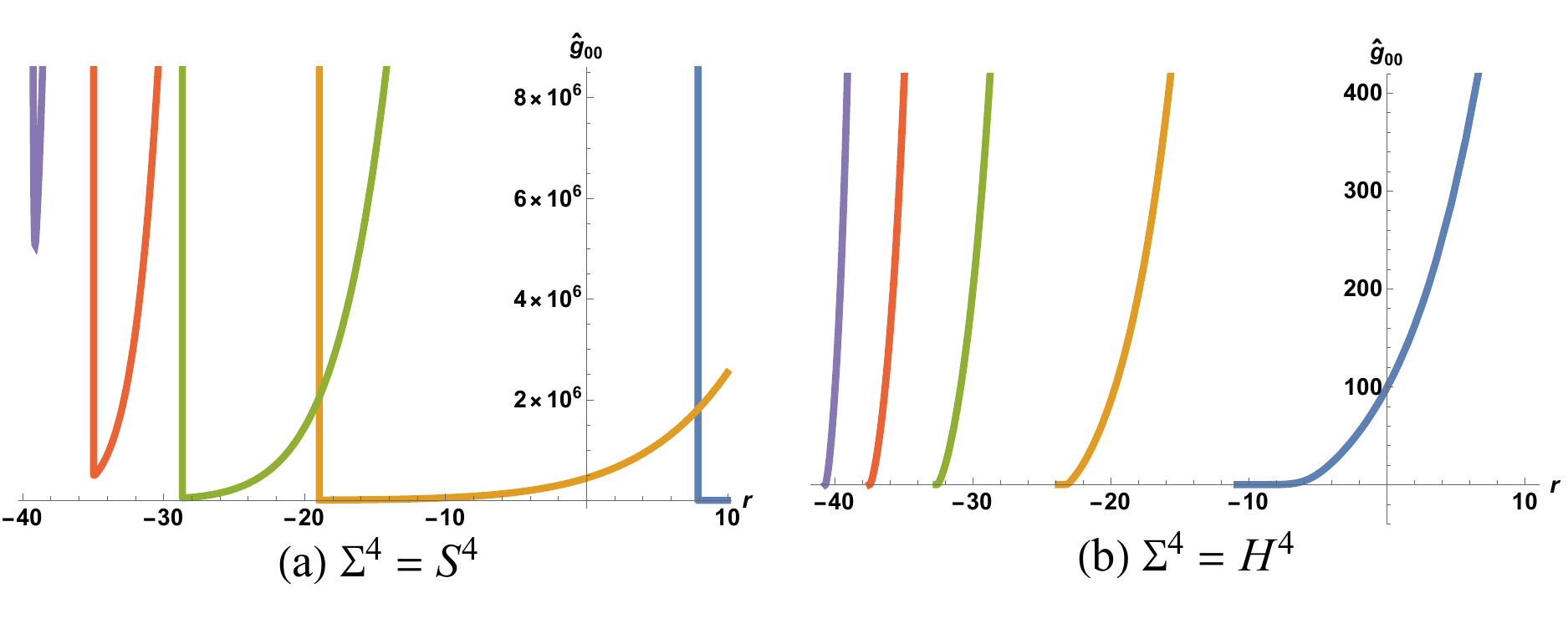}
\caption{The behavior of $\hat{g}_{00}$ along the RG flows given in figures \ref{15_S4_SO(4)_special_SO(5)gg_flows} and \ref{15_H4_SO(4)_special_SO(5)gg_flows} respectively for $\Sigma_1^4=S^4$ and $\Sigma_{-1}^4=H^4$ in $SO(5)$ gauge group.}
\label{g00fig1to3}
\end{figure}

As usually the case for holographic solutions dual to non-conformal field theories, all solutions flow to singular gemetries in the IR. To determine whether these singularities lead to any physical interpretrations, we can uplift the solutions to type IIA theory and look for the behavior of the ten-dimensional metric component $\hat{g}_{00}$. We have put a hat on the fields in ten dimensions. Using the formulae given in appendix \ref{g00}, we find that for the case of $SO(5)$ gauge group, the $(00)$-component of type IIA metric takes the form
\begin{equation}\label{15_g00}
\hat{g}_{00}=e^{2U+\varphi}\Delta^{3/8}\, .
\end{equation}
The warp factor $\Delta$ is given by
\begin{equation}
\Delta=e^{-4 \phi}[(\mu^1)^2+(\mu^2)^2+(\mu^3)^2+(\mu^4)^2]+\lambda ^2 e^{8 \phi}(\mu^5)^2
\end{equation}
in which $\mu^a$, $a=1,2,\ldots, 5$, are coordinates on $S^4$ satisfying $\mu^a\mu^a=1$.
\\
\indent According to the criterion of \cite{Maldacena_nogo}, $\hat{g}_{00}$ should not diverge when approaching the singularity in order for the IR singularity to be physically acceptable. For all the solutions shown in figures \ref{15_S4_SO(4)_special_SO(5)gg_flows} and \ref{15_H4_SO(4)_special_SO(5)gg_flows}, we give the behavior of $\hat{g}_{00}$ in figure \ref{g00fig1to3}. From this figure, we readily see that all the solutions with $t\times H^4$ slices are uplifted to type IIA theory solutions with physically acceptable singularities since $\hat{g}_{00}\rightarrow 0$ near the singularities. According to the DW/QFT correspondence, these solutions give holographic descriptions of twisted compactifications of the maximal five-dimensional SYM theory on $H^4$ leading to supersymmetric quantum mechanics preserving two supercharges. Moreover, the uplifted type IIA solutions can be interpreted as D4-branes wrapped on $H^4$. On the other hand, all solutions with $t\times S^4$ slices lead to unphysical type IIA solutions.

We now move to another class of solutions with asymptotic behaviors given by a locally flat domain wall with $SO(4)$ symmetry of the form
\begin{equation}\label{15_SO(4)_flat_DW_asym}
U\sim V\sim -\frac{1}{5} \ln \left[-\frac{5 g \rho }{\sqrt{2}}\right],\qquad\varphi\sim-\phi\sim \frac{1}{25} \ln \left[-\frac{5 g \rho }{\sqrt{2}}\right].
\end{equation}
The new radial coordinate $\rho$ is defined by $\frac{d\rho}{dr}=e^{-6V}$. We will choose the gauge coupling constant $g <0$ in order to identify the UV limit with $\rho\rightarrow +\infty$. It turns out that, in this case, all of the numerical solutions are uplifted to type IIA solutions with unphysical IR singularities since $\hat{g}_{00}$ diverges near the singularities. Accordingly, we will not present these solutions here.
\\
\indent For convenience and the sake of comparison with other cases to be given in subsequent analyses, we summarize physical acceptabilities of the IR singularities according to the criterion of \cite{Maldacena_nogo} in table \ref{tab1}. Note also that solutions with $t\times \mathbb{R}^4$ slices are not included in the table since these are untwisted flat domain walls with $p=k=0$ along the entire flow solutions. In this table and subsequent ones, we use the $\times$ symbol to mark the cases in which the criterion of \cite{Maldacena_nogo} is violated. For the solutions with physically acceptable singularities, the possible values of the independent parameter, in this case $g$, are also given. 

\begin{table}[h!]
\centering
\begin{tabular}{| c | c | c | c |}
\hline
UV flat & gauge & Riemannian & physical IR \\
domain wall& group & four-manifold &  singularity \\\hline
$SO(5)$&$SO(5)$&$S^4$ & $\times$ \\
&&$H^4$ & any $g$ \\\hline
$SO(4)$&$SO(5)$& $S^4$, $H^4$ & $\times$\\\cline{2-4}
&$SO(4,1)$ & $S^4$, $H^4$ & $\times$\\\hline
\end{tabular}
\caption{Summary of satisfaction of the physical criterion given in \cite{Maldacena_nogo} for the IR singularities of the curved domain wall solutions obtained from $SO(4)$ twist on a Riemannian four-manifold in $SO(5)$ and $SO(4,1)$ gauge groups.}\label{tab1}
\end{table}

%%%%%%%%%%%%%%%%%%%%%%%%%%%%%%%%%%%%%%%%%%%
\subsection{D4-branes wrapped on a product of two Riemann surfaces}\label{15_Sig2xSig2_Sec}
We now carry out a similar analysis for supersymmetric solutions describing D4-branes wrapped on a product of two Riemann surfaces $\Sigma^2_{k_1}\times\Sigma^2_{k_2}$. In this case, the ansatz for the metric takes the form of
\begin{equation}\label{SO(2)xSO(2)6Dmetric}
ds_6^2=-e^{2U(r)}dt^2+dr^2+e^{2V_1(r)}ds^2_{\Sigma^{2}_{k_1}}+e^{2V_2(r)}ds^2_{\Sigma^{2}_{k_2}}
\end{equation}
in which the metrics on $\Sigma^2_{k_1}$ and $\Sigma^2_{k_2}$ are given by
\begin{equation}\label{Sigma2metric}
 ds^2_{\Sigma^2_{k_i}}=d\theta_i^2+f_{k_i}(\theta_i)^2d\zeta_i^2,\qquad i=1,2\, .
\end{equation}
The function $f_{k_i}(\theta_i)$ is defined as in \eqref{fFn} with $k_i=0,\pm 1$ characterizing the constant curvature of $\Sigma^2_{k_i}$. We will split the six-dimensional coordinates as $x^\mu=(t,r,x^{i_1},x^{i_2})$ with $x^{i_1}=(x^{\theta_1},x^{\zeta_1})$ and $x^{i_2}=(x^{\theta_2},x^{\zeta_2})$.
\\
\indent Using the following choice of the vielbein
\begin{eqnarray}
e^{\hat{t}}&=&e^{U}dt, \qquad\, e^{\hat{r}}=dr,\qquad e^{\hat{\theta}_1}=e^{V_1}d\theta_1,\qquad e^{\hat{\theta}_2}=e^{V_2}d\theta_2,\nonumber \\ 
e^{\hat{\zeta}_1}&=&e^{V_1}f_{k_1}(\theta_1)d\zeta_1, \qquad e^{\hat{\zeta}_2}=e^{V_2}f_{k_2}(\theta_2)d\zeta_2,\label{Sigma2xSigma2bein}
\end{eqnarray}
we obtain all non-vanishing components of the spin connection as follows
\begin{eqnarray}
{\omega_{\hat{t}}}^{\hat{t}\hat{r}}&=&U', \qquad {\omega_{\hat{i}_1}}^{\hat{j}_1\hat{r}}= V'_1\,\delta^{\hat{j}_1}_{\hat{i}_1}, \qquad {\omega_{\hat{i}_2}}^{\hat{j}_2\hat{r}}= V'_2\,\delta^{\hat{j}_2}_{\hat{i}_2},\nonumber \\
{\omega_{\hat{\zeta}_1}}^{\hat{\zeta}_1\hat{\theta}_1}&=&e^{-V_1}\frac{f_{k_1}'(\theta_1)}{f_{k_1}(\theta_1)},\qquad {\omega_{\hat{\zeta}_2}}^{\hat{\zeta}_2\hat{\theta}_2}=e^{-V_2}\frac{f_{k_2}'(\theta_2)}{f_{k_2}(\theta_2)}\label{AdS3xSigma2xSigma2SpinCon}
\end{eqnarray}
with $\hat{i}_1=\hat{\theta}_1, \hat{\zeta}_1$ and $\hat{i}_2=\hat{\theta}_2, \hat{\zeta}_2$ being flat indices on $\Sigma^2_{k_1}$ and $\Sigma^2_{k_2}$, respectively.
\\
\indent To perform the topological twist, we turn on $SO(2)\times SO(2)$ gauge fields corresponding to $X_{12}$ and $X_{34}$ gauge generators of the form
\begin{eqnarray}
{A_{\hat{\zeta}_1}}^{12}&=&e^{-V_1}\frac{p_{11}}{4k_1}\frac{f_{k_1}'(\theta_1)}{f_{k_1}(\theta_1)},\qquad{A_{\hat{\zeta}_2}}^{12}=e^{-V_2}\frac{p_{12}}{4k_2}\frac{f_{k_2}'(\theta_2)}{f_{k_2}(\theta_2)},\nonumber \\
{A_{\hat{\zeta}_1}}^{34}&=&e^{-V_1}\frac{p_{21}}{4k_1}\frac{f_{k_1}'(\theta_1)}{f_{k_1}(\theta_1)},\qquad{A_{\hat{\zeta}_2}}^{34}=e^{-V_2}\frac{p_{22}}{4k_2}\frac{f_{k_2}'(\theta_2)}{f_{k_2}(\theta_2)}.\label{15_SO(2)xSO(2)_gauge_fields}
\end{eqnarray} 
The resulting two-form field strengths have the following non-vanishing components
\begin{eqnarray}
{\mathcal{H}_{\hat{\theta}_1\hat{\zeta}_1}}^{12}&=&-\frac{1}{4}p_{11}e^{-2V_1},\qquad{\mathcal{H}_{\hat{\theta}_2\hat{\zeta}_2}}^{12}=-\frac{1}{4}p_{12}e^{-2V_2},\nonumber\\{\mathcal{H}_{\hat{\theta}_1\hat{\zeta}_1}}^{34}&=&-\frac{1}{4}p_{21}e^{-2V_1},
\qquad{\mathcal{H}_{\hat{\theta}_2\hat{\zeta}_2}}^{34}=-\frac{1}{4}p_{22}e^{-2V_2}.
\end{eqnarray}
\indent As in the previous case, we will consider solutions in all gauge groups with an $SO(2)\times SO(2)$ subgroup. The embedding tensor for these gauge groups are characterized by the symmetric tensor
\begin{equation}\label{SO(2)xSO(2)Y}
Y_{mn}=\textrm{diag}(1,1,\kappa,\kappa,\lambda)
\end{equation} 
for $\lambda=0,\pm 1$ and $\kappa=\pm 1$. All possible gauge groups are then given explicitly by $SO(5)$ ($\kappa=\lambda=1$), $SO(4,1)$ ($\kappa=-\lambda=1$), $SO(3,2)$ ($\kappa=-\lambda=-1$), $CSO(4,0,1)$ ($\kappa=1,\lambda=0$), and $CSO(2,2,1)$ ($\kappa=-1,\lambda=0$).
\\
\indent Under $SO(2)\times SO(2)$ generated by $X_{12}$ and $X_{34}$, there are five singlet scalars. One of them is the dilaton $\varphi(r)$ corresponding to the $SO(1,1)$ generator $\boldsymbol{d}$. The remaining four are denoted by $\phi_1(r)$, $\phi_2(r)$, $\varsigma_1(r)$, and $\varsigma_2(r)$ and correspond respectively to the following non-compact generators
\begin{equation}
\mathcal{Y}_1=\hat{\boldsymbol{t}}_{1\dot{1}}+\hat{\boldsymbol{t}}_{2\dot{2}}-2\,\hat{\boldsymbol{t}}_{5\dot{5}},\qquad\mathcal{Y}_2=\hat{\boldsymbol{t}}_{3\dot{3}}+\hat{\boldsymbol{t}}_{4\dot{4}}-2\,\hat{\boldsymbol{t}}_{5\dot{5}},\qquad\mathcal{Y}_3=\boldsymbol{s}_{12},\qquad\mathcal{Y}_4=\boldsymbol{s}_{34}\, .
\end{equation}
The coset representative is then given by
\begin{equation}\label{fullSO(2)xSO(2)singlet_coset}
V=e^{\varphi\boldsymbol{d}+\phi_1\mathcal{Y}_1+\phi_2\mathcal{Y}_2+\varsigma_1\mathcal{Y}_3+\varsigma_2\mathcal{Y}_4}\, .
\end{equation}
\indent As in the solutions with $SO(2)\times SO(2)$ twist on a single Riemann surface considered in \cite{6D_twist_I}, the $SO(2)\times SO(2)$ gauge field strengths lead to the following non-vanishing terms in the two-/four-form duality, given in \eqref{2_4SD},
\begin{eqnarray}
g{\theta^A}_mM_{AB}\mathbb{T}^{B}_{np}{\mathcal{H}_{\hat{\theta}_1\hat{\zeta}_1}}^{np}&=&-\frac{g\lambda}{\sqrt{2}}e^{-2(V_1-\varphi-2\phi_1-2\phi_2)}(p_{21}e^{4\phi_2}\varsigma_1+p_{11}e^{4\phi_1}\varsigma_2)\delta_m^5,\nonumber\\g{\theta^A}_mM_{AB}\mathbb{T}^{B}_{np}{\mathcal{H}_{\hat{\theta}_2\hat{\zeta}_2}}^{np}&=&-\frac{g\lambda}{\sqrt{2}}e^{-2(V_2-\varphi-2\phi_1-2\phi_2)}(p_{22}e^{4\phi_2}\varsigma_1+p_{12}e^{4\phi_1}\varsigma_2)\delta_m^5.\qquad\label{Sig2xSig2_prob_terms}
\end{eqnarray}
Accordingly, we would need to turn on the following components of the four-form fields $\mathcal{H}_{tr\theta_1\zeta_1,A}$ and $\mathcal{H}_{tr\theta_2\zeta_2,A}$. In addition, this leads to non-vanishing components ${\mathcal{H}_{tr}}^{mn}$ and algebraic constraints on the magnetic charges and the warp factors $V_1$ and $V_2$. To avoid these complications as well as unnecessary restrictions on the solutions, we will set the two terms in \eqref{Sig2xSig2_prob_terms} to zero. As in \cite{6D_twist_I}, there are two possibilities namely vanishing shift scalars $\varsigma_1=\varsigma_2=0$ or imposing some additional constraints on the magnetic charges with non-vanishing shift scalars. 

\subsubsection{Solutions with $\varsigma_1=\varsigma_2=0$}
Although setting $\varsigma_1=\varsigma_2=0$ solves the two-/four-form duality \eqref{2_4SD}, there is an additional term in the Bianchi identity \eqref{DefBianchi2} of the form
\begin{equation}\label{Sig2xSig2_gamFF}
(\Gamma_m)_{AB}\mathbb{T}^{A}_{np}\mathbb{T}^{B}_{qr}{\mathcal{H}_{\hat{\theta}_1\hat{\zeta}_1}}^{np}{\mathcal{H}_{\hat{\theta}_2\hat{\zeta}_2}}^{qr}=\frac{1}{12\sqrt{2}}(p_{11}p_{22}+p_{12}p_{21})e^{-2 V_1-2 V_2}\delta_m^5\, .
\end{equation}
This can be canceled by turning on the following magnetic two-form potential
\begin{equation}\label{Sig2xSig2_mag_two_pot}
{B_{\hat{t}\hat{r}}}^5=-\frac{1}{8 g^2}(p_{11}p_{22}+p_{12}p_{21})  e^{-2 (V_1+V_2+3\varphi+4\phi_1+4\phi_2)}
\end{equation}
which leads to another contribution to the two-form field strengths as
\begin{equation}\label{another_SO(4)2form}
\mathcal{H}_{\hat{t}\hat{r},5}=\frac{\lambda}{4 \sqrt{2}g}(p_{11}p_{22}+p_{12}p_{21})  e^{-2 (V_1+V_2+3\varphi+4\phi_1+4\phi_2)}\, .
\end{equation}
This in turn gives rise to, via the Bianchi identity \eqref{DefBianchi2} and the two-/four-form duality \eqref{2_4SD}, the four-form field strength
\begin{equation}\label{Sig2xSig2_H4}
{\theta^A}_m\mathcal{H}_{\hat{\theta}_1\hat{\zeta}_1\hat{\theta}_2\hat{\zeta}_2, A}=-{\theta^A}_mM_{AB}\mathbb{T}^{Bn}{\mathcal{H}^{\hat{t}\hat{r}}}_n=\frac{\lambda^2}{4\sqrt{2}g}(p_{11}p_{22}+p_{12}p_{21})e^{-2V_1-2V_2}\delta_m^5\, .
\end{equation}
For gauge groups with $\lambda=\pm1$, the two terms in \eqref{Sig2xSig2_gamFF} and \eqref{Sig2xSig2_H4} cancel each other recovering the Bianchi identity \eqref{DefBianchi2}. 
\\
\indent On the other hand, for $CSO(4,0,1)$ and $CSO(2,2,1)$ gauge groups with $\lambda=0$, the problematic term \eqref{Sig2xSig2_gamFF} within the Bianchi identity \eqref{DefBianchi2} cannot be canceled since the corresponding four-form field strengths do not appear. To satisfy the Bianchi identity \eqref{DefBianchi2} in this case, we further impose the following condition between the magnetic charges
\begin{equation}\label{15_SO2xSO2_4p_con}
p_{11}p_{22}+p_{12}p_{21}=0\, .
\end{equation}
\indent Implementing the relevant topological twist by imposing the projection condition $\hat{\gamma}_r\epsilon_\pm=\epsilon_\mp$ as well as
\begin{eqnarray}
\hat{\gamma}_{\hat{\zeta}_1\hat{\theta}_1}\epsilon_{+\alpha}&=&\hat{\gamma}_{\hat{\zeta}_2\hat{\theta}_2}\epsilon_{+\alpha}={(\gamma_{12})_\alpha}^\beta\epsilon_{+\beta}={(\gamma_{34})_\alpha}^\beta\epsilon_{+\beta}\label{SO(2)xSO(2)ProjCon}\\
\text{and}\qquad\hat{\gamma}_{\hat{\zeta}_1\hat{\theta}_1}\epsilon_{-\dot{\alpha}}&=&\hat{\gamma}_{\hat{\zeta}_2\hat{\theta}_2}\epsilon_{-\dot{\alpha}}={(\gamma_{\dot{1}\dot{2}})_{\dot{\alpha}}}^{\dot{\beta}}\epsilon_{-\dot{\beta}}={(\gamma_{\dot{3}\dot{4}})_{\dot{\alpha}}}^{\dot{\beta}}\epsilon_{-\dot{\beta}}\label{SO(2)xSO(2)ProjCon-}
\end{eqnarray}
together with the twist conditions 
\begin{equation}\label{15_SO(2)xSO(2)_twist_con}
g(p_{11}+\kappa p_{21})=k_1\qquad \textrm{and} \qquad g(p_{12}+\kappa p_{22})=k_2,
\end{equation}
we find the following set of BPS equations 
\begin{eqnarray}
U'&=&\frac{ge^{\varphi}}{4\sqrt{2}}(2e^{-4\phi_1}+2\kappa e^{-4\phi_2}+\lambda e^{8(\phi_1+\phi_2)})+\frac{3\lambda(p_{11}p_{22}+p_{12}p_{21})}{16 \sqrt{2}ge^{2V_1+2V_2+3\varphi+4\phi_1+4\phi_2}}\nonumber\\
&&-\frac{e^{-\varphi}}{8\sqrt{2}}\left[e^{-2V_1}(p_{11}e^{4\phi_1}+p_{21}e^{4\phi_2})+e^{-2V_2}(p_{12}e^{4\phi_1}+p_{22}e^{4\phi_2})\right],\label{Sig2xSig2SO(2)xSO(2)BPS1}\\
V'_1&=&\frac{ge^{\varphi}}{4\sqrt{2}}(2e^{-4\phi_1}+2\kappa e^{-4\phi_2}+\lambda e^{8(\phi_1+\phi_2)})-\frac{\lambda(p_{11}p_{22}+p_{12}p_{21})}{16 \sqrt{2}ge^{2V_1+2V_2+3\varphi+4\phi_1+4\phi_2}}\nonumber\\
&&+\frac{e^{-\varphi}}{8\sqrt{2}}\left[3e^{-2V_1}(p_{11}e^{4\phi_1}+p_{21}e^{4\phi_2})-e^{-2V_2}(p_{12}e^{4\phi_1}+p_{22}e^{4\phi_2})\right],\\
V'_2&=&\frac{ge^{\varphi}}{4\sqrt{2}}(2e^{-4\phi_1}+2\kappa e^{-4\phi_2}+\lambda e^{8(\phi_1+\phi_2)})-\frac{\lambda(p_{11}p_{22}+p_{12}p_{21})}{16 \sqrt{2}ge^{2V_1+2V_2+3\varphi+4\phi_1+4\phi_2}}\nonumber\\
&&-\frac{e^{-\varphi}}{8\sqrt{2}}\left[e^{-2V_1}(p_{11}e^{4\phi_1}+p_{21}e^{4\phi_2})-3e^{-2V_2}(p_{12}e^{4\phi_1}+p_{22}e^{4\phi_2})\right],\\
\varphi'&=&-\frac{ge^{\varphi}}{20\sqrt{2}}(2e^{-4\phi_1}+2\kappa e^{-4\phi_2}+\lambda e^{8(\phi_1+\phi_2)})-\frac{3\lambda(p_{11}p_{22}+p_{12}p_{21})}{80 \sqrt{2}ge^{2V_1+2V_2+3\varphi+4\phi_1+4\phi_2}}\nonumber\\
&&+\frac{e^{-\varphi}}{40\sqrt{2}}\left[e^{-2V_1}(p_{11}e^{4\phi_1}+p_{21}e^{4\phi_2})+e^{-2V_2}(p_{12}e^{4\phi_1}+p_{22}e^{4\phi_2})\right],\\
\phi'_1&=&\frac{ge^{\varphi}}{5\sqrt{2}}(3e^{-4\phi_1}-2\kappa e^{-4\phi_2}-\lambda e^{8(\phi_1+\phi_2)})-\frac{\lambda(p_{11}p_{22}+p_{12}p_{21})}{40 \sqrt{2}ge^{2V_1+2V_2+3\varphi+4\phi_1+4\phi_2}}\nonumber\\
&&-\frac{e^{-\varphi}}{20\sqrt{2}}\left[e^{-2V_1}(3p_{11}e^{4\phi_1}-2p_{21}e^{4\phi_2})+e^{-2V_2}(3p_{12}e^{4\phi_1}-2p_{22}e^{4\phi_2})\right],\\
\phi'_2&=&\frac{ge^{\varphi}}{5\sqrt{2}}(3\kappa e^{-4\phi_2}-2e^{-4\phi_1}-\lambda e^{8(\phi_1+\phi_2)})-\frac{\lambda(p_{11}p_{22}+p_{12}p_{21})}{40 \sqrt{2}ge^{2V_1+2V_2+3\varphi+4\phi_1+4\phi_2}}\nonumber\\
&&+\frac{e^{-\varphi}}{20\sqrt{2}}\left[e^{-2V_1}(2p_{11}e^{4\phi_1}-3p_{21}e^{4\phi_2})+e^{-2V_2}(2p_{12}e^{4\phi_1}-3p_{22}e^{4\phi_2})\right]\qquad\quad\label{Sig2xSig2SO(2)xSO(2)BPS6}
\end{eqnarray}
with the Killing spinors given by \eqref{DW_Killing_spinor}. As in the previous case, we will numerically solve these equations for solutions interpolating between locally asymptotically flat domain walls and $t\times \Sigma^2_{k_1}\times \Sigma^2_{k_2}$-sliced domain walls. We will consider the solutions from gauge groups with $\lambda\neq0$ and $\lambda=0$ separately.

\begin{figure}[h!]
  \centering
    \includegraphics[width=\linewidth]{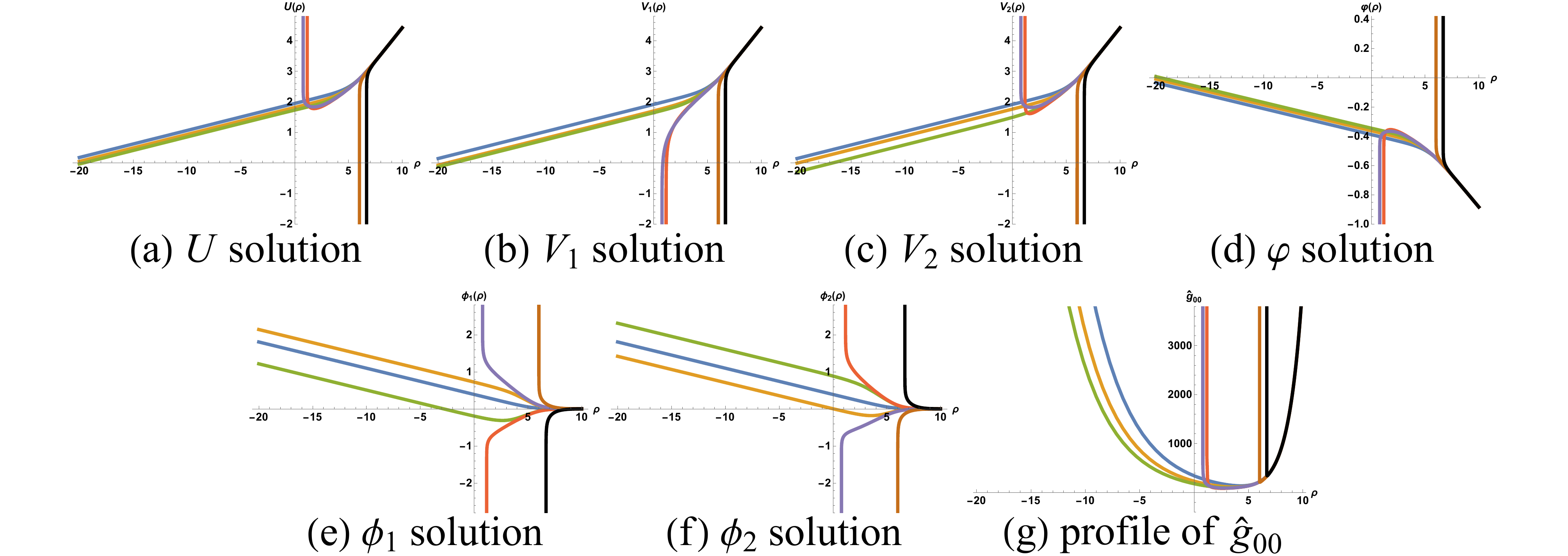}
\caption{Interpolating solutions between the locally $SO(5)$ flat domain wall as $\rho\rightarrow+\infty$ and $t\times S^2\times S^2$-sliced curved domain walls for $SO(2)\times SO(2)$ twist in $SO(5)$ gauge group. The blue, orange, green, red, purple, brown, and black curves refer to $(z_1,z_2)=(0,0), (1.7,0), (0,-1.9), (-2,0), (3,-1), (4,3), (0,-16)$.}
\label{15_SS_special_SO(2)xSO(2)_SO(5)gg_flows}
\end{figure}

\begin{figure}[h!]
  \centering
    \includegraphics[width=\linewidth]{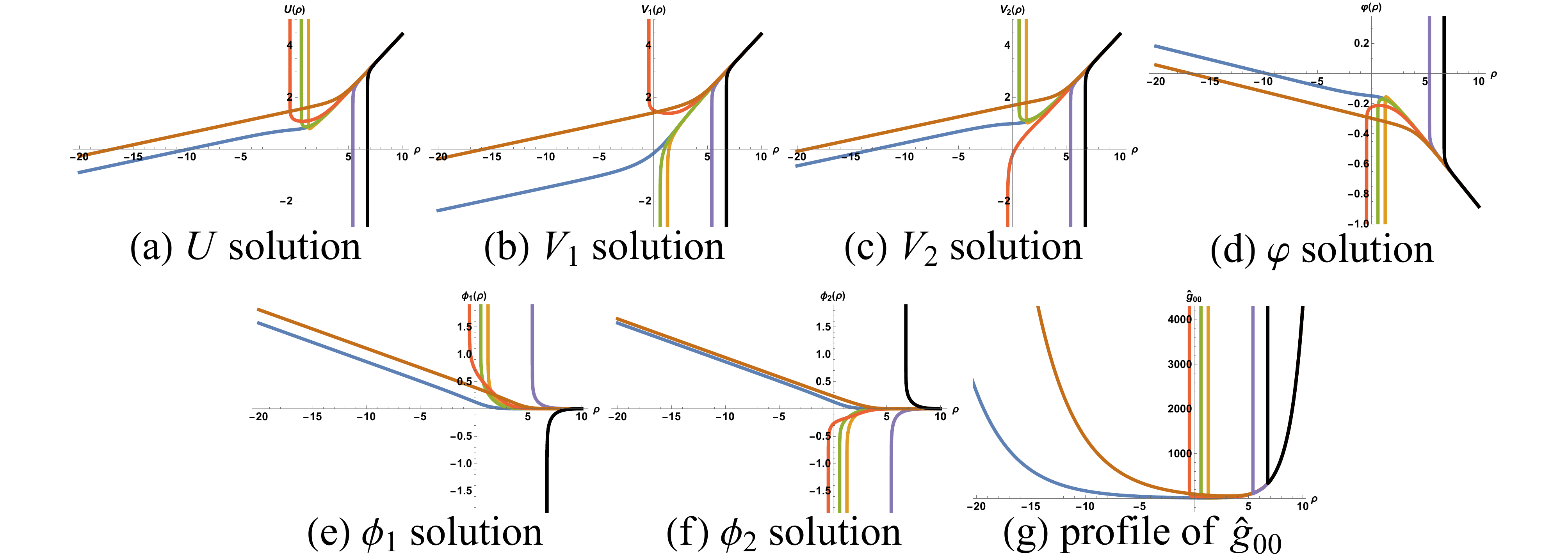}
\caption{Interpolating solutions between the locally $SO(5)$ flat domain wall as $\rho\rightarrow+\infty$ and $t\times S^2\times H^2$-sliced curved domain walls for $SO(2)\times SO(2)$ twist in $SO(5)$ gauge group. The blue, orange, green, red, purple, brown, and black curves refer to $(z_1,z_2)=(0,0), (0.01,0), (0.50,-0.50), (-2,2), (3,-1), (8,-8), (0,-16)$.}
\label{15_SH_special_SO(2)xSO(2)_SO(5)gg_flows}
\end{figure}

\begin{figure}[h!]
  \centering
    \includegraphics[width=\linewidth]{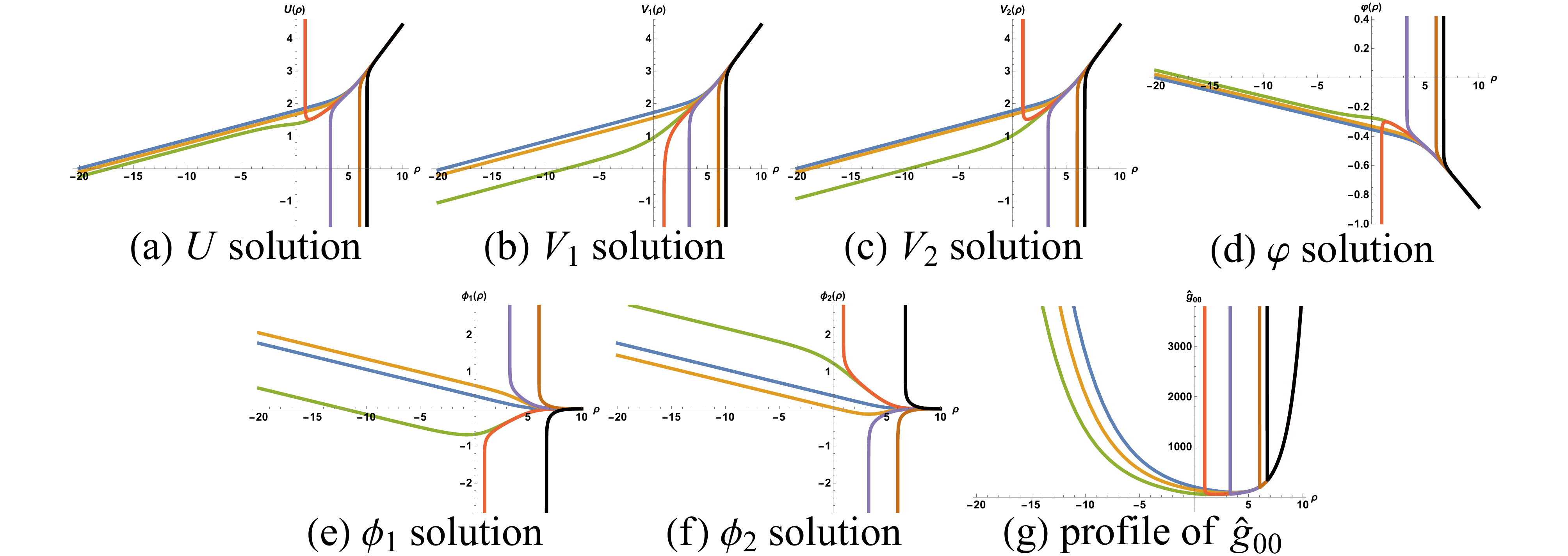}
\caption{Interpolating solutions between the locally $SO(5)$ flat domain wall as $\rho\rightarrow+\infty$ and $t\times S^2\times \mathbb{R}^2$-sliced curved domain walls for $SO(2)\times SO(2)$ twist in $SO(5)$ gauge group. The blue, orange, green, red, purple, brown, and black curves refer to $(z_1,z_2)=(0,0), (0.8,0), (0,-1), (-1,0), (0.5,0.6), (3,3), (0,-16)$.}
\label{15_SR_special_SO(2)xSO(2)_SO(5)gg_flows}
\end{figure}

\begin{figure}[h!]
  \centering
    \includegraphics[width=\linewidth]{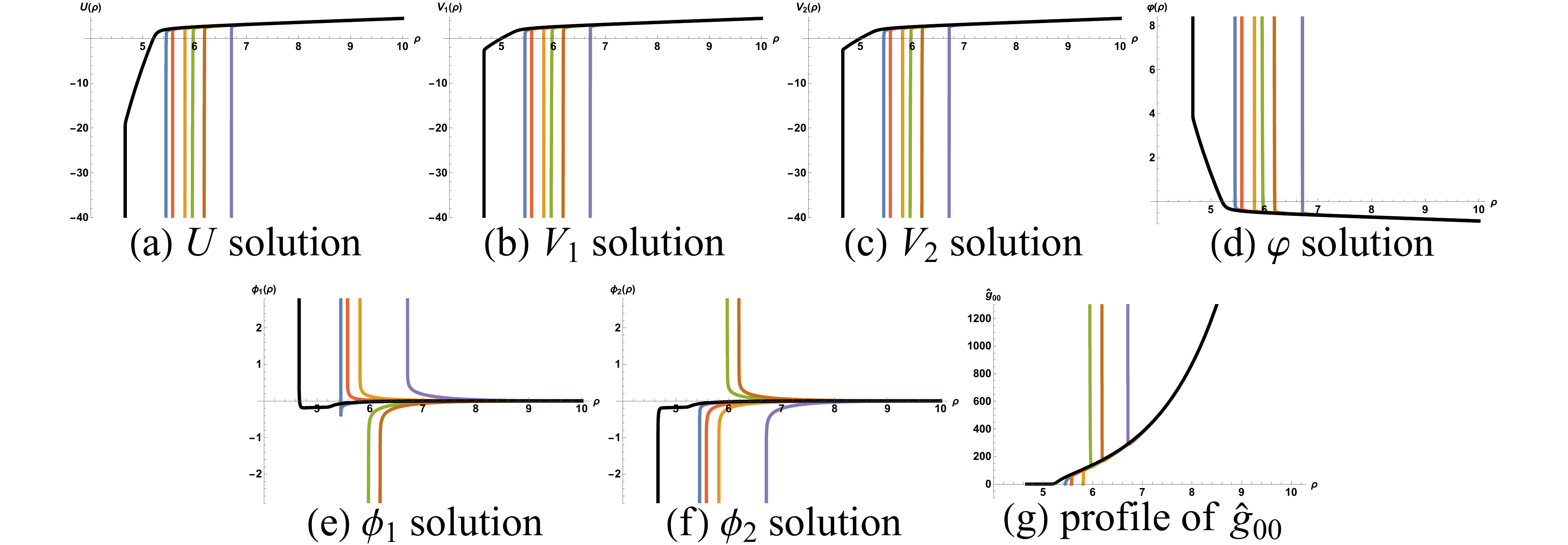}
\caption{Interpolating solutions between the locally $SO(5)$ flat domain wall as $\rho\rightarrow+\infty$ and $t\times H^2\times H^2$-sliced curved domain walls for $SO(2)\times SO(2)$ twist in $SO(5)$ gauge group. The blue, orange, green, red, purple, brown, and black curves refer to $(z_1,z_2)=(0,0), (1.5,0), (0,-2.5), (-3,3.5), (7,5), (-9,4.5), (8,-8)$.}
\label{15_HH_special_SO(2)xSO(2)_SO(5)gg_flows}
\end{figure}

\begin{figure}[h!]
  \centering
    \includegraphics[width=\linewidth]{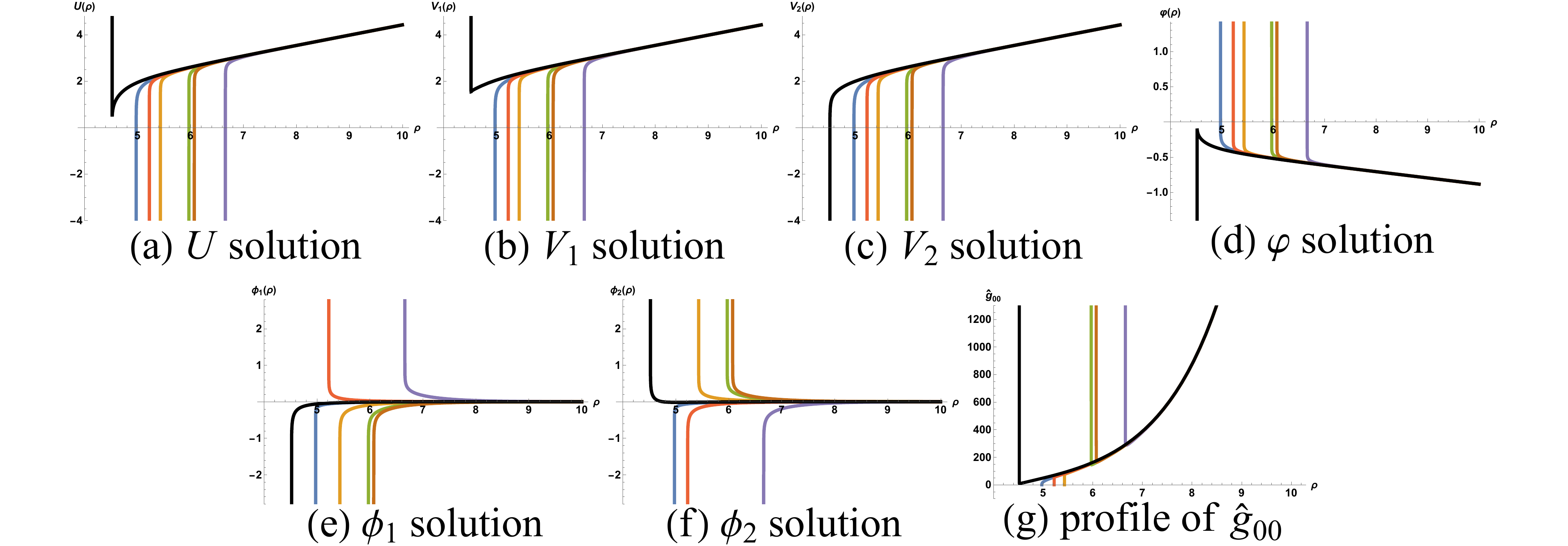}
\caption{Interpolating solutions between the locally $SO(5)$ flat domain wall as $\rho\rightarrow+\infty$ and $t\times H^2\times \mathbb{R}^2$-sliced curved domain walls for $SO(2)\times SO(2)$ twist in $SO(5)$ gauge group. The blue, orange, green, red, purple, brown, and black curves refer to $(z_1,z_2)=(0,0), (0.5,-1.5), (-1.5,-2), (-2.5,3), (7,5), (-9,4.5), (8,-8)$.}
\label{15_HR_special_SO(2)xSO(2)_SO(5)gg_flows}
\end{figure}

\begin{figure}[h!]
  \centering
    \includegraphics[width=\linewidth]{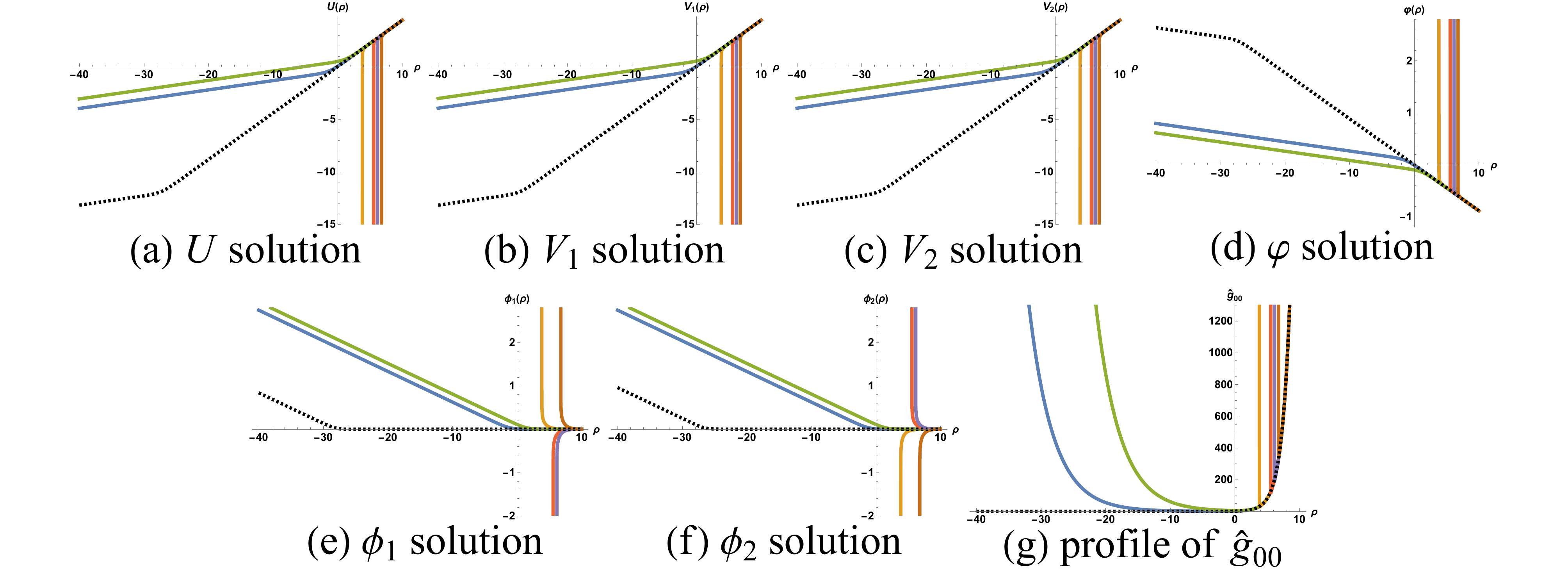}
\caption{Interpolating solutions between the locally $SO(5)$ flat domain wall as $\rho\rightarrow+\infty$ and $t\times \mathbb{R}^2\times \mathbb{R}^2$-sliced curved domain walls for $SO(2)\times SO(2)$ twist in $SO(5)$ gauge group. The blue, orange, green, red, purple, and brown curves refer to $(z_1,z_2)=(0.1,-0.1), (0.2,0), (0.8,-0.8), (-1,-1.5), (3,-3), (0,16)$. The dashed curve is the $SO(2)\times SO(2)$ flat domain wall with $z_1=z_2=0$.}
\label{15_RR_special_SO(2)xSO(2)_SO(5)gg_flows}
\end{figure}

\paragraph{Solutions from $SO(5)$, $SO(4,1)$, and $SO(3,2)$ gauge groups:}
We first consider semisimple gauge groups $SO(5)$, $SO(4,1)$, and $SO(3,2)$ with $\lambda\neq0$. We begin with solutions from $SO(5)$ gauge group that, as $\rho\rightarrow \infty$, are asymptotically flat domain walls with $SO(5)$ symmetry given by
\begin{equation}\label{SO(5)_flat_DW_asym}
U\sim V_1\sim V_2\sim\frac{5g \lambda \rho}{4\sqrt{2}},\qquad \varphi\sim-\frac{g \lambda \rho}{4\sqrt{2}},\qquad \phi_1\sim\phi_2\sim0
\end{equation}
for the new radial coordinate $\rho$ defined by $\frac{d\rho}{dr}=e^{\varphi+8(\phi_1+\phi_2)}$.
\\
\indent Before giving examples of numerical solutions, we note that the solutions with $t\times \Sigma^2_{k_2}\times \Sigma^2_{k_1}$ slices are equivalent to the solutions with $t\times \Sigma^2_{k_1}\times \Sigma^2_{k_2}$ slices with all subscripts $1$ and $2$ interchanged.
For the completely flat case with $k_1=0$ and $k_2=0$, the topological twist is not needed since there is no spin connection on $\mathbb{R}^2\times \mathbb{R}^2$ to be canceled. As a result, these solutions preserve eight supercharges due to the $\hat{\gamma}_r$ projector and the last projector in \eqref{SO(2)xSO(2)ProjCon} and \eqref{SO(2)xSO(2)ProjCon-}. Although the solutions take the form of a flat $t\times\mathbb{R}^2\times \mathbb{R}^2$-sliced domain wall, the gauge fields are non-vanishing with the magnetic charges related by $p_{11}=-\kappa p_{21}$ and $p_{12}=-\kappa p_{22}$. Therefore, we will also give numerical solutions in this case. For solutions with an $\mathbb{R}^2$ factor corresponding to either $k_1=0$ or $k_2=0$, the unbroken supersymmetry consists of four supercharges due to the two projectors on $\epsilon_\pm$ and the $\hat{\gamma}_r$ projector. Finally, for the case of $k_1\neq 0$ and $k_2\neq 0$, both $\epsilon_+$ and $\epsilon_-$ are subject to three independent projectors together with the $\hat{\gamma}_r$ projector relating the two chiralities, so the corresponding solutions preserve only two supercharges. 
\\
\indent To characterize the solutions, it is also useful to define the following two parameters 
\begin{equation}\label{two_z_define}
z_1=g(p_{11}-\kappa p_{21})\qquad\text{and}\qquad z_2=g(p_{12}-\kappa p_{22})\, .
\end{equation}
According to the twist conditions given in \eqref{15_SO(2)xSO(2)_twist_con}, the solutions with $z_1=z_2=0$ are special in the sense that $p_{11}=\kappa p_{21}=\frac{k_1}{2g}$ and $p_{12}=\kappa p_{22}=\frac{k_2}{2g}$. In particular, the solution with $z_1=z_2=0$ and $k_1=k_2=0$ corresponds to the untwisted flat domain wall solution preserving sixteen supercharges. Examples of numerical solutions asymptotic to the locally flat domain wall \eqref{SO(5)_flat_DW_asym} with $SO(5)$ symmetry are given in figures \ref{15_SS_special_SO(2)xSO(2)_SO(5)gg_flows} to \ref{15_RR_special_SO(2)xSO(2)_SO(5)gg_flows}. In these figures, we have chosen $g=0.5$ and also given the behavior of the type IIA metric component $\hat{g}_{00}$.

\begin{figure}[h!]
  \centering
    \includegraphics[width=0.94\linewidth]{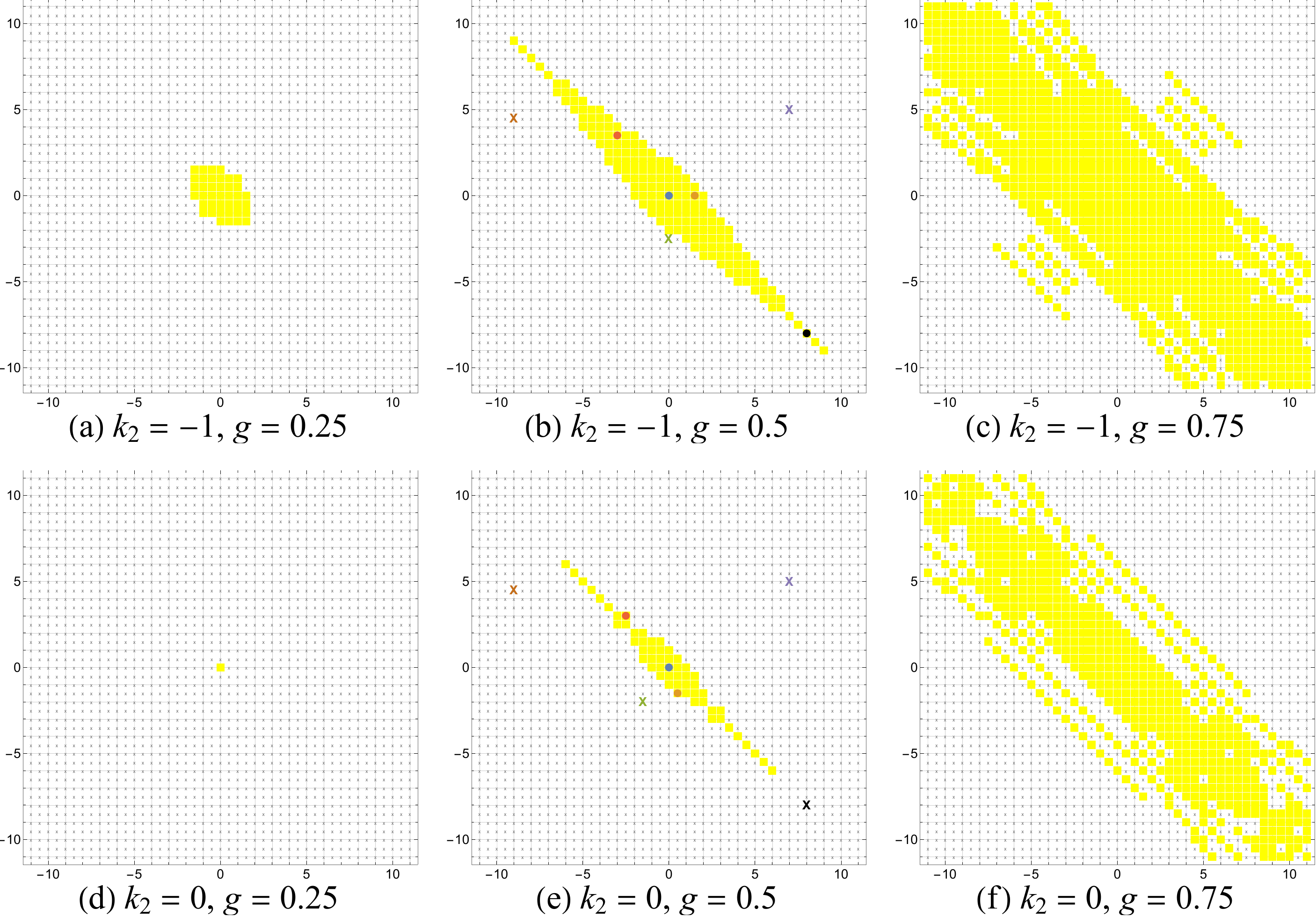}
\caption{Regions (yellow) in the parameter space ($z_1$,  $z_2$) where the physical IR singularities exist from $t\times H^2\times H^2$- and $t\times H^2\times \mathbb{R}^2$-sliced curved domain walls with $k_1=-1$ and $g=0.25, 0.50, 0.75$. In the middle figures, we also show the colored dots corresponding to numerical solutions with physical IR singularities given in figures \ref{15_HH_special_SO(2)xSO(2)_SO(5)gg_flows} and \ref{15_HR_special_SO(2)xSO(2)_SO(5)gg_flows} while the $\times$ symbol refer to solutions with $\hat{g}_{00}\rightarrow+\infty$ near the IR singularities.}
\label{15_SigSig_special_Regions_SO(5)gg_flows}
\end{figure}

From these figures, we readily see that only the solutions with $t\times H^2\times H^2$ slices in figure \ref{15_HH_special_SO(2)xSO(2)_SO(5)gg_flows} and solutions with $t\times H^2\times \mathbb{R}^2$ slices in figure \ref{15_HR_special_SO(2)xSO(2)_SO(5)gg_flows} are uplifted to type IIA solutions with physically acceptable IR singularities for specific values of $z_1$ and $z_2$. We further examine these cases and find regions in the parameter space $(z_1, z_2)$ for the IR singularities to be physically acceptable. The result is shown in figure \ref{15_SigSig_special_Regions_SO(5)gg_flows}, in which the physical regions (represented by the yellow areas) appear along the line $z_1+z_2\approx0$ for small $z_1$ and $z_2$ in all cases. We also point out that these regions expand as $g$ increases.
\vfil
\clearpage\newpage
\vfil
\vfil
\vfil
\clearpage\newpage
\vfil
\vfil
\vfil
\clearpage\newpage
\vfil
\vfil
\vfil
\clearpage\newpage
\vfil

\begin{figure}[h!]
  \centering
    \includegraphics[width=\linewidth]{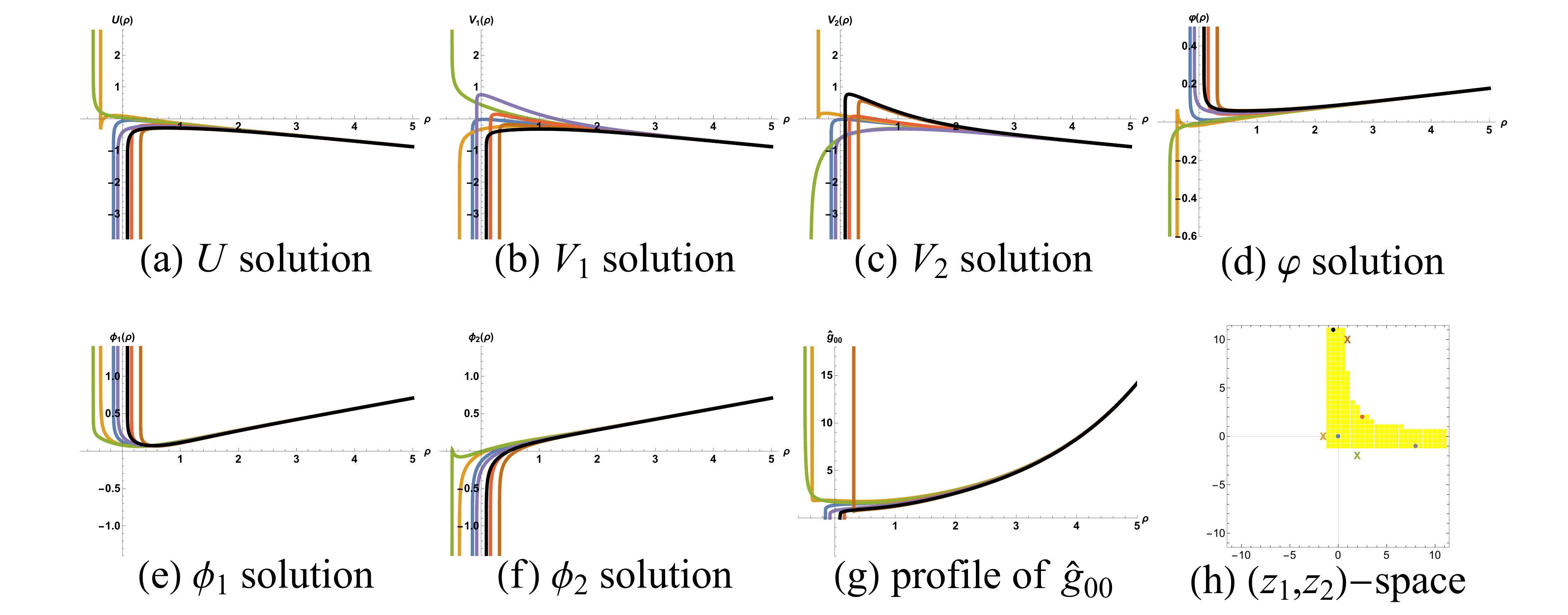}
\caption{Interpolating solutions between the locally $SO(2)\times SO(2)$ flat domain wall as $\rho\rightarrow+\infty$ and $t\times S^2\times S^2$-sliced curved domain walls for $SO(2)\times SO(2)$ twist in $SO(3,2)$ gauge group. The blue, orange, green, red, purple, brown, and black curves refer to $(z_1,z_2)=(0,0), (-1.5,0), (2,-2), (2.5,2), (8,-1), (1,10), (-0.5,11)$. The yellow region in ($z_1$,  $z_2$)-space shows the area of physical IR singularities.}
\label{15_SS_SO(2)xSO(2)_SO(32)gg_flows}
\end{figure}

\begin{figure}[h!]
  \centering
    \includegraphics[width=\linewidth]{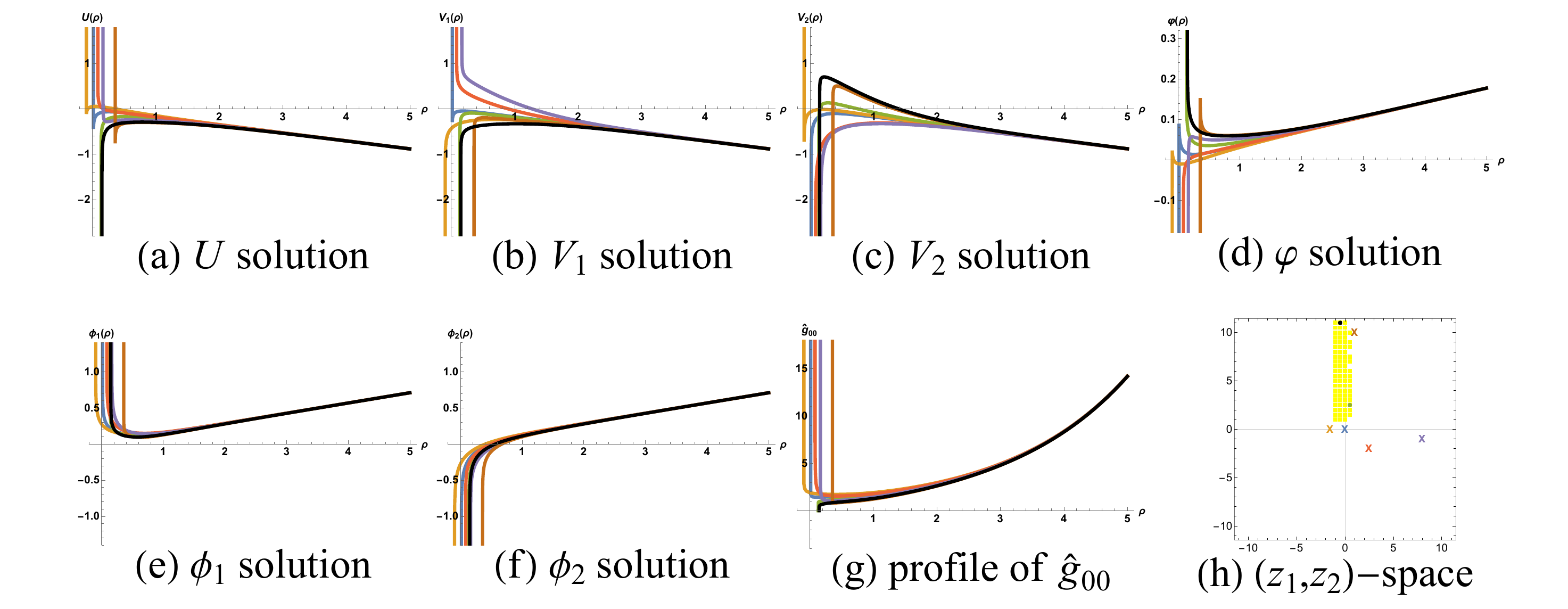}
\caption{Interpolating solutions between the locally $SO(2)\times SO(2)$ flat domain wall as $\rho\rightarrow+\infty$ and $t\times S^2\times H^2$-sliced curved domain walls for $SO(2)\times SO(2)$ twist in $SO(3,2)$ gauge group. The blue, orange, green, red, purple, brown, and black curves refer to $(z_1,z_2)=(0,0), (-1.5,0), (2,-2), (2.5,2), (8,-1), (1,10), (-0.5,11)$. The yellow region in ($z_1$,  $z_2$)-space shows the area of physical IR singularities.}
\label{15_SH_SO(2)xSO(2)_SO(32)gg_flows}
\end{figure}

\begin{figure}[h!]
  \centering
    \includegraphics[width=\linewidth]{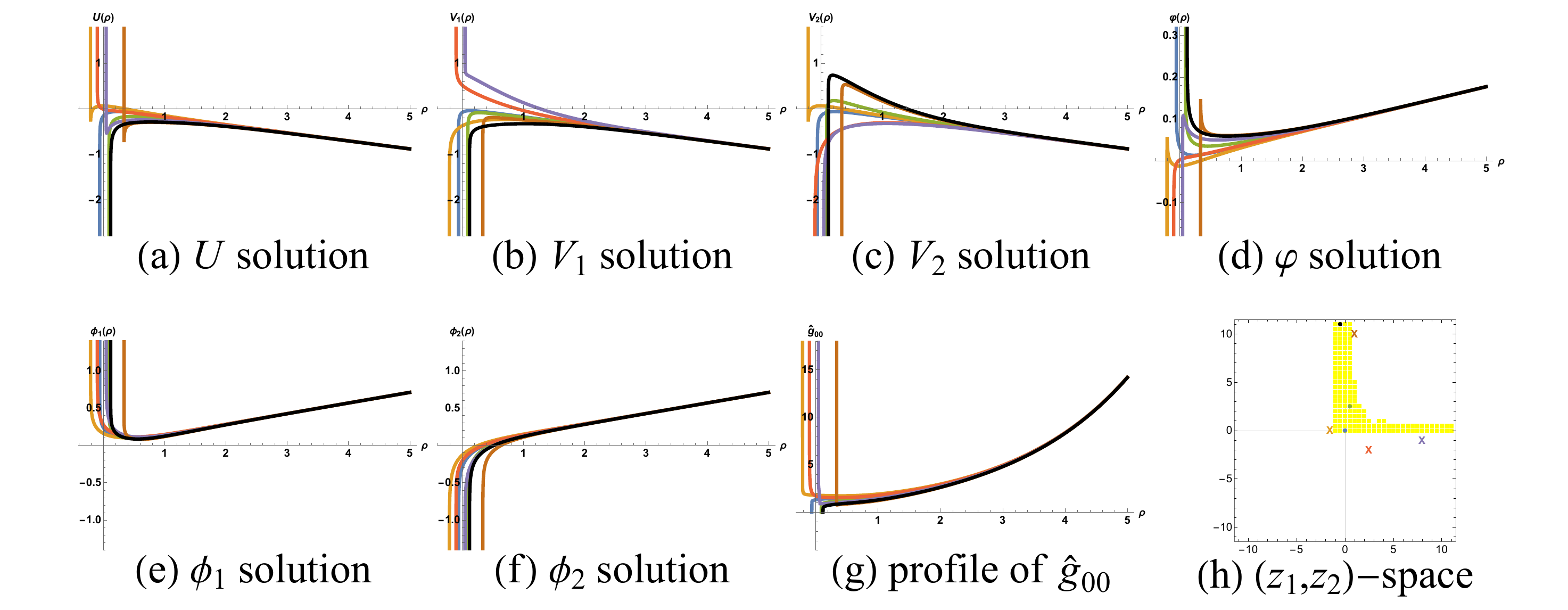}
\caption{Interpolating solutions between the locally $SO(2)\times SO(2)$ flat domain wall as $\rho\rightarrow+\infty$ and $t\times S^2\times \mathbb{R}^2$-sliced curved domain walls for $SO(2)\times SO(2)$ twist in $SO(3,2)$ gauge group. The blue, orange, green, red, purple, brown, and black curves refer to $(z_1,z_2)=(0,0), (-1.5,0), (2,-2), (2.5,2), (8,-1), (1,10), (-0.5,11)$. The yellow region in ($z_1$,  $z_2$)-space shows the area of physical IR singularities.}
\label{15_SR_SO(2)xSO(2)_SO(32)gg_flows}
\end{figure}

\begin{figure}[h!]
  \centering
    \includegraphics[width=\linewidth]{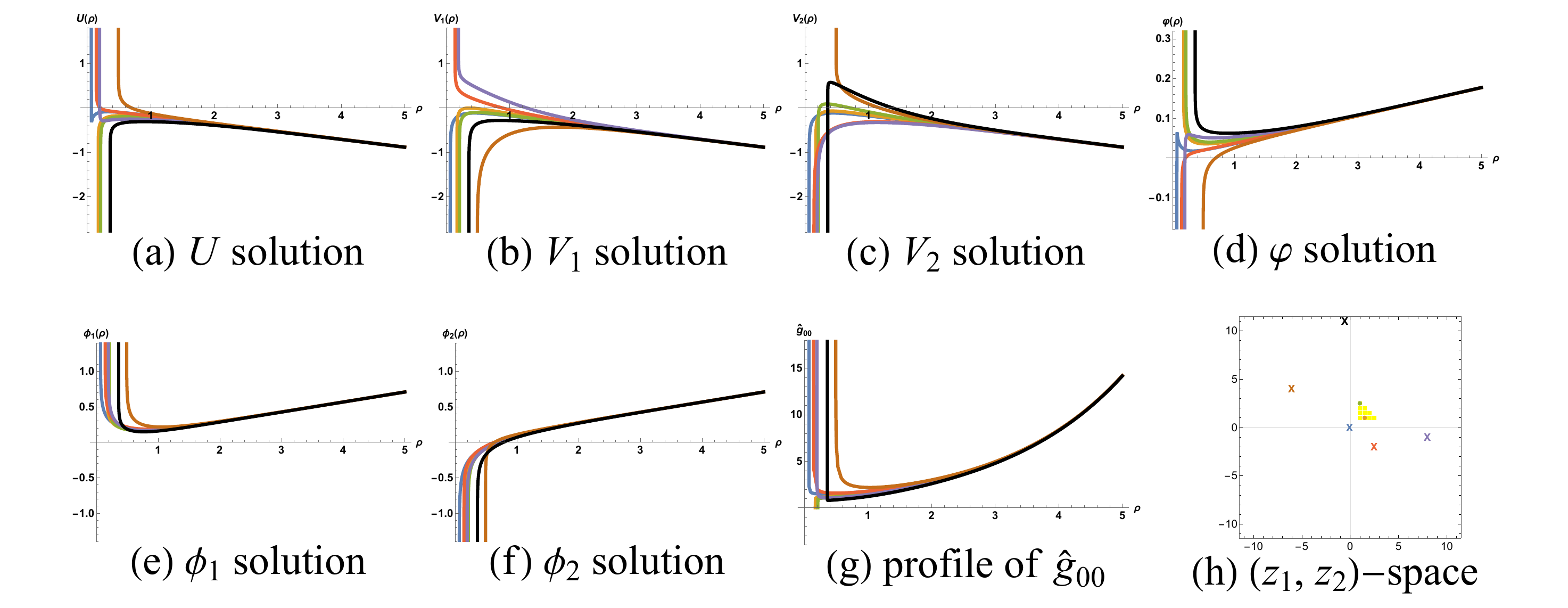}
\caption{Interpolating solutions between the locally $SO(2)\times SO(2)$ flat domain wall as $\rho\rightarrow+\infty$ and $t\times H^2\times H^2$-sliced curved domain walls for $SO(2)\times SO(2)$ twist in $SO(3,2)$ gauge group. The blue, orange, green, red, purple, brown, and black curves refer to $(z_1,z_2)=(0,0), (-1.5,0), (2,-2), (2.5,2), (8,-1), (1,10), (-0.5,11)$. The yellow region in ($z_1$,  $z_2$)-space shows the area of physical IR singularities.}
\label{15_HH_SO(2)xSO(2)_SO(32)gg_flows}
\end{figure}

\begin{figure}[h!]
  \centering
    \includegraphics[width=\linewidth]{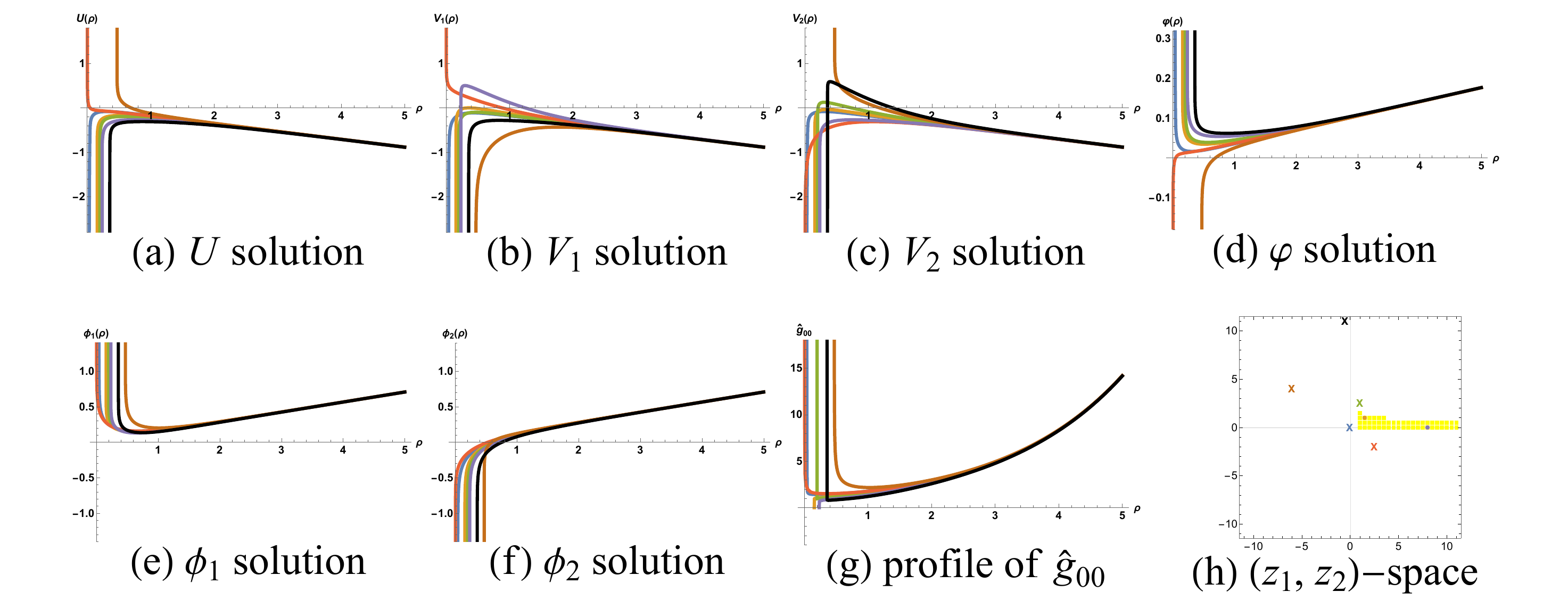}
\caption{Interpolating solutions between the locally $SO(2)\times SO(2)$ flat domain wall as $\rho\rightarrow+\infty$ and $t\times H^2\times \mathbb{R}^2$-sliced curved domain walls for $SO(2)\times SO(2)$ twist in $SO(3,2)$ gauge group. The blue, orange, green, red, purple, brown, and black curves refer to $(z_1,z_2)=(0,0), (-1.5,0), (2,-2), (2.5,2), (8,-1), (1,10), (-0.5,11)$. The yellow region in ($z_1$,  $z_2$)-space shows the area of physical IR singularities.}
\label{15_HR_SO(2)xSO(2)_SO(32)gg_flows}
\end{figure}

\begin{figure}[h!]
  \centering
    \includegraphics[width=\linewidth]{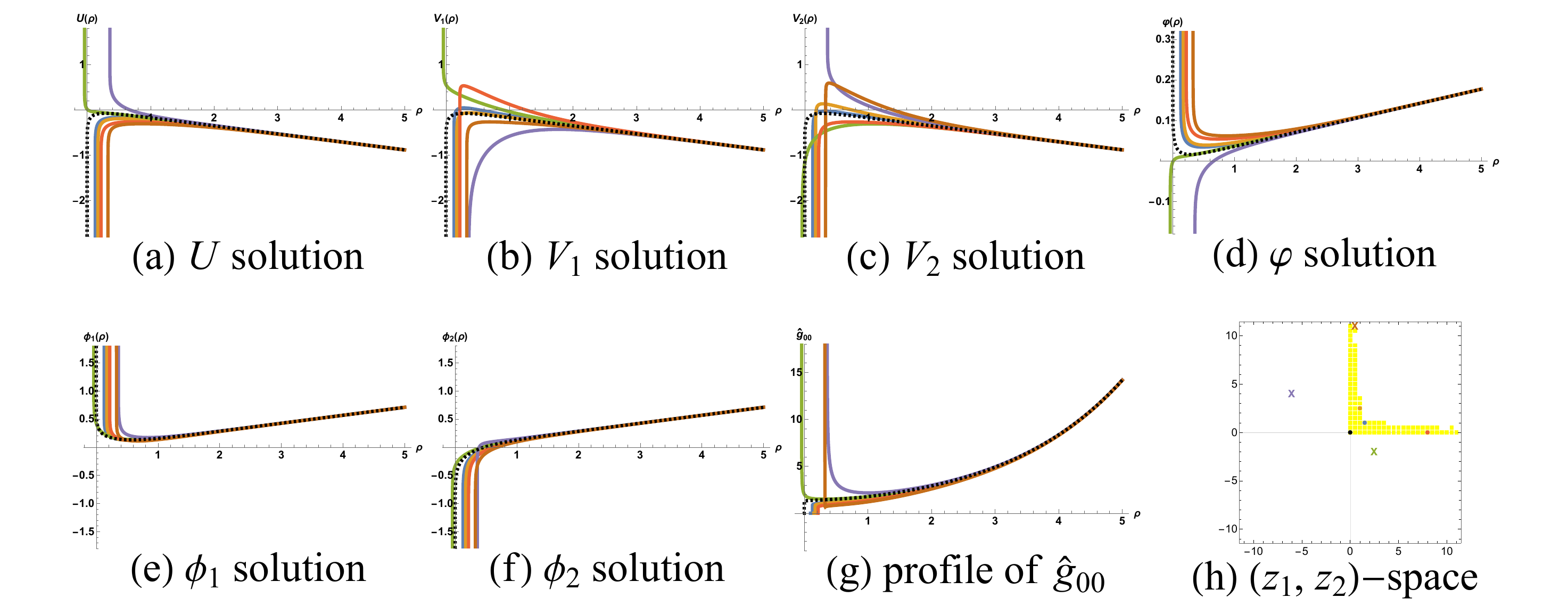}
\caption{Interpolating solutions between the locally $SO(2)\times SO(2)$ flat domain wall as $\rho\rightarrow+\infty$ and $t\times \mathbb{R}^2\times \mathbb{R}^2$-sliced curved domain walls for $SO(2)\times SO(2)$ twist in $SO(3,2)$ gauge group. The blue, orange, green, red, purple, and brown curves refer to $(z_1,z_2)=(-1.5,0), (2,-2), (2.5,2), (8,-1), (1,10), (-0.5,11)$. The dashed curve is the $SO(2)\times SO(2)$ flat domain wall with $z_1=z_2=0$. The yellow region in ($z_1$,  $z_2$)-space shows the area of physical IR singularities.}
\label{15_RR_SO(2)xSO(2)_SO(32)gg_flows}
\end{figure}

We now turn to solutions with an asymptotic behavior given by $SO(2)\times SO(2)$ symmetric flat domain wall of the form
\begin{equation}\label{SO(2)xSO(2)_flat_DW_asym}
U\sim V_1\sim V_2\sim\frac{g \lambda  \rho }{4\sqrt{2}},\qquad \varphi\sim-\frac{g \lambda  \rho }{20 \sqrt{2}},\qquad \phi_1\sim\phi_2\sim-\frac{g \lambda  \rho }{5 \sqrt{2}}\, .
\end{equation}
We have chosen $g\lambda <0$ in order to identify the UV limit with $\rho\rightarrow +\infty$. From the numerical analysis, it turns out that all of the solutions in $SO(5)$ gauge group are uplifted to type IIA solutions with unphysical IR singularities. For $SO(4,1)$ gauge group with $g=1$, only solutions with $t\times H^2\times \mathbb{R}^2$ and $t\times \mathbb{R}^2\times \mathbb{R}^2$ slices admit physical IR singularities for $-1\leq z_1\leq1$ and $z_2=0$, and $z_1=z_2=0$, respectively. The former corresponds to the $Mkw_3\times H^2$-sliced curved domain walls obtained from $SO(2)\times SO(2)$ twist on a Riemann surface previously studied in \cite{6D_twist_I} while the latter, with $z_1=z_2=k_1=k_2=0$, is the $SO(2)\times SO(2)$ symmetric flat domain wall. Accordingly, we will not give all of these solutions and those with unphysical IR singularities here.
\\
\indent For $SO(3,2)$ gauge group, examples of numerical solutions with $g=-1$ are given in figures \ref{15_SS_SO(2)xSO(2)_SO(32)gg_flows} to \ref{15_RR_SO(2)xSO(2)_SO(32)gg_flows}. In these figures, we have shown the regions in the parameter space ($z_1$,  $z_2$) where physical IR singularities exist in each case. Furthermore, we have also located solutions with physical IR singularities in the yellow regions by dots with the colors corresponding to the numerical curves of these solutions. As in the previous case, $\times$ marks that lie outside the yellow regions represent solutions with unphysical IR singularities. We end this section by giving a summary of all the solutions from $SO(5)$, $SO(4,1)$ and $SO(3,2)$ gauge groups in table \ref{tab2}. 

\begin{table}[h!]
\centering
\begin{tabular}{| c | c | c | c |}
\hline
UV flat & gauge & two Riemann & physical IR \\
domain wall& group & surfaces &  singularity \\\hline
$SO(5)$&$SO(5)$&$S^2\times\Sigma^2$ & $\times$ \\
&&$H^2\times H^2$ & figure \ref{15_SigSig_special_Regions_SO(5)gg_flows}(a,b,c) \\
&&$H^2\times \mathbb{R}^2$ & figure \ref{15_SigSig_special_Regions_SO(5)gg_flows}(d,e,f) \\
&&$\mathbb{R}^2\times\mathbb{R}^2$ &  $z_1=z_2=0$\\\hline
$SO(2)\times SO(2)$&$SO(5)$& $\Sigma^2\times\Sigma^2$ & $\times$\\\cline{2-4}
&$SO(4,1)$& $S^2\times\Sigma^2$, $H^2\times H^2$ & $\times$\\
&& $H^2\times\mathbb{R}^2$ & $-1\leq z_1\leq1$, $z_2=0$\\
&& $\mathbb{R}^2\times \mathbb{R}^2$ & $z_1=z_2=0$\\\cline{2-4}
&$SO(3,2)$ & $S^2\times S^2$ & figure \ref{15_SS_SO(2)xSO(2)_SO(32)gg_flows}(h)\\
& & $S^2\times H^2$ & figure \ref{15_SH_SO(2)xSO(2)_SO(32)gg_flows}(h)\\
& & $S^2\times \mathbb{R}^2$ & figure \ref{15_SR_SO(2)xSO(2)_SO(32)gg_flows}(h)\\
& & $H^2\times H^2$ & figure \ref{15_HH_SO(2)xSO(2)_SO(32)gg_flows}(h)\\
& & $H^2\times \mathbb{R}^2$ & figure \ref{15_HR_SO(2)xSO(2)_SO(32)gg_flows}(h)\\
& & $\mathbb{R}^2\times \mathbb{R}^2$ & figure \ref{15_RR_SO(2)xSO(2)_SO(32)gg_flows}(h)\\\hline
\end{tabular}
\caption{Summary of satisfaction of the criterion \cite{Maldacena_nogo} for the IR singularities of the solutions with $t\times \Sigma_{k_1}^2\times \Sigma_{k_2}^2$ slices obtained from $SO(2)\times SO(2)$ twist with $\varsigma_1=\varsigma_2=0$ in $SO(5)$, $SO(4,1)$, and $SO(3,2)$ gauge groups.}\label{tab2}
\end{table}
\pagebreak
\vfil
\vfil
\vfil
\pagebreak
\vfil
\vfil
\vfil
\pagebreak
\vfil
\vfil

\begin{figure}[h!]
  \centering
    \includegraphics[width=\linewidth]{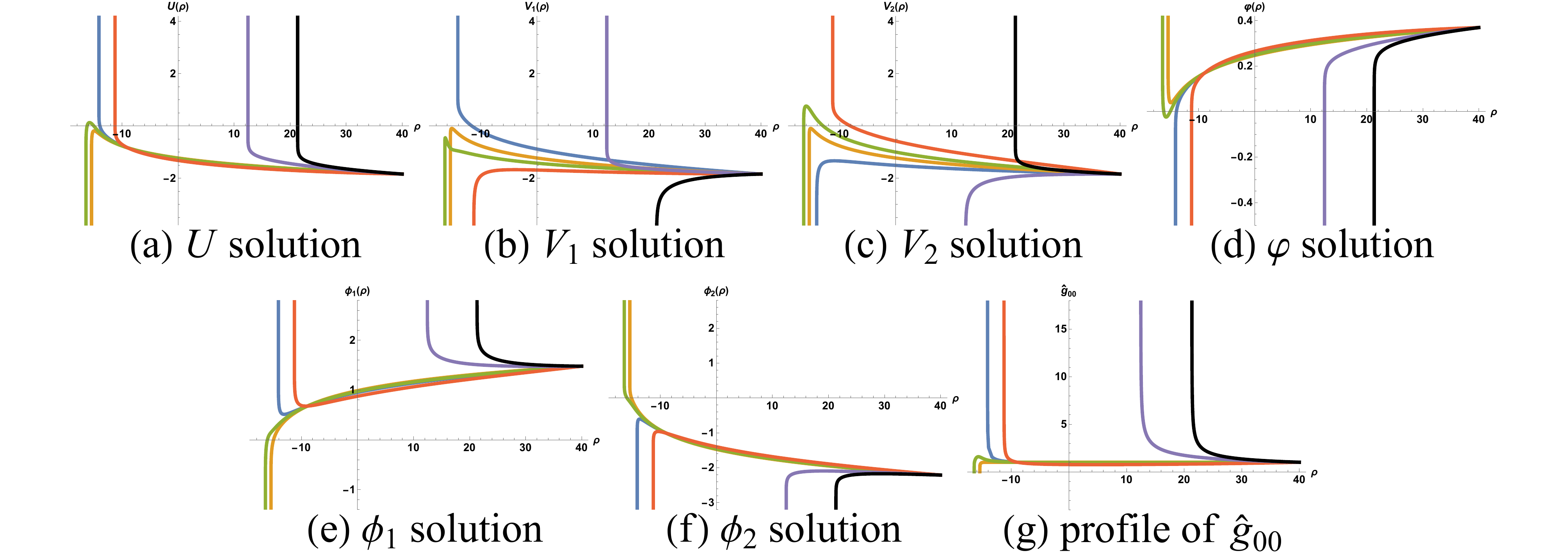}
\caption{Numerical solutions interpolating between the locally $SO(2)\times SO(2)$ flat domain wall as $\rho\rightarrow+\infty$ and $t\times S^2\times S^2$-sliced curved domain walls for $SO(2)\times SO(2)$ twist in $CSO(2,2,1)$ gauge group. The blue, orange, green, red, purple, and black curves, respectively, refer to $z_1=-2.80, -1, -0.50, -0.15, 0.50, 4$.}
\label{15_SS_SO(2)xSO(2)_CSO(221)gg_flows}
\end{figure}

\begin{figure}[h!]
  \centering
    \includegraphics[width=\linewidth]{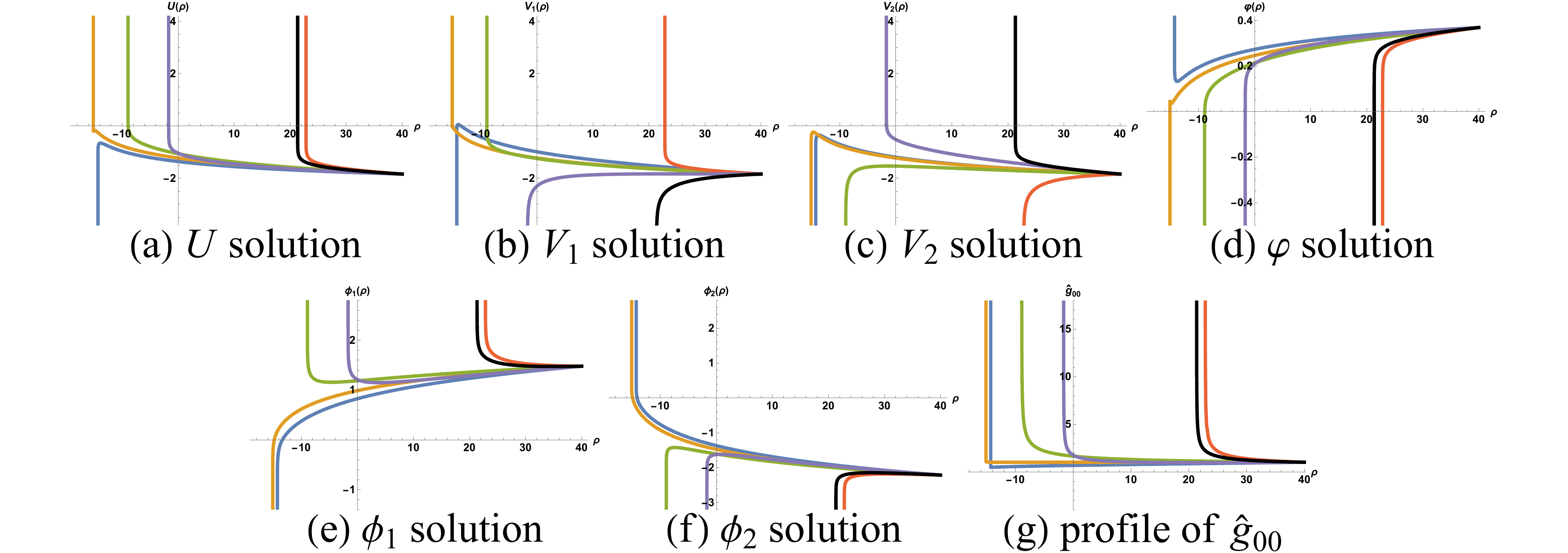}
\caption{Interpolating solutions between the locally $SO(2)\times SO(2)$ flat domain wall as $\rho\rightarrow+\infty$ and $t\times S^2\times H^2$-sliced curved domain walls for $SO(2)\times SO(2)$ twist in $CSO(2,2,1)$ gauge group. The blue, orange, green, red, purple, and black curves, respectively, refer to $z_1=-2.80, -1, -0.50, -0.15, 0.50, 4$.}
\label{15_SH_SO(2)xSO(2)_CSO(221)gg_flows}
\end{figure}

\begin{figure}[h!]
  \centering
    \includegraphics[width=\linewidth]{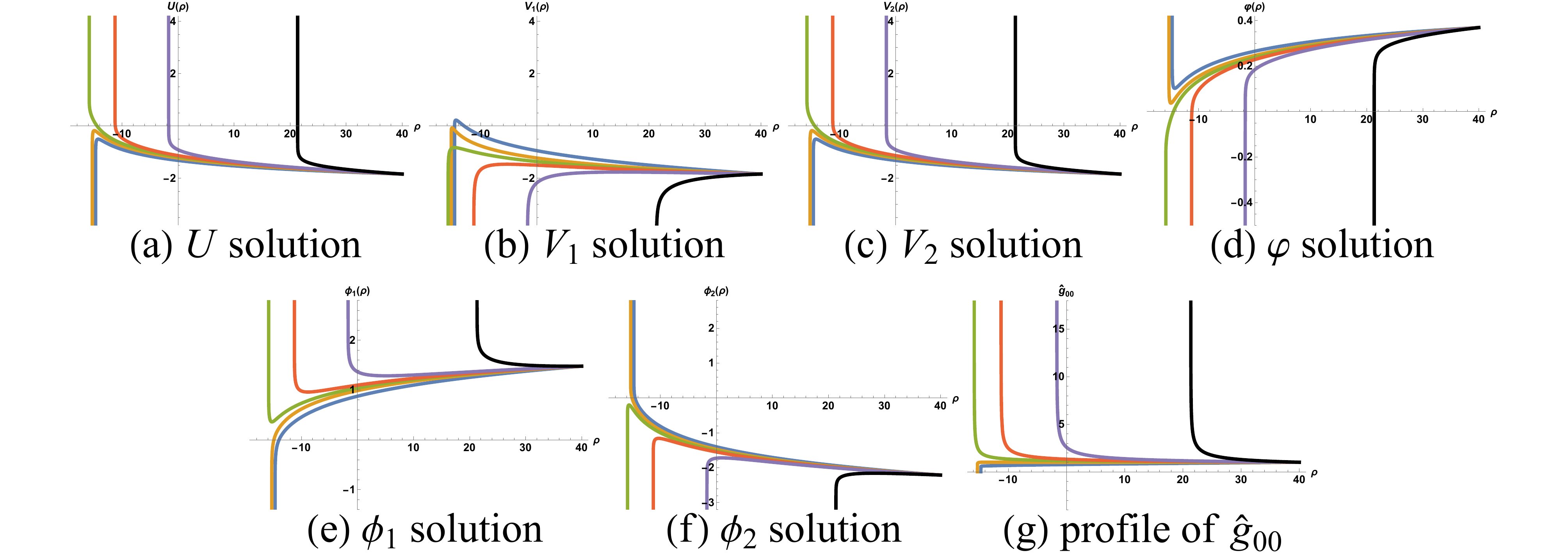}
\caption{Interpolating solutions between the locally $SO(2)\times SO(2)$ flat domain wall as $\rho\rightarrow+\infty$ and $t\times S^2\times \mathbb{R}^2$-sliced curved domain walls for $SO(2)\times SO(2)$ twist in $CSO(2,2,1)$ gauge group. The blue, orange, green, red, purple, and black curves, respectively, refer to $z_1=-2.80, -1, -0.50, -0.15, 0.50, 4$.}
\label{15_SR_SO(2)xSO(2)_CSO(221)gg_flows}
\end{figure}

\begin{figure}[h!]
  \centering
    \includegraphics[width=\linewidth]{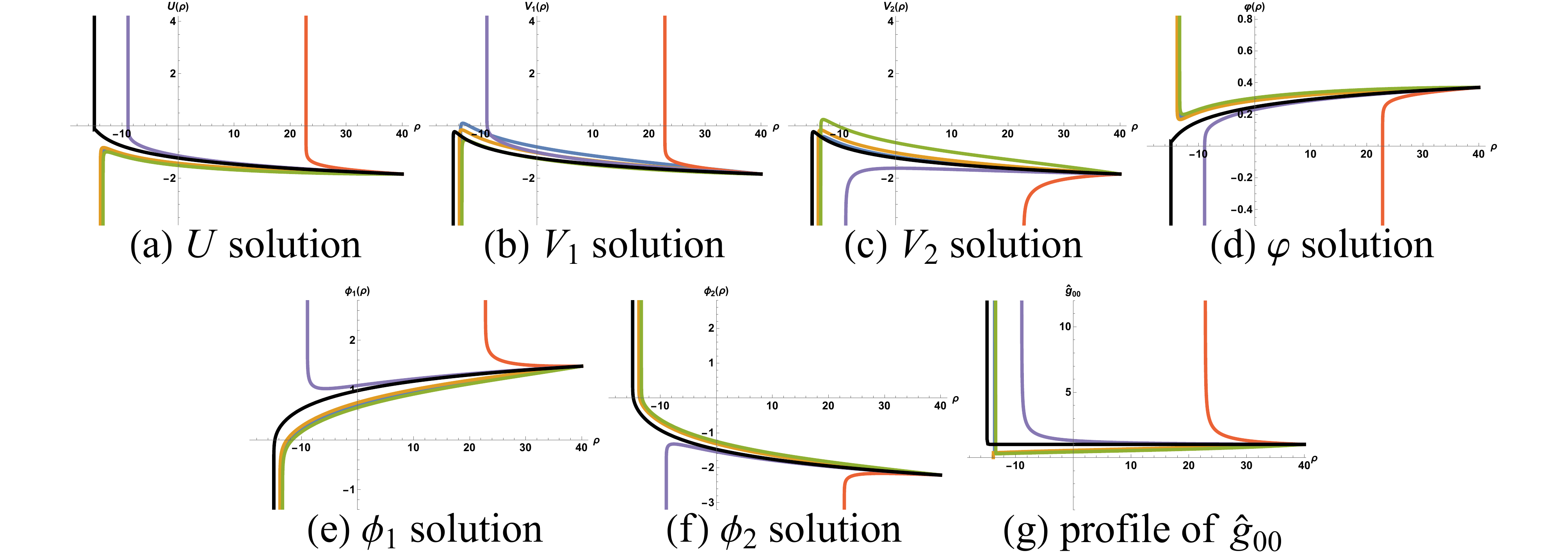}
\caption{Interpolating solutions between the locally $SO(2)\times SO(2)$ flat domain wall as $\rho\rightarrow+\infty$ and $t\times H^2\times H^2$-sliced curved domain walls for $SO(2)\times SO(2)$ twist in $CSO(2,2,1)$ gauge group. The blue, orange, green, red, purple, and black curves, respectively, refer to $z_1=-3.50, -1, -0.15, 0.15, 0.50, 1$.}
\label{15_HH_SO(2)xSO(2)_CSO(221)gg_flows}
\end{figure}

\begin{figure}[h!]
  \centering
    \includegraphics[width=\linewidth]{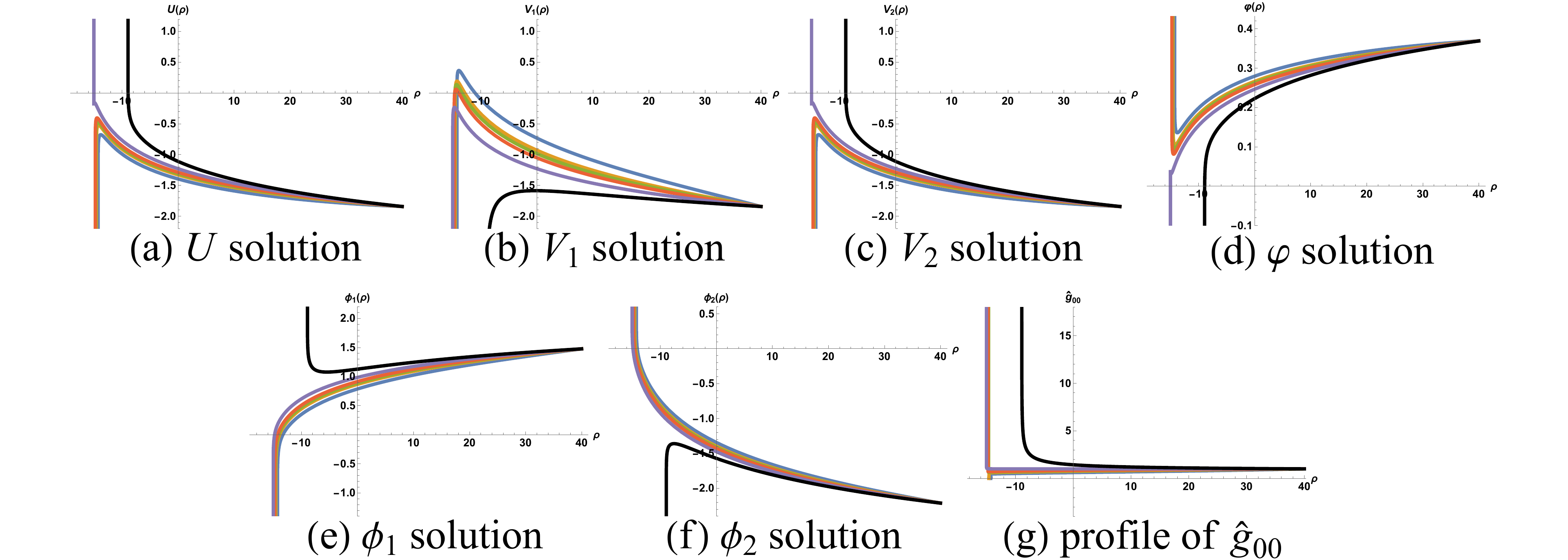}
\caption{Interpolating solutions between the locally $SO(2)\times SO(2)$ flat domain wall as $\rho\rightarrow+\infty$ and $t\times H^2\times \mathbb{R}^2$-sliced curved domain walls for $SO(2)\times SO(2)$ twist in $CSO(2,2,1)$ gauge group. The blue, orange, green, red, purple, and black curves, respectively, refer to $z_1=-3.50, -1, -0.50, 0.01, 1, 2$.}
\label{15_HR_SO(2)xSO(2)_CSO(221)gg_flows}
\end{figure}

\begin{figure}[h!]
  \centering
    \includegraphics[width=\linewidth]{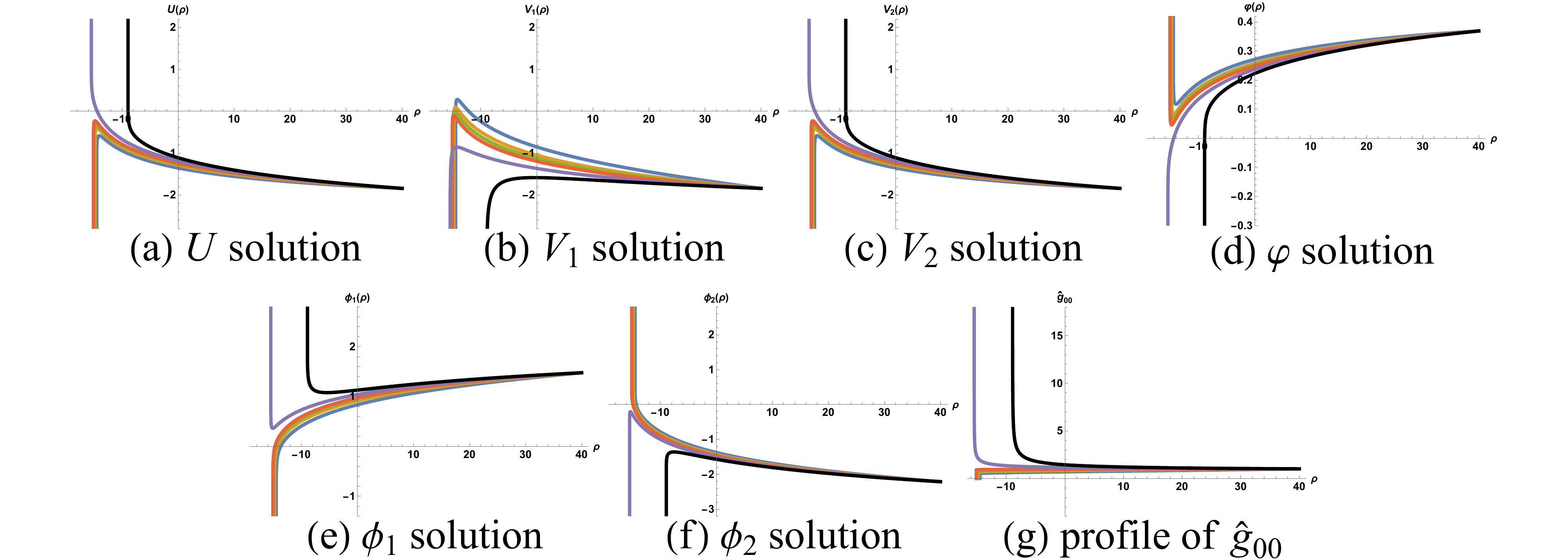}
\caption{Interpolating solutions between the locally $SO(2)\times SO(2)$ flat domain wall as $\rho\rightarrow+\infty$ and $t\times \mathbb{R}^2\times \mathbb{R}^2$-sliced curved domain walls for $SO(2)\times SO(2)$ twist in $CSO(2,2,1)$ gauge group. The blue, orange, green, red, purple, and black curves, respectively, refer to $z_1=-2.80, -1, -0.50, -0.15, 0.50, 1$.}
\label{15_RR_SO(2)xSO(2)_CSO(221)gg_flows}
\end{figure}

\paragraph{Solutions from $CSO(4,0,1)$ and $CSO(2,2,1)$ gauge groups:}
We now consider the case of non-semisimple gauge groups with $\lambda=0$. In this case, we define a new radial coordinate $\rho$ by the relation $\frac{d\rho}{dr}=e^{-\frac{5}{2}\phi_1-\frac{15}{2}\phi_2}$. As $\rho\rightarrow+\infty$, we require that the solutions are asymptotic to the $SO(2)\times SO(2)$ symmetric flat domain wall
\begin{eqnarray}
U&\sim&V_1\sim V_2\sim\frac{3}{4}\ln \rho-\frac{5}{4}\ln(-g\kappa\rho) ,\qquad \varphi\sim\frac{1}{4}\ln(-g\kappa\rho)-\frac{3}{20}\ln \rho,\nonumber\\\phi_1&\sim&\frac{9}{10}\ln \rho-\frac{1}{2}\ln(-g\kappa\rho),\qquad\qquad\qquad \phi_2\sim-\frac{3}{5}\ln\rho\, .\label{15_SO(2)xSO(2)_fDW_lam0}
\end{eqnarray}
All numerical solutions in $CSO(4,0,1)$ gauge group lead to unphysical IR singularities, so we will omit these solutions here. 
\\
\indent Examples of numerical solutions from $CSO(2,2,1)$ gauge group are given in figures \ref{15_SS_SO(2)xSO(2)_CSO(221)gg_flows} to \ref{15_RR_SO(2)xSO(2)_CSO(221)gg_flows}. We also note that, in this case, the parameters $z_1$ and $z_2$ are not independent from each other due to the constraint $p_{11}p_{22}+p_{12}p_{21}=0$ required by the Bianchi identity \eqref{DefBianchi2} for $\lambda=0$. This leads to the condition 
\begin{equation}
z_1z_2=k_1k_2\, .
\end{equation}
In particular, solutions with $z_1=0$ for $k_1k_2\neq0$ do not exist. In the figures, we have set $g=-\kappa=1$ and given the solutions for different values of $z_1$. We collect all possible values of $z_1$ giving rise to type IIA solutions with physical IR singularities in table \ref{tab3}. 
\vfil
\vfil\pagebreak
\vfil
\vfil
\vfil\pagebreak
\vfil
\vfil
\clearpage\newpage

\begin{table}[h!]
\centering
\begin{tabular}{| c | c | c |}
\hline
gauge & two Riemann & physical IR \\
group & surfaces & singularity \\\hline
$CSO(4,0,1)$& $\Sigma^2\times\Sigma^2$ & $\times$\\\hline
$CSO(2,2,1)$& $S^2\times S^2$ & $-1\leq z_1\leq-0.5$\\
& $S^2\times H^2$ & $\times$\\
& $S^2\times \mathbb{R}^2$, $H^2\times \mathbb{R}^2$ & $z_1\leq-1$\\
& $H^2\times H^2$ & $z_1=-1$\\
& $\mathbb{R}^2\times \mathbb{R}^2$ & $z_1\leq-0.15$\\\hline
\end{tabular}
\caption{Summary of satisfaction of the criterion \cite{Maldacena_nogo} for the IR singularities of the solutions interpolating between the locally $SO(2)\times SO(2)$ flat domain wall and  $t\times \Sigma_{k_1}^2\times \Sigma_{k_2}^2$-sliced curved domain walls obtained from $SO(2)\times SO(2)$ twist with $\varsigma_1=\varsigma_2=0$ in $CSO(4,0,1)$ and $CSO(2,2,1)$ gauge groups.}\label{tab3}
\end{table}

\subsubsection{Solutions with non-vanishing $\varsigma_1$ and $\varsigma_2$}
We now repeat a similar analysis for solutions with non-vanishing shift scalars. In order to keep $\varsigma_1$ and $\varsigma_2$ non-vanishing, we need to impose the following conditions on the magnetic charges 
\begin{equation}\label{Sig2xSig2_SO(2)diag_con}
p_{21}=\kappa p_{11}\qquad\text{ and }\qquad p_{22}=\kappa p_{12}
\end{equation}
together with an algebraic constraint
\begin{equation}\label{15_Sig2xSig2_keep_shift_con}
\kappa e^{4\phi_2}\varsigma_1+e^{4\phi_1}\varsigma_2=0\, .
\end{equation}
\indent With all these and the magnetic two-form potential \eqref{Sig2xSig2_mag_two_pot} as well as the twist conditions
\begin{equation}\label{Sig2xSig2_SO(2)diag_twistcon}
2gp_{11}=k_1\qquad \textrm{and} \qquad 2gp_{12}=k_2,
\end{equation}
we find the following BPS equations 
\begin{eqnarray}
U'&=&\frac{ge^{\varphi}}{4\sqrt{2}}(2e^{-4\phi_1}+2\kappa e^{-4\phi_2}+\lambda e^{8(\phi_1+\phi_2)})+\frac{3\kappa\lambda p_{11}p_{12}}{8 \sqrt{2}ge^{2V_1+2V_2+3\varphi+4\phi_1+4\phi_2}}\nonumber\\
&&-\frac{e^{-\varphi}}{8\sqrt{2}}(p_{11}e^{-2V_1}+p_{12}e^{-2V_2})(e^{4\phi_1}+\kappa e^{4\phi_2}),\\
V'_1&=&\frac{ge^{\varphi}}{4\sqrt{2}}(2e^{-4\phi_1}+2\kappa e^{-4\phi_2}+\lambda e^{8(\phi_1+\phi_2)})-\frac{\kappa\lambda p_{11}p_{12}}{8 \sqrt{2}ge^{2V_1+2V_2+3\varphi+4\phi_1+4\phi_2}}\nonumber\\
&&+\frac{e^{-\varphi}}{8\sqrt{2}}(3p_{11}e^{-2V_1}-p_{12}e^{-2V_2})(e^{4\phi_1}+\kappa e^{4\phi_2}),\\
V'_2&=&\frac{ge^{\varphi}}{4\sqrt{2}}(2e^{-4\phi_1}+2\kappa e^{-4\phi_2}+\lambda e^{8(\phi_1+\phi_2)})-\frac{\kappa\lambda p_{11}p_{12}}{8 \sqrt{2}ge^{2V_1+2V_2+3\varphi+4\phi_1+4\phi_2}}\nonumber\\
&&-\frac{e^{-\varphi}}{8\sqrt{2}}(p_{11}e^{-2V_1}-3p_{12}e^{-2V_2})(e^{4\phi_1}+\kappa e^{4\phi_2}),\\
\varphi'&=&-\frac{ge^{\varphi}}{20\sqrt{2}}(2e^{-4\phi_1}+2\kappa e^{-4\phi_2}+\lambda e^{8(\phi_1+\phi_2)})-\frac{3\kappa\lambda p_{11}p_{12}}{40 \sqrt{2}ge^{2V_1+2V_2+3\varphi+4\phi_1+4\phi_2}}\nonumber\\
&&+\frac{e^{-\varphi}}{40\sqrt{2}}(p_{11}e^{-2V_1}+p_{12}e^{-2V_2})(e^{4\phi_1}+\kappa e^{4\phi_2}),\\
\phi'_1&=&\frac{ge^{\varphi}}{5\sqrt{2}}(3e^{-4\phi_1}-2\kappa e^{-4\phi_2}-\lambda e^{8(\phi_1+\phi_2)})-\frac{\kappa\lambda p_{11}p_{12}}{20 \sqrt{2}ge^{2V_1+2V_2+3\varphi+4\phi_1+4\phi_2}}\nonumber\\
&&-\frac{e^{-\varphi}}{20\sqrt{2}}(p_{11}e^{-2V_1}+p_{12}e^{-2V_2})(3e^{4\phi_1}-2\kappa e^{4\phi_2}),\\
\phi'_2&=&\frac{ge^{\varphi}}{5\sqrt{2}}(3\kappa e^{-4\phi_2}-2e^{-4\phi_1}-\lambda e^{8(\phi_1+\phi_2)})-\frac{\kappa\lambda p_{11}p_{12}}{20 \sqrt{2}ge^{2V_1+2V_2+3\varphi+4\phi_1+4\phi_2}}\nonumber\\
&&+\frac{e^{-\varphi}}{20\sqrt{2}}(p_{11}e^{-2V_1}+p_{12}e^{-2V_2})(2e^{4\phi_1}-3\kappa e^{4\phi_2}),\\
\varsigma'_1&=&\frac{ge^{\varphi}}{\sqrt{2}}(2e^{-4\phi_1}-2\kappa e^{-4\phi_2}-\lambda e^{8(\phi_1+\phi_2)})\varsigma_1-\frac{\kappa\lambda p_{11}p_{12}\varsigma_1}{2 \sqrt{2}ge^{2V_1+2V_2+3\varphi+4\phi_1+4\phi_2}}\nonumber\\
&&-\frac{e^{-\varphi}}{2\sqrt{2}}(p_{11}e^{-2V_1}+p_{12}e^{-2V_2})(e^{4\phi_1}-\kappa e^{4\phi_2})\varsigma_1\, .\qquad\quad
\end{eqnarray}
It should be pointed out that these are the same as those given in \eqref{Sig2xSig2SO(2)xSO(2)BPS1} to \eqref{Sig2xSig2SO(2)xSO(2)BPS6} subject to the conditions \eqref{Sig2xSig2_SO(2)diag_con} together with an additional equation for $\varsigma_1$. 
\\
\indent For non-vanishing $\varsigma_1$ and $\varsigma_2$, the Killing spinors generally take the form 
\begin{equation}
\epsilon_+=e^{\frac{1}{2}(U+W\gamma_{12})}\epsilon^0_+\qquad \textrm{and}\qquad \epsilon_-=e^{\frac{1}{2}(U+W\gamma_{\dot{1}\dot{2}})}\epsilon^0_-
\end{equation}
with the $r$-dependent function $W$ determined by the composite connection via
\begin{equation}\label{15_Sig2_W_Im}
W'={Q_r}^{12}+{Q_r}^{34}=\varsigma'_1+\varsigma'_2-4\varphi'(\varsigma_1+\varsigma_2)-4 \varsigma_1\phi'_1-4 \varsigma_2\phi'_2\, .
\end{equation}
However, it turns out that the BPS equations given above imply that $W'=0$. Without loss of generality, we can choose the constant $W$ to be zero leading to the Killing spinors of the form \eqref{DW_Killing_spinor}. 
\\
\indent For $SO(5)$, $SO(4,1)$, and $SO(3,2)$ gauge groups with $\lambda\neq 0$, these BPS equations are compatible with all the bosonic field equations. On the other hand, for non-semisimple gauge groups $CSO(4,0,1)$ and $CSO(2,2,1)$ with $\lambda=0$, the four-form field strengths given in \eqref{Sig2xSig2_H4} do not contribute to the Bianchi identity \eqref{DefBianchi2}. Accordingly, the contribution of $SO(2)\times SO(2)$ gauge field strengths to the Bianchi identity \eqref{DefBianchi2} must vanish identically. This leads to the following condition 
\begin{equation}
(\Gamma_m)_{AB}\mathbb{T}^{A}_{np}\mathbb{T}^{B}_{qr}{\mathcal{H}_{\hat{\theta}_1\hat{\zeta}_1}}^{np}{\mathcal{H}_{\hat{\theta}_2\hat{\zeta}_2}}^{qr}=\frac{\kappa}{6\sqrt{2}}p_{11}p_{12}e^{-2 V_1-2 V_2}\delta_m^5=0\, .
\end{equation}
However, this implies that either $p_{11}=0$ or $p_{12}=0$ incompatible with the twist conditions for $k_1\neq 0$ and $k_2\neq 0$. Accordingly, we will not consider solutions from non-semisimple gauge groups $CSO(4,0,1)$ and $CSO(2,2,1)$ in the following analysis. 
\\
\indent Similar to the previous case, we will look for numerical solutions with asymptotic behaviors given by flat domain walls. For $SO(5)$ gauge group, all solutions asymptotic to $SO(5)$ symmetric domain wall in \eqref{SO(5)_flat_DW_asym} have $\varsigma_1=0$ and $\varsigma_2=0$ along the entire flow solutions. Moreover, imposing the conditions \eqref{Sig2xSig2_SO(2)diag_con} and the twist conditions \eqref{Sig2xSig2_SO(2)diag_twistcon} is equivalent to setting $z_1=z_2=0$ in the previous case. Therefore, the solutions are similar to those already found in the previous section. Accordingly, we will not give the numerical solutions for this case to avoid repetition. 
\\
\indent We now consider solutions with the asymptotic behavior given by $SO(2)\times SO(2)$ symmetric flat domain wall of the form
\begin{equation}\label{SO(2)d_flat_DW_asym}
U\sim V_1\sim V_2\sim\frac{g \lambda  \rho }{4\sqrt{2}},\qquad \varphi\sim-\frac{g \lambda  \rho }{20 \sqrt{2}},\qquad \phi_1\sim\phi_2\sim-\frac{g \lambda  \rho }{5 \sqrt{2}},\qquad\varsigma_1\sim e^{-\frac{g \lambda  \rho }{\sqrt{2}}}
\end{equation}
with the new radial coordinate $\rho$ defined by $\frac{d\rho}{dr}=e^{\varphi+8(\phi_1+\phi_2)}$. We will choose $g\lambda<0$ in order to identify the UV limit with $\rho\rightarrow+\infty$. 
\\
\indent As in the previous case with $\varsigma_1=\varsigma_2=0$, all of the solutions in $SO(5)$ gauge group are uplifted to type IIA solutions with unphysical IR singularities, so we also neglect these solutions. For $SO(4,1)$ and $SO(3,2)$ gauge groups, examples of numerical solutions are given in figures \ref{15_SS_SO(2)d_SO(41)gg_flows} to \ref{15_HR_SO(2)d_SO(32)gg_flows} for different values of $g$ in Appendix \ref{AppC_1_1}. We also recall that, in this case, all the four magnetic charges have been fixed in terms of the gauge coupling constant by the four conditions given in \eqref{Sig2xSig2_SO(2)diag_con} and \eqref{Sig2xSig2_SO(2)diag_twistcon}. In addition, we have not given $t\times \mathbb{R}^2\times \mathbb{R}^2$-sliced solutions since, in these solutions, we have $k_1=k_2=0$ and $p_{11}=p_{12}=p_{21}=p_{22}=0$. Therefore, the solutions are essentially flat domain walls. From the figures, we find that all solutions from $SO(3,2)$ gauge group lead to physical IR singularities for any values of $k_1$ and $k_2$. On the other hand, for $SO(4,1)$ gauge group, only solutions with an $\mathbb{R}^2$ factor, $k_1=0$ or $k_2=0$, admit physical IR singularities. We collect all possible values of the gauge coupling constant $g$ giving rise to type IIA solutions with physical IR singularities in table \ref{tab4}. 
\begin{table}[h!]
\centering
\begin{tabular}{| c | c | c |}
\hline
gauge & \multirow{2}{*}{\centering two Riemann surfaces} & physical IR \\
group &  &  singularity \\\hline
$SO(5)$& $\Sigma^2\times \Sigma^2$ & $\times$\\\hline
$SO(4,1)$ & $S^2\times S^2$, $S^2\times H^2$, $H^2\times H^2$& $\times$\\
 & $S^2\times\mathbb{R}^2$& $g\geq2$\\
 & $H^2\times\mathbb{R}^2$& any $g$\\\hline
$SO(3,2)$ & $S^2\times S^2$, $S^2\times \mathbb{R}^2$ & $g\leq-0.6$\\
& $S^2\times H^2$, $H^2\times H^2$, $H^2\times \mathbb{R}^2$ & $g\leq-2$\\\hline
\end{tabular}
\caption{Summary of satisfaction of the criterion \cite{Maldacena_nogo} for the IR singularities of the solutions interpolating between the locally $SO(2)\times SO(2)$ flat domain wall and  $t\times \Sigma_{k_1}^2\times \Sigma_{k_2}^2$-sliced curved domain walls with shift scalars in $SO(5)$, $SO(4,1)$, and $SO(3,2)$ gauge groups.}\label{tab4}
\end{table}

%%%%%%%%%%%%%%%%%%%%%%%%%%%%%%%%%%%%%%%%%%%%%%%%
\subsection{D4-branes wrapped on a Kahler four-cycle}\label{15_K4_Sec}
In this section, we perform a similar analysis for D4-branes wrapped on a Kahler four-cycle $K^4_k$. As in the previous cases, the constant $k$ characterizes the constant curvature of $K^4_k$ with $k=1,0,-1$ corresponding to a two-dimensional complex projective space $CP^2$, a four-dimensional flat space $\mathbb{R}^4$, and a two-dimensional complex hyperbolic space $CH^2$, respectively. The manifold $K^4_k$ has a $U(2)\sim U(1)\times SU(2)$ spin connection. Therefore, we can perform a twist by turning on $SO(2)\sim U(1)$ or $SO(3)\sim SU(2)$ gauge fields to cancel the $U(1)$ or $SU(2)$ parts of the spin connection. 
\\
\indent A general ansatz for the six-dimensional metric takes the form of
\begin{equation}\label{Kahler7Dmetric}
ds_6^2=-e^{2U(r)}dt^2+dr^2+e^{2V(r)}ds^2_{K^{4}_{k}}
\end{equation}
in which the explicit form of the metric on a Kahler four-cycle will be specified in each case.

\subsubsection{Solutions with $SO(3)$ twist}\label{15_K4_SO(3)_section}
We begin with a twist along the $SU(2)\sim SO(3)$ part of the spin connection and choose the following form of the metric on a Kahler four-cycle
\begin{equation}\label{Kahlermetric}
ds^2_{K^4_k}=d\psi^2+f_k(\psi)^2(\tau_1^2+\tau_2^2+\tau_3^2)
\end{equation}
with $\psi\in[0,\frac{\pi}{2}]$. $f_k(\psi)$ is a function defined as in \eqref{fFn} while $\tau_i$, $i=1,2,3$, are $SU(2)$ left-invariant one-forms satisfying 
\begin{equation}
d\tau_i=\varepsilon_{ijl}\tau_j\wedge\tau_l\, . 
\end{equation}
The explicit forms of $\tau_i$ can be chosen to be
\begin{eqnarray}
\tau_1&=&\frac{1}{2}(-\sin\vartheta d\theta+\cos\vartheta\sin\theta d \zeta),\nonumber \\
\tau_2&=&\frac{1}{2}(\cos\vartheta d\theta+\sin\vartheta\sin\theta d \zeta),\nonumber \\
\tau_3&=&\frac{1}{2}(d\vartheta+\cos\theta d \zeta)\label{SU(2)Inv1form}
\end{eqnarray}
with the ranges of the angular coordinates given by $\theta\in[0,\pi]$, $\zeta\in[0,2\pi]$, and $\vartheta\in[0,4\pi]$.
\\
\indent Using the following choice of vielbein
\begin{eqnarray}
e^{\hat{t}}&=&e^{U}dt, \qquad\qquad\  e^{\hat{r}}=dr,  \qquad\qquad\quad\ e^{\hat{\psi}}=e^{V}d\psi,\nonumber \\ e^{\hat{1}}&=&e^{V}f_k(\psi)\tau_1,\qquad e^{\hat{2}}=e^{V}f_k(\psi)\tau_2, \qquad e^{\hat{3}}=e^{V}f_k(\psi)\tau_3,   \label{Kahler4bein}
\end{eqnarray}
we find the following non-vanishing components of the spin connection
\begin{equation}\label{K4_SU(2)_spincon}
{\omega_{\hat{t}}}^{\hat{t}\hat{r}}=U', \qquad {\omega_{\hat{I}}}^{\hat{J}\hat{r}}= V'\delta^{\hat{J}}_{\hat{I}},\qquad
{\omega_{\hat{i}}}^{\hat{j}\hat{\psi}}=e^{-V}\frac{f'_k(\psi)}{f_k(\psi)}\delta^{\hat{j}}_{\hat{i}},\qquad {\omega_{\hat{i}}}^{\hat{j}\hat{l}}=e^{-V}\frac{1}{f_k(\psi)}{\varepsilon_{\hat{i}}}^{\hat{j}\hat{l}}\, .
\end{equation}
We have split the six-dimensional coordinates as $x^\mu=(t,r,x^I)$, $I=(\psi,i)$ for $i=1,2,3$, and used $\hat{I}=\{\hat{\psi},\hat{i}\}$ as a flat index on $K^4_k$ with $\hat{i},\hat{j},\hat{l}=\hat{1}, \hat{2}, \hat{3}$. The $SU(2)$ spin connections on $K^4_k$ are given by ${\omega_{\hat{i}}}^{\hat{j}\hat{\psi}}$ and ${\omega_{\hat{i}}}^{\hat{j}\hat{l}}$. To perform a topological twist, we turn on $SO(3)$ gauge fields with the following ansatz
\begin{equation}\label{15_K4_SO(3)_gauge_fields}
{A_{\hat{1}}}^{23}={A_{\hat{2}}}^{31}={A_{\hat{3}}}^{12}=e^{-V}\frac{p}{4k}\frac{[f'_k(\psi)-1]}{f_k(\psi)}\, .
\end{equation} 
\indent We then consider all gauge groups containing an $SO(3)$ subgroup. By choosing the $SO(3)$ gauge generators to be $X_{12}$, $X_{13}$, and $X_{23}$, we can describe all possible gauge groups using the embedding tensor characterized by the following symmetric tensor
\begin{equation}\label{SO(3)_Ytensor}
Y_{mn}=\text{diag}(1,1,1,\kappa,\lambda).
\end{equation}
There are six possible gauge groups given by $SO(5)$ ($\kappa=\lambda=1$), $SO(4,1)$ ($\kappa=-\lambda=1$), $SO(3,2)$ ($\kappa=\lambda=-1$), $CSO(4,0,1)$ ($\kappa=1$, $\lambda=0$), $CSO(3,1,1)$ ($\kappa=-1$, $\lambda=0$), and $CSO(3,0,2)$ ($\kappa=\lambda=0$). 
\\
\indent In addition to the dilaton, there are four $SO(3)$ singlet scalars corresponding to the following non-compact generators
\begin{eqnarray}
\overline{\mathcal{Y}}_1&=&2\,\hat{\boldsymbol{t}}_{1\dot{1}}+2\,\hat{\boldsymbol{t}}_{2\dot{2}}+2\,\hat{\boldsymbol{t}}_{3\dot{3}}-3\,\hat{\boldsymbol{t}}_{4\dot{4}}-3\,\hat{\boldsymbol{t}}_{5\dot{5}},\nonumber \\
\overline{\mathcal{Y}}_2&=&\hat{\boldsymbol{t}}_{4\dot{5}},\qquad \overline{\mathcal{Y}}_3\,=\,\hat{\boldsymbol{t}}_{4\dot{4}}-\hat{\boldsymbol{t}}_{5\dot{5}},\qquad\quad \overline{\mathcal{Y}}_4\,=\,\boldsymbol{s}_{45}\label{15_SO(3)_non_com}
\end{eqnarray}
leading to the coset representative of the form
\begin{equation}\label{15_SO(3)_coset}
V=e^{\varphi\boldsymbol{d}+\phi_1\overline{\mathcal{Y}}_1+\phi_2\overline{\mathcal{Y}}_2+\phi_3\overline{\mathcal{Y}}_3+\varsigma\overline{\mathcal{Y}}_4}.
\end{equation}
\indent With $B_{\mu\nu,m}=0$, the only non-vanishing components of the modified gauge field strengths are given by
\begin{equation}
{\mathcal{H}_{\hat{\psi}\hat{1}}}^{23}={\mathcal{H}_{\hat{2}\hat{3}}}^{23}={\mathcal{H}_{\hat{\psi}\hat{2}}}^{31}={\mathcal{H}_{\hat{3}\hat{1}}}^{31}={\mathcal{H}_{\hat{\psi}\hat{3}}}^{12}={\mathcal{H}_{\hat{1}\hat{2}}}^{12}=-\frac{e^{-2V}p}{4}.
\end{equation}
These satisfy all the Bianchi identities without any further conditions. However, there is a non-vanishing term in the two-/four-form duality given in \eqref{2_4SD} of the form
\begin{eqnarray}
\mathbb{T}^{Am}M_{AB}\mathbb{T}^{B}_{np}{\mathcal{H}_{\hat{\psi}\hat{i}}}^{np}&=&-\frac{1}{\sqrt{2}}e^{-2(V-\varphi-2\phi_1)}p\varsigma\delta^m_{\hat{i}},\\
\mathbb{T}^{Am}M_{AB}\mathbb{T}^{B}_{np}{\mathcal{H}_{\hat{i}\hat{j}}}^{np}&=&-\frac{1}{\sqrt{2}}e^{-2(V-\varphi-2\phi_1)}p\varsigma\varepsilon_{\hat{i}\hat{j}\hat{k}}\delta^m_{\hat{k}}\, .
\end{eqnarray}
In order to keep $p\neq0$, we will set $\varsigma=0$ such that the two-/four-form duality is identically satisfied. 
\\
\indent We now impose the twist condition $gp=k$ and the projectors $\hat{\gamma}_r\epsilon_\pm=\epsilon_\mp$ as well as
\begin{equation}
\hat{\gamma}_{\hat{1}\hat{2}}\epsilon_{+\alpha}=-{(\gamma_{12})_\alpha}^\beta\epsilon_{+\beta}, \qquad\hat{\gamma}_{\hat{2}\hat{3}}\epsilon_{+\alpha}=-{(\gamma_{23})_\alpha}^\beta\epsilon_{+\beta},\qquad \hat{\gamma}_{\hat{\psi}\hat{1}}\epsilon_{+\alpha}=\hat{\gamma}_{\hat{2}\hat{3}}\epsilon_{+\alpha},\label{SO(3)_K4_Projcon}
\end{equation}
together with
\begin{equation}
\hat{\gamma}_{\hat{1}\hat{2}}\epsilon_{-\dot{\alpha}}=-{(\gamma_{\dot{1}\dot{2}})_{\dot{\alpha}}}^{\dot{\beta}}\epsilon_{-\dot{\beta}}, \qquad\hat{\gamma}_{\hat{2}\hat{3}}\epsilon_{-\dot{\alpha}}=-{(\gamma_{\dot{2}\dot{3}})_{\dot{\alpha}}}^{\dot{\beta}}\epsilon_{-\dot{\beta}},\qquad \hat{\gamma}_{\hat{\psi}\hat{1}}\epsilon_{-\dot{\alpha}}=\hat{\gamma}_{\hat{2}\hat{3}}\epsilon_{-\dot{\alpha}}\, . \label{SO(3)_K4_Projcon-}
\end{equation}
The Killing spinors take the form
\begin{equation}\label{15_sig2_full_Killing}
\epsilon_+(r)=e^{\frac{1}{2}\left[U(r)+ W(r)\gamma_{45}\right]}\epsilon_+^{0}\qquad\text{ and }\qquad\epsilon_-(r)=e^{\frac{1}{2}\left[U(r)- W(r)\gamma_{\dot{4}\dot{5}}\right]}\epsilon_-^{0}
\end{equation}
for an $r$-dependent function $W(r)$. With all these, the resulting BPS equations read
\begin{eqnarray}
U'&=&\frac{ge^{\varphi-8\phi_1}}{4\sqrt{2}}\left[3+e^{20\phi_1}\left((\kappa+\lambda)\cosh{2\phi_2}\cosh{4\phi_3}-(\kappa-\lambda)\sinh{4\phi_3}\right)\right]\nonumber\\&&-\frac{3p}{4\sqrt{2}}e^{-2V-\varphi+8\phi_1},\\
V'&=&\frac{ge^{\varphi-8\phi_1}}{4\sqrt{2}}\left[3+e^{20\phi_1}\left((\kappa+\lambda)\cosh{2\phi_2}\cosh{4\phi_3}-(\kappa-\lambda)\sinh{4\phi_3}\right)\right]\nonumber\\&&+\frac{3p}{4\sqrt{2}}e^{-2V-\varphi+8\phi_1},\\
\varphi'&=&-\frac{ge^{\varphi-8\phi_1}}{20\sqrt{2}}\left[3+e^{20\phi_1}\left((\kappa+\lambda)\cosh{2\phi_2}\cosh{4\phi_3}-(\kappa-\lambda)\sinh{4\phi_3}\right)\right]\nonumber\\&&+\frac{3p}{20\sqrt{2}}e^{-2V-\varphi+8\phi_1}, \\
\phi'_1&=&\frac{ge^{\varphi-8\phi_1}}{10\sqrt{2}}\left[2-e^{20\phi_1}\left((\kappa+\lambda)\cosh{2\phi_2}\cosh{4\phi_3}-(\kappa-\lambda)\sinh{4\phi_3}\right)\right]\nonumber\\&&-\frac{p}{5\sqrt{2}}e^{-2V-\varphi+8\phi_1}\\
\phi'_2&=&-\frac{ge^{\varphi+12\phi_1}}{\sqrt{2}}(\kappa+\lambda)\sinh{2\phi_2}\,\text{sech }{4\phi_3},\label{15_SO(3)_phi2_BPS}\\
\phi'_3&=&\frac{ge^{\varphi+12\phi_1}}{2\sqrt{2}}\left((\kappa-\lambda)\cosh{4\phi_3}-(\kappa+\lambda)\cosh{2\phi_2}\sinh{4\phi_3}\right)\label{15_SO(3)_phi3_BPS}
\end{eqnarray}
together with
\begin{equation}\label{15_Sig2_W_Im}
W'=-\frac{ge^{\varphi+12\phi_1}}{2\sqrt{2}}(\kappa+\lambda)\sinh{2\phi_2}\tanh{4\phi_3}\, .
\end{equation}
It can be verified that these BPS equations imply all the second-ordered field equations. We also note that for $k=0$, the twist condition implies $p=0$. This leads to untwisted solutions of standard flat domain walls preserving sixteen supercharges. As in the previous cases, we will not further consider this type of solutions. On the other hand, for $k=\pm 1$, the above projection conditions imply that the solutions preserve only two supercharges. 
\\
\indent For $CSO(3,0,2)$ gauge group with $\kappa=\lambda=0$, the BPS equations can be analytically solved. In this case, the BPS equations lead to $\phi'_2=\phi'_3=W'=0$, and we can choose $\phi_2=\phi_3=W=0$ without loss of generality. The other BPS equations give $U'=-5\varphi'$ and $\phi'_1=-\frac{4}{3}\varphi'$ with the solutions given by 
\begin{equation}
U=-5\varphi \qquad  \textrm{and}\qquad  \phi_1=C_1-\frac{4}{3}\varphi
\end{equation} 
for an integration constant $C_1$ . We have neglected an additive integration constant for $U$ which can be absorbed by rescaling the time coordinate.
\\
\indent Taking a linear combination $V'+5\varphi'$ and changing to a new radial coordinate $\rho$ given by $\frac{d\rho}{dr}=e^{4 \varphi+8\phi_1-V}$, we find
\begin{equation}
V= \ln \left[\frac{3p \rho }{2\sqrt{2}}+C\right]-5 \varphi\, .
\end{equation}
The integration constant $C$ can also be set to zero by shifting the coordinate $\rho$. With all these results and $C=0$, the equation for $\varphi'$ gives
\begin{equation}
\varphi=\frac{6}{5}C_1-\frac{3}{40} \ln \left[\frac{9(p\rho ^{10/3}+C_0)}{20g \rho^{4/3} }\right]
\end{equation}
in which $C_0$ is another integration constant. 
\\
\indent In the UV limit as $\rho\rightarrow+\infty$, we find the following asymptotic behavior
\begin{eqnarray}\label{15_UV_Asy_SO(3)}
U\sim \frac{3}{4} \ln \rho,\qquad V\sim\frac{7}{4} \ln \rho,\qquad \varphi \sim -\frac{3}{20} \ln \rho,\qquad \phi_1\sim\frac{1}{5}\ln \rho
\end{eqnarray}
which leads to the six-dimensional metric of the form
\begin{equation}\label{15_UV_metric_SO(3)}
ds^2=\rho^{\frac{3}{2}}(-dt^2+d\rho^2+\rho^2ds^2_{K^4_k}).
\end{equation}
As $\rho\rightarrow 0$ and $C_0\neq 0$, the solution becomes  
\begin{equation}
U\sim-\frac{1}{2} \ln\rho,\qquad V\sim\frac{1}{2}\ln \rho,\qquad\varphi\sim \frac{1}{10}\ln \rho,\qquad\phi_1\sim-\frac{2}{15} \ln \rho
\end{equation}
leading to the six-dimensional metric
\begin{equation}
ds^2=-\rho^{-1}dt^2+\rho^{\frac{7}{3}}d\rho^2+\rho ds^2_{K^4_k}.
\end{equation}
By uplifting the solution to type IIA theory, we can find the behavior of the ten-dimensional metric component $\hat{g}_{00}$ near this IR singularity as
\begin{equation}
\hat{g}_{00}= e^{2U+\varphi-3\phi_1}\sim \frac{1}{\sqrt{\rho}}\rightarrow +\infty
\end{equation}  
which implies that the singularity is unphysical.
\\
\indent For $C_0=0$, we find the same asymptotic behavior \eqref{15_UV_Asy_SO(3)} as $\rho\rightarrow 0$. The six-dimensional metric is also given by \eqref{15_UV_metric_SO(3)}, and the behavior of $\hat{g}_{00}$ near this IR singularity is 
\begin{equation}
\hat{g}_{00}= e^{2U+\varphi-3\phi_1}\sim \rho^{3/4}\rightarrow 0\, .
\end{equation}
Accordingly, in this case, the singularity is physical, and the solution describes D4-branes wrapped on $K^4_k$ giving rise to $N=2$ supersymmetric quantum mechanics in the IR.
\\
\indent In other gauge groups, we need to solve the BPS equations numerically. 
We begin with solutions that are asymptotic to a locally flat domain walls with $SO(5)$ symmetry given by
\begin{equation}
U\sim V\sim\frac{5g\rho}{4\sqrt{2}},\qquad \varphi\sim-\frac{g\rho}{4\sqrt{2}},\qquad \phi_1\sim\phi_2\sim\phi_3\sim0
\end{equation}
with the radial coordinate $\rho$ defined by $\frac{d\rho}{dr}=e^{\varphi-8\phi_1}$. Examples of numerical solutions interpolating between this asymptotic behavior and $t\times K^4_k$-sliced curved domain walls in $SO(5)$ gauge group for different values of $g$ are given in figures \ref{15_CP2_special_SO(3)_SO(5)gg_flows} and \ref{15_CH2_special_SO(3)_SO(5)gg_flows} in Appendix \ref{AppC_2_1}. In all of these solutions, we find that $\phi_2$ and $\phi_3$ vanish identically along the entire flows, so we have neglected the corresponding numerical solutions for $\phi_2$ and $\phi_3$ for convenience. From the behavior of the ten-dimensional metric component $\hat{g}_{00}$, we find that all $t\times CH^2$-sliced solutions are uplifted to type IIA solutions with physically acceptable IR singularities. On the other hand, the $t\times CP^2$-sliced solutions admit physical IR singularities only for $g\geq0.3$. 

Another class of solutions consists of solutions that are asymptotically locally $SO(3)$ symmetric flat domain wall of the form
\begin{equation}\label{15_SO(3)_flat_DW_asym}
U\sim V\sim\frac{3 g \rho }{4 \sqrt{2}},\qquad \varphi\sim-\frac{3 g \rho }{20 \sqrt{2}},\qquad \phi_1\sim\frac{g \rho }{5 \sqrt{2}},\qquad \phi_2\sim\phi_3\sim0\, .
\end{equation}
Choosing $g<0$ in order to identify the UV limit with $\rho\rightarrow +\infty$, we find a number of numerical solutions interpolating between this locally $SO(3)$ symmetric flat domain wall and singular geometries in the IR. It turns out that all solutions in $SO(5)$, $SO(4,1)$, $CSO(4,0,1)$, and $CSO(3,1,1)$ gauge groups are uplifted to type IIA solutions with unphysical IR singularities. We will not give these solutions here. In the case of $SO(3,2)$ gauge group, examples of numerical solutions are shown in figures \ref{15_CP2_SO(32)_SO(5)gg_flows} and \ref{15_CH2_SO(3)_SO(32)gg_flows} in Appendix \ref{AppC_2_1}. We also note that $\phi_2=\phi_3=0$ in all of the solutions, so we have omitted the numerical solutions for $\phi_2$ and $\phi_3$ in the figures. From the figures, we see that all $t\times CP^2$-sliced solutions are uplifted to ten dimensions with physical IR singularities for any values of $g$. On the other hand, the $t\times CH^2$-sliced solutions admit physical IR singularities only for $g<-2$. All of these $t\times K_k^4$-sliced curved domain walls obtained from $SO(3)$ twist are summarized in table \ref{tab5}. We also note that these results are very similar to solutions with an $SO(3)$ twist on a Riemannian three-manifold studied recently in \cite{6D_twist_I}.

\begin{table}[h!]
\centering
\begin{tabular}{| c | c | c  | c |}
\hline
UV flat&  \multirow{2}{*}{gauge group} & Kahler  & physical IR \\
domain wall& & four-cycle  & singularity \\\hline
$SO(5)$ & $SO(5)$& $CP^2$& $g\geq0.3$ \\
&& $CH^2$ & any $g$ \\\hline
$SO(3)$ &$SO(5)$& $CP^2$, $CH^2$  & $\times$ \\\cline{2-4}
&$SO(4,1)$ & $CP^2$, $CH^2$& $\times$ \\\cline{2-4}
&$SO(3,2)$ & $CP^2$ & any $g$ \\
&& $CH^2$ & $g\leq-2$ \\\cline{2-4}
&$CSO(4,0,1)$& $CP^2$, $CH^2$ & $\times$ \\\cline{2-4}
&$CSO(3,1,1)$ & $CP^2$, $CH^2$& $\times$ \\\cline{2-4}
&$CSO(3,0,2)$ & $CP^2$, $CH^2$& $C_0=0$ \\
&(analytic) &  &   \\\hline
\end{tabular}
\caption{Summary of satisfaction of the criterion \cite{Maldacena_nogo} for the IR singularities of the $t\times K_k^4$-sliced curved domain walls obtained from $SO(3)$ twist in $SO(5)$, $SO(4,1)$, $SO(3,2)$, $CSO(4,0,1)$, $CSO(3,1,1)$, and $CSO(3,0,2)$ gauge groups.}\label{tab5}
\end{table}

\subsubsection{Solutions with $SU(2)$ twist}\label{15_K4_SO(3)sd_section}
We now move to the topological twist obtained by canceling the $SU(2)$ part of the spin connection with the self-dual $SU(2)_+$ subgroup of $SU(2)_+\times SU(2)_-\sim SO(4)\subset SO(5)$ generated by $X_{12}+X_{34}$, $X_{13}+X_{24}$, and $X_{23}+X_{14}$. We then turn on the following gauge fields
\begin{eqnarray}
{A_{\hat{1}}}^{23}&=&{A_{\hat{2}}}^{31}={A_{\hat{3}}}^{12}=e^{-V}\frac{p}{8k}\frac{[f'_k(\psi)-1]}{f_k(\psi)}\\
\text{and}\qquad{A_{\hat{1}}}^{14}&=&{A_{\hat{2}}}^{24}={A_{\hat{3}}}^{34}=e^{-V}\frac{p}{8k}\frac{[f'_k(\psi)-1]}{f_k(\psi)}\, .
\end{eqnarray}
The gauge fields can also be written compactly as
\begin{equation}
{A_{\hat{i}}}^{i}=\frac{1}{2}\varepsilon^{ijl}{A_{\hat{i}}}^{jl}+ {A_{\hat{i}}}^{i4}=e^{-V}\frac{p}{4k}\frac{[f'_k(\psi)-1]}{f_k(\psi)}\delta^i_{\hat{i}}\, .
\end{equation}
In this case, we perform the twist by imposing the twist condition $gp=k$ and the projections given in \eqref{SO(3)_K4_Projcon} and \eqref{SO(3)_K4_Projcon-} as well as two additional projectors 
\begin{equation}\label{GammaSDProj}
{(\gamma_{12})_\alpha}^\beta\epsilon_{+\beta}={(\gamma_{34})_\alpha}^\beta\epsilon_{+\beta}\quad\text{ and }\quad{(\gamma_{\dot{1}\dot{2}})_{\dot{\alpha}}}^{\dot{\beta}}\epsilon_{-\dot{\beta}}={(\gamma_{\dot{3}\dot{4}})_{\dot{\alpha}}}^{\dot{\beta}}
\epsilon_{-\dot{\beta}}\, .
\end{equation}
Accordingly, the resulting solutions preserve only one supercharges.
\\  
\indent In addition, we also need to turn on the magnetic two-form potential
\begin{equation}\label{sdSO(3)Btr}
{B_{\hat{t}\hat{r}}}^5=-\frac{3p^2}{16 g^2}  e^{-2 (2V+3\varphi+8\phi)}\, .
\end{equation}
Recall that the vector representation of $SO(4)$ is identified with $(\mathbf{2},\mathbf{2})$ representation of $SU(2)_+\times SU(2)_-$, we see that the $SU(2)_+$ is embedded in $SO(4)$ in such a way that the $SU(2)_+$ singlets are essentially $SO(4)$ singlets. Accordingly, we need to consider gauge groups with an $SO(4)$ subgroup and singlet scalars under $SO(4)$. These gauge groups are $SO(5)$, $SO(4,1)$, and $CSO(4,0,1)$ with the embedding tensor given by \eqref{SO(4)Ytensor}. As in the case of $SO(4)$ twist, we will consider only $SO(5)$ and $SO(4,1)$ gauge group with $\lambda\neq 0$ in order to satisfy the Bianchi identities and the two-/four-form duality.
\\
\indent Using the $SO(4)$ invariant coset representative \eqref{15_SO(4)_coset} and the $\hat{\gamma}_r$ projection \eqref{15_DW_Proj}, we find the following BPS equations
\begin{eqnarray}
U'&=&\frac{g}{4\sqrt{2}}e^{\varphi-4\phi}(4+\lambda e^{20\phi})-\frac{3p}{4\sqrt{2}}e^{-2V-\varphi+4\phi}+\frac{9\lambda p^2}{32\sqrt{2}g}e^{-4V-3\varphi-8\phi},\\
V'&=&\frac{g}{4\sqrt{2}}e^{\varphi-4\phi}(4+\lambda e^{20\phi})+\frac{3p}{4\sqrt{2}}e^{-2V-\varphi+4\phi}-\frac{3\lambda p^2}{32\sqrt{2}g}e^{-4V-3\varphi-8\phi},\\
\varphi'&=&-\frac{g}{20\sqrt{2}}e^{\varphi-4\phi}(4+\lambda e^{20\phi})+\frac{3p}{20\sqrt{2}}e^{-2V-\varphi+4\phi}-\frac{9\lambda p^2}{160\sqrt{2}g}e^{-4V-3\varphi-8\phi},\qquad\quad \\
\phi'&=&\frac{g}{5\sqrt{2}}e^{\varphi-4\phi}(1-\lambda e^{20\phi})-\frac{3p}{20\sqrt{2}}e^{-2V-\varphi+4\phi}-\frac{3\lambda p^2}{80\sqrt{2}g}e^{-4V-3\varphi-8\phi}\, .
\end{eqnarray}
We then look for numerical solutions asymptotic to locally flat domain walls with $SO(5)$ and $SO(4)$ symmetries given in \eqref{SO(4)_SO(5)_flat_DW_asym} and \eqref{15_SO(4)_flat_DW_asym}, respectively. We first note that the resulting BPS equations take a very similar form as those in section \ref{15_Sig4_Sec}. The corresponding numerical solutions are also very similar to those obtained from the $SO(4)$ twist on a Riemannian four-manifold in section \ref{15_Sig4_Sec}. We will not present these solutions here to avoid repetition but simply give a summary of the physical IR singularities in table \ref{tab6}. 
\clearpage\newpage
\begin{table}[h!]
\centering
\begin{tabular}{| c | c | c | c |}
\hline
UV flat & gauge & Kahler & physical IR \\
domain wall& group & four-cycle & singularity \\\hline
$SO(5)$&$SO(5)$&$CP^2$ & $\times$ \\
&&$CH^2$ & any $g$ \\\hline
$SO(4)$&$SO(5)$& $CP^2$, $CH^2$ & $\times$\\\cline{2-4}
&$SO(4,1)$ & $CP^2$, $CH^2$ & $\times$\\\hline
\end{tabular}
\caption{Summary of satisfaction of the criterion \cite{Maldacena_nogo} for the IR singularities of the curved domain wall solutions obtained from $SU(2)$ twist on a Kahler four-cycle in $SO(5)$ and $SO(4,1)$ gauge groups.}\label{tab6}
\end{table}

\subsubsection{Solutions with $SO(2)\times SO(2)$ twist}\label{15_K4_SO(2)xSO(2)_section}
Another possibility to perform a topological twist on a Kahler four-cycle is achieved by canceling the $U(1)$ part of the spin connection. In this case, it is convenient to write the metric on $K^4_k$ as
\begin{equation}\label{U(1)Kahlermetric}
ds^2_{K^4_k}=\frac{d\psi^2}{(k\psi^2+1)^2}+\frac{\psi^2\tau_3^2}{(k\psi^2+1)^2}+\frac{\psi^2}{(k\psi^2+1)}(\tau_1^2+\tau_2^2)
\end{equation}
with $\tau_i$ being the $SU(2)$ left-invariant one-forms given in \eqref{SU(2)Inv1form}. The six-dimensional metric is still given by \eqref{Kahler7Dmetric} with $ds^2_{K^4_k}$ given by \eqref{U(1)Kahlermetric}.
\\
\indent With the vielbein of the form
\begin{eqnarray}
e^{\hat{t}}&=&e^{U}dt, \qquad\qquad\quad\, \ e^{\hat{r}}=dr, \qquad\qquad\qquad\, \ e^{\hat{\psi}}=\frac{e^{V}d\psi}{(k\psi^2+1)} ,\nonumber \\ e^{\hat{1}}&=&\frac{e^{V}\psi}{\sqrt{k\psi^2+1}}\tau_1,\qquad e^{\hat{2}}=\frac{e^{V}\psi}{\sqrt{k\psi^2+1}}\tau_2, \qquad e^{\hat{3}}=\frac{e^{V}\psi}{(k\psi^2+1)}\tau_3,\label{U(1)Kahler4bein}
\end{eqnarray}
all non-vanishing components of the spin connection are given by
\begin{eqnarray}
{\omega_{\hat{t}}}^{\hat{t}\hat{r}}&=&U', \qquad {\omega_{\hat{I}}}^{\hat{J}\hat{r}}= V'\delta^{\hat{J}}_{\hat{I}},\qquad  \hat{I},\hat{J}=\hat{\psi}, \hat{1}, \hat{2}, \hat{3},\nonumber \\
{\omega_{\hat{1}}}^{\hat{1}\hat{\psi}}&=&{\omega_{\hat{1}}}^{\hat{2}\hat{3}}={\omega_{\hat{2}}}^{\hat{3}\hat{1}}={\omega_{\hat{2}}}^{\hat{2}\hat{\psi}}=\frac{e^{-V}}{\psi},\nonumber \\
{\omega_{\hat{3}}}^{\hat{3}\hat{\psi}}&=&e^{-V}\frac{(1-k\psi^2)}{\psi},\qquad {\omega_{\hat{3}}}^{\hat{1}\hat{2}}=e^{-V}\frac{(2k\psi^2+1)}{\psi}.\label{AdS3xU(1)Kahler4SpinCon}
\end{eqnarray}
Imposing the projector 
\begin{equation}
\hat{\gamma}_{\hat{\psi}\hat{1}}\epsilon_{\pm}=\hat{\gamma}_{\hat{2}\hat{3}}\epsilon_{\pm},\label{SO2SO2_twist_Kahler4_pro}
\end{equation}
we find that ${\omega_{\hat{1}}}^{\hat{1}\hat{\psi}}$ and ${\omega_{\hat{1}}}^{\hat{2}\hat{3}}$ as well as ${\omega_{\hat{2}}}^{\hat{3}\hat{1}}$ and ${\omega_{\hat{2}}}^{\hat{2}\hat{\psi}}$ cancel each other in the covariant derivatives of the supersymmetry parameters. The remaining $U(1)$ part takes the form of
\begin{equation}
{\omega_{\hat{3}}}^{\hat{3}\hat{\psi}}-{\omega_{\hat{3}}}^{\hat{1}\hat{2}}=e^{-V}k\psi\, .
\end{equation}
\indent We perform a twist to cancel this by turning on $SO(2)\times SO(2)$ gauge fields 
\begin{equation}\label{15_K4_SO(2)xSO(2)_gauge_fields}
{A_{\hat{3}}}^{12}=\frac{3p_1}{4}e^{-V}\psi\qquad \textrm{and}\qquad {A_{\hat{3}}}^{34}=\frac{3p_2}{4}e^{-V}\psi\, .
\end{equation}  
We also need to impose the following projection conditions on the Killing spinors
\begin{equation}\label{SO(2)xSO(2)_K4Projcon+}
\hat{\gamma}_{\hat{1}\hat{2}}\epsilon_{+\alpha}={(\gamma_{12})_\alpha}^\beta\epsilon_{+\beta}={(\gamma_{34})_\alpha}^\beta\epsilon_{+\beta}
\end{equation}
and
\begin{equation}\label{SO(2)xSO(2)_K4Projcon-}
\hat{\gamma}_{\hat{1}\hat{2}}\epsilon_{-\dot{\alpha}}={(\gamma_{\dot{1}\dot{2}})_{\dot{\alpha}}}^{\dot{\beta}}\epsilon_{-\dot{\beta}}={(\gamma_{\dot{3}\dot{4}})_{\dot{\alpha}}}^{\dot{\beta}}\epsilon_{-\dot{\beta}}
\end{equation}
together with the twist condition 
\begin{equation}\label{15_SO(2)xSO(2)_K4_twist_con}
g(p_1+\kappa p_2)=k\, .
\end{equation}
Including the $\hat{\gamma}_r$ projector in \eqref{15_DW_Proj}, the solutions perserve two supercharges.
\\
\indent As in the case of $SO(2)\times SO(2)$ twist on two Riemann surfaces, the embedding tensor and scalar coset representative are given respectively by \eqref{SO(2)xSO(2)Y} and \eqref{fullSO(2)xSO(2)singlet_coset}. The two-form gauge field strengths take the form
\begin{equation}
{\mathcal{H}_{\hat{1}\hat{2}}}^{12}={\mathcal{H}_{\hat{\psi}\hat{3}}}^{12}=\frac{3p_1}{2}e^{-2V}\qquad \textrm{and} \qquad
{\mathcal{H}_{\hat{1}\hat{2}}}^{34}={\mathcal{H}_{\hat{\psi}\hat{3}}}^{34}=\frac{3p_2}{2}e^{-2V}\, .
\end{equation}
In order to satisfy the Bianchi identity \eqref{DefBianchi2}, we also need to turn on the magnetic two-form potential of the form
\begin{equation}\label{Sig2xSig2_K4_mag_two_pot}
{B_{\hat{t}\hat{r}}}^5=-\frac{9}{g^2} p_1p_2 e^{-2 (2V+3\varphi+4\phi_1+4\phi_2)}\, .
\end{equation}
As in the previous cases, for the gauge groups with $\lambda=0$, the Bianchi identity implies either $p_1=0$ or $p_2=0$. Accordingly, to keep both of the $SO(2)$ gauge fields non-vanishing, we will consider only gauge groups with $\lambda\neq0$. Furthermore, consistency of the BPS equations also requires either $\varsigma_1=\varsigma_2=0$ or $p_2=p_1/\kappa$ with $\varsigma_1\neq 0$ and $\varsigma_2\neq 0$. We consider these two possibilities separately.

\paragraph{Solutions with $\varsigma_1=\varsigma_2=0$:}
For $\varsigma_1=\varsigma_2=0$, the corresponding BPS equations read
\begin{eqnarray}
U'&=&\frac{ge^{\varphi}}{4\sqrt{2}}(2e^{-4\phi_1}+2\kappa e^{-4\phi_2}+\lambda e^{8(\phi_1+\phi_2)})-\frac{3e^{-2V-\varphi}}{2\sqrt{2}}(p_1e^{4\phi_1}+p_2e^{4\phi_2})\nonumber\\&&+\frac{27\lambda}{2\sqrt{2} g}p_1p_2 e^{-4V-3\varphi-4\phi_1+4\phi_2},\label{15_SO(2)xSO(2)_K4_BPS1}\\
V'&=&\frac{ge^{\varphi}}{4\sqrt{2}}(2e^{-4\phi_1}+2\kappa e^{-4\phi_2}+\lambda e^{8(\phi_1+\phi_2)})+\frac{3e^{-2V-\varphi}}{2\sqrt{2}}(p_1e^{4\phi_1}+p_2e^{4\phi_2})\nonumber\\&&-\frac{9\lambda}{2\sqrt{2} g}p_1p_2 e^{-4V-3\varphi-4\phi_1+4\phi_2},\label{15_SO(2)xSO(2)_K4_BPS2}\\
\varphi'&=&-\frac{ge^{\varphi}}{20\sqrt{2}}(2e^{-4\phi_1}+2\kappa e^{-4\phi_2}+\lambda e^{8(\phi_1+\phi_2)})+\frac{3e^{-2V-\varphi}}{10\sqrt{2}}(p_1e^{4\phi_1}+p_2e^{4\phi_2})\nonumber\\&&-\frac{27\lambda}{10\sqrt{2} g}p_1p_2 e^{-4V-3\varphi-4\phi_1+4\phi_2},\qquad\label{15_SO(2)xSO(2)_K4_BPS3}\\
\phi'_1&=&\frac{ge^{\varphi}}{5\sqrt{2}}(3e^{-4\phi_1}-2\kappa e^{-4\phi_2}-\lambda e^{8(\phi_1+\phi_2)})-\frac{3e^{-2V-\varphi}}{5\sqrt{2}}(3p_1e^{4\phi_1}-2p_2e^{4\phi_2})\nonumber\\&&-\frac{9\lambda}{5\sqrt{2} g}p_1p_2 e^{-4V-3\varphi-4\phi_1+4\phi_2},\label{15_SO(2)xSO(2)_K4_BPS4}\\
%\end{eqnarray}
%\begin{eqnarray}
\phi'_2&=&\frac{ge^{\varphi}}{5\sqrt{2}}(3\kappa e^{-4\phi_2}-2e^{-4\phi_1}-\lambda e^{8(\phi_1+\phi_2)})+\frac{3e^{-2V-\varphi}}{5\sqrt{2}}(2p_1e^{4\phi_1}-3p_2e^{4\phi_2})\nonumber\\&&-\frac{9\lambda}{5\sqrt{2} g}p_1p_2 e^{-4V-3\varphi-4\phi_1+4\phi_2}.\label{15_SO(2)xSO(2)_K4_BPS5}
\end{eqnarray}
We begin with solutions that are asymptotic to a locally $SO(5)$ symmetric flat domain wall with the following asymptotic behavior,
as $\rho\rightarrow+\infty$,
\begin{equation}\label{15_K4_SO(2)xSO(2)_full_IR_fDW}
U\sim V\sim\frac{5g \lambda \rho}{4\sqrt{2}},\qquad \varphi\sim-\frac{g \lambda \rho}{4\sqrt{2}},\qquad \phi_1\sim\phi_2\sim0\, .
\end{equation}
The new radial coordinate $\rho$ is defined by $\frac{d\rho}{dr}=e^{\varphi+8(\phi_1+\phi_2)}$. Examples of numerical solutions from $SO(5)$ gauge group are given in figures \ref{15_CP2_special_SO(2)xSO(2)_SO(5)gg_flows}, \ref{15_CH2_special_SO(2)xSO(2)_SO(5)gg_flows}, and \ref{15_R4_special_SO(2)xSO(2)_SO(5)gg_flows} in Appendix \ref{AppC_2_2}. In these figures, we have set $g=1$ and different values of $p_1$. All $t\times CP^2$-sliced domain walls are uplifted to type IIA theory with unphysical IR singularities. On the other hand, for $t\times CH^2$- or $t\times \mathbb{R}^4$-sliced domain walls, all solutions lead to physical solutions in type IIA theory. We also point out that the solutions with $t\times \mathbb{R}^4$ slices do not need a topological twits and preserve eight supercharges due to the projectors \eqref{15_DW_Proj} and \eqref{SO2SO2_twist_Kahler4_pro}. However, the $SO(2)\times SO(2)$ gauge fields are non-vanishing in these solutions. For the sake of comparison, we also give the flat domain wall solutions with $p_1=p_2=k=0$ represented by the dashed line in figure \ref{15_R4_special_SO(2)xSO(2)_SO(5)gg_flows}.

Another class of solutions consists of solutions with the asymptotic geometry given by an $SO(2)\times SO(2)$ symmetric flat domain wall of the form
\begin{equation}
U\sim V\sim\frac{g \lambda  \rho }{4\sqrt{2}},\qquad \varphi\sim-\frac{g \lambda  \rho }{20 \sqrt{2}},\qquad \phi_1\sim\phi_2\sim-\frac{g \lambda  \rho }{5 \sqrt{2}}\, .
\end{equation}
We will choose $g\lambda <0$ in order to identify the UV limit with $\rho\rightarrow +\infty$. For $SO(5)$ gauge group, all solutions are uplifted to type IIA theory with unphysical IR singularities. For $SO(4,1)$ and $SO(3,2)$ gauge groups, examples of numerical solutions are shown in figures \ref{15_CP2_SO(2)xSO(2)_SO(41)gg_flows} to \ref{15_R4_SO(2)xSO(2)_SO(32)gg_flows} in Appendix \ref{AppC_2_2}. From these figures, we see that only the solutions with specific values of the magnetic charge $p_1$ can be uplifted to ten-dimensional solutions with physical IR singularities. The conditions for obtaining solutions with physical IR singularities are summarized in table \ref{tab7}.

\begin{table}[h!]
\centering
\begin{tabular}{| c | c | c | c |}
\hline
UV flat & gauge & Kahler & physical IR \\
domain wall& group & four-cycle &  singularity \\\hline
$SO(5)$&$SO(5)$&$CP^2$ & $\times$ \\
&&$CH^2$, $\mathbb{R}^4$ &  any $p_1$ \\\hline
$SO(2)\times SO(2)$&$SO(5)$& $K^4$ &  $\times$ \\\cline{2-4}
&$SO(4,1)$ & $CP^2$ &  $\times$ \\
& & $CH^2$ &  $-1\leq p_1\leq0$ \\
& & $\mathbb{R}^4$ &  $p_1=0$ \\\cline{2-4}
&$SO(3,2)$ & $CP^2$ & $-0.8\leq p_1\leq-1.1$ \\
& & $CH^2$, $\mathbb{R}^4$ & $p_1=0$ \\\hline
\end{tabular}
\caption{Summary of satisfaction of the criterion \cite{Maldacena_nogo} for the IR singularities of the solutions with $t\times K_k^4$ slices obtained from $SO(2)\times SO(2)$ twist with $\varsigma_1=\varsigma_2=0$ in $SO(5)$, $SO(4,1)$, and $SO(3,2)$ gauge groups.}\label{tab7}
\end{table}

\paragraph{Solutions with $\varsigma_1$ and $\varsigma_2$ scalars:}
To include non-vanishing shift scalars $\varsigma_1$ and $\varsigma_2$ in the BPS equations, we have to impose the condition $p_2=p_1/\kappa$ and an algebraic condition
\begin{equation}
\kappa e^{4\phi_2}\varsigma_1+e^{4\phi_1}\varsigma_2=0\, .
\end{equation}
The resulting BPS equations are the same as those given in \eqref{15_SO(2)xSO(2)_K4_BPS1} to \eqref{15_SO(2)xSO(2)_K4_BPS5} with $p_2=p_1/\kappa$ together with an additional equation 
\begin{eqnarray}
\varsigma'_1&=&\frac{ge^{\varphi}}{\sqrt{2}}(2e^{-4\phi_1}-2\kappa e^{-4\phi_2}-\lambda e^{8(\phi_1+\phi_2)})\varsigma_1-\frac{9 \sqrt{2}\lambda p_{1}^2\varsigma_1}{g\kappa}e^{-4V-3\varphi-4\phi_1-4\phi_2}\nonumber\\
&&-3\sqrt{2}p_1e^{-2V-\varphi}(e^{4\phi_1}-\kappa e^{4\phi_2})\varsigma_1\, .\qquad\quad
\end{eqnarray}
It turns out that all solutions asymptotic to an $SO(5)$ symmetric flat domain wall from $SO(5)$ gauge group have $\varsigma_1=\varsigma_2=0$ along the entire solutions. Since these solutions are qualitatively the same as those considered in the previous case, we will not give the corresponding numerical solutions here.
\\
\indent We now move to solutions with the asymptotic geometry given by an $SO(2)\times SO(2)$ symmetric flat domain wall of the form
\begin{equation}\label{15_K4_SO(2)d_flat_DW_asym}
U\sim V\sim\frac{g \lambda  \rho }{4\sqrt{2}},\qquad \varphi\sim-\frac{g \lambda  \rho }{20 \sqrt{2}},\qquad \phi_1\sim\phi_2\sim-\frac{g \lambda  \rho }{5 \sqrt{2}},\qquad\varsigma_1\sim e^{-\frac{g \lambda  \rho }{\sqrt{2}}}
\end{equation}
with the new radial coordinate $\rho$ defined by $\frac{d\rho}{dr}=e^{\varphi+8(\phi_1+\phi_2)}$. By choosing $g\lambda<0$, we can identify the UV limit with $\rho\rightarrow+\infty$. From the numerical analysis, we find that all solutions from $SO(5)$ and $SO(4,1)$ gauge groups are uplifted to ten-dimensional solutions with unphysical IR singularities. Accordingly, we will omit all of these solutions here. For $SO(3,2)$ gauge group, examples of numerical solutions for different values of $g$ are given in figures \ref{15_CP2_SO(2)d_SO(32)gg_flows} and \ref{15_CH2_SO(2)d_SO(32)gg_flows} in Appendix \ref{AppC_2_2}. We have omitted the $t\times \mathbb{R}^4$-sliced solutions with $k=0$ and $p_2=p_1=0$ since, in this case, the solutions are essentially flat domain walls. From these figures, we find that only $t\times CP^2$-sliced domain walls with $g\leq-1.5$ (an example of these solutions is given by the purple line in figure \ref{15_CP2_SO(2)d_SO(32)gg_flows}) are uplifted to solutions of type IIA theory with physical IR singularities. All of the solutions are summarized in table \ref{tab8}.

\begin{table}[h!]
\centering
\begin{tabular}{| c | c | c |}
\hline
gauge & Kahler & physical IR \\
group & four-cycles &  singularity \\\hline
$SO(5)$& $CP^2$, $CH^2$ & $\times$\\\hline
$SO(4,1)$& $CP^2$, $CH^2$ & $\times$\\\hline
$SO(3,2)$ & $CP^2$ & $g\leq-1.5$\\
& $CH^2$ & $\times$\\\hline
\end{tabular}
\caption{Summary of satisfaction of the criterion \cite{Maldacena_nogo} for the IR singularities of the solutions interpolating between the locally $SO(2)\times SO(2)$ flat domain wall and  $t\times K_k^4$-sliced curved domain walls with shift scalars in $SO(5)$, $SO(4,1)$, and $SO(3,2)$ gauge groups.}\label{tab8}
\end{table}
%%%%%%%%%%%%%%%%%%%%%%%%%%%%%%%%%%%%%%%%%%%%%%%%%%%%%%%%%%%%%%%%%%%%%%%%%%%%%%%%%%%%%%%%%%%%%%%%%%%%%%%%%%%%%%%%%%%%%%%%%%%%%%%%%%%%%%%%%
\section{Supersymmetric quantum mechanics from $CSO(p,q,4-p-q)\ltimes \mathbb{R}^4$ gauged supergravity}\label{40_Sec}
In this section, we repeat the same analysis for solutions describing D4-branes wrapped on a four-manifold from six-dimensional gauged supergravity with $CSO(p,q,4-p-q)\ltimes \mathbb{R}^4$. This gauge group arises from the embedding tensor in $\overline{\mathbf{40}}^{-1}$ representation of $GL(5)$. The explicit form of the embedding tensor is given by
\begin{equation}\label{40_rep_theta}
\theta^{Am}=\mathbb{T}^A_{np}U^{np,m}\qquad\text{and}\qquad{\theta^{A}}_m\ = \ 0\, .
\end{equation}
We also recall that the tensor $U^{mn,p}$ satisfies the conditions $U^{mn,p}=U^{[mn],p}$ and $U^{[mn,p]}=0$. The quadratic constraint can be solved by taking 
\begin{equation}
U^{mn,p}\ =\ v^{[m}w^{n]p}
\end{equation}
in which $v^m$ is a $GL(5)$ vector, and $w^{mn}$ is a symmetric tensor, $w^{mn}=w^{(mn)}$. 
\\
\indent Using $SL(5)$ symmetry, we can choose the vector $v^m=\delta^m_5$ and split the $SL(5)$ index $m=(i,5)$ with $i=1,\ldots,4$. In addition, we also restrict the discussion to cases with $w^{i5}=w^{55}=0$. With a proper $SL(4)\subset SL(5)$ transformation that leaves $v^m=\delta^m_5$ invariant, we can write $w^{ij}$ as
\begin{equation}
w^{ij}\ =\ \text{diag}(\underbrace{1,..,1}_p,\underbrace{-1,..,-1}_q,\underbrace{0,..,0}_r)
\end{equation}
with $p+q+r=4$. The non-vanishing gauge generators are given by 
\begin{equation}\label{GenCSO(p,q,4-p-q)X}
X_{ij}=\frac{1}{\sqrt{2}}\varepsilon_{ijl_1l_2}w^{l_1l_3}{\boldsymbol{t}^{l_2}}_{l_3}\qquad\textrm{and}\qquad X^i=w^{ij}\boldsymbol{s}_{5j}\, .
\end{equation}
These generators satisfy the following Lie algebra
\begin{equation}
[X^i,X^j]=0,\qquad [X_{ij},X^l]={(X_{ij})_{l'}}^lX^{l'},\qquad [X_{i_1i_2},X_{i_3i_4}]={(X_{i_1i_2})_{i_3i_4}}^{i_5i_6}X_{i_5i_6}
\end{equation}
with ${(X_{i_1i_2})_{i_3i_4}}^{i_5i_6}=2{(X_{i_1i_2})_{[i_3}}^{[i_5}\delta_{i_4]}^{i_6]}$. The gauge group is then given by
\begin{equation}\label{40_gauge_group}
G_0=CSO(p,q,4-p-q)\ltimes \mathbb{R}^{4}=SO(p,q) \ltimes \left(\mathbb{R}^{(p+q)(4-p-q)}\times\mathbb{R}^{4}\right).
\end{equation}
The $CSO(p,q,4-p-q)\subset SL(4)\subset SL(5)$ factor and the four-dimensional translation group $\mathbb{R}^{4}$ are respectively generated by $X_{ij}$ and $X^i$. 
\\
\indent In this case, it is convenient to describe scalar fields in the $SL(4)$ basis. In accordance with the branching $SL(5)\rightarrow SL(4)\times \mathbb{R}^+$, we split the $SO(5)\times SO(5)$ vector indices as $a=(i,5)$ and $\dot{a}=(\dot{i},\dot{5})$. The $SL(5)$ non-compact generators then decompose as
\begin{equation}
\hat{\boldsymbol{t}}_{a\dot{b}}\rightarrow\left(\hat{\boldsymbol{t}}_{i\dot{j}},\,\hat{\boldsymbol{t}}_{i\dot{5}},\,\hat{\boldsymbol{t}}_{5\dot{5}}\right).
\end{equation}
The $\mathbb{R}^+\sim SO(1,1)$ generator is given by
\begin{equation}\label{40repDil}
\widetilde{\mathcal{Y}}_0=\hat{\boldsymbol{t}}_{1\dot{1}}+\hat{\boldsymbol{t}}_{2\dot{2}}+\hat{\boldsymbol{t}}_{3\dot{3}}+\hat{\boldsymbol{t}}_{4\dot{4}}-4\,\hat{\boldsymbol{t}}_{5\dot{5}}
\end{equation}
while the non-compact generators of $SL(4)$ can be written as
\begin{equation}
\overline{\boldsymbol{t}}_{i\dot{j}}=\hat{\boldsymbol{t}}_{i\dot{j}}+\frac{1}{4}\hat{\boldsymbol{t}}_{5\dot{5}}\delta_{i\dot{j}}
\end{equation}
with $\hat{\boldsymbol{t}}_{1\dot{1}}+\hat{\boldsymbol{t}}_{2\dot{2}}+\hat{\boldsymbol{t}}_{3\dot{3}}+\hat{\boldsymbol{t}}_{4\dot{4}}
=-\hat{\boldsymbol{t}}_{5\dot{5}}$ due to the tracelessness of $SL(5)$ generators. The nine scalar fields corresponding to $\overline{\boldsymbol{t}}_{i\dot{j}}$ generators parametrize an $SL(4)/SO(4)\subset SL(5)/SO(5)$ coset. The remaining four scalars associated with $\hat{\boldsymbol{t}}_{i\dot{5}}$ are nilpotent scalars and will be denoted by $b_i$. In addition to the dilaton associated with generator $\boldsymbol{d}$, there are also ten axions corresponding to the antisymmetric shift generators $\boldsymbol{s}_{mn}$ as in the previous section. 
\\
\indent The gauge generators couple to ten gauge fields denoted by ${A_\mu}^{ij}={A_\mu}^{[ij]}$ and $A_{\mu,i}$ in the gauge covariant derivatives 
\begin{equation}
D_\mu=\nabla_\mu-gA_{\mu,i} X^i-g{A_\mu}^{ij} X_{ij}\, .
\end{equation}
We will perform topological twists by turning on gauge fields ${A_\mu}^{ij}$ associated with compact residual symmetries that are subgroups of the $CSO(p,q,4-p-q)$ factor and truncate out all the gauge fields $A_{\mu,i}$. Moreover, since gaugings in $\overline{\mathbf{40}}^{-1}$ representation are purely electric, the three-forms $C_{\mu\nu\rho, A}$ and the magnetic two-forms ${B_{\mu\nu}}^m$ do not appear in the Lagrangian. 
\\
\indent As in the previous section, the gauge fields arising from the topological twist on a four-manifold usually lead to a non-vanishing term in the Bianchi identity \eqref{DefBianchi2} of the form $(\Gamma_m)_{AB}\mathbb{T}^{A}_{i_1i_2}\mathbb{T}^{B}_{i_3i_4}{\mathcal{H}_{[\mu\nu}}^{i_1i_2}{\mathcal{H}_{\rho\sigma]}}^{i_3i_4}$. However, unlike the case of $CSO(p,q,5-p-q)$ gauge group, there are no magnetic two-form fields ${B_{\mu\nu}}^m$ to cancel this term in such a way that the Bianchi identity is satisfied. On the other hand, the Bianchi identity could be recovered by turning on the electric two-form fields $B_{\mu\nu,m}$ to make
\begin{equation}
\mc{D}_{[\mu}\mc{H}_{\nu\rho\sigma],m}=\frac{3}{2\sqrt{2}}(\Gamma_m)_{AB}\mathbb{T}^{A}_{i_1i_2}\mathbb{T}^{B}_{i_3i_4}{\mathcal{H}_{[\mu\nu}}^{i_1i_2}{\mathcal{H}_{\rho\sigma]}}^{i_3i_4}
\end{equation}
However, we are not able to find a consistent set of BPS equations with non-vanishing three-form field strengths. Therefore, in this section, we will set all the tensor fields to zero namely
\begin{equation}
B_{\mu\nu,m}=0,\qquad {B_{\mu\nu}}^m=0,\qquad C_{\mu\nu\rho,A}=0
\end{equation}
and solve the problematic Bianchi identity by requiring the term $(\Gamma_m)_{AB}\mathbb{T}^{A}_{i_1i_2}\mathbb{T}^{B}_{i_3i_4}{\mathcal{H}_{[\mu\nu}}^{i_1i_2}{\mathcal{H}_{\rho\sigma]}}^{i_3i_4}$ to vanish identically. This condition is rather restrictive. In particular, a topological twist on a Riemannian four-manifold $\Sigma_k^4$ by turning on $SO(4)$ gauge fields is not possible. In the following analysis, we will then consider topological twists only on four-dimensional manifolds given by a product of two Riemann surfaces $\Sigma_{k_1}^2\times \Sigma_{k_2}^2$ and a Kahler four-cycle $K_k^4$ in this section. 
%%%%%%%%%%%%%%%%%%%%%%%%%%%%%%%%%%%%%%%%%%%%%%%%%%%%%%%%%%%%%%%%
\subsection{D4-branes wrapped on a product of two Riemann surfaces}\label{40_Sig2xSig2_SO(2)xSO(2)_section}
For solutions describing D4-branes wrapped on a product of two Riemann surfaces, the ansatz for the metric is still given in \eqref{SO(2)xSO(2)6Dmetric}. To cancel the spin connection on $\Sigma^2_{k_1}\times\Sigma^2_{k_2}$, we turn on the following $SO(2)\times SO(2)$ gauge fields
\begin{eqnarray}
{A_{\hat{\zeta}_1}}^{12}&=&-\frac{e^{-V_1}p_{21}}{2\sqrt{2}k_1}\frac{f_{k_1}'(\theta_1)}{f_{k_1}(\theta_1)},\qquad{A_{\hat{\zeta}_2}}^{12}=-\frac{e^{-V_2}p_{22}}{2\sqrt{2}k_2}\frac{f_{k_2}'(\theta_2)}{f_{k_2}(\theta_2)},\nonumber \\
{A_{\hat{\zeta}_1}}^{34}&=&-\frac{e^{-V_1}p_{11}}{2\sqrt{2}k_1}\frac{f_{k_1}'(\theta_1)}{f_{k_1}(\theta_1)},\qquad{A_{\hat{\zeta}_2}}^{34}=-\frac{e^{-V_2}p_{12}}{2\sqrt{2}k_2}\frac{f_{k_2}'(\theta_2)}{f_{k_2}(\theta_2)}\label{40_SO(2)xSO(2)gaugeAnt}
\end{eqnarray} 
with the corresponding two-form field strengths
\begin{eqnarray}
{\mathcal{H}_{\hat{\theta}_1\hat{\zeta}_1}}^{12}&=&\frac{e^{-2V_1}p_{21}}{2\sqrt{2}},\qquad{\mathcal{H}_{\hat{\theta}_2\hat{\zeta}_2}}^{12}=\frac{e^{-2V_2}p_{22}}{2\sqrt{2}},\nonumber\\{\mathcal{H}_{\hat{\theta}_1\hat{\zeta}_1}}^{34}&=&\frac{e^{-2V_1}p_{11}}{2\sqrt{2}},
\qquad{\mathcal{H}_{\hat{\theta}_2\hat{\zeta}_2}}^{34}=\frac{e^{-2V_2}p_{12}}{2\sqrt{2}}\, .
\end{eqnarray}
These lead to a non-vanishing term in the Bianchi identity \eqref{DefBianchi2} of the form 
\begin{equation}\label{40_Sig2xSig2_gamFF}
(\Gamma_m)_{AB}\mathbb{T}^{A}_{i_1i_2}\mathbb{T}^{B}_{i_3i_4}{\mathcal{H}_{\hat{\theta}_1\hat{\zeta}_1}}^{i_1i_2}{\mathcal{H}_{\hat{\theta}_2\hat{\zeta}_2}}^{i_3i_4}=\frac{1}{6\sqrt{2}}(p_{11}p_{22}+p_{12}p_{21})e^{-2 V_1-2 V_2}\delta_m^5\, .
\end{equation}
As previously mentioned, in order for the Bianchi identity to hold, we require that the four magnetic charges satisfy the following relation
\begin{equation}\label{40_Sig2xSig2_ppcon}
p_{11}p_{22}+p_{12}p_{21}=0\, .
\end{equation}
\indent There are two gauge groups that contain an $SO(2)\times SO(2)$ subgroup. These are $SO(4)\ltimes \mathbb{R}^{4}$ and $SO(2,2)\ltimes \mathbb{R}^{4}$ obtained from the embedding tensor with
\begin{equation}\label{SO(2)xSO(2)wij}
w^{ij}=\text{diag}(1,1,\kappa,\kappa)
\end{equation}
for $\kappa=1,-1$, respectively. Under $SO(2)\times SO(2)$ residual symmetry generated by $X_{12}$ and $X_{34}$, there are five singlet scalars with the coset representative of the form
\begin{equation}\label{40_fullSO(2)xSO(2)singlet_coset}
V=e^{\varphi\boldsymbol{d}+\phi_0\widetilde{\mathcal{Y}}_0+\phi\widehat{\mathcal{Y}}_1
+\varsigma_1\widehat{\mathcal{Y}}_2+\varsigma_2\widehat{\mathcal{Y}}_3}\, .
\end{equation}
$\varphi(r)$ is the dilaton corresponding to the $SO(1,1)$ generator $\boldsymbol{d}$ while $\phi_0(r)$ is another dilatonic scalar  associated with the $SO(1,1)$ factor in $SL(5)\rightarrow SL(4)\times SO(1,1)$ branching. The remaining three scalars correspond to the following $SO(5,5)$ non-compact generators
\begin{equation}
\widehat{\mathcal{Y}}_1=\overline{\boldsymbol{t}}_{1\dot{1}}+\overline{\boldsymbol{t}}_{2\dot{2}}-\overline{\boldsymbol{t}}_{3\dot{3}}-\overline{\boldsymbol{t}}_{4\dot{4}},\qquad
\widehat{\mathcal{Y}}_2=\boldsymbol{s}_{12},\qquad
\widehat{\mathcal{Y}}_3=\boldsymbol{s}_{34}\, .
\end{equation}
\indent We now perform an $SO(2)\times SO(2)$ twist by imposing the twist condition \eqref{15_SO(2)xSO(2)_twist_con} together with the projection conditions \eqref{SO(2)xSO(2)ProjCon}, \eqref{SO(2)xSO(2)ProjCon-}, and 
\begin{equation}\label{pureZProj}
\hat{\gamma}_r\epsilon_\pm=\gamma^5\epsilon_\mp
\end{equation}
on the Killing spinors \eqref{DW_Killing_spinor}. Similar to the $SO(2)\times SO(2)$ twist performed in $CSO(p,q,5-p-q)$ gauge group, consistency requires either vanishing shift scalars or additional conditions on the magnetic charges $p_{21}=\kappa p_{11}$ and $p_{22}=\kappa p_{12}$ with non-vanishing shift scalars. However, the latter together with the aforementioned algebraic constraint \eqref{40_Sig2xSig2_ppcon}, leads to $p_{11}p_{12}=0$ which is not compatible with the topological twist. Accordingly, we will not consider the solutions with non-vanishing $\varsigma_1$ and $\varsigma_2$ in the following analysis.
\\
\indent With $\varsigma_1=\varsigma_2=0$, we find the following BPS equations
\begin{eqnarray}
U'&=&\frac{g}{4}e^{\varphi-4(\phi_0+\phi)}(e^{8\phi}+\kappa)\nonumber\\
&&-\frac{1}{8}e^{-\varphi+4\phi_0}\left[e^{-2V_1}(p_{11}e^{-4\phi}+p_{21}e^{4\phi})+e^{-2V_2}(p_{12}e^{-4\phi}+p_{22}e^{4\phi})\right],\qquad\\
V_1'&=&\frac{g}{4}e^{\varphi-4(\phi_0+\phi)}(e^{8\phi}+\kappa)\nonumber\\
&&+\frac{1}{8}e^{-\varphi+4\phi_0}\left[3e^{-2V_1}(p_{11}e^{-4\phi}+p_{21}e^{4\phi})-e^{-2V_2}(p_{12}e^{-4\phi}+p_{22}e^{4\phi})\right],\qquad\\
V_2'&=&\frac{g}{4}e^{\varphi-4(\phi_0+\phi)}(e^{8\phi}+\kappa)\nonumber\\
&&-\frac{1}{8}e^{-\varphi+4\phi_0}\left[e^{-2V_1}(p_{11}e^{-4\phi}+p_{21}e^{4\phi})-3e^{-2V_2}(p_{12}e^{-4\phi}+p_{22}e^{4\phi})\right],\qquad\\
\varphi'&=&-\frac{g}{20}e^{\varphi-4(\phi_0+\phi)}(e^{8\phi}+\kappa)\nonumber\\
&&+\frac{1}{40}e^{-\varphi+4\phi_0}\left[e^{-2V_1}(p_{11}e^{-4\phi}+p_{21}e^{4\phi})+e^{-2V_2}(p_{12}e^{-4\phi}+p_{22}e^{4\phi})\right],\qquad\\
\phi'_0&=&\frac{g}{20}e^{\varphi-4(\phi_0+\phi)}(e^{8\phi}+\kappa)\nonumber\\
&&-\frac{1}{40}e^{-\varphi+4\phi_0}\left[e^{-2V_1}(p_{11}e^{-4\phi}+p_{21}e^{4\phi})+e^{-2V_2}(p_{12}e^{-4\phi}+p_{22}e^{4\phi})\right],\qquad\\
\phi'&=&-\frac{g}{4}e^{\varphi-4(\phi_0+\phi)}(e^{8\phi}-\kappa)\nonumber\\
&&+\frac{1}{8}e^{-\varphi+4\phi_0}\left[e^{-2V_1}(p_{11}e^{-4\phi}-p_{21}e^{4\phi})+e^{-2V_2}(p_{12}e^{-4\phi}-p_{22}e^{4\phi})\right].\qquad
\end{eqnarray}
\indent As in the previous section, we will solve these equations numerically. For $SO(4)\ltimes \mathbb{R}^{4}$ gauge group with $\kappa=1$, we first look for solutions with an asymptotic behavior, as $r\rightarrow+\infty$, given by a locally $SO(4)$ symmetric flat domain wall
\begin{equation}\label{40_SO(2)xSO(2)_full_IR_fDW}
U\sim V_1\sim V_2\sim  \ln \left[\frac{g r}{2}\right],\qquad \varphi\sim-\phi_0\sim-\frac{1}{5}\ln \left[\frac{g r}{2}\right],\qquad\phi\sim \frac{16}{g^4 r^4}\, .
\end{equation}
This behavior implies that $\phi\rightarrow0$ as $r\rightarrow+\infty$, so the $SO(4)$ symmetry is unbroken in this limit. In this case, the parameters $z_1$ and $z_2$, defined in \eqref{two_z_define}, are not independent due to the algebraic condition $p_{11}p_{22}+p_{12}p_{21}=0$ or, equivalently, in terms of $z_1$ and $z_2$, $z_1z_2=k_1k_2$. From the behavior of the ten-dimensional metric component $\hat{g}_{00}$, we find that all solutions involving $S^2$ factors are uplifted to type IIA solutions with unphysical IR singularities. Accordingly, we have omitted the corresponding numerical solutions in these cases. For $t\times \mathbb{R}^2\times \mathbb{R}^2$-sliced domain walls, all solutions but for $z_1=0$ lead to unphysical solutions in type IIA theory. The solutions with $z_1=0$ corresponds to a flat domain wall with $p_{11}=p_{12}=p_{21}=p_{22}=k_1=k_2=0$ due to the twist conditions, so all $t\times \mathbb{R}^2\times \mathbb{R}^2$-sliced solutions have been omitted as well. 
\\
\indent  Examples of numerical solutions for the remaining two cases are given in figures \ref{40_HH_SO(2)xSO(2)_special_SO(4)gg_flows} and \ref{40_HR_SO(2)xSO(2)_special_SO(4)gg_flows} in Appendix \ref{App_D_1} for $g=1$ and different values of $z_1$. For $t\times H^2\times H^2$-sliced domain walls, the solutions with $z_1\approx\pm1$ lead to solutions in type IIA theory with physical IR singularities. On the other hand, solutions with $t\times H^2\times \mathbb{R}^2$ slices admit physical IR singularities for $-1\leq z_1\leq 1$.

Another class of solutions consists of those with an asymptotic behavior taking the form of an $SO(2)\times SO(2)$ symmetric flat domain wall
\begin{equation}\label{40_SO(2)xSO(2)_flat_DW_asym}
U\sim V_1\sim V_2\sim-\phi\sim-\ln \left[-\frac{g \rho}{4}\right]\qquad\text{and}\qquad \varphi\sim-\phi_0\sim\frac{1}{5}\ln \left[-\frac{g \rho}{4}\right]
\end{equation}
with the radial coordinate $\rho$ defined by $\frac{d\rho}{dr}=e^{5 \phi-5 \phi_0}$. We will also choose $g<0$ in order to identify the UV limit with $\rho\rightarrow +\infty$. Solutions of these type can be found in both $SO(4)\ltimes \mathbb{R}^{4}$ and $SO(2,2)\ltimes \mathbb{R}^{4}$ gauge groups. For $SO(4)\ltimes \mathbb{R}^4$ gauge group, all solutions turn out to be unphysical upon uplifted to type IIA theory, so we will not present the corresponding numerical solutions here. For $SO(2,2)\ltimes \mathbb{R}^4$ gauge group, examples of numerical solutions are shown in figures \ref{40_SS_SO(2)xSO(2)_SO(22)gg_flows} to \ref{40_RR_SO(2)xSO(2)_SO(22)gg_flows} in Appendix \ref{App_D_1}. From these figures, we see that many solutions can be uplifted to ten-dimensional solutions with physical IR singularities. We have summarized the conditions on $z_1$ in order for the IR singularities to be physically acceptable in table \ref{tab9}.

\begin{table}[h!]
\centering
\begin{tabular}{| c | c | c | c |}
\hline
UV flat & gauge & two Riemann & physical IR \\
domain wall& group & surfaces &  singularity \\\hline
$SO(4)$&$SO(4)\ltimes \mathbb{R}^{4}$&$S^2\times\Sigma^2$ & $\times$ \\
&&$H^2\times H^2$ & $z_1\approx\pm1$ \\
&&$H^2\times \mathbb{R}^2$ & $-1\leq z_1\leq1$ \\
&&$\mathbb{R}^2\times\mathbb{R}^2$ &  $z_1\approx0$\\\hline
$SO(2)\times SO(2)$&$SO(4)\ltimes \mathbb{R}^{4}$& $\Sigma^2\times\Sigma^2$ & $\times$\\\cline{2-4}
&$SO(2,2)\ltimes \mathbb{R}^{4}$& $S^2\times S^2$ & $0.48\leq z_1\leq2$\\
&& $S^2\times H^2$ & $\times$\\
&& $S^2\times \mathbb{R}^2$ & $z_1\geq0.54$\\
&& $H^2\times H^2$, $H^2\times \mathbb{R}^2$, $\mathbb{R}^2\times \mathbb{R}^2$ & $z_1>0$\\\hline
\end{tabular}
\caption{Summary of satisfaction of the criterion \cite{Maldacena_nogo} for the IR singularities of the solutions with $t\times \Sigma_{k_1}^2\times \Sigma_{k_2}^2$ slices obtained from $SO(2)\times SO(2)$ twist in $SO(4)\ltimes \mathbb{R}^{4}$ and $SO(2,2)\ltimes \mathbb{R}^{4}$ gauge groups.}\label{tab9}
\end{table}

%%%%%%%%%%%%%%%%%%%%%%%%%%%%%%%%%%%%%%%%%%%%
\subsection{D4-branes wrapped on a Kahler four-cycle}\label{40_K4_SO(3)_section}
We now consider D4-branes wrapped on a Kahler four-cycle. Unlike the previous case of $CSO(p,q,5-p-q)$ gauge group, we can not perform an $SU(2)$ twist on a Kahler four-cycle. As previously mentioned, in the absence of tensor fields, the Bianchi identity \eqref{DefBianchi2} requires the $SU(2)_\pm\subset SO(4)$ gauge fields to vanish. 

\subsubsection{Solutions with $SO(3)$ twist}
We first consider an $SO(3)$ twist from $SO(4)\ltimes \mathbb{R}^{4}$, $SO(3,1)\ltimes \mathbb{R}^{4}$, and $CSO(3,0,1)\ltimes \mathbb{R}^{4}$ gauge groups which can be collectively characterized by a symmetric tensor
\begin{equation}\label{40_SO3_w}
w^{ij}=\textrm{diag}(1,1,1,\kappa)
\end{equation}
for $\kappa=1,-1,0$, respectively. The metric ansatz is still given by \eqref{Kahler7Dmetric} with the metric on $K^4_k$ given in \eqref{Kahlermetric}. To cancel the $SU(2)$ part of the spin connection, we turn on $SO(3)$ gauge fields of the form
\begin{equation}\label{40_KahlerSO(3)Ant}
{A_{\hat{1}}}^{14}={A_{\hat{2}}}^{24}={A_{\hat{3}}}^{34}=-\frac{e^{-V}p}{2\sqrt{2}k}\frac{[f'_k(\psi)-1]}{f_k(\psi)}\, .
\end{equation} 
\indent There are five $SO(3)$ singlet scalars consisting of the two dilatons and three additional scalars corresponding to the following non-compact generators 
\begin{equation}
\widetilde{\mathcal{Y}}_1=\overline{\boldsymbol{t}}_{1\dot{1}}+\overline{\boldsymbol{t}}_{2\dot{2}}+\overline{\boldsymbol{t}}_{3\dot{3}}-3\,\overline{\boldsymbol{t}}_{4\dot{4}},\qquad \widetilde{\mathcal{Y}}_2=\hat{\boldsymbol{t}}_{4\dot{5}},\qquad \widetilde{\mathcal{Y}}_3=\boldsymbol{s}_{45}\, .
\end{equation}
The coset representative then takes the form
\begin{equation}\label{40_SO(3)_coset}
V=e^{\varphi\boldsymbol{d}+\phi_0\widetilde{\mathcal{Y}}_0+\phi\widetilde{\mathcal{Y}}_1+b\widetilde{\mathcal{Y}}_2
+\varsigma\widetilde{\mathcal{Y}}_3}\, .
\end{equation}
The gauge two-form field strengths are given by
\begin{equation}
{\mathcal{H}_{\hat{\psi}\hat{1}}}^{14}={\mathcal{H}_{\hat{2}\hat{3}}}^{14}={\mathcal{H}_{\hat{\psi}\hat{2}}}^{24}={\mathcal{H}_{\hat{3}\hat{1}}}^{24}={\mathcal{H}_{\hat{\psi}\hat{3}}}^{34}={\mathcal{H}_{\hat{1}\hat{2}}}^{34}=\frac{e^{-2V}p}{2\sqrt{2}}\, .
\end{equation}
We also note that these non-vanishing components lead to $\mc{H}^{ij}\wedge \mc{H}^{kl}=0$ such that the Bianchi identities are satisfied identically. However, this is not the case for $SU(2)$ or $SO(4)$ twists.
\\
\indent We perform the $SO(3)$ twist by imposing a twist condition $gp=k$ together with the projections \eqref{SO(3)_K4_Projcon}, \eqref{SO(3)_K4_Projcon-}, and \eqref{pureZProj} on the Killing spinors \eqref{DW_Killing_spinor}. However, it turns out that consistency of the BPS equations requires the axionic scalar $b$ to vanish. With $b=0$, the resulting BPS equations are given by
\begin{eqnarray}
U'&=&\frac{g}{8}e^{\varphi-4(\phi_0+3\phi)}(3e^{16\phi}+\kappa)-\frac{3p}{4}e^{-2V-\varphi+4(\phi_0-\phi)},\\
V'&=&\frac{g}{8}e^{\varphi-4(\phi_0+3\phi)}(3e^{16\phi}+\kappa)+\frac{3p}{4}e^{-2V-\varphi+4(\phi_0-\phi)},\\
\varphi'&=&-\frac{g}{40}e^{\varphi-4(\phi_0+3\phi)}(3e^{16\phi}+\kappa)+\frac{3p}{20}e^{-2V-\varphi+4(\phi_0-\phi)},\\
\phi'_0&=&\frac{g}{40}e^{\varphi-4(\phi_0+3\phi)}(3e^{16\phi}+\kappa)-\frac{3p}{20}e^{-2V-\varphi+4(\phi_0-\phi)},\\
\phi'&=&-\frac{g}{8}e^{\varphi-4(\phi_0+3\phi)}(e^{16\phi}-\kappa)+\frac{p}{4}e^{-2V-\varphi+4(\phi_0-\phi)},\\
\varsigma'&=&-g\kappa e^{\varphi-4(\phi_0+3\phi)}\varsigma\, .
\end{eqnarray}
\indent For $CSO(3,0,1)\ltimes \mathbb{R}^{4}$ gauge group with $\kappa=0$, we can analytically solve the BPS equations. In this case, the BPS equations imply $\varsigma'=0$. We will choose the constant value of $\varsigma$ to be zero for simplicity. The other equations give the following solutions for $U$, $\varphi$, and $\phi_0$ as functions of $\phi$
\begin{equation}
U=-3\phi,\qquad \varphi=C+\frac{3}{5}\phi,\qquad \phi_0=C_0-\frac{3}{5}\phi\, .\label{40_K4_SO(3)_CSO301_simsoln}
\end{equation}
As usual, we have neglected an additive integration constant for the warp factor $U$. 
\\
\indent We then consider a linear combination of the BPS equations of the form
\begin{equation}
V'+3\phi'=\frac{3}{2}pe^{-2V-\varphi+4(\phi_0-\phi)}\, .
\end{equation}
Using the results in \eqref{40_K4_SO(3)_CSO301_simsoln} and changing to a new radial coordinate $\rho$ defined by $\frac{d\rho}{dr}=e^{-\phi}$, we find
\begin{equation}
V=\frac{1}{2}\ln \left[3p\rho e^{4C_0}\right]-3\phi-\frac{C}{2}\, .
\end{equation}
With all these results, we finally find the solution for $\phi(\rho)$ of the form
\begin{equation}
\phi=\frac{C_0}{2}-\frac{1}{8}\ln \left[\frac{3p}{5k}e^C \rho +\frac{C_1}{\rho^{2/3}}\right]
\end{equation}
in which $C_1$ is another integration constant. 
\\
\indent As $\rho\rightarrow+\infty$, the asymptotic behavior of the solution is given by 
\begin{eqnarray}\label{40_CP2_CSO301_asym}
U\sim \frac{3}{8} \ln \rho,\qquad V\sim\frac{7}{8} \ln \rho,\qquad \varphi \sim-\phi_0\sim -\frac{3}{40} \ln \rho,\qquad \phi\sim-\frac{1}{8}\ln \rho\, .
\end{eqnarray}
This leads to the six-dimensional metric of the form 
\begin{equation}
ds^2=\rho^{\frac{3}{4}}(-dt^2+\rho^{-1}d\rho^2+\rho ds^2_{K^4_k}).
\end{equation}
\indent As $\rho\rightarrow 0$ and $C_1\neq 0$, we find
\begin{equation}
U\sim-\frac{1}{4} \ln\rho,\qquad V\sim\frac{1}{4}\ln \rho,\qquad \varphi \sim-\phi_0\sim \frac{1}{20} \ln \rho,\qquad\phi\sim\frac{1}{12} \ln \rho
\end{equation}
with the six-dimensional metric given by
\begin{equation}
ds^2=-\rho^{\frac{1}{2}}dt^2+\rho^{\frac{1}{6}}d\rho^2+\rho^{\frac{1}{2}} ds^2_{K^4_k}\, .
\end{equation}
Upon uplifted to type IIA theory, the solution gives the behavior of the ten-dimensional metric component $\hat{g}_{00}$ near this IR singularity of the form
\begin{equation}
\hat{g}_{00}= e^{2 U+2 \varphi-\frac{3}{4}\phi_0-\phi}\sim \frac{1}{\rho ^{107/240}}\rightarrow +\infty
\end{equation}  
which implies that the singularity is unphysical.
\\
\indent On the other hand, for $C_1=0$ and $\rho\rightarrow 0$, we find the asymptotic behavior as given in \eqref{40_CP2_CSO301_asym}. In this case, the behavior of $\hat{g}_{00}$ near this IR singularity is 
\begin{equation}
\hat{g}_{00}= e^{2 U+2 \varphi-\frac{3}{4}\phi_0-\phi}\sim \rho ^{107/160}\rightarrow 0
\end{equation}
indicating that the singularity is physical. 
\\
\indent For other gauge groups, we look for numerical solutions to the BPS equations. For $SO(4)\ltimes \mathbb{R}^{4}$ gauge group with $\kappa=1$, there are solutions with an asymptotic geometry, as $r\rightarrow +\infty$, given by a locally $SO(4)$ symmetric flat domain wall given by
\begin{equation}
U\sim V\sim\ln \left[\frac{g r}{2}\right],\qquad \varphi\sim-\phi_0\sim-\frac{1}{5}\ln \left[\frac{g r}{2}\right],\qquad \phi\sim\frac{16}{g^4r^4},\qquad \varsigma\sim \frac{1}{r^2}\, .\label{40_SO(5)_fDW_asym_fromSO(3)}
\end{equation}
Examples of numerical solutions for this type of solution for different values of $g$ are shown in figures \ref{40_CP2_special_SO(3)_SO(4)gg_flows} and \ref{40_CH2_special_SO(3)_SO(4)gg_flows} in Appendix \ref{App_D_2_1}. We find that all $t\times CH^2$-sliced solutions admit physical IR singularities upon uplifted to type IIA theory. For $t\times CP^2$-sliced domain walls, only solutions with $g\geq1.5$ can lead to physical solutions in type IIA theory. 

We then move to solutions with an asymptotic geometry given by an $SO(3)$ symmetric flat domain wall of the form  
\begin{equation}
U\sim V\sim\phi\sim -\ln \left[-\frac{ g\kappa \rho}{8}\right],\qquad \varphi\sim-\phi_0\sim\frac{1}{5} \ln \left[-\frac{ g\kappa \rho}{8}\right],\qquad\ \varsigma\sim\rho^8\label{40_SO(3)_flat_DW_asym}
\end{equation}
with the new radial coordinate $\rho$ defined by $\frac{d\rho}{dr}=e^{-V-13\phi}$. We will choose $g\kappa<0$ to identify the UV limit with $\rho\rightarrow +\infty$. For $SO(4)\ltimes \mathbb{R}^4$ gauge group, all solutions are uplifted to ten-dimensional solutions with unphysical IR singularities, so we will not give the numerical solutions in this case. For $SO(3,1)\ltimes \mathbb{R}^4$ gauge group, examples of numerical solutions for different values of $g$ are shown in figures \ref{40_S3_SO(3)_SO(31)gg_flows} and \ref{40_H3_SO(3)_SO(31)gg_flows} in Appendix \ref{App_D_2_1}. In this case, all $t\times CH^2$-sliced domain walls can be uplifted to type IIA solutions with physical IR singularities. For $t\times CP^2$-sliced domain walls, only the solutions represented by the purple and brown curves in figure \ref{40_S3_SO(3)_SO(31)gg_flows} lead to physical ten-dimensional solutions. All of the solutions in this case are summarized in table \ref{tab10}.

\begin{table}[h!]
\centering
\begin{tabular}{| c | c | c  | c |}
\hline
UV flat&  \multirow{2}{*}{gauge group} & Kahler  & physical IR \\
domain wall& & four-cycle  &  singularity \\\hline
$SO(4)$ & $SO(4)\ltimes \mathbb{R}^{4}$& $CP^2$& $g\geq1.5$ \\
&& $CH^2$ & any $g$ \\\hline
$SO(3)$ &$SO(4)\ltimes \mathbb{R}^{4}$& $CP^2$, $CH^2$  & $\times$ \\\cline{2-4}
&$SO(3,1)\ltimes \mathbb{R}^{4}$ & $CP^2$ & $g\geq2.56$ \\
&& $CH^2$ & any $g$ \\\cline{2-4}
&$CSO(3,0,1)\ltimes \mathbb{R}^{4}$ & $CP^2$, $CH^2$& $C_1=0$ \\
&(analytic) &  &   \\\hline
\end{tabular}
\caption{Summary of satisfaction of the criterion \cite{Maldacena_nogo} for the IR singularities of the $t\times K_k^4$-sliced curved domain walls obtained from $SO(3)$ twist in $SO(4)\ltimes \mathbb{R}^{4}$, $SO(3,1)\ltimes \mathbb{R}^{4}$, and $CSO(3,0,1)\ltimes \mathbb{R}^{4}$ gauge groups.}\label{tab10}
\end{table}
%%%%%%%%%%%%%%%%%%%%%%%%%%%%%%%%%%%%%%%%%%%%
\subsubsection{Solutions with $SO(2)$ twist}\label{40_K4_SO(3)sd_section}
We end this paper by considering D4-branes wrapped on a Kahler four-cycle with an $SO(2)$ twist. To cancel the $U(1)$ part of the spin connection on $K^4_k$ with the metric given by \eqref{U(1)Kahlermetric}, we turn on an $SO(2)$ gauge field of the form 
\begin{equation}\label{40_KahlerSO(2)gaugeAnt}
{A_{\hat{3}}}^{12}=-\frac{3e^{-V}p}{2\sqrt{2}}\psi\, .
\end{equation}  
All gauge groups with an $SO(2)$ subgroup are collectively characterized by the following component of the embedding tensor
\begin{equation}\label{40_SO(2)w}
w^{ij}=\text{diag}(1,1,\kappa,\lambda).
\end{equation}
\indent Apart from the two dilatons, there are additional nine $SO(2)$ singlet scalars. Three scalars are in the $SL(4)/SO(4)$ coset with non-compact generators
\begin{equation}
\overline{\mathcal{Y}}_1=\overline{\boldsymbol{t}}_{1\dot{1}}+\overline{\boldsymbol{t}}_{2\dot{2}}-\overline{\boldsymbol{t}}_{3\dot{3}}-\overline{\boldsymbol{t}}_{4\dot{4}},\qquad
\overline{\mathcal{Y}}_2=\overline{\boldsymbol{t}}_{3\dot{4}},\qquad
\overline{\mathcal{Y}}_3=\overline{\boldsymbol{t}}_{3\dot{3}}-\overline{\boldsymbol{t}}_{4\dot{4}}\, .
\end{equation}
The other two scalars correspond to two nilpotent generators 
\begin{equation}
\overline{\mathcal{Y}}_4=\hat{\boldsymbol{t}}_{3\dot{5}},\qquad
\overline{\mathcal{Y}}_5=\hat{\boldsymbol{t}}_{4\dot{5}},
\end{equation}
and the remaining four scalars are associated with the following generators
\begin{equation}
\overline{\mathcal{Y}}_6=\boldsymbol{s}_{12},\qquad
\overline{\mathcal{Y}}_7=\boldsymbol{s}_{35},\qquad
\overline{\mathcal{Y}}_8=\boldsymbol{s}_{45},\qquad
\overline{\mathcal{Y}}_9=\boldsymbol{s}_{34}\, .
\end{equation}
The analysis turns out to be highly complicated, so we will perform a subtruncation by setting the shift and nilpotent scalars corresponding to $\boldsymbol{s}_{12}$, $\hat{\boldsymbol{t}}_{3\dot{5}}$, and $\hat{\boldsymbol{t}}_{4\dot{5}}$ to zero. This is a consistent truncation and still gives interesting results. In addition, consistency of the BPS equations also requires the vanishing of the shift scalar corresponding to $\boldsymbol{s}_{35}$. We then end up with seven scalars and the coset representative
\begin{equation}\label{40SO(2)FullCoset}
V=e^{\varphi\boldsymbol{d}+\phi_0\widetilde{\mathcal{Y}}_0+\phi_1\overline{\mathcal{Y}}_1
+\phi_2\overline{\mathcal{Y}}_2+\phi_3\overline{\mathcal{Y}}_3+\varsigma_1\overline{\mathcal{Y}}_8+\varsigma_2\overline{\mathcal{Y}}_9}\, .
\end{equation} 
\indent With a twist condition $gp=k$, and the projector \eqref{pureZProj} together with
\begin{equation}\label{SO(2)KahlerProj}
\hat{\gamma}_{\hat{\psi}\hat{3}}\epsilon_{+\alpha}=\hat{\gamma}_{\hat{1}\hat{2}}\epsilon_{+\alpha}={(\gamma_{12})_\alpha}^\beta\epsilon_{+\beta}\quad\text{ and }\quad \hat{\gamma}_{\hat{\psi}\hat{3}}\epsilon_{-\dot{\alpha}}=\hat{\gamma}_{\hat{1}\hat{2}}\epsilon_{-\dot{\alpha}}={(\gamma_{\dot{1}\dot{2}})_{\dot{\alpha}}}^{\dot{\beta}}\epsilon_{-\dot{\beta}}
\end{equation}
on the Killing spinors
\begin{equation}\label{40_SO(2)_Killing_spinors}
\epsilon_+=e^{\frac{U(r)}{2}+W(r)\gamma_{34}}\epsilon_{+}^{0}\qquad\text{ and }\qquad\epsilon_-=e^{\frac{U(r)}{2}+W(r)\gamma_{\dot{3}\dot{4}}}\epsilon_{-}^{0},	
\end{equation}
the resulting BPS equations read
\begin{eqnarray}
U'&=&\frac{g}{8}e^{\varphi-4(\phi_0+\phi_1)}\left[2e^{8\phi_1}+(\kappa+\lambda)\cosh{2\phi_2}\cosh{4\phi_3}+(\kappa-\lambda)\sinh{4\phi_3}\right]\nonumber\\&&-\frac{3p}{2}e^{-2V-\varphi+4(\phi_0-\phi_1)},\\
V'&=&\frac{g}{8}e^{\varphi-4(\phi_0+\phi_1)}\left[2e^{8\phi_1}+(\kappa+\lambda)\cosh{2\phi_2}\cosh{4\phi_3}+(\kappa-\lambda)\sinh{4\phi_3}\right]\nonumber\\&&+\frac{3p}{2}e^{-2V-\varphi+4(\phi_0-\phi_1)},\\
\varphi'&=&-\frac{g}{40}e^{\varphi-4(\phi_0+\phi_1)}\left[2e^{8\phi_1}+(\kappa+\lambda)\cosh{2\phi_2}\cosh{4\phi_3}+(\kappa-\lambda)\sinh{4\phi_3}\right]\nonumber\\&&+\frac{3p}{10}e^{-2V-\varphi+4(\phi_0-\phi_1)},\qquad\ \\
\phi'_0&=&\frac{g}{40}e^{\varphi-4(\phi_0+\phi_1)}\left[2e^{8\phi_1}+(\kappa+\lambda)\cosh{2\phi_2}\cosh{4\phi_3}+(\kappa-\lambda)\sinh{4\phi_3}\right]\nonumber\\&&-\frac{3p}{10}e^{-2V-\varphi+4(\phi_0-\phi_1)},\\
\phi_1'&=&-\frac{g}{8}e^{\varphi-4(\phi_0+\phi_1)}\left[2e^{8\phi_1}-(\kappa+\lambda)\cosh{2\phi_2}\cosh{4\phi_3}-(\kappa-\lambda)\sinh{4\phi_3}\right]\nonumber\\&&+\frac{3p}{2}e^{-2V-\varphi+4(\phi_0-\phi_1)},\\
\phi_2'&=&-\frac{g}{2}e^{\phi_0-4(\phi_0+\phi_1)}(\kappa+\lambda)\sinh{2\phi_2}\,\text{sech}\,{4\phi_3},\label{40SO(2)GenBPS6}\\
\phi_3'&=&-\frac{g}{4}e^{\phi_0-4(\phi_0+\phi_1)}\left((\kappa+\lambda)\cosh{2\phi_2}\sinh{4\phi_3}+(\kappa-\lambda)\cosh{4\phi_3}\right)\label{40SO(2)GenBPS7}
\end{eqnarray}
together with 
\begin{eqnarray}
\varsigma'_1&=&-\frac{ge^{\varphi+4\phi_3}}{2e^{4(\phi_0+\phi_1)}}\left[\varsigma_1\left(\kappa-\lambda+(\kappa+\lambda)\cosh{2\phi_2}\right)+\varsigma_2(\kappa+\lambda)\sinh{2\phi_2}\,\text{sech}\,{4\phi_3}\right],\nonumber\\ \label{40SO(2)GenBPS8}\\
\varsigma'_2&=&-\frac{ge^{\varphi-4\phi_3}}{2e^{4(\phi_0+\phi_1)}}\left[\varsigma_1(\kappa+\lambda)\sinh{2\phi_2}\,\text{sech}\,{4\phi_3}-\varsigma_2\left(\kappa-\lambda-(\kappa+\lambda)\cosh{2\phi_2}\right)\right],\nonumber\\\label{40SO(2)GenBPS9}\\
W'&=&-\frac{g}{4}e^{\phi_0-4(\phi_0+\phi_1)}(\kappa+\lambda)\sinh{2\phi_2}\tanh{4\phi_3}\, .\label{40SO(2)GenBPS10}
\end{eqnarray}
We will not consider solutions with $k=0$ since the twist condition essentially implies $p=0$ leading to flat domain wall solutions. 
\\
\indent For $CSO(2,0,2)\ltimes \mathbb{R}^4$ gauge group with $\kappa=\lambda=0$, we can analytically solve the BPS equations by the same procedure as in the case of $SO(3)$ twist in $CSO(3,0,1)\ltimes \mathbb{R}^{4}$ gauge group. With the new radial coordinate $\rho$ defined by $\frac{d\rho}{dr}=e^{-3\phi_1}$, the resulting solution is given by
\begin{eqnarray}
U&=&-\phi_1,\qquad V=\frac{1}{2}\ln \left[6p\rho e^{4C_0}\right]-\phi_1-\frac{C}{2},\qquad \varphi=C+\frac{\phi_1}{5},\nonumber\\ 
\phi_0&=&C_0-\frac{\phi_1}{5},\qquad \phi_1=\frac{C_0}{2}-\frac{1}{8}\ln \left[\frac{2e^Cg \rho}{3} +\frac{C_1}{\rho^2}\right]
\end{eqnarray}
with $\phi_2=\phi_3=W=\varsigma_1=\varsigma_2=0$. Apart from some numerical factors, this solution is very similar to that of $CSO(3,0,1)\ltimes \mathbb{R}^{4}$ gauge group, so we will only look at the behavior near the IR singularities. 
\\
\indent As $\rho\rightarrow 0$ and $C_1\neq 0$, the solution becomes
\begin{equation}
U\sim-\frac{1}{4} \ln\rho,\qquad V\sim\frac{1}{4}\ln \rho,\qquad \varphi \sim-\phi_0\sim \frac{1}{20} \ln \rho,\qquad\phi_1\sim\frac{1}{4} \ln \rho\, .
\end{equation}
We find the ten-dimensional metric component 
\begin{equation}
\hat{g}_{00}= e^{2 U+2 \varphi-\frac{3}{4}\phi_0-\phi_1}\sim \frac{1}{\rho ^{49/80}}\rightarrow +\infty\, .
\end{equation}  
Therefore, the solution has an unphysical singularity.
\\
\indent As $\rho\rightarrow 0$ and $C_1=0$, we find 
\begin{eqnarray}
U\sim \frac{1}{8} \ln \rho,\qquad V\sim\frac{5}{8} \ln \rho,\qquad \varphi \sim-\phi_0\sim -\frac{1}{40} \ln \rho,\qquad \phi_1\sim-\frac{1}{8}\ln \rho
\end{eqnarray}
and
\begin{equation}
\hat{g}_{00}= e^{2 U+2 \varphi-\frac{3}{4}\phi_0-\phi_1}\sim \rho ^{49/160}\rightarrow 0\, .
\end{equation}
Accordingly, in this case, the IR singularity is physically acceptable, and the solution can be interpreted as a twisted compactification of five-dimensional Yang-Mills theory on $K^4_k$ to supersymmetric quantum mechanics in the IR.
\\
\indent For other gauge groups, we will look for numerical solutions to the BPS equations. All solutions with an asymptotic geometry given by an $SO(4)$ symmetric flat domain wall from $SO(4)\ltimes \mathbb{R}^4$ gauge group lead to uplifted ten-dimensional solutions with unphysical IR singularities. Therefore, we will not present these solutions here. Another class of solutions is given by those asymptotic to a locally $SO(2)$ symmetric flat domain wall of the form
\begin{equation}\label{40_SO(2)_flat_DW_IR_asym}
U\sim V\sim-5\varphi\sim5\phi_0\sim\frac{1}{2}\ln \rho-\ln \left[-\frac{g\rho}{2}\right],\qquad  \phi_1\sim\frac{1}{2}\ln \rho
\end{equation}
with $\phi_2=\phi_3=\varsigma_1=\varsigma_2=0$ and the new radial coordinate $\rho$ defined by $\frac{d\rho}{dr}=e^{5\phi_1-10\phi_0}$. It turns out that, in this case, all solutions have $\phi_2=\varsigma_1=\varsigma_2=0$ along the entire flow solutions. Moreover, all solutions in $SO(4)\ltimes \mathbb{R}^4$ and $CSO(3,0,1)\ltimes \mathbb{R}^4$ gauge groups are uplifted to type IIA solutions with unphysical IR singularities, so we will omit numerical solutions for all of these solutions. 
\\
\indent For $SO(3,1)\ltimes \mathbb{R}^4$, $SO(2,2)\ltimes \mathbb{R}^4$, and $CSO(2,1,1)\ltimes \mathbb{R}^4$ gauge groups, examples of numerical solutions are given in figures \ref{40_CP2_SO(2)_SO(31)gg_flows} to \ref{40_CH2_SO(2)_CSO(211)gg_flows} in Appendix \ref{App_D_2_2}. We find that all $t\times CP^2$-sliced solutions can be uplifted to solutions of type IIA theory with physical IR singularities. On the other hand, $t\times CH^2$-sliced solutions lead to ten-dimensional solutions with physical IR singularities only for large $|g|$. We collect all these conditions in tables \ref{tab11}. We also point out that the result is very similar to solutions with $SO(2)$ twist on a Riemann surface studied in \cite{6D_twist_I}.

\begin{table}[h!]
\centering
\begin{tabular}{| c | c | c | c |}
\hline
UV flat&  \multirow{2}{*}{gauge group} & Kahler  & physical IR \\
domain wall& & four-cycle  &  singularity \\\hline
$SO(4)$ & $SO(4)\ltimes\mathbb{R}^4$& $CP^2$, $CH^2$ & $\times$ \\\hline
$SO(2)$ &$SO(4)\ltimes\mathbb{R}^4$& $CP^2$, $CH^2$ & $\times$ \\\cline{2-4}
&$SO(3,1)\ltimes\mathbb{R}^4$& $CP^2$ & any $g$ \\
& & $CH^2$ & $g<-2.5$ \\\cline{2-4}
&$SO(2,2)\ltimes\mathbb{R}^4$& $CP^2$ & any $g$ \\
& & $CH^2$ & $g<-5$ \\\cline{2-4}
&$CSO(3,0,1)\ltimes\mathbb{R}^4$& $CP^2$, $CH^2$ & $\times$ \\\cline{2-4}
&$CSO(2,1,1)\ltimes\mathbb{R}^4$& $CP^2$ & any $g$ \\
& & $CH^2$ & $g<-2.5$ \\\cline{2-4}
&$CSO(2,0,2)\ltimes\mathbb{R}^4$ & $CP^2$, $CH^2$ & $C_1=0$ \\
&(analytic) &  & \\\hline
\end{tabular}
\caption{Summary of satisfaction of the criterion \cite{Maldacena_nogo} for the IR singularities of the $t\times K_k^4$-sliced curved domain walls obtained from $SO(3)$ twist in $SO(4)\ltimes\mathbb{R}^4$, $SO(3,1)\ltimes\mathbb{R}^4$, $SO(2,2)\ltimes\mathbb{R}^4$, $CSO(3,0,1)\ltimes\mathbb{R}^4$, $CSO(2,1,1)\ltimes\mathbb{R}^4$, and $CSO(2,0,2)\ltimes\mathbb{R}^4$  gauge groups.}\label{tab11}
\end{table}
%%%%%%%%%%%%%%%%%%%%%%%%%%%%%%%%%%%%%%%%%%%%%%%%%%%%%%%%%%%%%%%%%%%%%%%%%%%%%%%%%%%%%%%%%%%%%%%%%%%%%%%%%%%%%%%%%%%%%%%%%%%%%%%%%%%%%%%%%
\section{Conclusions and discussions}\label{conclusion_sec}
We have studied a large number of supersymmetric solutions interpolating between locally flat domain walls and $t\times\mc{M}_4$-sliced domain walls from maximal gauged supergravity in six dimensions with $CSO(p,q,5-p-q)$ and $CSO(p,q,4-p-q)\ltimes \mathbb{R}^4$ gauged groups. The four-manifold with constant curvature $\mc{M}_4$ under consideration here is given by a Riemannian four-manifold, a product of two Riemann surfaces and a Kahler four-cycle. By performing different types of topological twists, various amounts of supersymmetry are preserved along the entire solutions. These solutions are expected to give holographic duals of RG flows across dimensions from the maximal SYM theories in five dimensions dual to the locally flat domain wall to supersymmetric quantum mechanics in the IR. 
\\
\indent Upon uplifted to type IIA theory, many of these solutions possess physically acceptable IR singularities. Therefore, these solutions could be interpreted as D4-branes wrapped on $\mc{M}_4$ giving rise to supersymmetric quantum mechanics via twisted compactifications of the maximal five-dimensional SYM on the world-volume of D4-branes. Due to recent interests in the holographic study of matrix quantum mechanics, we hope the results given in this paper would be useful along this line. Rather than considering supersymmetric quantum mechanics on the world-volume of D0-branes as in many previous works, we have given a new class of gravity duals to supersymmetric quantum mechanics from D4-branes wrapped on four-manifolds. This might lead to some insight into supersymmetric quantum mechanics via twisted compactifications of maximal SYM in five dimensions. With some of the solutions given analytically, we hope the present results could provide a very useful tool to further study along these and related directions.
\\
\indent Due to the relation between six-dimensional gauge supergravity considered here and the maximal gauged supergravity in seven dimensions via an $S^1$ reduction, it is natural to look for a possible relation between supersymmetric solutions with topological twists on $\mc{M}_4$ in both cases. In seven dimensions, a number of $AdS_3\times \Sigma_{k_1}\times \Sigma_{k_2}$ solutions have been found in \cite{2D_Bobev} from $SO(5)$ maximal gauged supergravity. These solutions describe RG flows across dimensions from $N=(2,0)$ SCFT in six dimensions dual to the supersymmetric $AdS_7$ vacuum to $N=(2,0)$ two-dimensional SCFTs in the IR via twisted compactifications on $\Sigma_{k_1}\times \Sigma_{k_2}$. In addition, similar solutions in $CSO(p,q,5-p-q)$ and $CSO(p,q,4-p-q)$ gauge groups have been considered in \cite{7D_max_twist}. Apart from $SO(5)$ gauge group, all of these gauge groups do not lead to supersymmetric $AdS_7$ vacua. The resulting RG flow solutions then interpolate between $N=(2,0)$ non-conformal field theories in six dimensions to conformal or non-conformal field theories in two dimensions. 
\\
\indent It turns out that all of the solutions with physical IR singularities obtained in this paper correspond to seven-dimensional solutions with $AdS_3\times \mc{M}_4$ fixed points. Indeed, from the results of this paper, we have not found solutions with physical singularities only for the case of $\mc{M}_4=S^2\times H^2$ in $SO(5)$ gauge group. However, as shown in \cite{7D_max_twist}, the $AdS_3$ vacua in this case exist only in a small region of the parameter space. The corresponding six-dimensional solutions with physical singularities might exist within a very small region of the parameter space, but we have not found these solutions due to the limitation of numerical analysis. According to the result of \cite{DW_from_AdS}, an $S^1$ compactification of $AdS_{d+1}$ spaces leads to domain walls in $d$ dimensions. We then expect that the flow solutions with physical IR singularities are those obtained from the corresponding flow solutions to $AdS_3$ fixed points in seven dimensions.           
\\
\indent It would be interesting to perform a holographic study of the resulting supersymmetric quantum mechanics as in \cite{2D_MQM_holo} and \cite{deformed_Matrix_henning} using holographic renormalization to compute some correlation functions among the operators dual to various scalar fields in the six-dimensional gauged supergravity. The quantum mechanics dual to some of the solutions found here might be related to certain deformations of the BFSS matrix model which plays an important role in string/M-theory, and it would be interesting to explicitly verify this. On the field theory side, it could be interesting to identify the maximal SYM theories in five dimensions dual to the flat domain walls in the presence of topological twists on $\mc{M}_4$. The dual quantum mechanics could be directly constructed by twisted compactifications of the maximal SYM theories in five dimensions, and it would be interesting to study both the field theory and supergravity sides of the gauge/gravity duality in this case. On the other hand, constructing the complete truncation ansatze of type IIA theory on both $H^{p,q}\times \mathbb{R}^{5-p-q}$ and $H^{p,q}\times \mathbb{R}^{4-p-q}\times S^1$ is also worth considering. This could be used to uplift the solutions found here and in \cite{6D_twist_I} to the complete solutions of type IIA theory in which an interpretation in terms of D4-branes wrapped on compact manifolds of different dimensions can be given.   

\begin{acknowledgments}
P. N. is supported by the faculty research grants from the Faculty of Science, Ramkhamhaeng University.
\end{acknowledgments}
%%%%%%%%%%%%%%%%%%%%%%%%%%%%%%%%%%%%%%%%%%%%%%%%%%%%%%%%%%%%%%%%%%%%%%%%%%%%%%%%%%%%%%%%%%%%%%%%%%%%%%%%%%%%%%%%%%%%%%%%%%%%%%%%%%%%%%%%%
\appendix
\section{Useful formulae}\label{formula}
In this appendix, we collect relevant formulae used in the main text for convenience. All of these relations have been thoroughly described in \cite{6D_DW_I}, so we will omit all the details here. Recall that the branching rule for an $SO(5,5)$ vector representation under $GL(5)$ takes the form
\begin{equation}\label{VecDec}
\mathbf{10}\ \rightarrow\ \mathbf{5}^{+2}\,\oplus\,\overline{\mathbf{5}}^{-2},
\end{equation}
we can decompose the $SO(5,5)$ generators as 
\begin{equation}
\boldsymbol{t}_{MN}\ \rightarrow\ (\boldsymbol{t}_{mn},{\boldsymbol{t}^m}_n,\boldsymbol{t}^{mn})
\end{equation}
with ${\boldsymbol{t}_m}^n=-{\boldsymbol{t}^n}_m$. 
\\
\indent $SO(5,5)$ generators in vector representation can be chosen as
\begin{equation}
{(\boldsymbol{t}_{MN})_P}^Q\ =\ 4\eta_{P[M}\delta^Q_{N]}
\end{equation}
with the corresponding Lie algebra of the form 
\begin{equation}\label{SO(5,5)algebra}
\left[\boldsymbol{t}_{MN},\boldsymbol{t}_{PQ}\right]\ =\ 4(\eta_{M[P}\boldsymbol{t}_{Q]N}-\eta_{N[P}\boldsymbol{t}_{Q]M})\, .
\end{equation}
In terms of the shift generators $\boldsymbol{s}_{mn}=\boldsymbol{t}_{mn}$ and the generators for the hidden symmetries $\boldsymbol{h}^{mn}=\boldsymbol{t}^{mn}$, the $SO(5,5)$ generators can be written as
\begin{equation}
{(\boldsymbol{t}_{MN})_P}^Q=\begin{pmatrix} {\boldsymbol{t}^m}_n & \boldsymbol{h}^{mn} \\
								\boldsymbol{s}_{mn} & -{\boldsymbol{t}^n}_m \end{pmatrix}.
\end{equation}
\indent The spinor and conjugate spinor representations of $SO(5,5)$ decompose as
\begin{equation}
\mathbf{16}_s\ \rightarrow\ \overline{\mathbf{5}}^{+3}\,\oplus\,\mathbf{10}^{-1}\,\oplus\,\mathbf{1}^{-5}\qquad\text{and}\qquad\mathbf{16}_c\ \rightarrow\ \mathbf{5}^{-3}\,\oplus\,\overline{\mathbf{10}}^{+1}\,\oplus\,\mathbf{1}^{+5}\, .
\end{equation}
To concretly realize this decomposition, we introduce the following transformation matrices 
\begin{eqnarray}
\mathbb{T}_{Am}&=&\frac{1}{2\sqrt{2}}(\Gamma_m)_{AB}\boldsymbol{p}^B_{\alpha\beta}\Omega^{\alpha\beta},\label{TranMatTDef1}\\
\mathbb{T}_{A}^{mn}&=&\frac{1}{4\sqrt{2}}{(\Gamma^{mn})_A}^B\boldsymbol{p}_B^{\alpha\beta}\Omega_{\alpha\beta},\label{TranMatTDef2}\\
\mathbb{T}_{A\ast}&=&\frac{1}{10}{({\Gamma^m}_m)_A}^B\boldsymbol{p}_B^{\alpha\beta}\Omega_{\alpha\beta}\, .\label{TranMatTDef3}
\end{eqnarray}
The matrices $\boldsymbol{p}_A^{\alpha\beta}$ and inverse matrices $\boldsymbol{p}^A_{\alpha\beta}$ are given by
\begin{eqnarray}\label{thepmatrix}
\boldsymbol{p}_A^{\alpha\beta}&=&\delta_A^\alpha\delta_1^{\beta}+\delta_A^{\alpha+4}\delta_2^{\beta}+\delta_A^{\alpha+8}
\delta_3^{\beta}+\delta_A^{\alpha+12}\delta_4^{\beta},\nonumber\\
\boldsymbol{p}^A_{\alpha\beta}&=&\delta^A_\alpha\delta^1_{\beta}+\delta^A_{\alpha+4}\delta^2_{\beta}
+\delta^A_{\alpha+8}\delta^3_{\beta}+\delta^A_{\alpha+12}\delta^4_{\beta}\, .
\end{eqnarray}
These matrices satisfy the relations 
\begin{equation}
\boldsymbol{p}_A^{\alpha\beta}\boldsymbol{p}^B_{\alpha\beta}\ =\ \delta_A^B\qquad\text{and}\qquad\boldsymbol{p}_A^{\alpha\delta}\boldsymbol{p}^A_{\beta\gamma}\ =\ \delta^{\alpha}_{\beta}\delta^{\delta}_{\gamma}\, .
\end{equation}
\indent The inverse matrices of $\mathbb{T}_A$ are simply given by their complex conjugation $\mathbb{T}^A=(\mathbb{T}_A)^{-1}=(\mathbb{T}_A)^*$ satisfying the relations
\begin{eqnarray}
\mathbb{T}^{Am}\mathbb{T}_{An}\ =\ \delta^m_n,\qquad\quad\mathbb{T}^{A}_{mn}\mathbb{T}_{A}^{pq}& =& \delta^{[p}_{m}\delta^{q]}_{n},\qquad\mathbb{T}^{A}_\ast\mathbb{T}_{A\ast}\ =\ 1,\nonumber\\
\mathbb{T}^{Am}\mathbb{T}_{Anp}\ =\ 0,\qquad\quad\,\mathbb{T}^{Am}\mathbb{T}_{A\ast}& =& 0,\qquad\quad\mathbb{T}^{A}_{mn}\mathbb{T}_{A\ast}\ =\ 0\label{TT_rel}
\end{eqnarray}
together with
\begin{equation}
\mathbb{T}^{Am}\mathbb{T}_{Bm}+\mathbb{T}^{A}_{mn}\mathbb{T}_{B}^{mn}+\mathbb{T}^{A}_\ast\mathbb{T}_{B\ast}=\delta^A_B\, .
\end{equation}
With all these, we can write the following relations for $\mathbf{16}_s$ and $\mathbf{16}_c$ representations
\begin{equation}
\Psi_A\ =\ \mathbb{T}_{Am}\Psi^m+\mathbb{T}_{A}^{mn}\Psi_{mn}+\mathbb{T}_{A\ast}\Psi_\ast
\end{equation}
with $\Psi_{mn}=\Psi_{[mn]}$ and
\begin{equation}
\Psi^A\ =\ \mathbb{T}^{Am}\Psi_m+\mathbb{T}^{A}_{mn}\Psi^{mn}+\mathbb{T}^{A}_{\ast}\Psi_\ast\, .
\end{equation}
\indent The $SO(5,5)$ gamma matrices $(\Gamma_M)_{AB}$ are related to $SO(5)\times SO(5)$ gamma matrices $\Gamma_{\ul{A}}=(\Gamma_a,\Gamma_{\dot{a}})$ according to the relation
\begin{equation}
(\Gamma_M)_{AB}={{\mathbb{M}}_M}^{\underline{A}}(\Gamma_{\underline{A}})_{AB}
\end{equation}
with
\begin{equation}\label{offDiagTrans}
\mathbb{M}\ =\ \frac{1}{\sqrt{2}}\begin{pmatrix} \mathds{1}_5 & \mathds{1}_5 \\ \mathds{1}_5 & -\mathds{1}_5 \end{pmatrix}.
\end{equation}
The $SO(5,5)$ spinor indices are raised and lowered by the charge conjugation matrix $\boldsymbol{c}_{AB}$ of the form 
\begin{equation}
\boldsymbol{c}_{AB}=\boldsymbol{p}_{A}^{\alpha\dot{\alpha}}\boldsymbol{p}_B^{\beta\dot{\beta}}
\Omega_{\alpha\beta}\Omega_{\dot{\alpha}\dot{\beta}}
\end{equation}
in which $\boldsymbol{p}_{A}^{\alpha\dot{\alpha}}$ and $\boldsymbol{p}^{A}_{\alpha\dot{\alpha}}$ are defined in parallel with \eqref{thepmatrix}. In particular, we have
\begin{equation}
(\Gamma_{\underline{A}})_{AB}={(\Gamma_{\underline{A}})_{A}}^{C}\boldsymbol{c}_{CB}\, .
\end{equation}
\indent On the other hand, the $SO(5)\times SO(5)$ gamma matrices can also be written in terms of $SO(5)$ gamma matrices ${(\gamma_a)_\alpha}^\beta$ and ${(\gamma_{\dot{a}})_{\dot{\alpha}}}^{\dot{\beta}}$ as 
\begin{equation}
{(\Gamma_{\underline{A}})_{A}}^{B}=\left({(\gamma_a)_{A}}^{B},{(\gamma_{\dot{a}})_{A}}^{B}\right)
\end{equation}
with
\begin{eqnarray}
{(\gamma_a)_{A}}^{B}&=&\boldsymbol{p}_A^{\alpha\dot{\alpha}}{(\gamma_a)_{\alpha\dot{\alpha}}}^{\beta\dot{\beta}}\boldsymbol{p}^B_{\beta\dot{\beta}}\ = \ \boldsymbol{p}_A^{\alpha\dot{\alpha}}{(\gamma_a)_\alpha}^\beta\delta_{\dot{\alpha}}^{\dot{\beta}}\boldsymbol{p}^B_{\beta\dot{\beta}},\\
{(\gamma_{\dot{a}})_{A}}^{B}&=&\boldsymbol{p}_A^{\alpha\dot{\alpha}}{(\gamma_{\dot{a}})_{\alpha\dot{\alpha}}}^{\beta\dot{\beta}}\boldsymbol{p}^B_{\beta\dot{\beta}}\ = \ \boldsymbol{p}_A^{\alpha\dot{\alpha}}\delta_\alpha^\beta{(\gamma_{\dot{a}})_{\dot{\alpha}}}^{\dot{\beta}}
\boldsymbol{p}^B_{\beta\dot{\beta}}\, .
\end{eqnarray}
All these relations lead to $(\Gamma_m)_{AB}$ matrices of the form
\begin{equation}\label{GammEx}
(\Gamma_m)_{AB}=\frac{1}{\sqrt{2}}\boldsymbol{p}_A^{\alpha\dot{\alpha}}\left[{(\gamma_m)_\alpha}^\beta\delta_{\dot{\alpha}}^{\dot{\beta}}
+\delta_\alpha^\beta{(\gamma_m)_{\dot{\alpha}}}^{\dot{\beta}}\right]\Omega_{\beta\gamma}\Omega_{\dot{\beta}\dot{\gamma}}
\boldsymbol{p}_B^{\gamma\dot{\gamma}}\, .
\end{equation}
\indent Finally, we can write the explicit form of ${(\Gamma^{mn})_A}^B$ and ${({\Gamma^m}_m)_A}^B$ matrices using the following definitions
\begin{equation}
{(\Gamma^{mn})_A}^B=\frac{1}{2}\left[{(\Gamma^m)_A}^{C}{(\Gamma^n)_{C}}^B-{(\Gamma^n)_A}^{C}{(\Gamma^m)_{C}}^B\right]
\end{equation}
and
\begin{equation}\label{Gammamm}
{({\Gamma^m}_m)_A}^B=\frac{1}{2}\left[{(\Gamma^m)_A}^{C}{(\Gamma_m)_{C}}^B-{(\Gamma_m)_A}^{C}{(\Gamma^m)_{C}}^B\right]
\end{equation}
with
\begin{eqnarray}\label{GammUpDown}
{(\Gamma_m)_A}^B&=&\frac{1}{\sqrt{2}}\boldsymbol{p}_A^{\alpha\dot{\alpha}}\left[{(\gamma_m)_\alpha}^\beta\delta_{\dot{\alpha}}^{\dot{\beta}}
+\delta_\alpha^\beta{(\gamma_m)_{\dot{\alpha}}}^{\dot{\beta}}\right]\boldsymbol{p}^B_{\beta\dot{\beta}},\\
{(\Gamma^m)_A}^B&=&\frac{1}{\sqrt{2}}\boldsymbol{p}_A^{\alpha\dot{\alpha}}\left[{(\gamma_m)_\alpha}^\beta\delta_{\dot{\alpha}}^{\dot{\beta}}
-\delta_\alpha^\beta{(\gamma_m)_{\dot{\alpha}}}^{\dot{\beta}}\right]\boldsymbol{p}^B_{\beta\dot{\beta}}\, .
\end{eqnarray}
\indent In this paper, we use the following representation of $SO(5)$ gamma matrices
\begin{eqnarray}
\gamma_1&=&-\sigma_2\otimes\sigma_2,\quad\, \gamma_2\ =\ \mathds{1}_2\otimes\sigma_1,\quad
\gamma_3\ =\ \mathds{1}_2\otimes\sigma_3,\nonumber\\ 
\gamma_4&=&\sigma_1\otimes\sigma_2,\qquad \gamma_5\ =\ \sigma_3\otimes\sigma_2
\end{eqnarray}
with the Pauli matrices given by
\begin{equation}
\sigma_1\ = \ \begin{pmatrix} 0 & 1\\ 1 & 0 \end{pmatrix},\qquad \sigma_2\ = \ \begin{pmatrix} 0 & -i\\ i & 0 \end{pmatrix},\qquad\sigma_3\ = \ \begin{pmatrix} 1 & 0\\ 0 & -1 \end{pmatrix}\, .
\end{equation}
%%%%%%%%%%%%%%%%%%%%%%%%%%%%%%%%%%%%%%%%%%%%%%%
\section{Type IIA uplift}\label{g00}
In this appendix, we give the ten-dimensional metric component $\hat{g}_{00}$ in type IIA theory relevant for determining whether a given IR singularity of six-dimensional solutions is physically acceptable upon uplifted to type IIA theory. Although, as previously mentioned, the complete truncations to six-dimensional gauged supergravity with $CSO(p,q,5-p-q)$ and $CSO(p,q,4-p-q)\ltimes \mathbb{R}^4$ gauge groups have not been worked out to date, we can at least find the explicit form of the $(00)$-component of the ten-dimensional metric from the partial results given in \cite{Malek_IIA_IIB} and \cite{Henning_KK}.  

\subsection{The embedding of $CSO(p,q,4-p-q)\ltimes \mathbb{R}^4$ gauged supergravity} 
We begin with a more straightforward case of gauged supergravity with $CSO(p,q,4-p-q)\ltimes \mathbb{R}^4$ gauge group. Using the consistent truncation to seven dimensions of type IIA theory on $H^{p,q}\times \mathbb{R}^{4-p-q}$ given in \cite{Malek_IIA_IIB}, we can perform the standard dimensional-reduction on $S^1$ using the ansatz  
\begin{equation}
ds^2_7=e^{2\varphi}ds^2+e^{-8\varphi}(dz+A)^2\, .\label{S1_reduction_7D_6D}
\end{equation}
The scalar $\varphi$ is the dilaton in the main text while $z$ is the seventh coordinate added to the six-dimensional metric $ds^2$. In the present notation, the $(00)$-component of the ten-dimensional metric is given by 
\begin{equation}
\hat{g}_{00}=e^{-3\phi_0}\Delta^{\frac{1}{4}} g^{(7)}_{00}\, .
\end{equation}
Recall that $\phi_0$ is the dilaton corresponding to the $SO(1,1)$ factor in the decomposition $SL(5)\rightarrow SL(4)\times SO(1,1)$. After a dimensional reduction on $S^1$, we can write the ten-dimensional metric in terms of the six-dimensional one as
\begin{equation}
\hat{g}_{00}=-e^{-3\phi_0}\Delta^{\frac{1}{4}}e^{2U+2\varphi}\, .\label{g00_40}
\end{equation} 
We have also used the metric ansatz with $g_{00}=-e^{2U}$. The warp factor $\Delta$ is defined by 
\begin{equation}
\Delta=m^{ij}\eta_{ik}\eta_{jl}\mu^k\mu^l,\qquad i,j,\ldots=1,2,3,4, 
\end{equation}
in which $m^{ij}$ is the inverse of the $SL(4)$ matrix $m_{ij}=(\widetilde{\mc{V}}\widetilde{\mc{V}}^T)_{ij}$. $\widetilde{\mc{V}}$ is the $SL(4)/SO(4)$ coset representative. The coordinates $\mu^i$ on $H^{p,q}\times \mathbb{R}^{4-p-q}$ satisfy $\mu^i\mu^j\eta_{ij}=1$ with $\eta_{ij}$ being the $CSO(p,q,4-p-q)$ invariant tensor. In all cases, the metric component $\hat{g}_{00}$ takes the same form as \eqref{g00_40} with different warp factor $\Delta$ due to different scalar sectors.
\\
\indent For $SO(2)\times SO(2)$ symmetric scalars, we have
\begin{equation}
\widetilde{\mc{V}}=\textrm{diag}(e^{2\phi},e^{2\phi},e^{-2\phi},e^{-2\phi})
\end{equation}
which gives
\begin{equation}
\Delta=e^{-4\phi}[(\mu^1)^2+(\mu^2)^2]+\kappa e^{4\phi}[(\mu^3)^2+(\mu^4)^2].
\end{equation}
In this equation, we have used the invariant tensor of $SO(4)$ and $SO(2,2)$ given by $\eta_{ij}=\textrm{diag}(1,1,\kappa,\kappa)$ for $\kappa=\pm 1$.
\\
\indent For $SO(3)$ symmetric scalar, the $SL(4)/SO(4)$ coset representative is given by
\begin{equation}
\widetilde{\mc{V}}=\textrm{diag}(e^{2\phi},e^{2\phi},e^{2\phi},e^{-6\phi})
\end{equation}
which gives
\begin{equation}
\Delta=e^{-4\phi}[(\mu^1)^2+(\mu^2)^2+(\mu^3)^2]+\kappa e^{12\phi}(\mu^4)^2
\end{equation}
for $\eta_{ij}=\textrm{diag}(1,1,1,\kappa)$ and $\kappa=0, \pm 1$.
\\
\indent In the case of $SO(2)$ singlet scalars, the coset representative $\widetilde{\mc{V}}$ is much more complicated, so we will restrict ourselves to the subtruncation considered in the main text. In this subtruncation, we have only $\phi_1$, $\phi_2$ and $\phi_3$ non-vanishing and 
\begin{equation}
\widetilde{\mc{V}}=\begin{pmatrix}
					e^{2\phi_1}& 0 &0 &0 \\
					0 & e^{2\phi_1} & 0 & 0 \\
					0&0 &e^{-2\phi_1+2\phi_3}\cosh\phi_2 & e^{-2\phi_1-2\phi_3}\sinh\phi_2 \\
					0&0 &e^{-2\phi_1+2\phi_3}\sinh\phi_2 & e^{-2\phi_1-2\phi_3}\cos\phi_2
					\end{pmatrix}.
\end{equation}
The warp factor is then given by
\begin{eqnarray}
\Delta&=&e^{-4\phi_1}[(\mu^1)^2+(\mu^2)^2]+e^{4\phi_1}\left\{\sinh4\phi_3[\lambda^2(\mu^4)^2-\kappa^2(\mu^3)^2]\right.\nonumber \\
& &\left.+e^{-2\phi_2}\cosh4\phi_3[\lambda^2(\mu^4)^2+\kappa^2(\mu^3)^2]\right\}
\end{eqnarray}
in which we have used $\eta_{ij}=\textrm{diag}(1,1,\kappa,\lambda)$ for $\kappa=0,\pm 1$ and $\lambda=0,\pm 1$.

\subsection{The embedding of $CSO(p,q,5-p-q)$ gauged supergravity}
To find the embedding in type IIA theory in this case, we first recall that the eleven-dimensional metric component $\hat{g}^{(11)}_{00}$ is given via a consistent truncation on $H^{p,q}\times \mathbb{R}^{5-p-q}$ by 
\begin{equation}
\hat{g}^{(11)}_{00}=\Delta^{\frac{1}{3}}g^{(7)}_{00}\, .
\end{equation}
The warp factor is defined by
\begin{equation}
\Delta=\mc{M}^{ab}\eta_{ac}\eta_{bd}\mu^c\mu^d
\end{equation}
with $\eta_{ab}$, $a,b=1,2,\ldots, 5$, being the $CSO(p,q,5-p-q)$ invariant tensor. $\mc{M}^{ab}$ is the inverse of the $SL(5)$ scalar matrix defined in terms of the $SL(5)/SO(5)$ coset representative as
\begin{equation}
\mc{M}_{ab}={\mc{V}_a}^{\underline{a}}{\mc{V}_b}^{\underline{a}}\, .
\end{equation}
Similar to the previous case, the coordinates $\mu^a$ satisfy $\mu^a\mu^b\eta_{ab}=1$. Using the $S^1$ reduction given in \eqref{S1_reduction_7D_6D}, we have 
\begin{equation}
g^{(7)}_{00}=e^{2\varphi}g_{00}
\end{equation}
which in turn gives
\begin{equation}
\hat{g}^{(11)}_{00}=\Delta^{\frac{1}{3}}e^{2\varphi}g_{00}\, .\label{11D_metric}
\end{equation}
\\
\indent As in the truncation on $S^4\times S^1$ leading to six-dimensional gauged supergravity with $SO(5)$ gauge group constructed in \cite{Pope_typeII_S3_S4}, we can perform a dimensional reduction of the eleven-dimensional metric on $S^1$ and find the $(00)$-component of the ten-dimensional metric of the form
\begin{equation}
\hat{g}_{00}=\Omega^{\frac{1}{8}}\Delta^{\frac{1}{3}}e^{2\varphi}g_{00}
\end{equation}
with
\begin{eqnarray}
\Omega=\Delta^{\frac{1}{3}}e^{-8\varphi}+\Delta^{-\frac{2}{3}}m_{ab}\chi^{ac}\chi^{bd}\eta_{ce}\eta_{df}\mu^e\mu^f\, .
\end{eqnarray}
It is also useful to note that the factor $\Omega$ is related to the type IIA dilaton via
\begin{equation}
e^{\Phi}=\Omega^{\frac{3}{4}}\, .
\end{equation}
The axionic scalars $\chi^{ab}=-\chi^{ba}$ arise from the dimensional reduction of the $CSO(p,q,5-p-q)$ gauge fields $A^{ab}_\mu$ in seven dimensions. These scalars transform in representation $\mathbf{10}$ of $SL(5)$ and can be identified with the shift scalars $\varsigma_1,\ldots,\varsigma_{10}$ appearing in \eqref{6D_scalar}.
\\
\indent For $SO(4)$ invariant scalars, we have 
\begin{equation}
\mc{V}=\textrm{diag}(e^{2\phi},e^{2\phi},e^{2\phi},e^{2\phi},e^{-8\phi})
\end{equation}
and $\eta_{ab}=\textrm{diag}(1,1,1,1,\lambda)$ for $\lambda=0,\pm 1$. In this case, all shift scalars vanish leading to 
\begin{equation}
\Omega=e^{-8\varphi}\Delta
\end{equation}
with the wrap factor $\Delta$ given by 
\begin{equation}
\Delta=e^{-4\phi}[(\mu^1)^2+(\mu^2)^2+(\mu^3)^2+(\mu^4)^2]+\lambda^2 e^{16\phi}(\mu^4)^2\, .
\end{equation}
This can also be used in the case of $SU(2)_+$ invariant scalars since all $SU(2)_+$ singlets are essentially $SO(4)$ singlets. 
\\
\indent For $SO(2)\times SO(2)$ singlet scalars, we have 
\begin{equation}
\mc{V}=\textrm{diag}(e^{2\phi_1},e^{2\phi_1},e^{2\phi_2},e^{2\phi_2},e^{-4\phi_1-4\phi_2}),
\end{equation}
and the non-vanishing shift scalars are identified as $\chi^{12}=\varsigma_1$ and $\chi^{34}=\varsigma_2$. These lead to 
\begin{equation}
\Omega=e^{-8\varphi}\Delta^{\frac{1}{3}}+\Delta^{-\frac{2}{3}}\left\{e^{4\phi_1}\varsigma_1^2[(\mu^1)^2+(\mu^2)^2]+e^{4\phi_2}\varsigma_2^2[(\mu^3)^2+(\mu^4)^2]\right\}
\end{equation}
with 
\begin{equation}
\Delta=e^{-4\phi_1}[(\mu^1)^2+(\mu^2)^2]+\kappa e^{-4\phi_2}[(\mu^3)^2+(\mu^4)^2]+\lambda^2e^{8\phi_1+8\phi_2}(\mu^5)^2\, .
\end{equation}
We have also used the $CSO(p,q,5-p-q)$ invariant tensor $\eta_{ab}=\textrm{diag}(1,1,\kappa,\kappa, \lambda)$ with $\kappa=\pm 1$ and $\lambda=0,\pm 1$.
\\
\indent Finally, in the case of $SO(3)$ singlet scalars, we have 
\begin{equation}
\mc{V}=\begin{pmatrix}
					e^{4\phi_1}& 0 &0 &0 & 0 \\
					0& e^{4\phi_1}& 0 &0 &0 \\
					0 & 0 & e^{4\phi_1} & 0 & 0 \\
					0&0 &0 &e^{-6\phi_1+2\phi_3}\cosh\phi_2 & e^{-6\phi_1-2\phi_3}\sinh\phi_2 \\
					0&0  & 0 &e^{-6\phi_1+2\phi_3}\sinh\phi_2 & e^{-6\phi_1-2\phi_3}\cos\phi_2
					\end{pmatrix}.
\end{equation}
For the solutions given in the main text, all shift scalars vanish, so we simply have the warp factor
\begin{equation}
\Omega=e^{-8\varphi}\Delta
\end{equation}
with 
\begin{eqnarray}
\Delta&=&e^{-8\phi_1}[(\mu^1)^2+(\mu^2)^2+(\mu^3)^2]+e^{12\phi_1}\left\{\sinh4\phi_3[\lambda^2(\mu^5)^2-\kappa^2(\mu^4)^2]\right.\nonumber \\
& &\left.+e^{-2\phi_2}\cosh4\phi_3[\lambda^2(\mu^5)^2+\kappa^2(\mu^4)^2]\right\}.
\end{eqnarray}
In the last relation, we have used the invariant tensor $\eta_{ab}=\textrm{diag}(1,1,1,\kappa,\lambda)$ with $\kappa,\lambda=0,\pm 1$.
%%%%%%%%%%%%%%%%%%%%%%%%%%%%%%%%%%%%%%%%%%%%%%%
\section{Numerical solutions from $CSO(p,q,5-p-q)$ gauged supergravity}\label{24_Numer_App}
In this appendix, we collect various numerical solutions from different $CSO(p,q,5-p-q)$ gauge groups.

\subsection{D4-branes wrapped on a product of two Riemann surfaces}
\subsubsection{Solutions with non-vanishing $\varsigma_1$ and $\varsigma_2$}\label{AppC_1_1}
\begin{figure}[h!]
  \centering
    \includegraphics[width=0.8\linewidth]{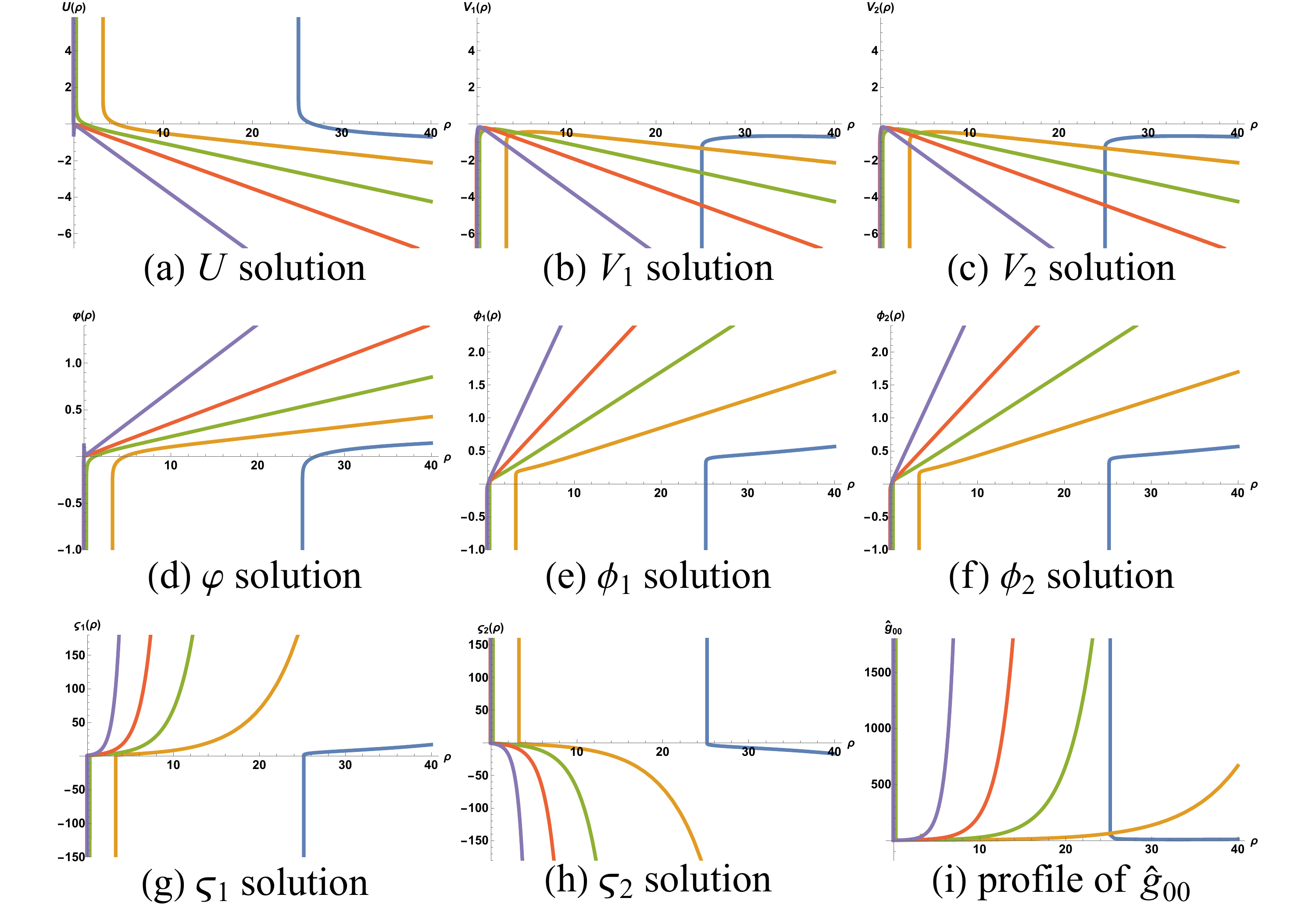}
\caption{Interpolating solutions between the locally $SO(2)\times SO(2)$ flat domain wall as $\rho\rightarrow+\infty$ and $t\times S^2\times S^2$-sliced curved domain walls with shift scalars in $SO(4,1)$ gauge group. The blue, orange, green, red, purple, and black curves refer to $g=0.1, 0.3, 0.6, 1, 2$.}
\label{15_SS_SO(2)d_SO(41)gg_flows}
\end{figure}

\begin{figure}[h!]
  \centering
    \includegraphics[width=0.8\linewidth]{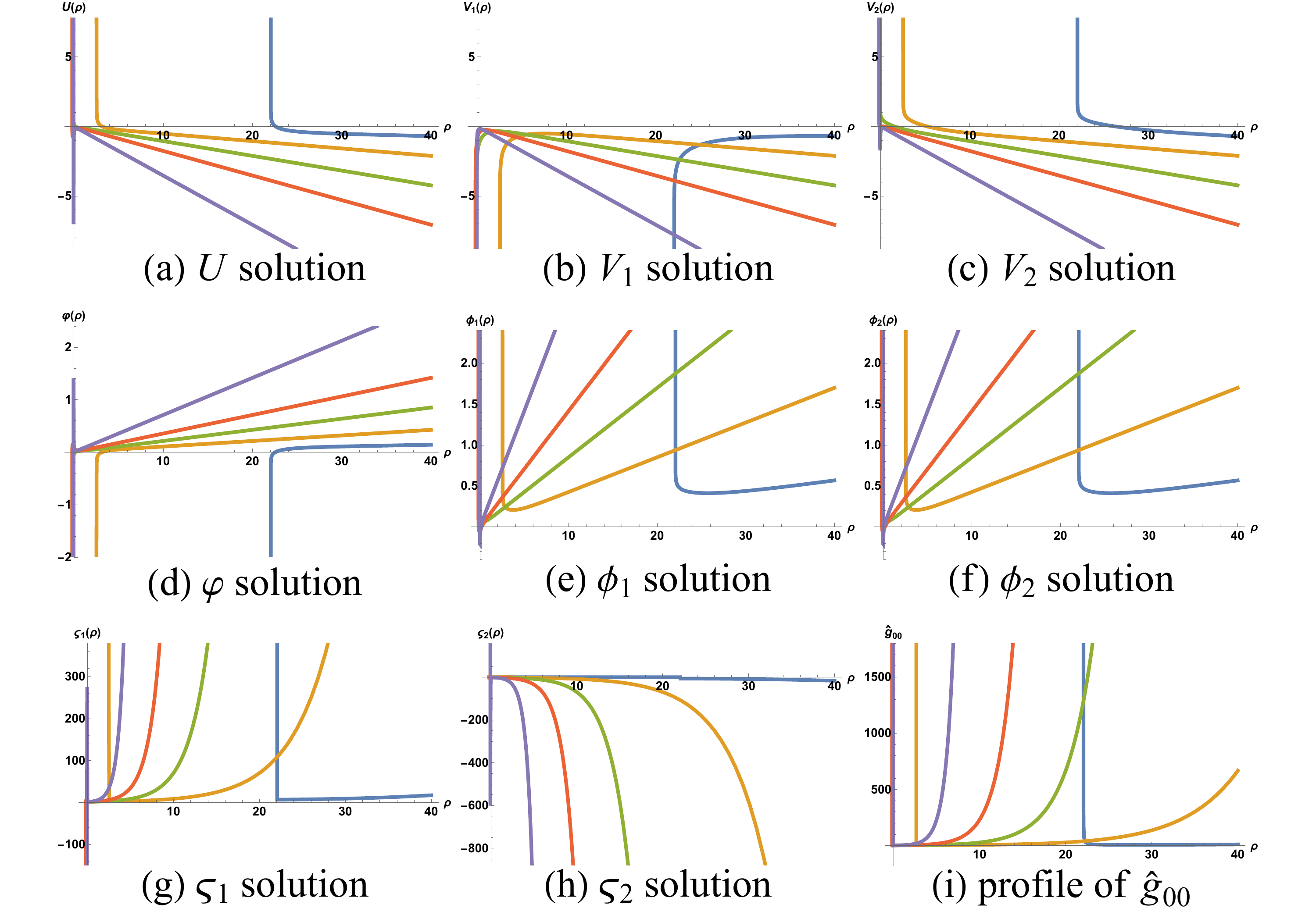}
\caption{Interpolating solutions between the locally $SO(2)\times SO(2)$ flat domain wall as $\rho\rightarrow+\infty$ and $t\times S^2\times H^2$-sliced curved domain walls with shift scalars in $SO(4,1)$ gauge group. The blue, orange, green, red, purple, and black curves refer to $g=0.1, 0.3, 0.6, 1, 2$.}
\label{15_SH_SO(2)d_SO(41)gg_flows}
\end{figure}

\begin{figure}[h!]
  \centering
    \includegraphics[width=0.7\linewidth]{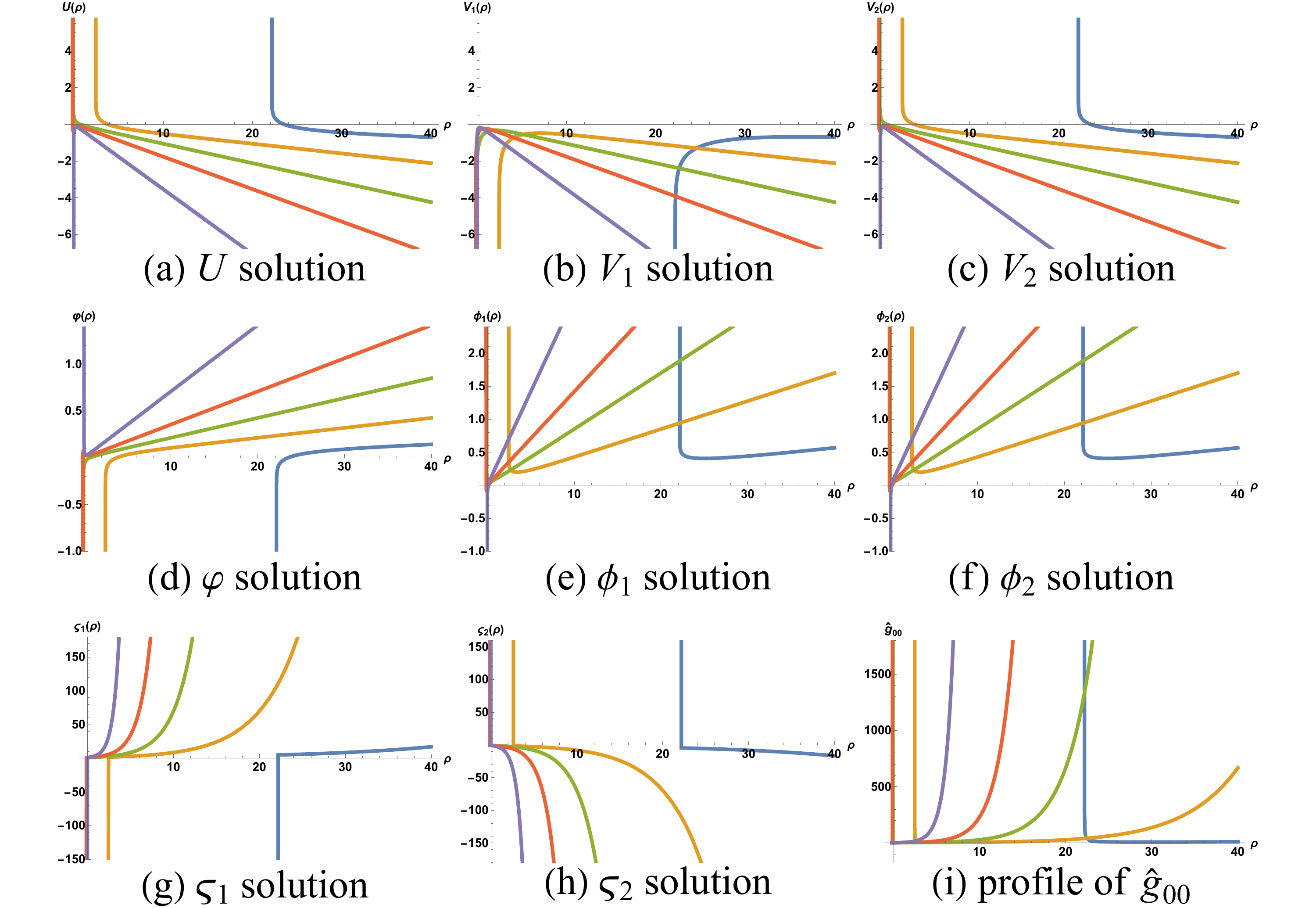}
\caption{Interpolating solutions between the locally $SO(2)\times SO(2)$ flat domain wall as $\rho\rightarrow+\infty$ and $t\times S^2\times \mathbb{R}^2$-sliced curved domain walls with shift scalars in $SO(4,1)$ gauge group. The blue, orange, green, red, purple, and black curves refer to $g=0.1, 0.3, 0.6, 1, 2$.}
\label{15_SR_SO(2)d_SO(41)gg_flows}
\end{figure}

\begin{figure}[h!]
  \centering
    \includegraphics[width=0.7\linewidth]{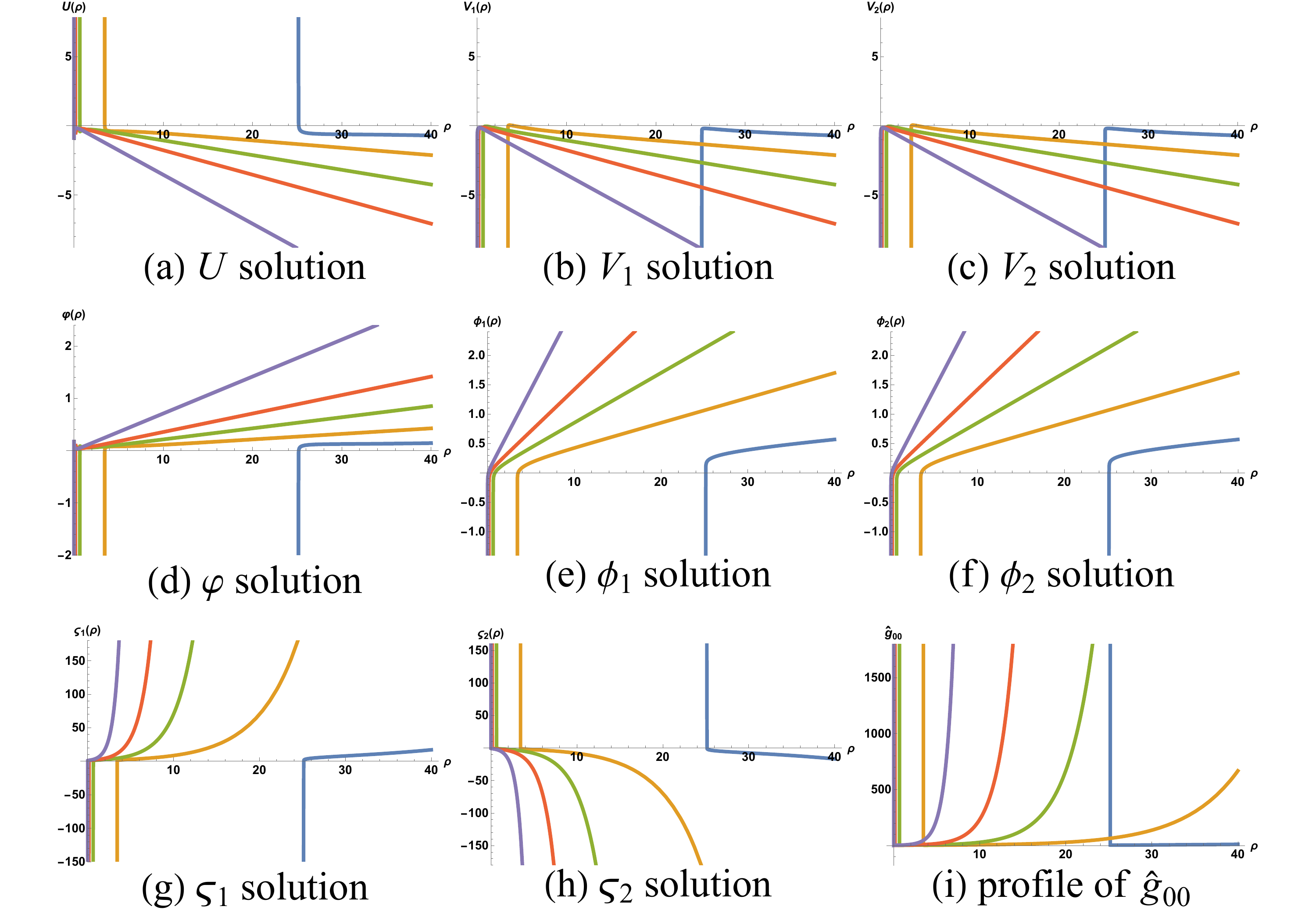}
\caption{Interpolating solutions between the locally $SO(2)\times SO(2)$ flat domain wall as $\rho\rightarrow+\infty$ and $t\times H^2\times H^2$-sliced curved domain walls with shift scalars in $SO(4,1)$ gauge group. The blue, orange, green, red, purple, and black curves refer to $g=0.1, 0.3, 0.6, 1, 2$.}
\label{15_HH_SO(2)d_SO(41)gg_flows}
\end{figure}

\begin{figure}[h!]
  \centering
    \includegraphics[width=0.7\linewidth]{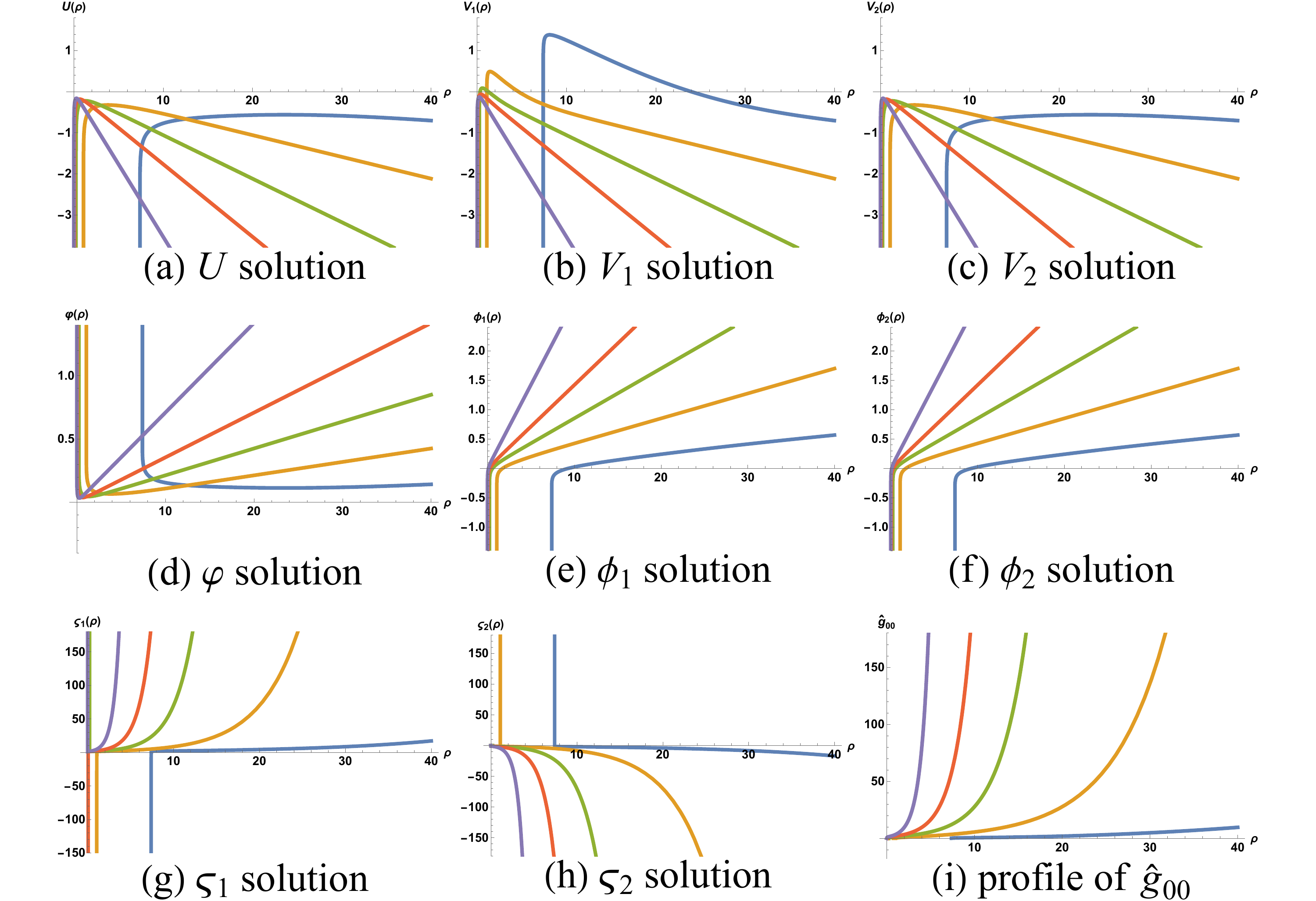}
\caption{Interpolating solutions between the locally $SO(2)\times SO(2)$ flat domain wall as $\rho\rightarrow+\infty$ and $t\times H^2\times \mathbb{R}^2$-sliced curved domain walls with shift scalars in $SO(4,1)$ gauge group. The blue, orange, green, red, purple, and black curves refer to $g=0.1, 0.3, 0.6, 1, 2$.}
\label{15_HR_SO(2)d_SO(41)gg_flows}
\end{figure}

\begin{figure}[p!]
  \centering
    \includegraphics[width=0.7\linewidth]{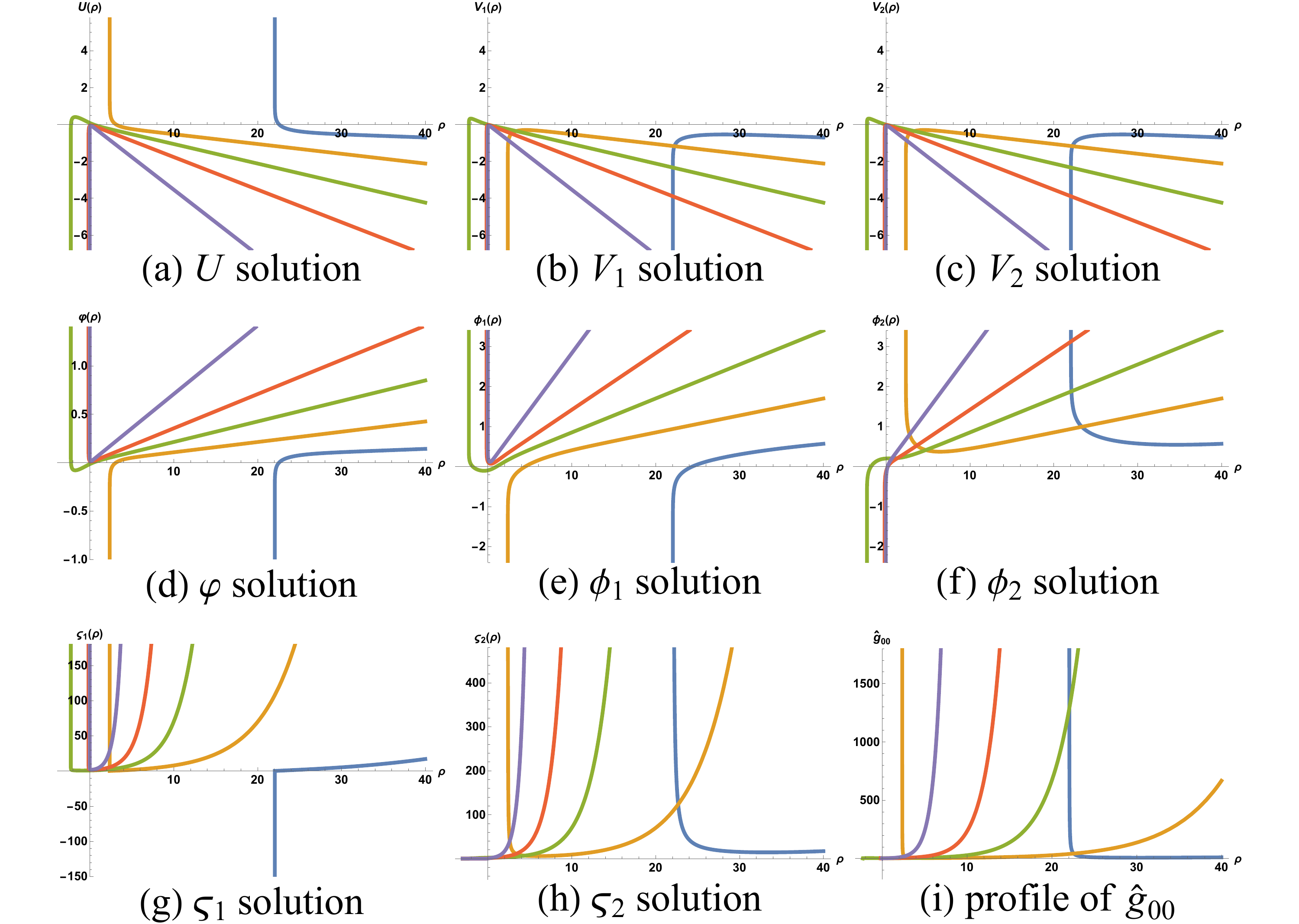}
\caption{Interpolating solutions between the locally $SO(2)\times SO(2)$ flat domain wall as $\rho\rightarrow+\infty$ and $t\times S^2\times S^2$-sliced curved domain walls with shift scalars in $SO(3,2)$ gauge group. The blue, orange, green, red, purple, and black curves refer to $g=-0.1, -0.3, -0.6, -1, -2$.}
\label{15_SS_SO(2)d_SO(32)gg_flows}
\end{figure}

\begin{figure}[p!]
  \centering
    \includegraphics[width=0.7\linewidth]{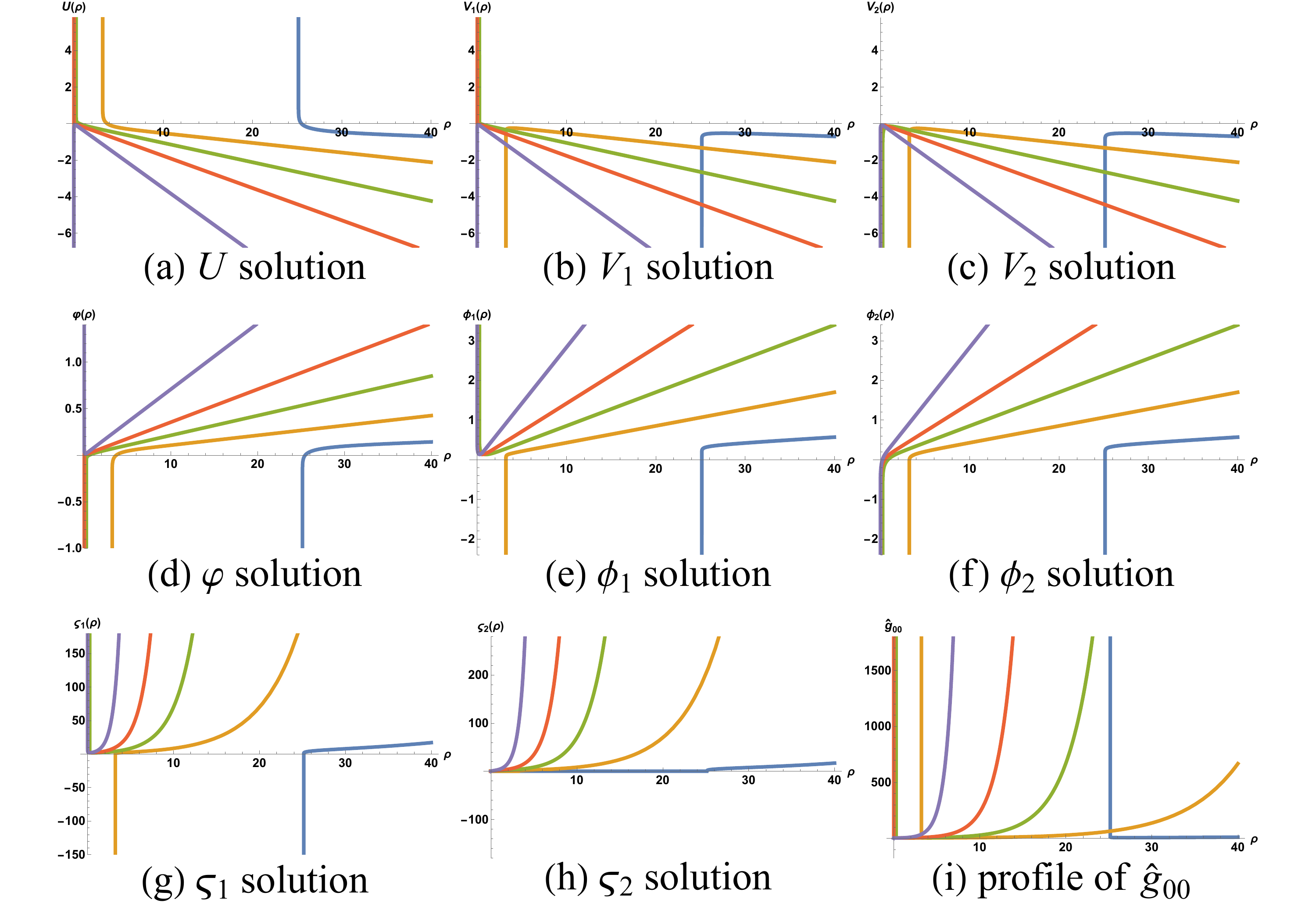}
\caption{Interpolating solutions between the locally $SO(2)\times SO(2)$ flat domain wall as $\rho\rightarrow+\infty$ and $t\times S^2\times H^2$-sliced curved domain walls with shift scalars in $SO(3,2)$ gauge group. The blue, orange, green, red, purple, and black curves refer to $g=-0.1, -0.3, -0.6, -1, -2$.}
\label{15_SH_SO(2)d_SO(32)gg_flows}
\end{figure}

\begin{figure}[h!]
  \centering
    \includegraphics[width=0.7\linewidth]{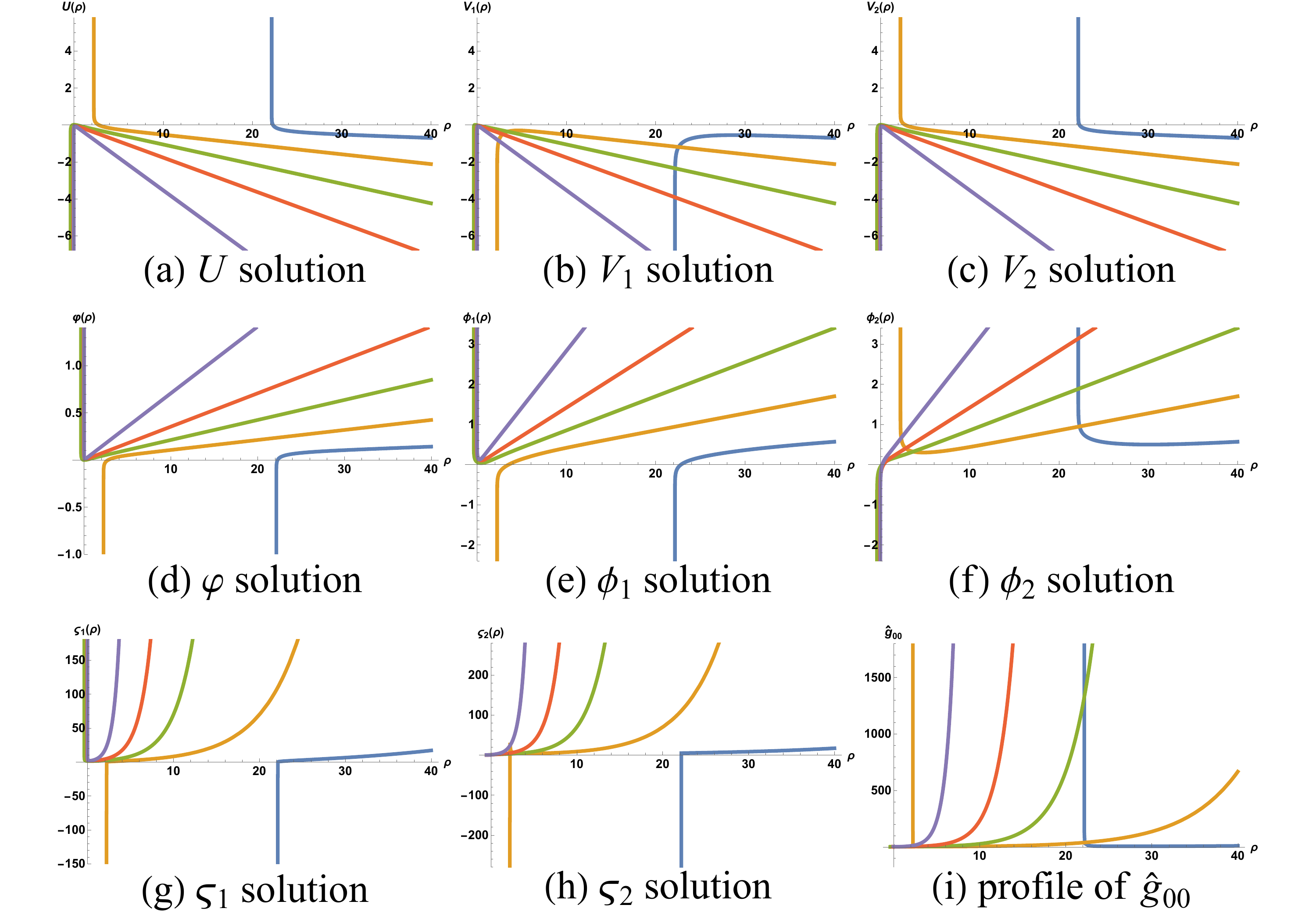}
\caption{Interpolating solutions between the locally $SO(2)\times SO(2)$ flat domain wall as $\rho\rightarrow+\infty$ and $t\times S^2\times \mathbb{R}^2$-sliced curved domain walls with shift scalars in $SO(3,2)$ gauge group. The blue, orange, green, red, purple, and black curves refer to $g=-0.1, -0.3, -0.6, -1, -2$.}
\label{15_SR_SO(2)d_SO(32)gg_flows}
\end{figure}

\begin{figure}[h!]
  \centering
    \includegraphics[width=0.7\linewidth]{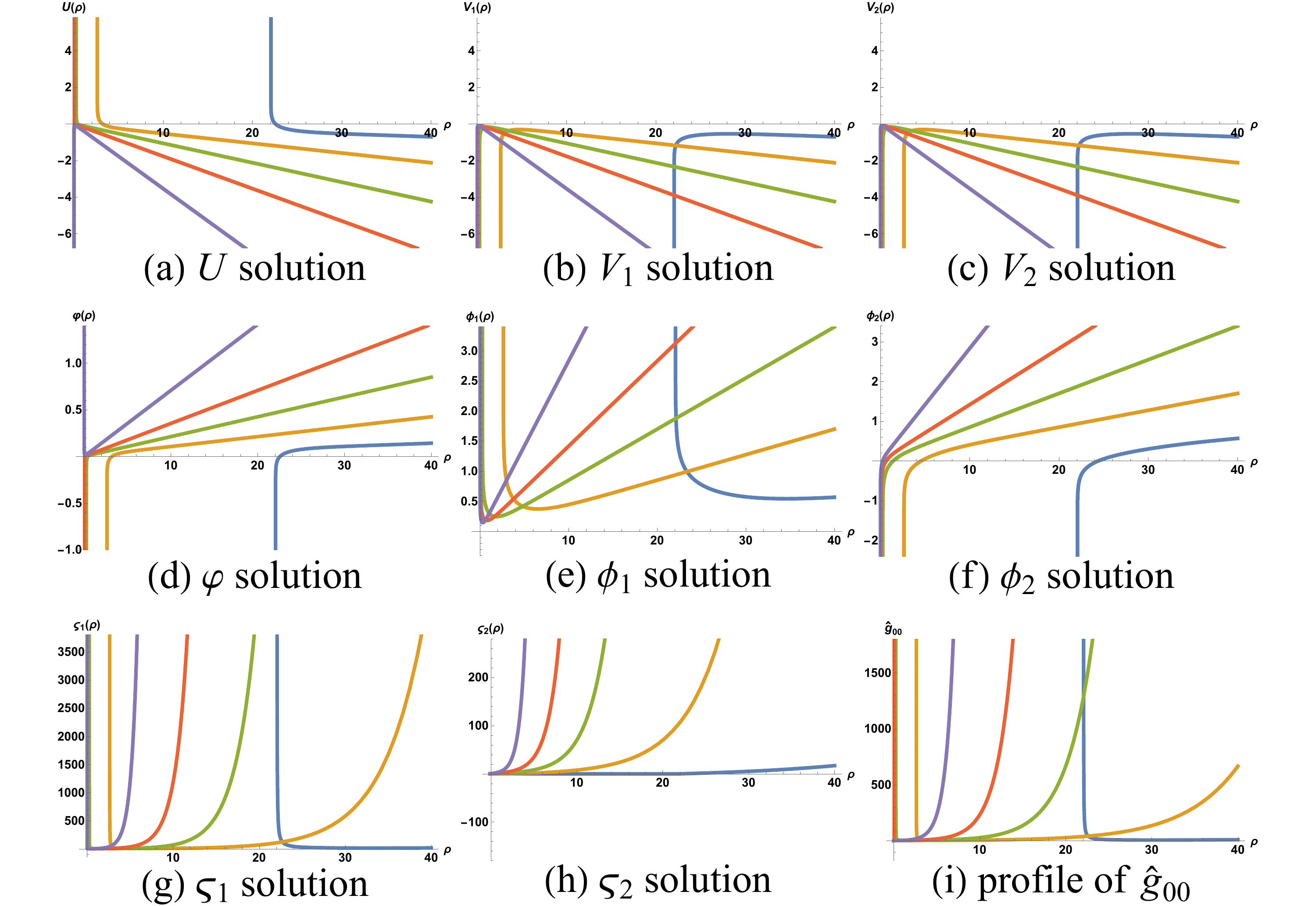}
\caption{Interpolating solutions between the locally $SO(2)\times SO(2)$ flat domain wall as $\rho\rightarrow+\infty$ and $t\times H^2\times H^2$-sliced curved domain walls with shift scalars in $SO(3,2)$ gauge group. The blue, orange, green, red, purple, and black curves refer to $g=-0.1, -0.3, -0.6, -1, -2$.}
\label{15_HH_SO(2)d_SO(32)gg_flows}
\end{figure}

\begin{figure}[h!]
  \centering
    \includegraphics[width=0.7\linewidth]{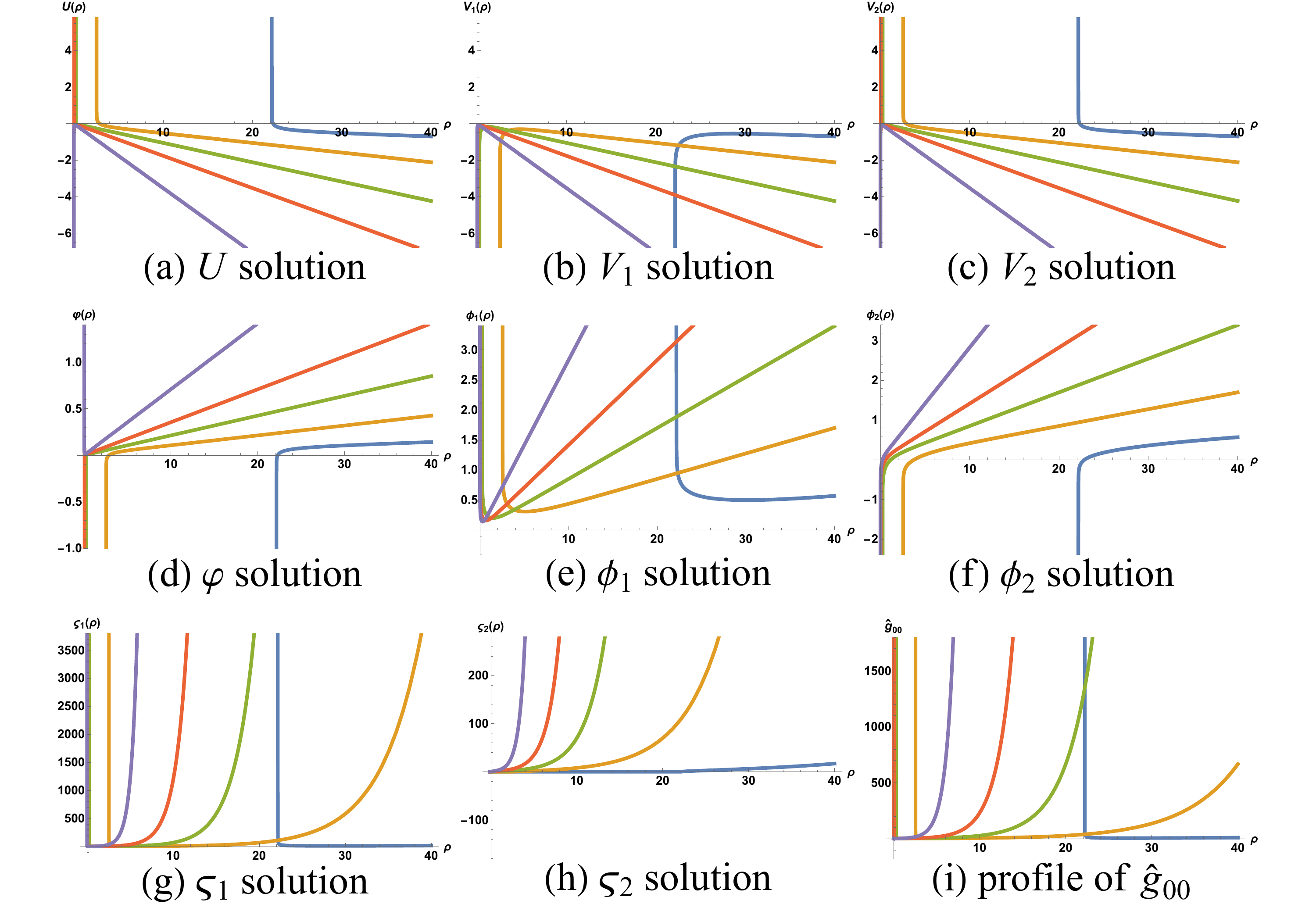}
\caption{Interpolating solutions between the locally $SO(2)\times SO(2)$ flat domain wall as $\rho\rightarrow+\infty$ and $t\times H^2\times \mathbb{R}^2$-sliced curved domain walls with shift scalars in $SO(3,2)$ gauge group. The blue, orange, green, red, purple, and black curves refer to $g=-0.1, -0.3, -0.6, -1, -2$.}
\label{15_HR_SO(2)d_SO(32)gg_flows}
\end{figure}

\clearpage\newpage
\subsection{D4-branes wrapped on a Kahler four-cycle}
\subsubsection{Solutions with $SO(3)$ twist}\label{AppC_2_1}
\begin{figure}[h!]
  \centering
    \includegraphics[width=\linewidth]{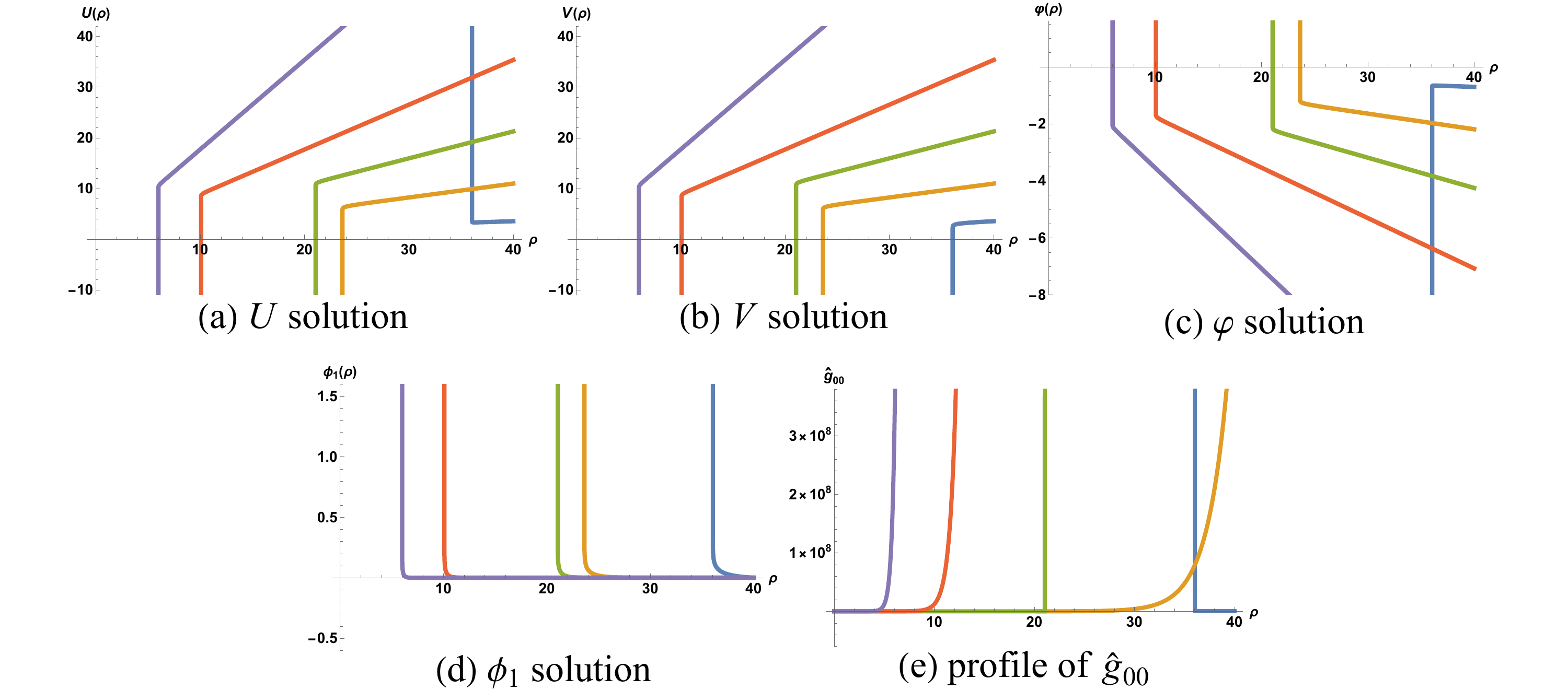}
\caption{Interpolating solutions between the locally $SO(5)$ flat domain wall as $\rho\rightarrow+\infty$ and $t\times CP^2$-sliced curved domain walls for $SO(3)$ twist in $SO(5)$ gauge group. The blue, orange, green, red, and purple curves refer to $g=0.1, 0.3, 0.6, 1$, and $2$, respectively.}
\label{15_CP2_special_SO(3)_SO(5)gg_flows}
\end{figure}

\begin{figure}[h!]
  \centering
    \includegraphics[width=\linewidth]{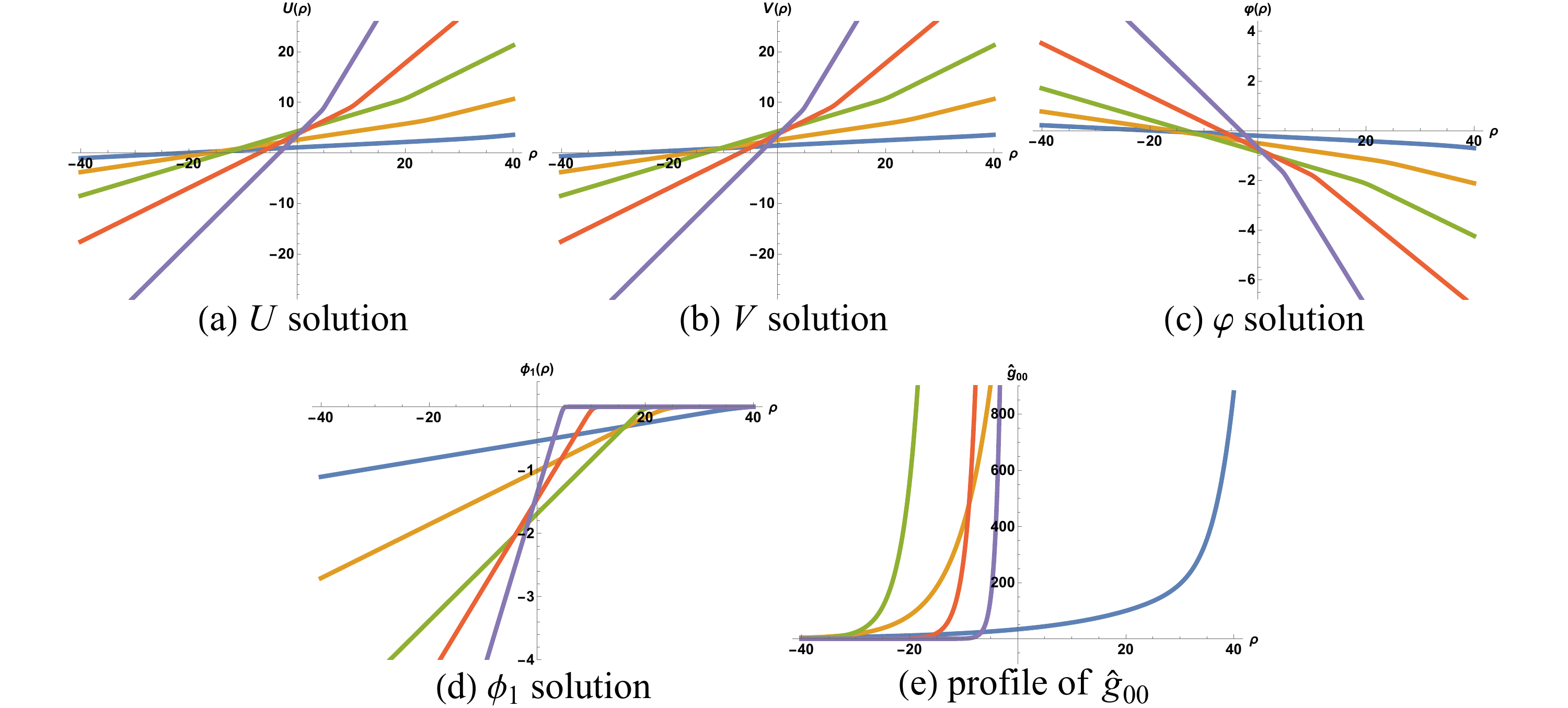}
\caption{Interpolating solutions between the locally $SO(5)$ flat domain wall as $\rho\rightarrow+\infty$ and $t\times CH^2$-sliced curved domain walls for $SO(3)$ twist in $SO(5)$ gauge group. The blue, orange, green, red, and purple curves refer to $g=0.1, 0.3, 0.6, 1$, and $2$, respectively.}
\label{15_CH2_special_SO(3)_SO(5)gg_flows}
\end{figure}

\begin{figure}[h!]
  \centering
    \includegraphics[width=\linewidth]{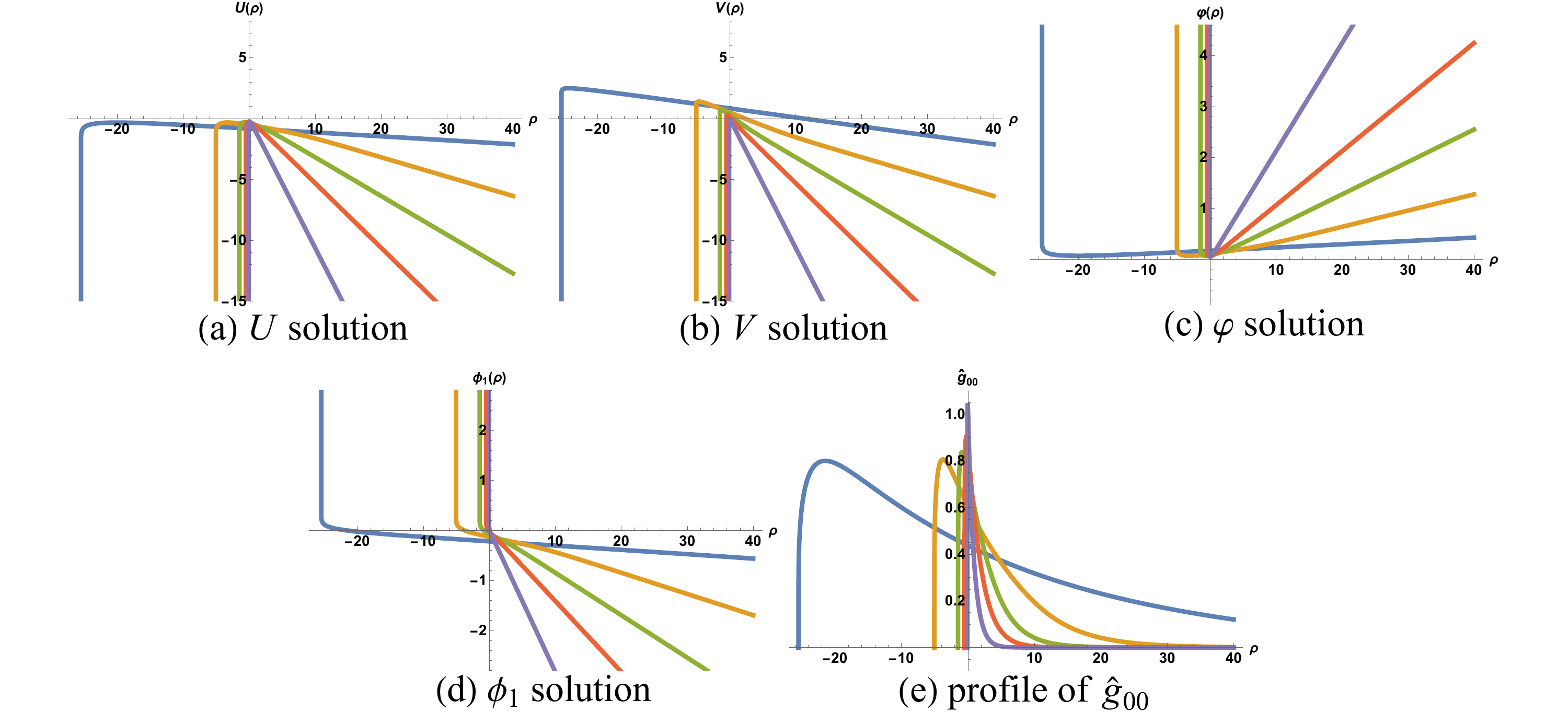}
\caption{Interpolating solutions between the locally $SO(3)$ flat domain wall as $\rho\rightarrow+\infty$ and $t\times CP^2$-sliced curved domain walls for $SO(3)$ twist in $SO(3,2)$ gauge group. The blue, orange, green, red, and purple curves refer to $g=-0.1, -0.3, -0.6, -1$, and $-2$, respectively.}
\label{15_CP2_SO(32)_SO(5)gg_flows}
\end{figure}

\begin{figure}[h!]
  \centering
    \includegraphics[width=\linewidth]{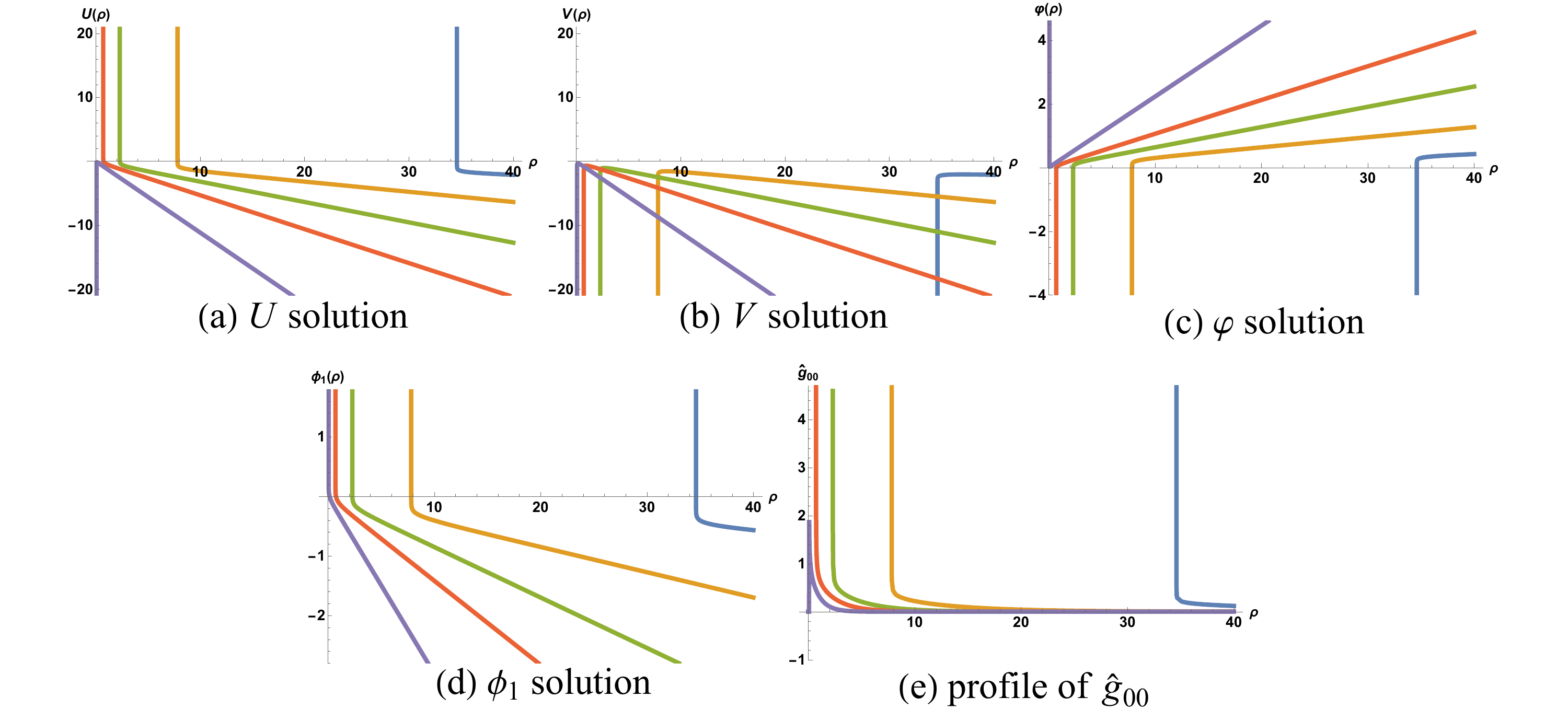}
\caption{Interpolating solutions between the locally $SO(3)$ flat domain wall as $\rho\rightarrow+\infty$ and $t\times CH^2$-sliced curved domain walls for $SO(3)$ twist in $SO(3,2)$ gauge group. The blue, orange, green, red, and purple curves refer to $g=-0.1, -0.3, -0.6, -1$, and $-2.1$, respectively.}
\label{15_CH2_SO(3)_SO(32)gg_flows}
\end{figure}

\clearpage\newpage
\subsubsection{Solutions with $SO(2)\times SO(2)$ twist}\label{AppC_2_2}
\begin{figure}[h!]
  \centering
    \includegraphics[width=\linewidth]{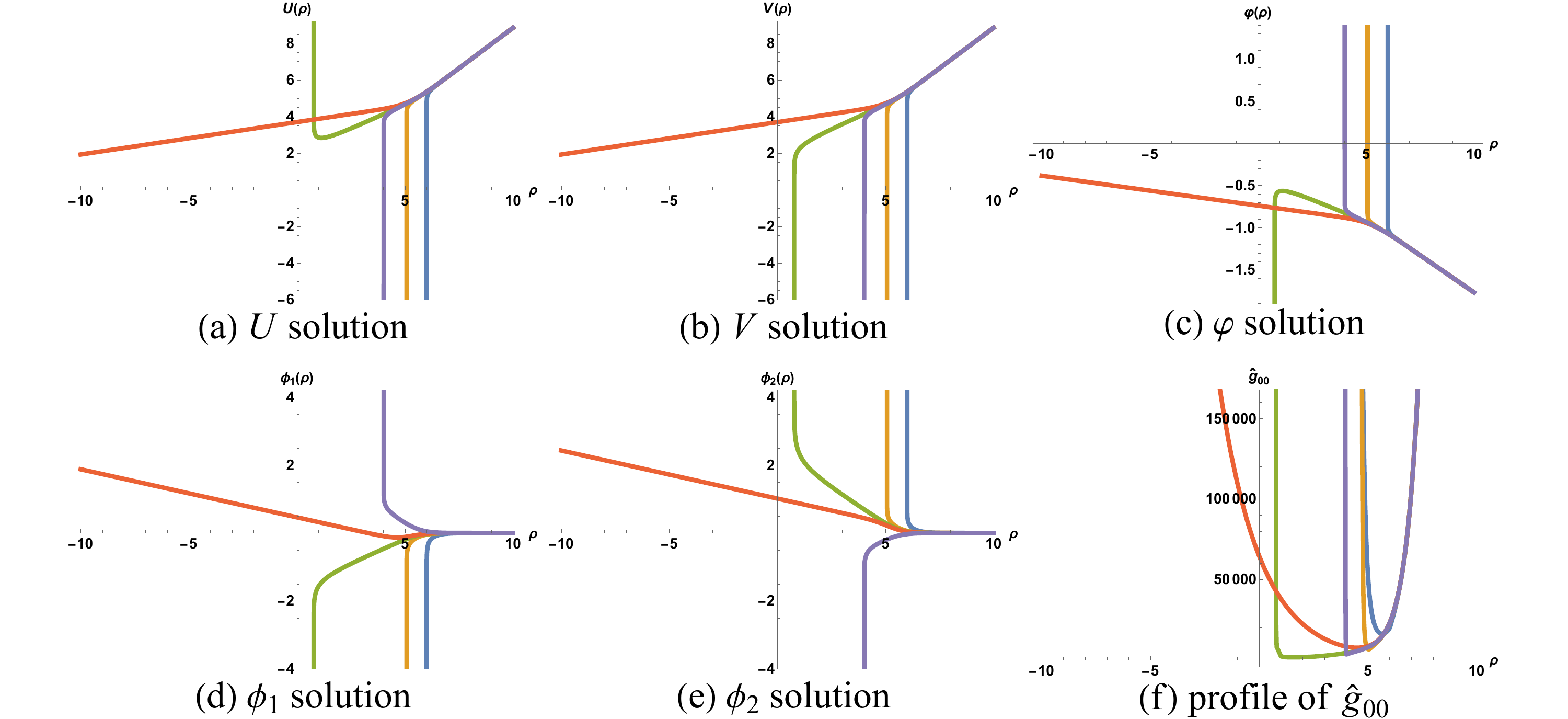}
\caption{Interpolating solutions between the locally $SO(5)$ flat domain wall as $\rho\rightarrow+\infty$ and $t\times CP^2$-sliced curved domain walls for $SO(2)\times SO(2)$ twist in $SO(5)$ gauge group with $\varsigma_1=\varsigma_2=0$. The blue, orange, green, red, and purple curves refer to $p_1=-2.90, -0.20, 0, 0.10$, and $1.01$, respectively. }
\label{15_CP2_special_SO(2)xSO(2)_SO(5)gg_flows}
\end{figure}

\begin{figure}[h!]
  \centering
    \includegraphics[width=\linewidth]{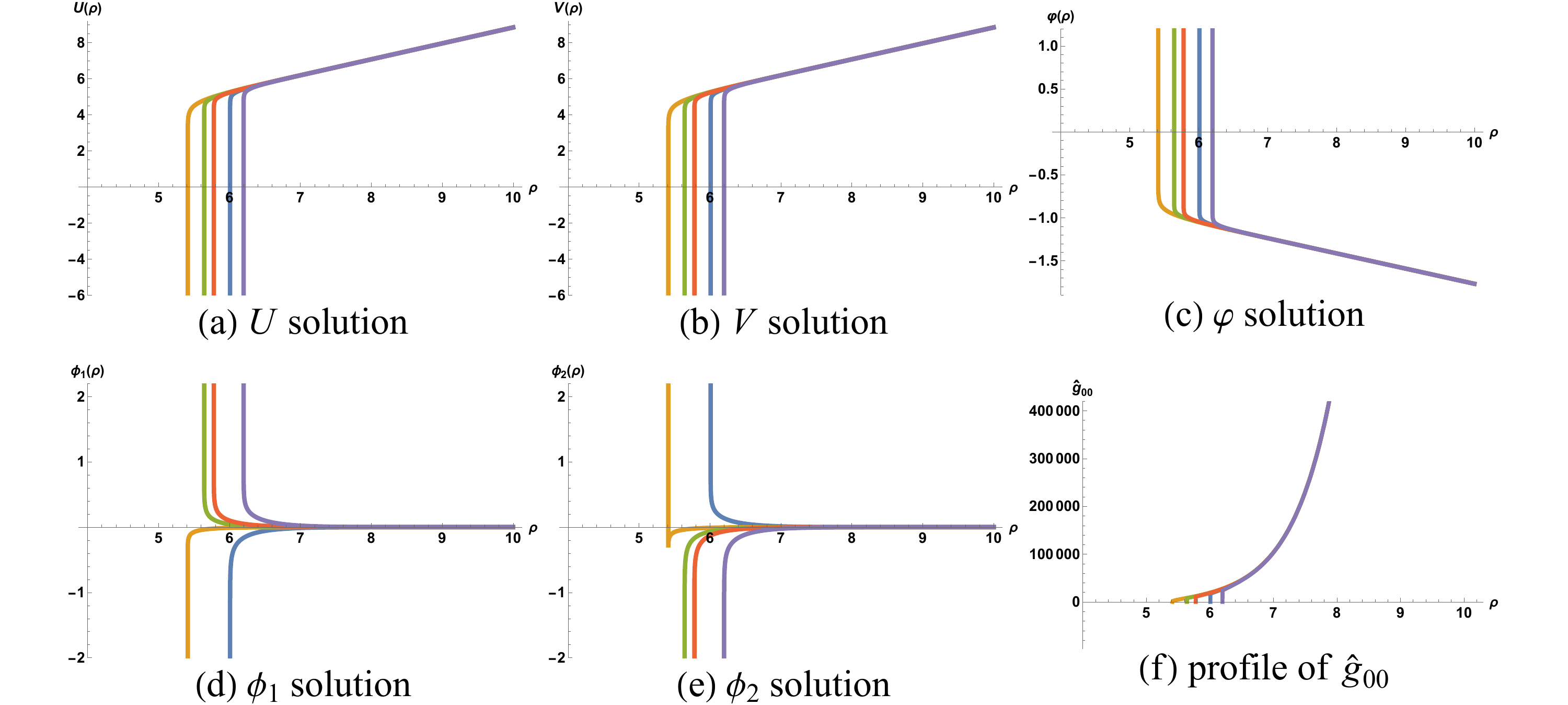}
\caption{Interpolating solutions between the locally $SO(5)$ flat domain wall as $\rho\rightarrow+\infty$ and $t\times CH^2$-sliced curved domain walls for $SO(2)\times SO(2)$ twist in $SO(5)$ gauge group with $\varsigma_1=\varsigma_2=0$. The blue, orange, green, red, and purple curves refer to $p_1=-2.9, -0.5, 0, 0.5$, and $4$, respectively. }
\label{15_CH2_special_SO(2)xSO(2)_SO(5)gg_flows}
\end{figure}

\begin{figure}[h!]
  \centering
    \includegraphics[width=\linewidth]{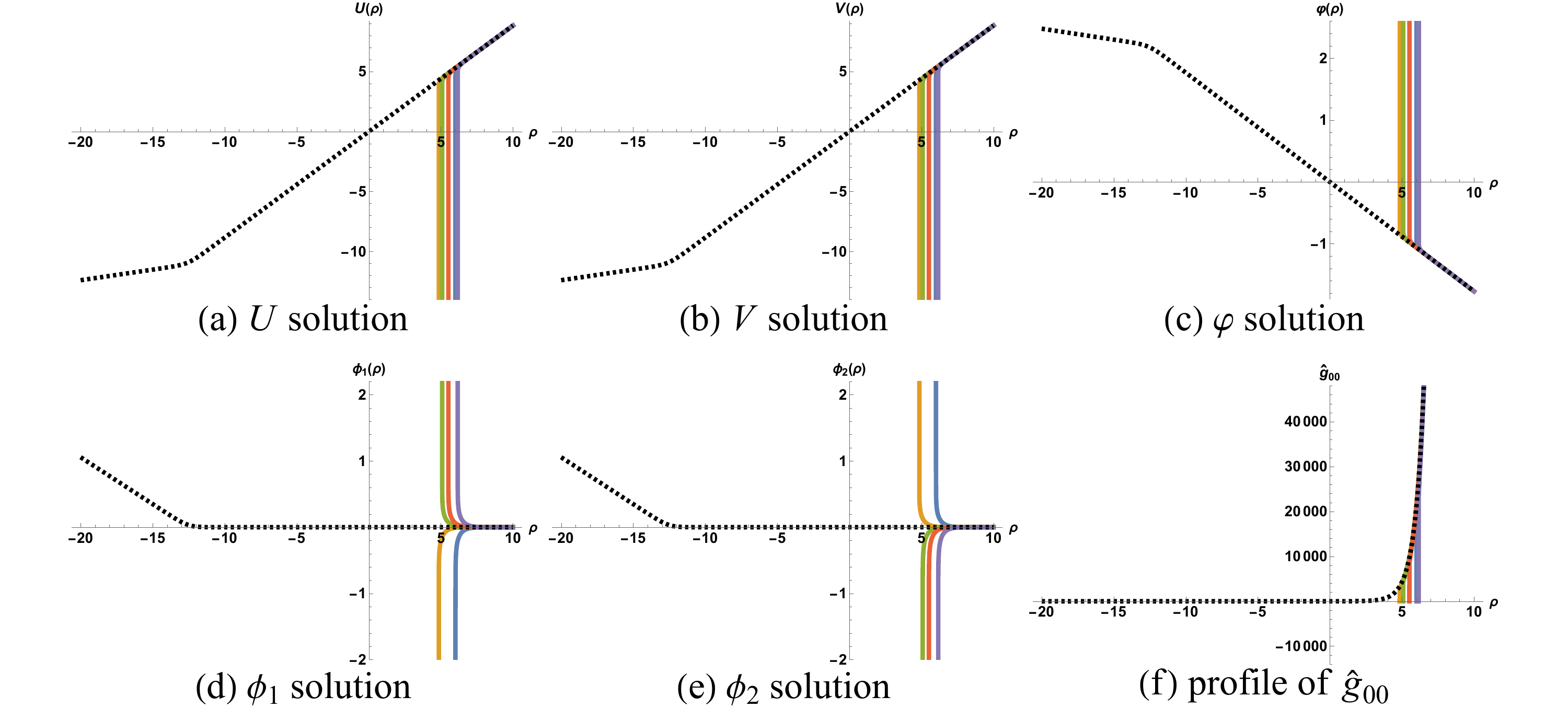}
\caption{Interpolating solutions between the locally $SO(5)$ flat domain wall as $\rho\rightarrow+\infty$ and $t\times \mathbb{R}^4$-sliced curved domain walls for $SO(2)\times SO(2)$ twist in $SO(5)$ gauge group with $\varsigma_1=\varsigma_2=0$. The blue, orange, green, red, and purple curves refer to $p_1=-2.9, -0.1, 0.2, 0.7$, and $4.5$, respectively. The dashed curve is the $SO(2)\times SO(2)$ flat domain wall obtained from $p_1=0$ which implies $p_2=k=0$.}
\label{15_R4_special_SO(2)xSO(2)_SO(5)gg_flows}
\end{figure}

\begin{figure}[h!]
  \centering
    \includegraphics[width=\linewidth]{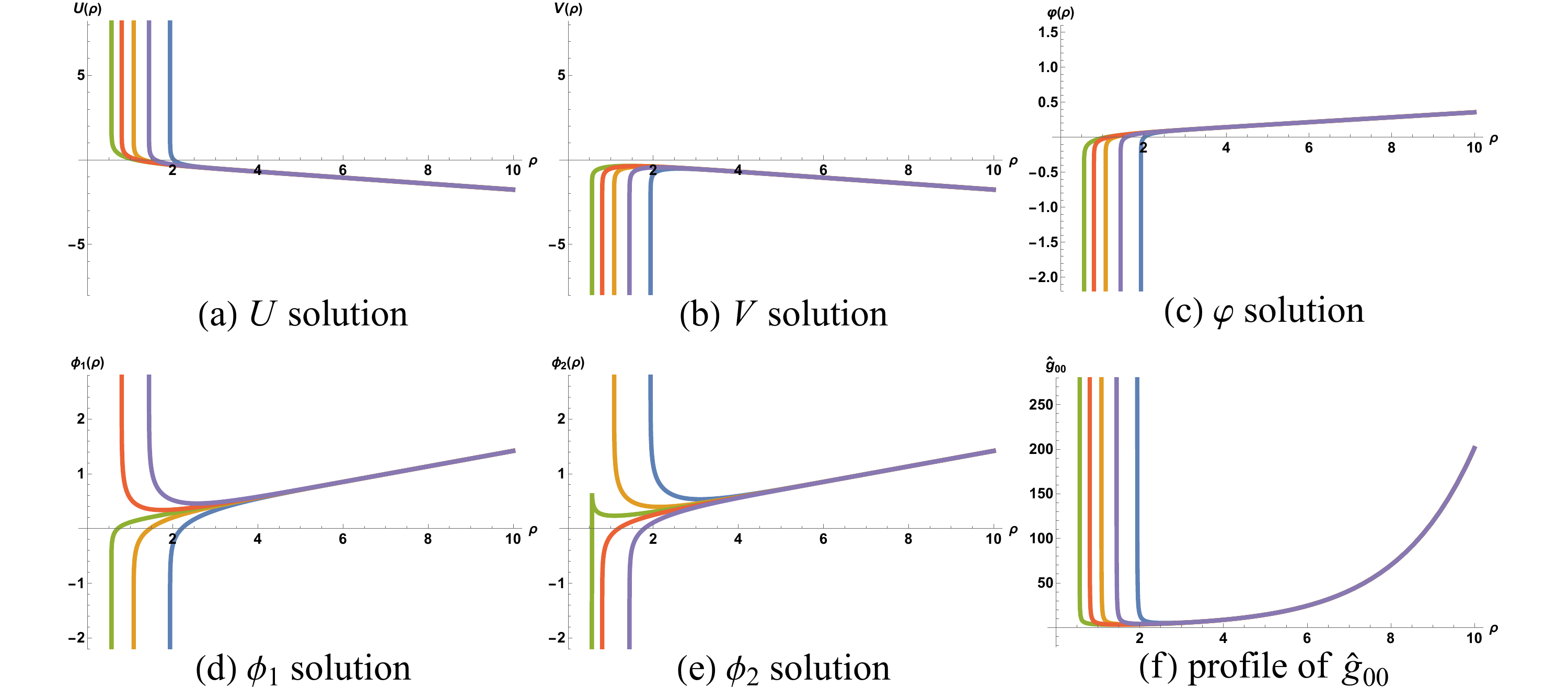}
\caption{Interpolating solutions between the locally $SO(2)\times SO(2)$ flat domain wall as $\rho\rightarrow+\infty$ and $t\times CP^2$-sliced curved domain walls for $SO(2)\times SO(2)$ twist in $SO(4,1)$ gauge group with $\varsigma_1=\varsigma_2=0$. The blue, orange, green, red, and purple curves refer to $p_1=-4, -0.5, 0.4, 1$, and $2.5$, respectively. }
\label{15_CP2_SO(2)xSO(2)_SO(41)gg_flows}
\end{figure}

\begin{figure}[h!]
  \centering
    \includegraphics[width=\linewidth]{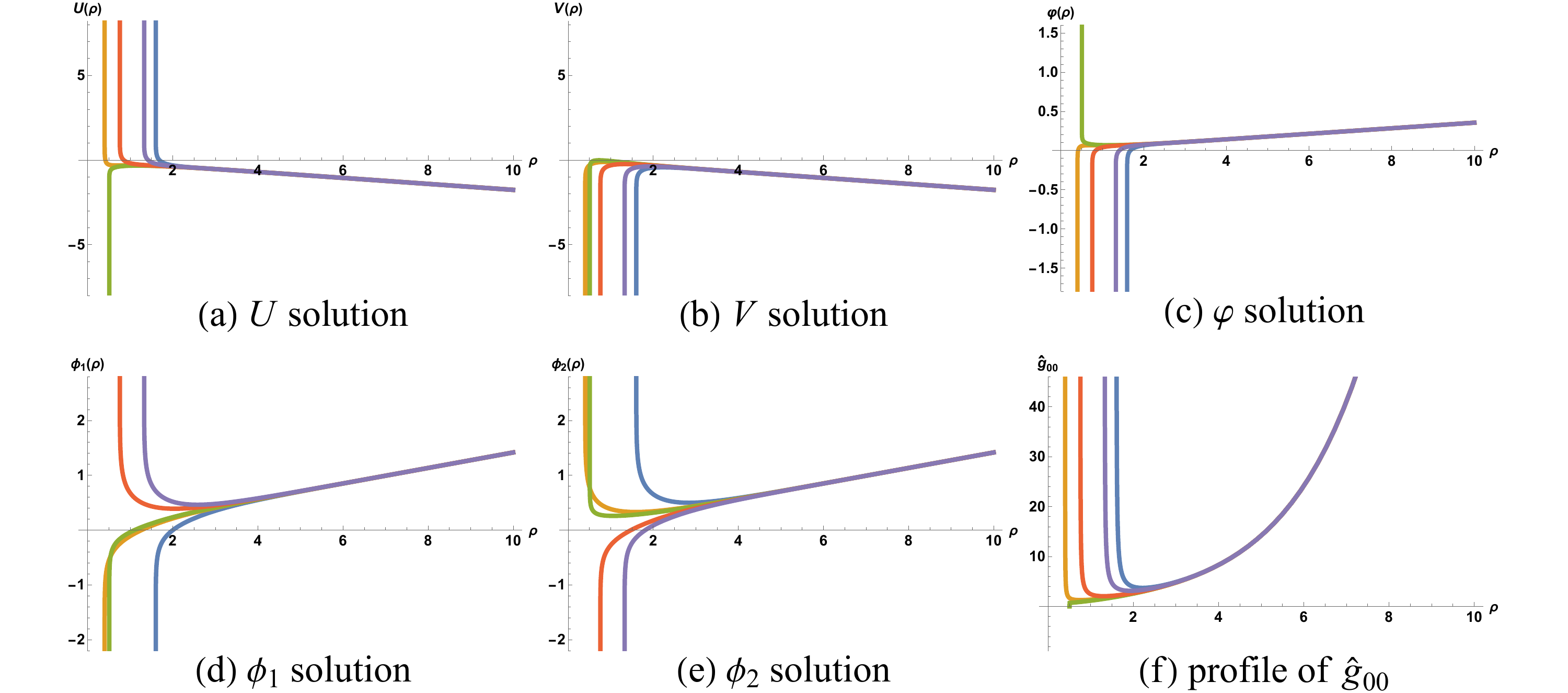}
\caption{Interpolating solutions between the locally $SO(2)\times SO(2)$ flat domain wall as $\rho\rightarrow+\infty$ and $t\times CH^2$-sliced curved domain walls for $SO(2)\times SO(2)$ twist in $SO(4,1)$ gauge group with $\varsigma_1=\varsigma_2=0$. The blue, orange, green, red, and purple curves refer to $p_1=-4, -1.3, -1, 0.8$, and $2$, respectively. }
\label{15_CH2_SO(2)xSO(2)_SO(41)gg_flows}
\end{figure}

\begin{figure}[h!]
  \centering
    \includegraphics[width=\linewidth]{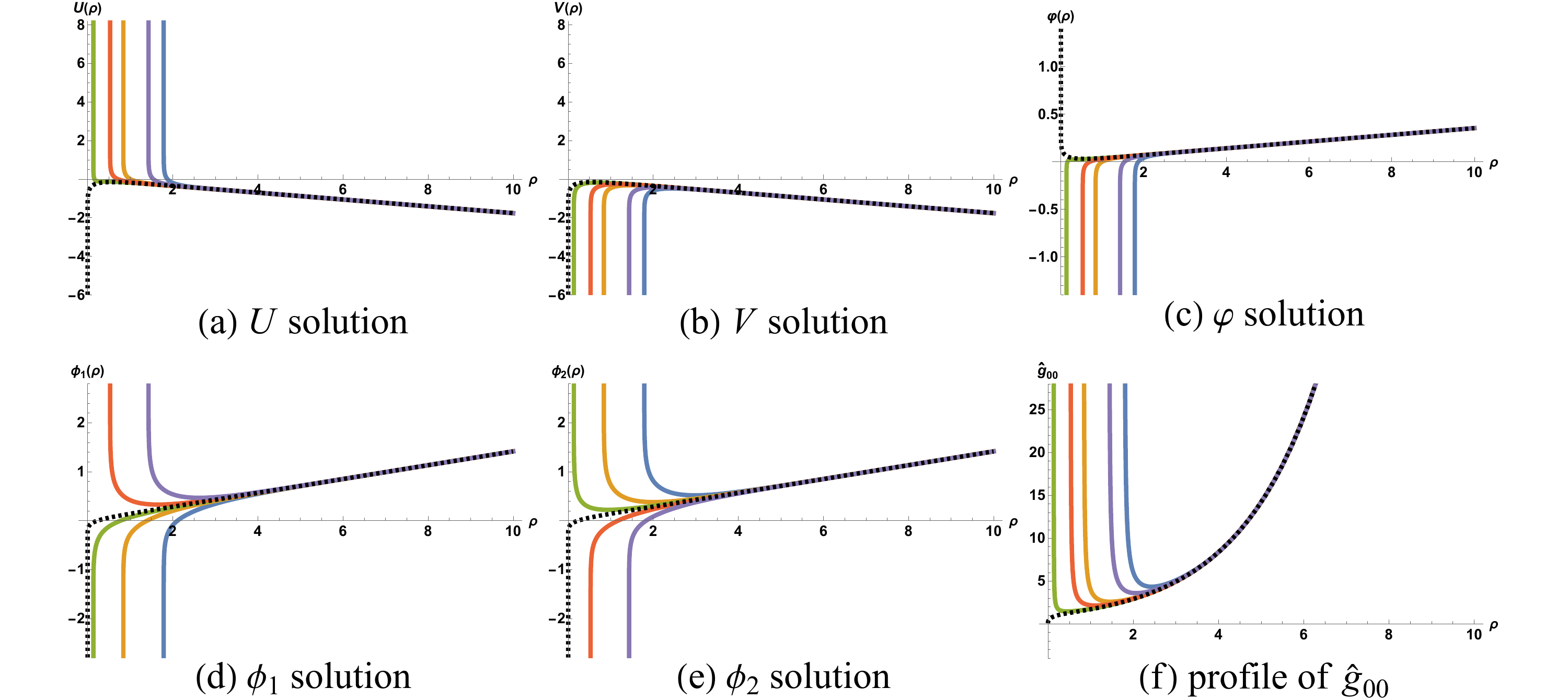}
\caption{Interpolating solutions between the locally $SO(2)\times SO(2)$ flat domain wall as $\rho\rightarrow+\infty$ and $t\times \mathbb{R}^4$-sliced curved domain walls for $SO(2)\times SO(2)$ twist in $SO(4,1)$ gauge group with $\varsigma_1=\varsigma_2=0$. The blue, orange, green, red, and purple curves refer to $p_1=-4, -1, -0.2, 0.6$, and $2.4$, respectively. The dashed curve is the $SO(2)\times SO(2)$ flat domain wall obtained from $p_1=0$ which implies $p_2=k=0$.}
\label{15_R4_SO(2)xSO(2)_SO(41)gg_flows}
\end{figure}

\begin{figure}[h!]
  \centering
    \includegraphics[width=\linewidth]{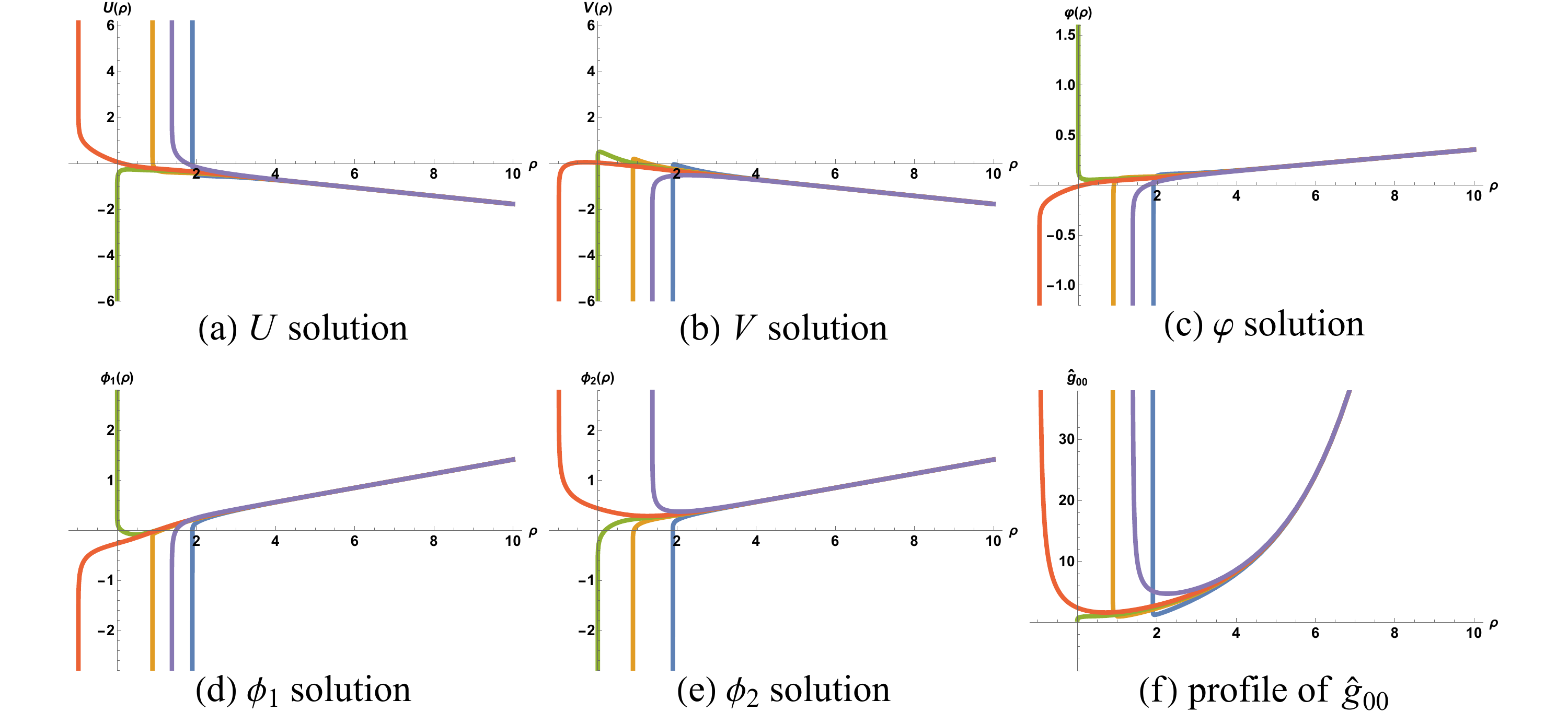}
\caption{Interpolating solutions between the locally $SO(2)\times SO(2)$ flat domain wall as $\rho\rightarrow+\infty$ and $t\times CP^2$-sliced curved domain walls for $SO(2)\times SO(2)$ twist in $SO(3,2)$ gauge group with $\varsigma_1=\varsigma_2=0$. The blue, orange, green, red, and purple curves refer to $p_1=-4, -1.5, -1, -0.7$, and $1.2$, respectively. }
\label{15_CP2_SO(2)xSO(2)_SO(32)gg_flows}
\end{figure}

\begin{figure}[h!]
  \centering
    \includegraphics[width=\linewidth]{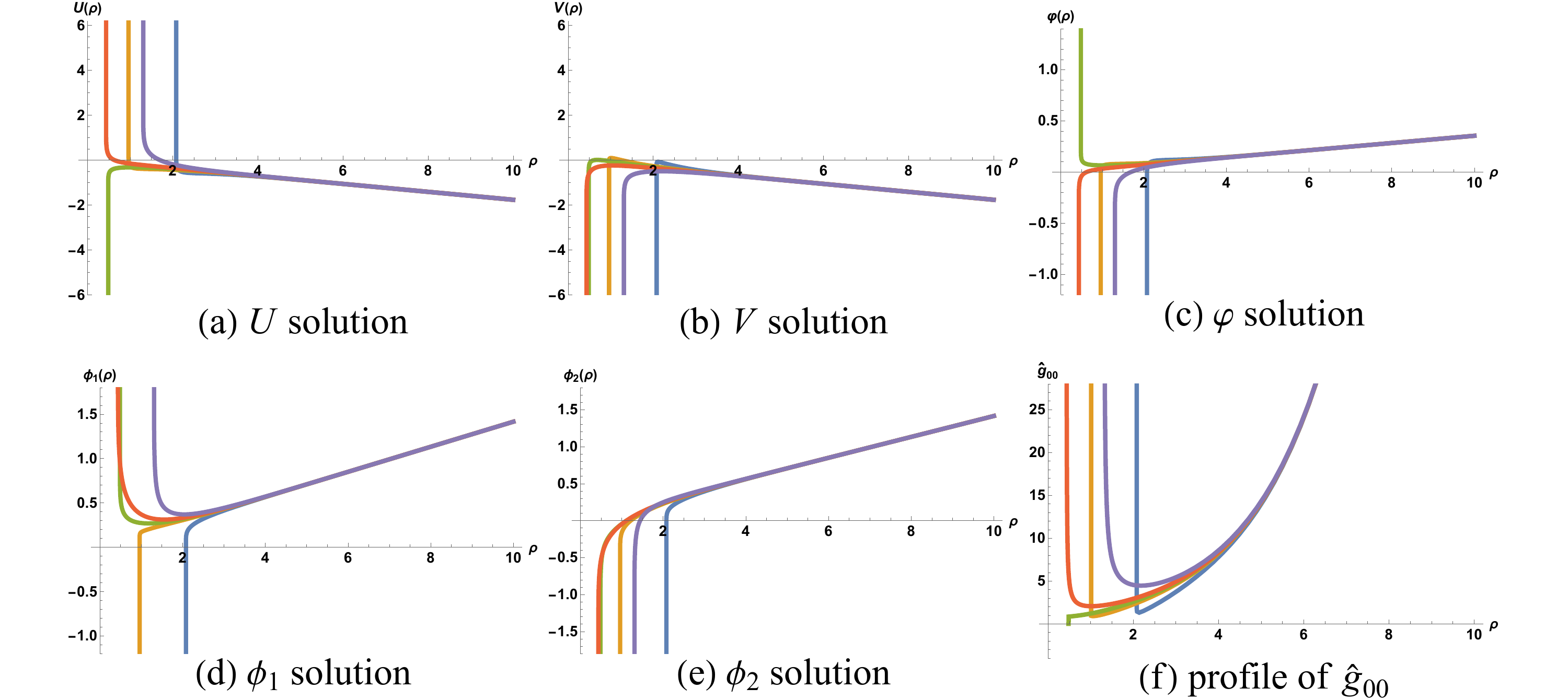}
\caption{Interpolating solutions between the locally $SO(2)\times SO(2)$ flat domain wall as $\rho\rightarrow+\infty$ and $t\times CH^2$-sliced curved domain walls for $SO(2)\times SO(2)$ twist in $SO(3,2)$ gauge group with $\varsigma_1=\varsigma_2=0$. The blue, orange, green, red, and purple curves refer to $p_1=-4, -0.5, 0, 0.5$, and $2$, respectively. }
\label{15_CH2_SO(2)xSO(2)_SO(32)gg_flows}
\end{figure}

\begin{figure}[h!]
  \centering
    \includegraphics[width=\linewidth]{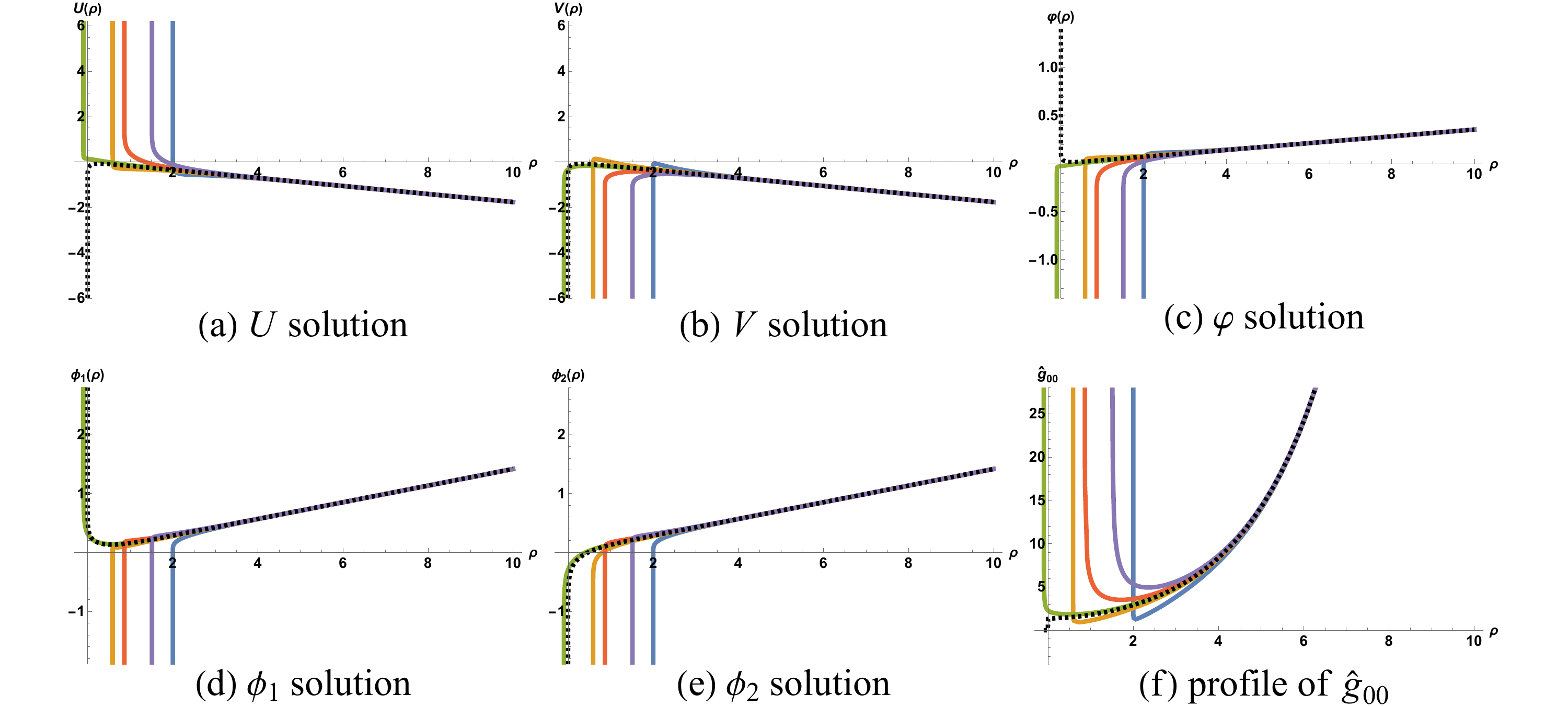}
\caption{Interpolating solutions between the locally $SO(2)\times SO(2)$ flat domain wall as $\rho\rightarrow+\infty$ and $t\times \mathbb{R}^4$-sliced curved domain walls for $SO(2)\times SO(2)$ twist in $SO(3,2)$ gauge group with $\varsigma_1=\varsigma_2=0$. The blue, orange, green, red, and purple curves refer to $p_1=-4, -0.5, 0.1, 0.8$, and $2$, respectively. The dashed curve is the $SO(2)\times SO(2)$ flat domain wall obtained from $p_1=0$ which implies $p_2=k=0$.}
\label{15_R4_SO(2)xSO(2)_SO(32)gg_flows}
\end{figure}

\begin{figure}[h!]
  \centering
    \includegraphics[width=0.98\linewidth]{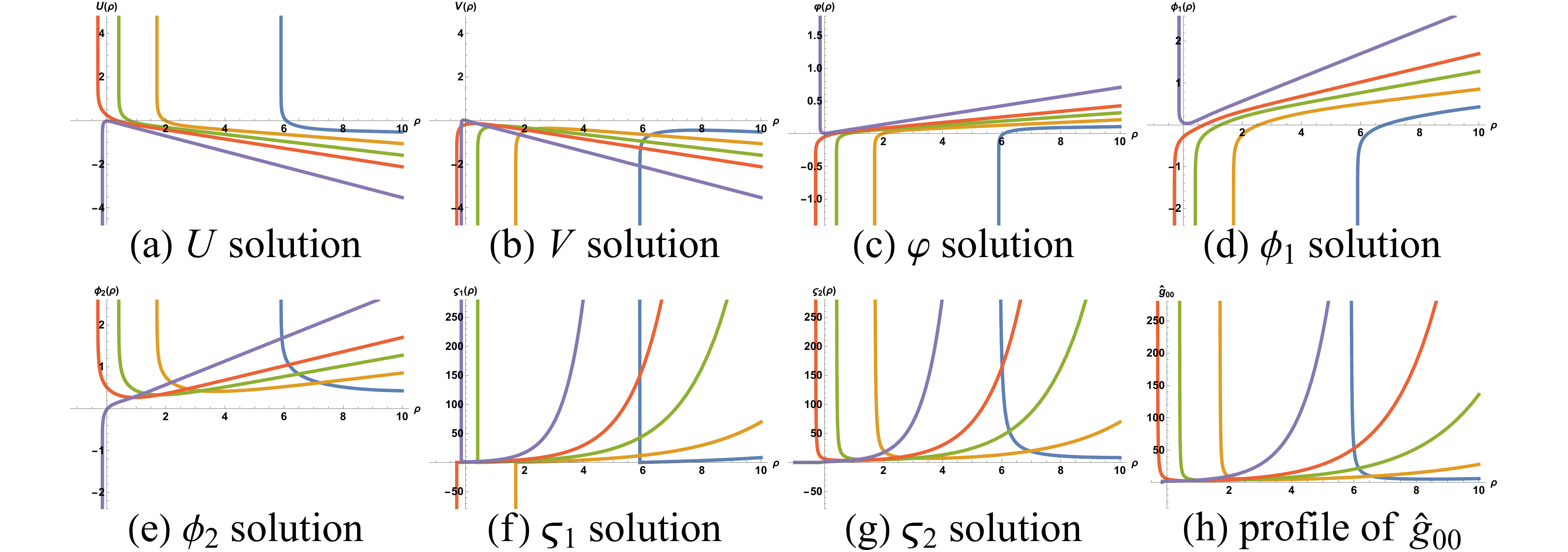}
\caption{Interpolating solutions between the locally $SO(2)\times SO(2)$ flat domain wall as $\rho\rightarrow+\infty$ and $t\times CP^2$-sliced curved domain walls for $SO(2)\times SO(2)$ twist with shift scalars in $SO(3,2)$ gauge group. The blue, orange, green, red, and purple curves refer to $p_1=-0.3, -0.6, -0.9, -1.2$, and $-2$, respectively. }
\label{15_CP2_SO(2)d_SO(32)gg_flows}
\end{figure}
\clearpage\newpage
\begin{figure}[h!]
  \centering
    \includegraphics[width=0.98\linewidth]{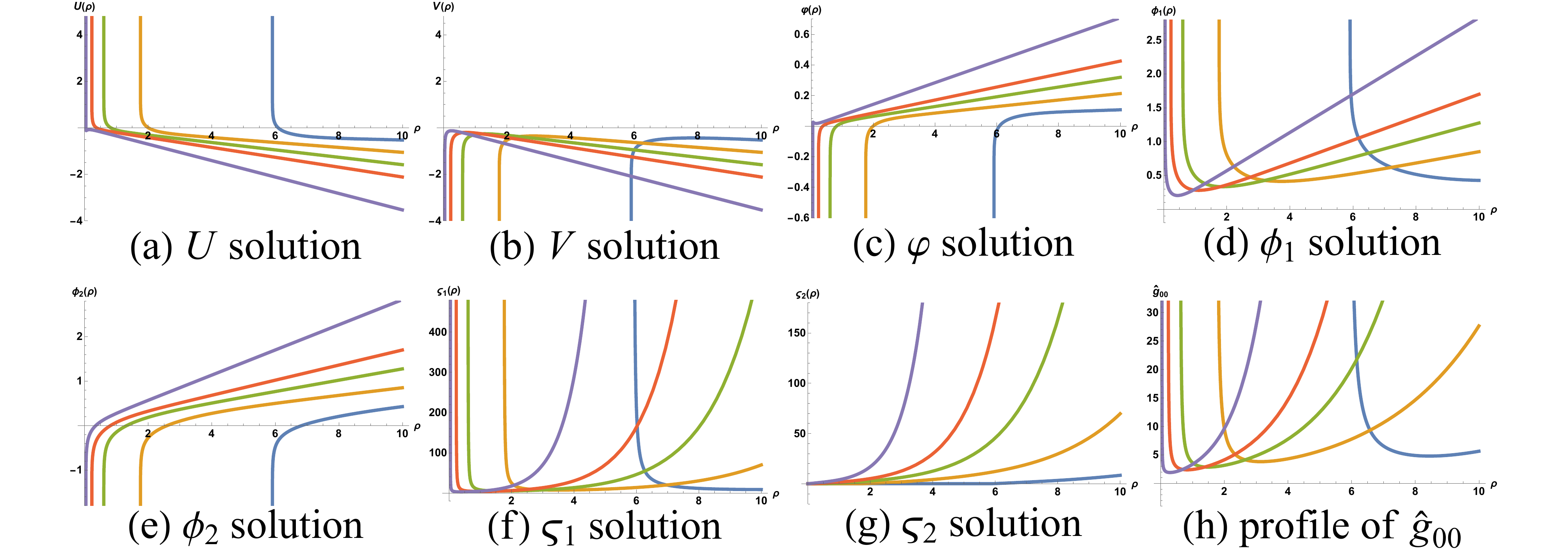}
\caption{Interpolating solutions between the locally $SO(2)\times SO(2)$ flat domain wall as $\rho\rightarrow+\infty$ and $t\times CH^2$-sliced curved domain walls for $SO(2)\times SO(2)$ twist with shift scalars in $SO(3,2)$ gauge group. The blue, orange, green, red, and purple curves refer to $p_1=-0.3, -0.6, -0.9, -1.2$, and $-2$, respectively. }
\label{15_CH2_SO(2)d_SO(32)gg_flows}
\end{figure}

%%%%%%%%%%%%%%%%%%%%%%%%%%%%%%%%%%%%%%%%%%%%%%%
\section{Numerical solutions from $CSO(p,q,4-p-q)\ltimes \mathbb{R}^4$ gauged supergravity}\label{40_Numer_App}
All numerical solutions from $CSO(p,q,4-p-q)\ltimes \mathbb{R}^4$ are given in this appendix.

\subsection{D4-branes wrapped on a product of two Riemann surfaces}\label{App_D_1}

\begin{figure}[h!]
  \centering
    \includegraphics[width=\linewidth]{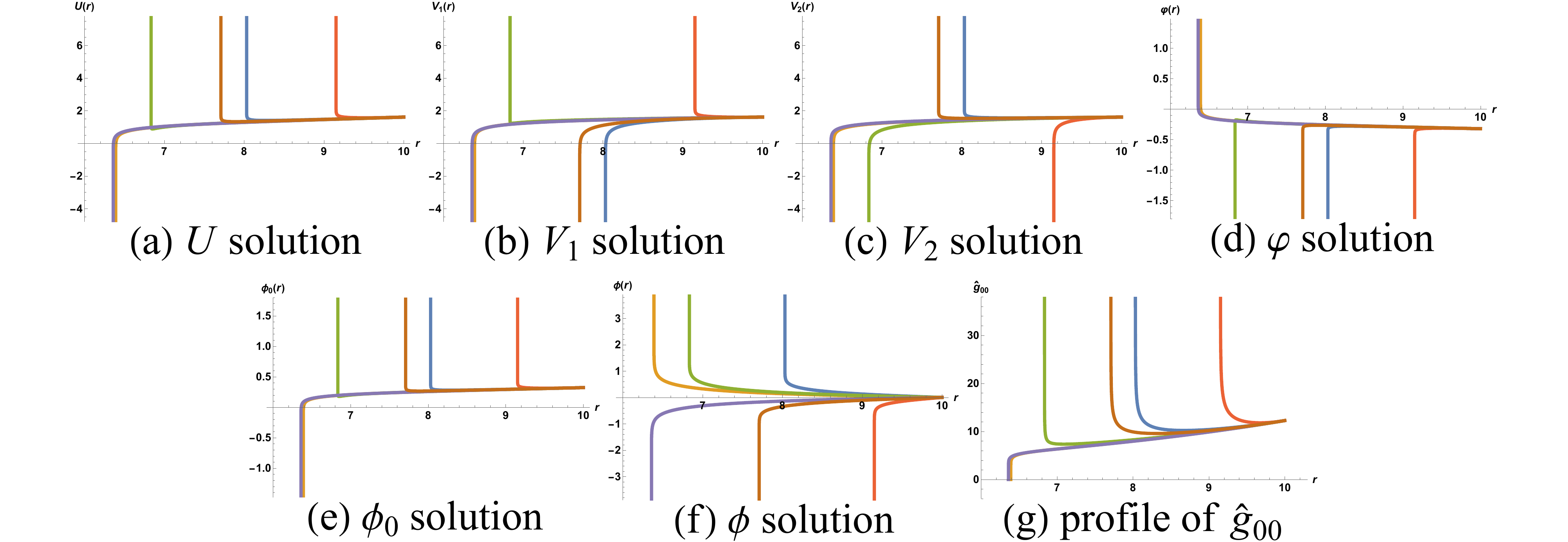}
\caption{Interpolating solutions between the locally $SO(4)$ flat domain wall as $r\rightarrow+\infty$ and $t\times H^2\times H^2$-sliced curved domain walls for $SO(2)\times SO(2)$ twist in $SO(4)\ltimes \mathbb{R}^{4}$ gauge group. The blue, orange, green, red, purple, and brown curves refer to  $z_1=-4.01, -1.01, -0.50, 0.10$, $1.12$, and $3.40$, respectively. }
\label{40_HH_SO(2)xSO(2)_special_SO(4)gg_flows}
\end{figure}

\begin{figure}[h!]
  \centering
    \includegraphics[width=\linewidth]{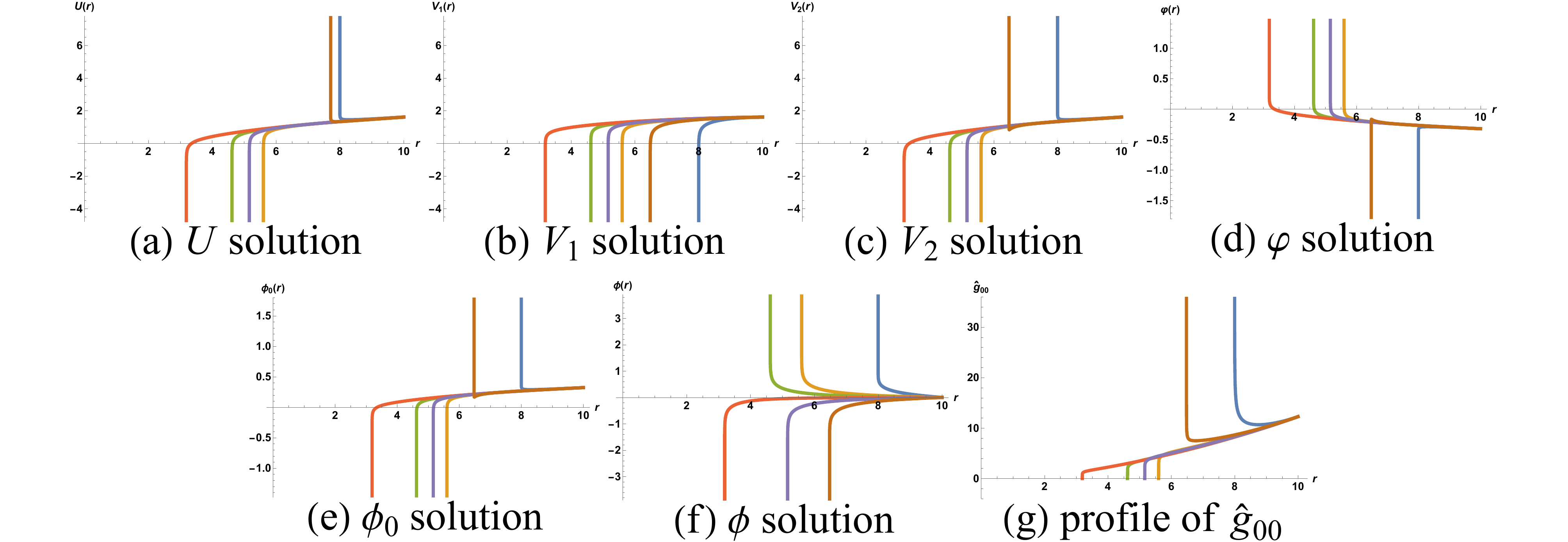}
\caption{Interpolating solutions between the locally $SO(4)$ flat domain wall as $r\rightarrow+\infty$ and $t\times H^2\times \mathbb{R}^2$-sliced curved domain walls for $SO(2)\times SO(2)$ twist in $SO(4)\ltimes \mathbb{R}^{4}$ gauge group. The blue, orange, green, red, purple, and brown curves refer to $z_1=-4, -1, -0.40, 0.10$, $0.75$, and $1.90$, respectively. }
\label{40_HR_SO(2)xSO(2)_special_SO(4)gg_flows}
\end{figure}

\begin{figure}[h!]
  \centering
    \includegraphics[width=\linewidth]{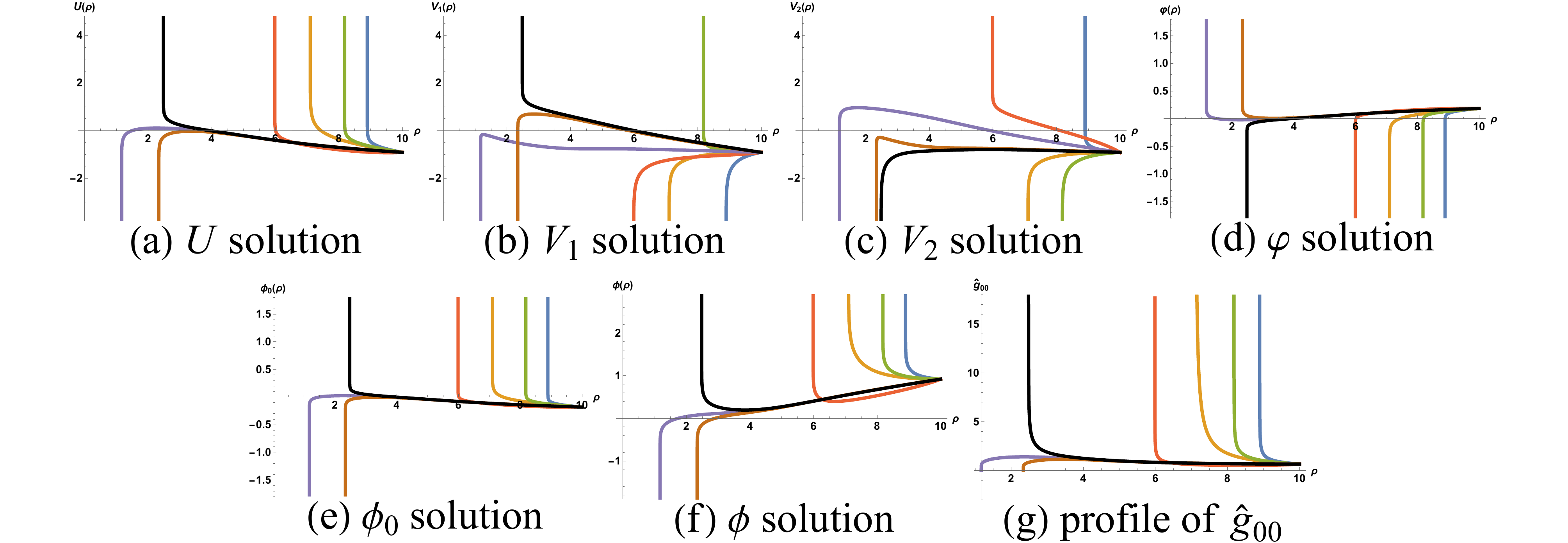}
\caption{Interpolating solutions between the locally $SO(2)\times SO(2)$ flat domain wall as $\rho\rightarrow+\infty$ and $t\times S^2\times S^2$-sliced curved domain walls for $SO(2)\times SO(2)$ twist in $SO(2,2)\ltimes \mathbb{R}^{4}$ gauge group. The blue, orange, green, red, purple, brown, and black curves refer to $z_1=-4, -1, -0.50, 0.10, 0.48$, $2$, and $2.17$, respectively. }
\label{40_SS_SO(2)xSO(2)_SO(22)gg_flows}
\end{figure}

\begin{figure}[h!]
  \centering
    \includegraphics[width=\linewidth]{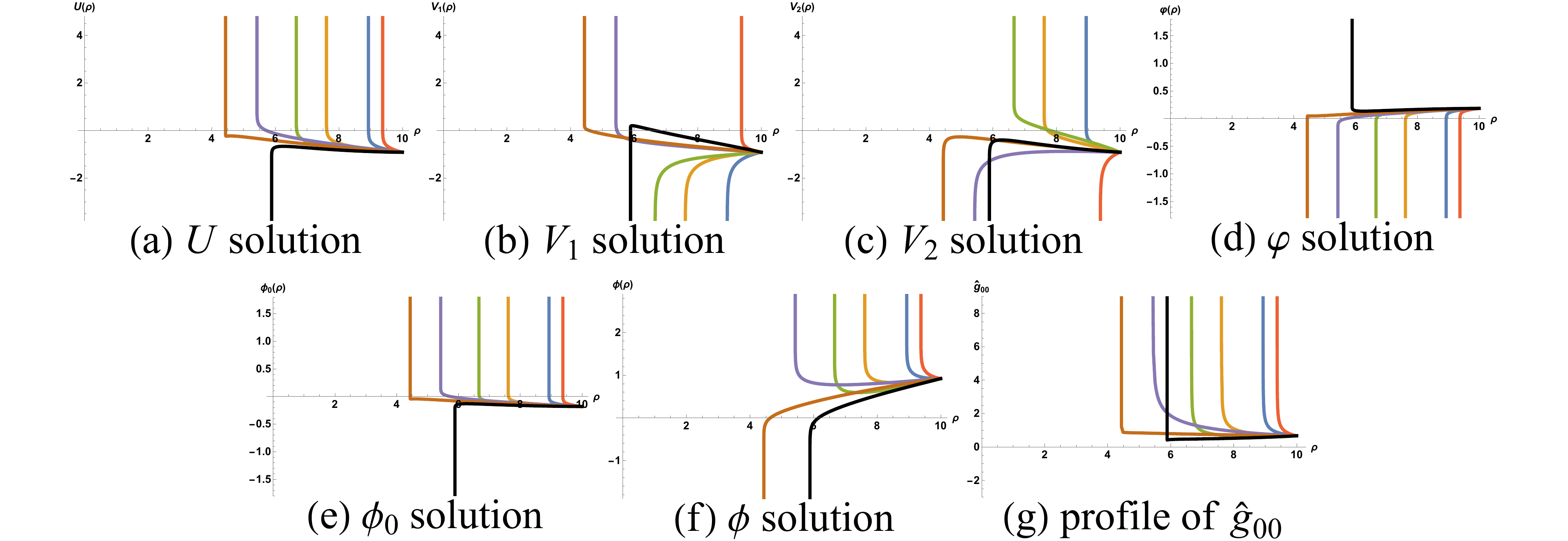}
\caption{Interpolating solutions between the locally $SO(2)\times SO(2)$ flat domain wall as $\rho\rightarrow+\infty$ and $t\times S^2\times H^2$-sliced curved domain walls for $SO(2)\times SO(2)$ twist in $SO(2,2)\ltimes \mathbb{R}^{4}$ gauge group. The blue, orange, green, red, purple, brown, and black curves refer to $z_1=-4, -1, -0.21, 0.10, 0.50$, $1.10$, and $3.90$, respectively. }
\label{40_SH_SO(2)xSO(2)_SO(22)gg_flows}
\end{figure}

\begin{figure}[h!]
  \centering
    \includegraphics[width=\linewidth]{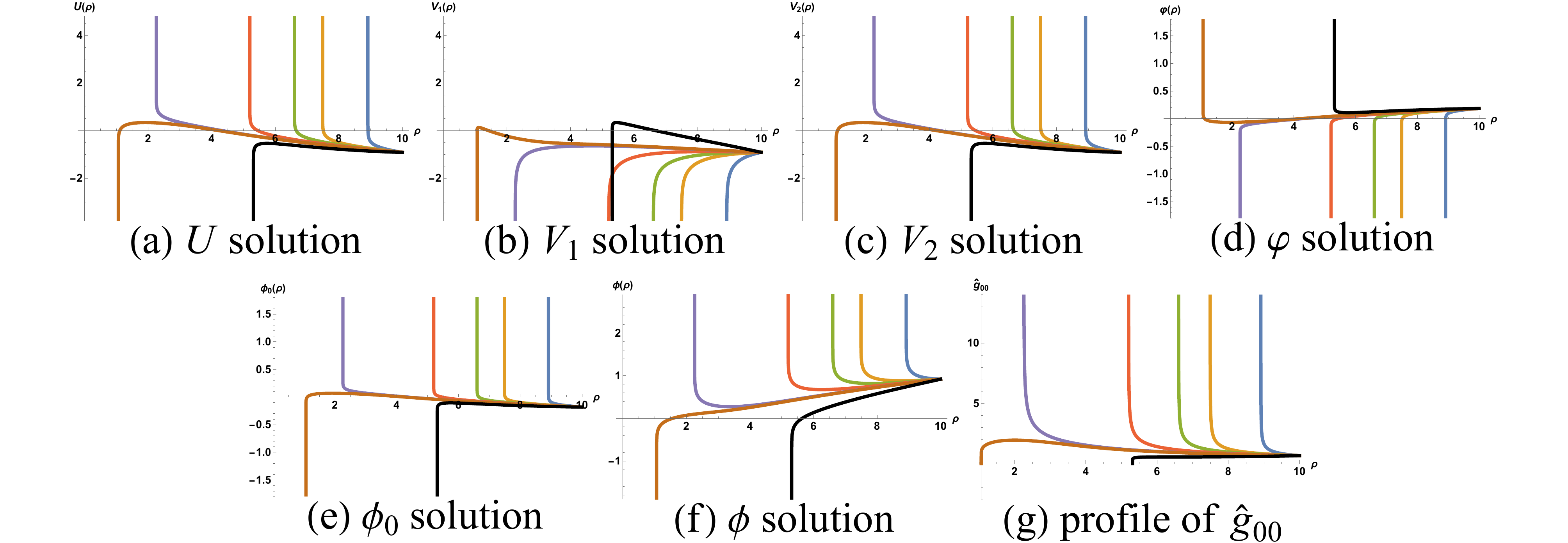}
\caption{Interpolating solutions between the locally $SO(2)\times SO(2)$ flat domain wall as $\rho\rightarrow+\infty$ and $t\times S^2\times \mathbb{R}^2$-sliced curved domain walls for $SO(2)\times SO(2)$ twist in $SO(2,2)\ltimes \mathbb{R}^{4}$ gauge group. The blue, orange, green, red, purple, brown, and black curves refer to $z_1=-4, -1, -0.40, 0.10, 0.50$, $0.54$, and $3.70$, respectively. }
\label{40_SR_SO(2)xSO(2)_SO(22)gg_flows}
\end{figure}

\begin{figure}[h!]
  \centering
    \includegraphics[width=\linewidth]{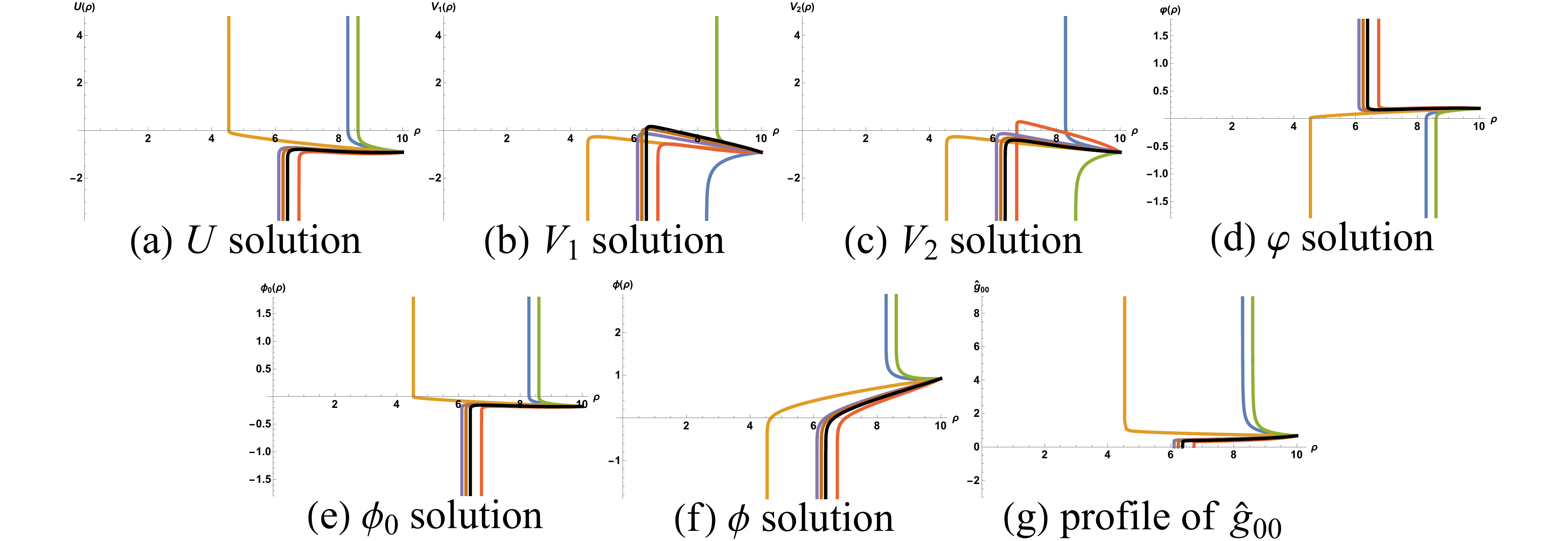}
\caption{Interpolating solutions between the locally $SO(2)\times SO(2)$ flat domain wall as $\rho\rightarrow+\infty$ and $t\times H^2\times H^2$-sliced curved domain walls for $SO(2)\times SO(2)$ twist in $SO(2,2)\ltimes \mathbb{R}^{4}$ gauge group. The blue, orange, green, red, purple, brown, and black curves refer to $z_1=-4, -1, -0.21, 0.11, 1$, $2.60$, and $4$, respectively. }
\label{40_HH_SO(2)xSO(2)_SO(22)gg_flows}
\end{figure}

\begin{figure}[h!]
  \centering
    \includegraphics[width=\linewidth]{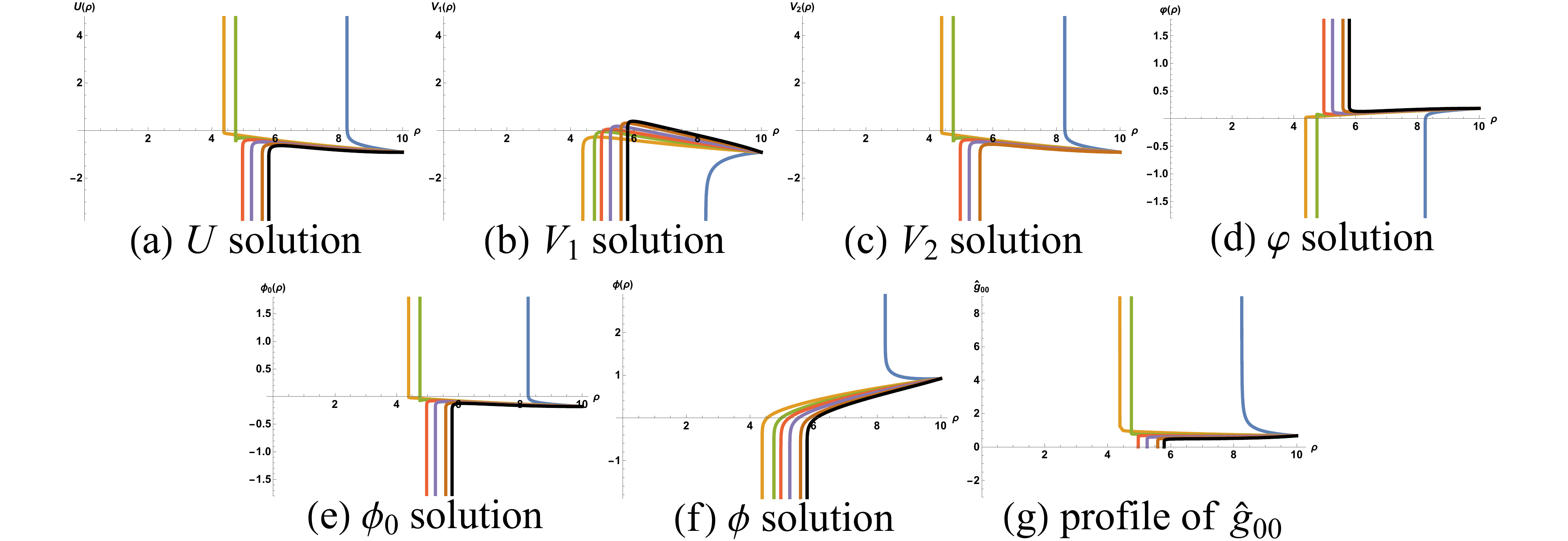}
\caption{Interpolating solutions between the locally $SO(2)\times SO(2)$ flat domain wall as $\rho\rightarrow+\infty$ and $t\times H^2\times \mathbb{R}^2$-sliced curved domain walls for $SO(2)\times SO(2)$ twist in $SO(2,2)\ltimes \mathbb{R}^{4}$ gauge group. The blue, orange, green, red, purple, brown, and black curves refer to $z_1=-4, -1, -0.40, 0.11, 1$, $2.60$, and $4$, respectively. }
\label{40_HR_SO(2)xSO(2)_SO(22)gg_flows}
\end{figure}
\clearpage\newpage
\begin{figure}[h!]
  \centering
    \includegraphics[width=\linewidth]{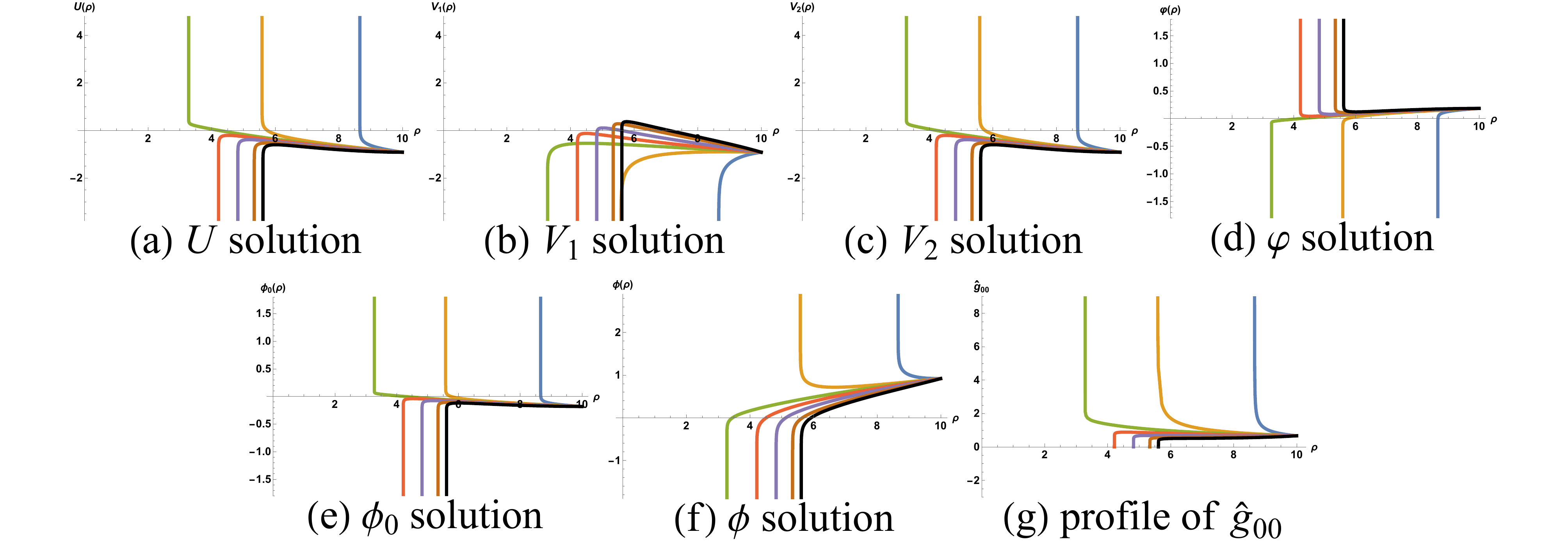}
\caption{Interpolating solutions between the locally $SO(2)\times SO(2)$ flat domain wall as $\rho\rightarrow+\infty$ and $t\times \mathbb{R}^2\times \mathbb{R}^2$-sliced curved domain walls for $SO(2)\times SO(2)$ twist in $SO(2,2)\ltimes \mathbb{R}^{4}$ gauge group. The blue, orange, green, red, purple, brown, and black curves refer to $z_1=-4, -1, -0.4, 0.1, 1$, $2.6$, and $4$, respectively. }
\label{40_RR_SO(2)xSO(2)_SO(22)gg_flows}
\end{figure}

\subsection{D4-branes wrapped on a Kahler four-cycle}
\subsubsection{Solutions with $SO(3)$ twist}\label{App_D_2_1}
\begin{figure}[h!]
  \centering
    \includegraphics[width=\linewidth]{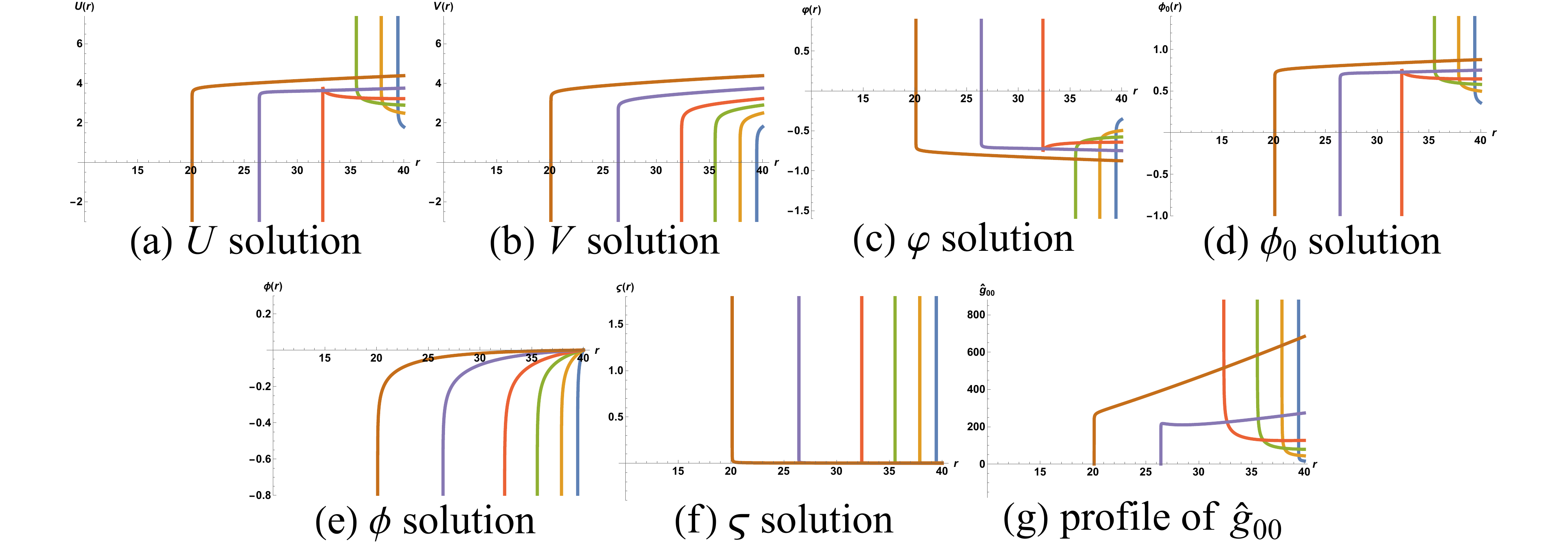}
\caption{Interpolating solutions between the locally $SO(4)$ flat domain wall as $\rho\rightarrow +\infty$ and $t\times CP^2$-sliced curved domain walls for $SO(3)$ twist in $SO(4)\ltimes \mathbb{R}^{4}$ gauge group. The blue, orange, green, red, purple, and brown curves refer to $g=0.30, 0.60, 0.90, 1.25, 2.12$, and $4$, respectively.}
\label{40_CP2_special_SO(3)_SO(4)gg_flows}
\end{figure}

\begin{figure}[h!]
  \centering
    \includegraphics[width=\linewidth]{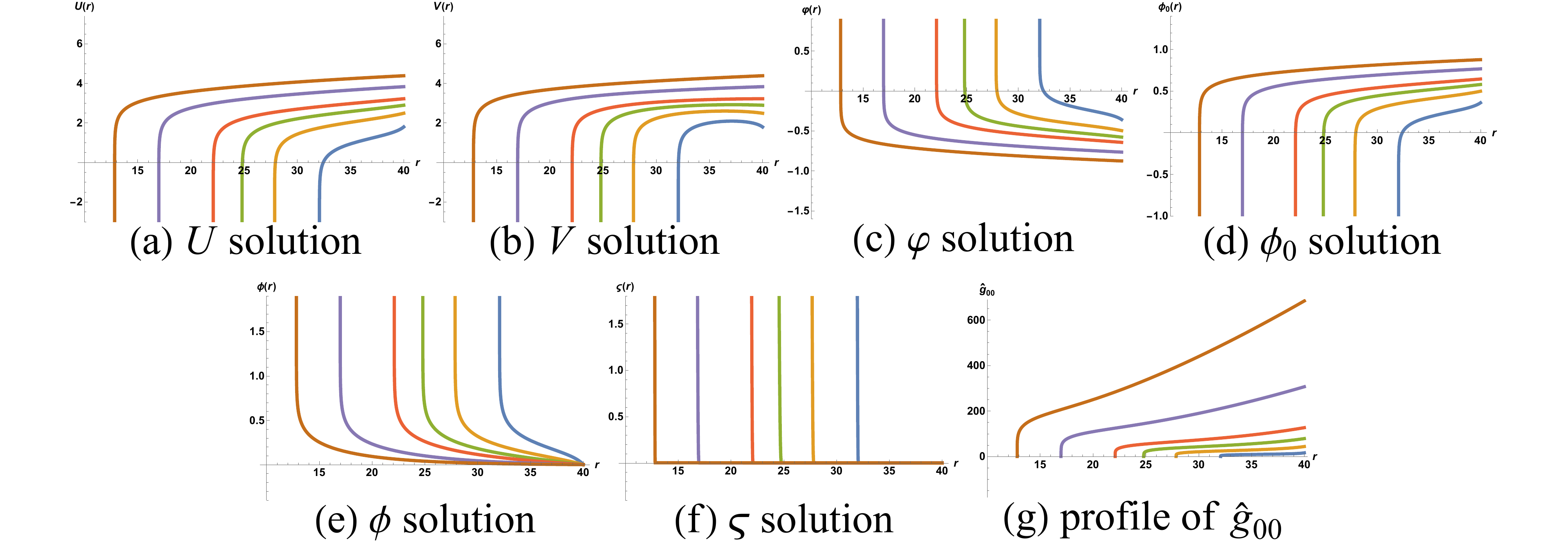}
\caption{Interpolating solutions between the locally $SO(4)$ flat domain wall as $\rho\rightarrow+\infty$ and $t\times CH^2$-sliced curved domain walls for $SO(3)$ twist in $SO(4)\ltimes \mathbb{R}^{4}$ gauge group. The blue, orange, green, red, purple, and brown curves refer to $g=0.30, 0.60, 0.90, 1.25, 2.30$, and $4$, respectively.}
\label{40_CH2_special_SO(3)_SO(4)gg_flows}
\end{figure}

\begin{figure}[h!]
  \centering
    \includegraphics[width=\linewidth]{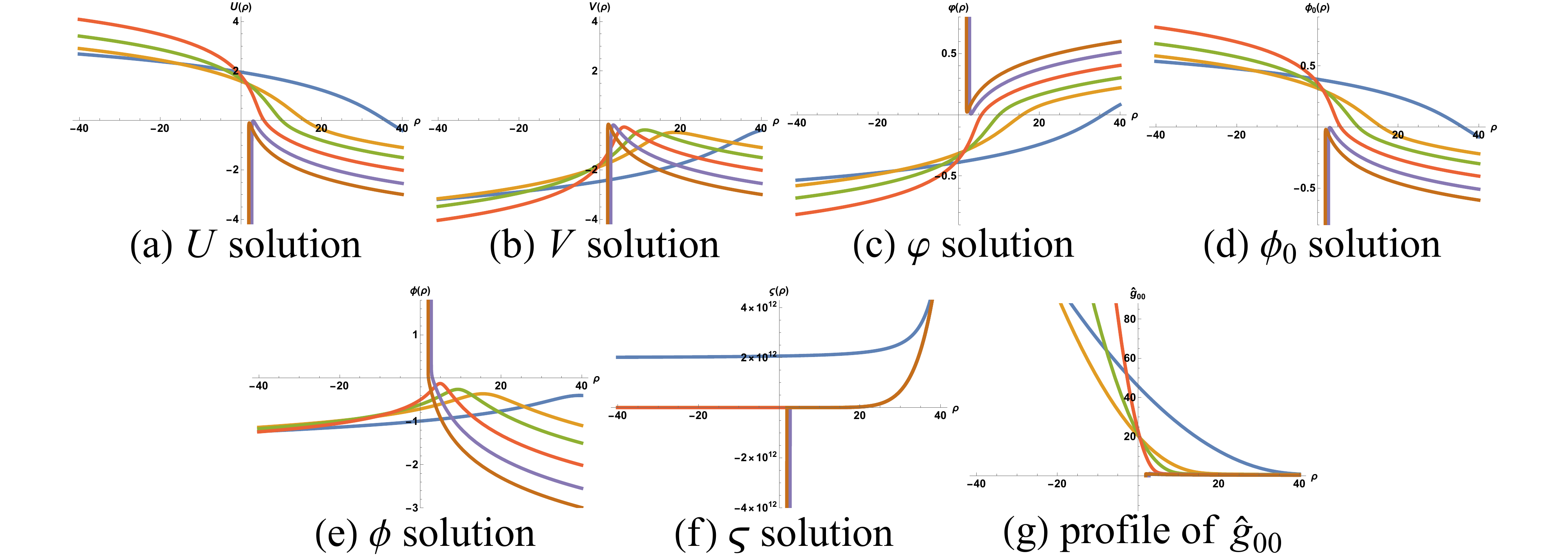}
\caption{Interpolating solutions between the locally $SO(3)$ flat domain wall as $\rho\rightarrow+\infty$ and $t\times CP^2$-sliced curved domain walls for $SO(3)$ twist in $SO(3,1)\ltimes \mathbb{R}^{4}$ gauge group. The blue, orange, green, red, purple, and brown curves refer to $g=0.3, 0.6, 0.9, 1.5, 2.56$, and $4$.}
\label{40_S3_SO(3)_SO(31)gg_flows}
\end{figure}

\clearpage\newpage
\begin{figure}[h!]
  \centering
    \includegraphics[width=\linewidth]{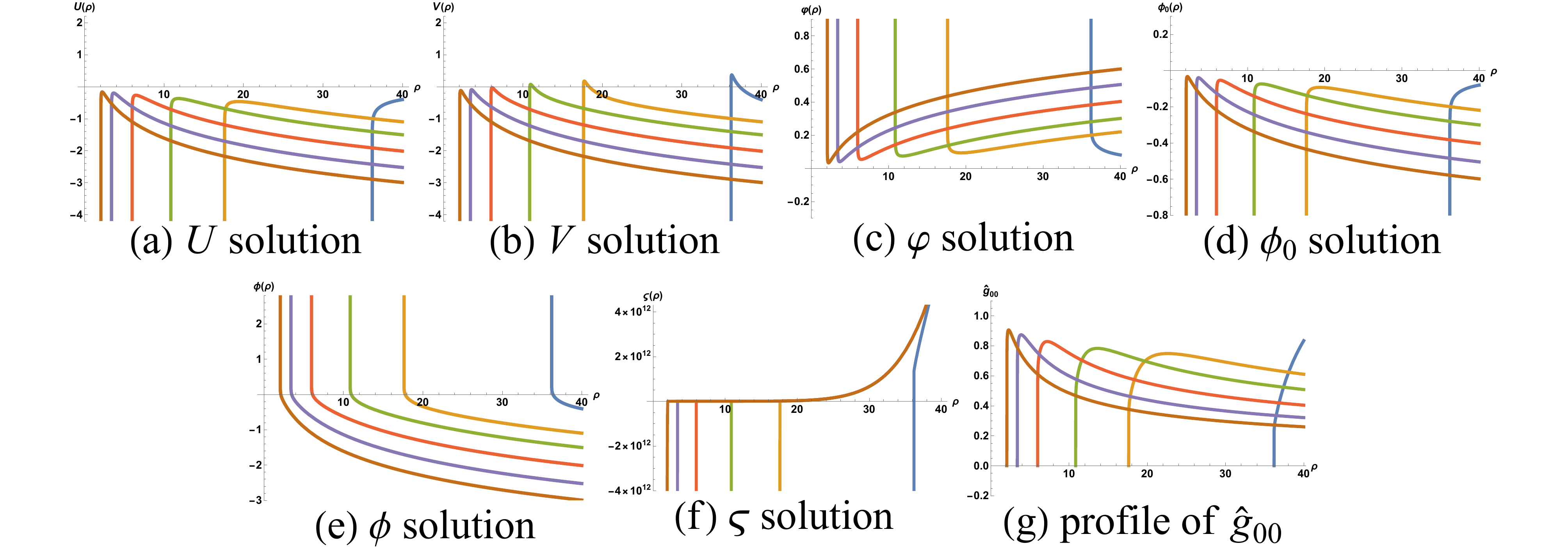}
\caption{Interpolating solutions between the locally $SO(3)$ flat domain wall as $\rho\rightarrow+\infty$ and $t\times CH^2$-sliced curved domain walls for $SO(3)$ twist in $SO(3,1)\ltimes \mathbb{R}^{4}$ gauge group. The blue, orange, green, red, purple, and brown curves refer to $g=0.3, 0.6, 0.9, 1.5, 2.5$, and $4$.}
\label{40_H3_SO(3)_SO(31)gg_flows}
\end{figure}

\subsubsection{Solutions with $SO(2)$ twist}\label{App_D_2_2}
\begin{figure}[h!]
  \centering
    \includegraphics[width=\linewidth]{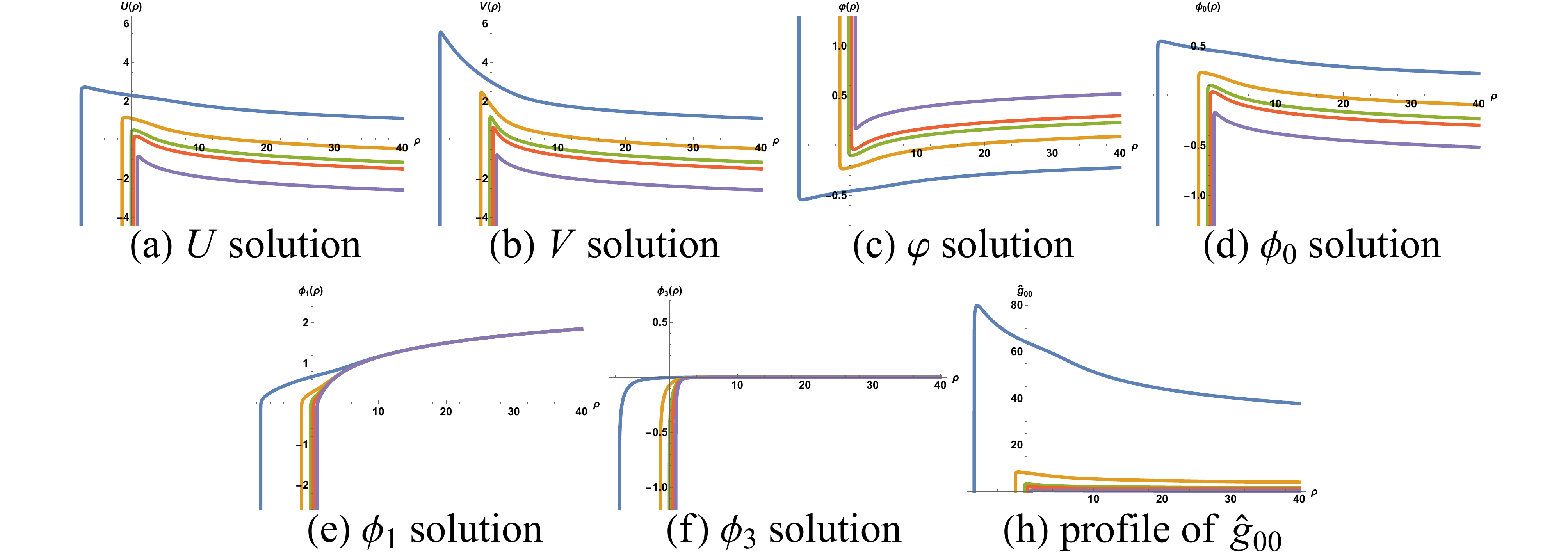}
\caption{Interpolating solutions between the locally $SO(2)$ flat domain wall as $\rho\rightarrow+\infty$ and $t\times CP^2$-sliced curved domain walls for $SO(2)$ twist in $SO(3,1)\ltimes \mathbb{R}^{4}$ gauge group. The blue, orange, green, red, and purple curves refer to $g=-0.1, -0.5, -1, -1.4$, and $-4.2$, respectively.}
\label{40_CP2_SO(2)_SO(31)gg_flows}
\end{figure}

\begin{figure}[h!]
  \centering
    \includegraphics[width=\linewidth]{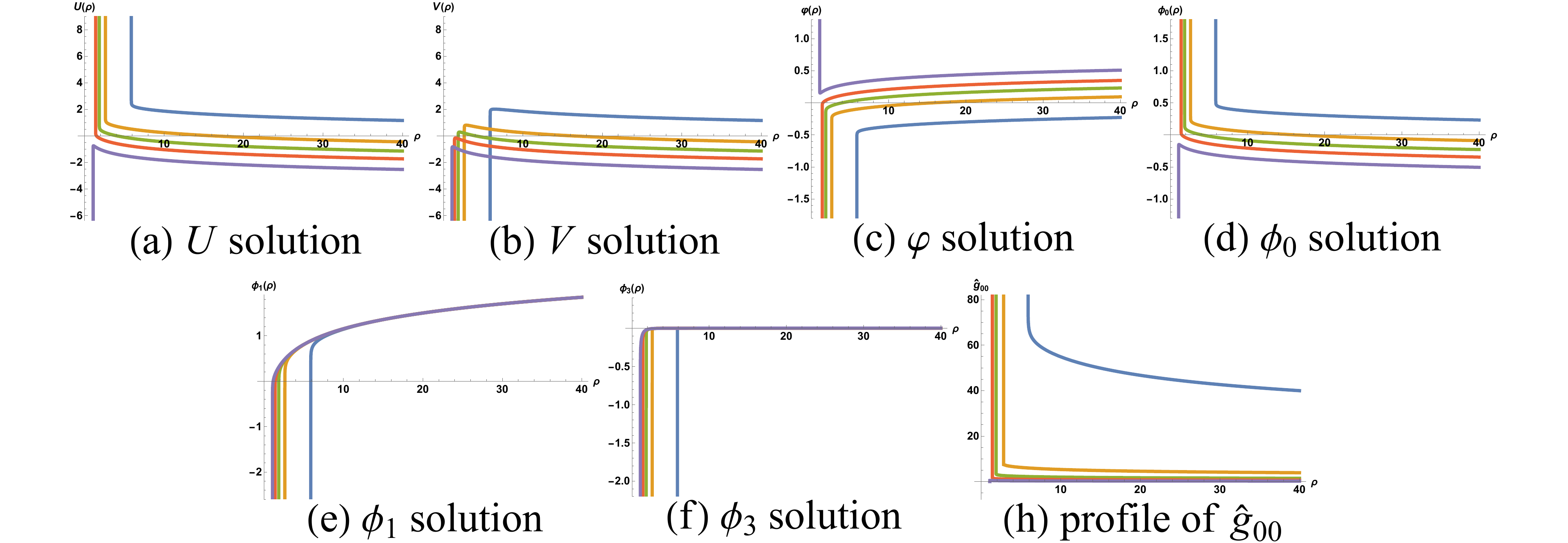}
\caption{Interpolating solutions between the locally $SO(2)$ flat domain wall as $\rho\rightarrow+\infty$ and $t\times CH^2$-sliced curved domain walls for $SO(2)$ twist in $SO(3,1)\ltimes \mathbb{R}^{4}$ gauge group. The blue, orange, green, red, and purple curves refer to $g=-0.1, -0.5, -1, -1.8$, and $-4$, respectively.}
\label{40_CH2_SO(2)_SO(31)gg_flows}
\end{figure}

\begin{figure}[h!]
  \centering
    \includegraphics[width=\linewidth]{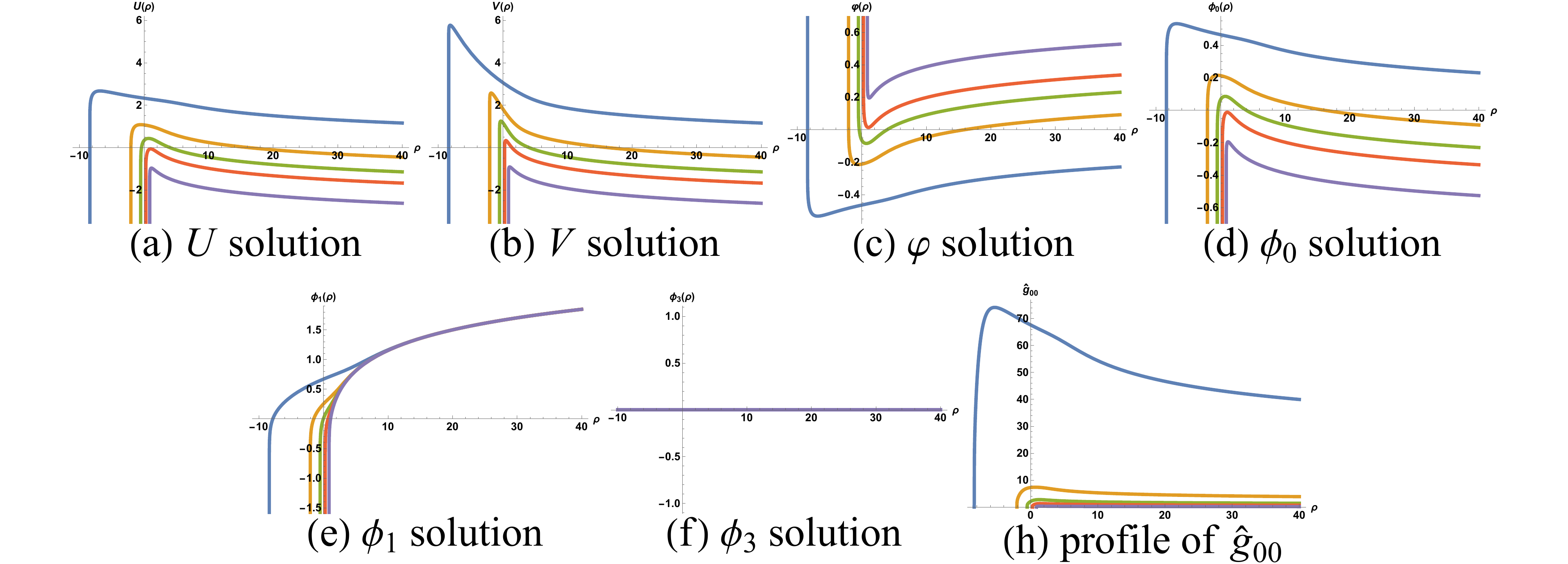}
\caption{Interpolating solutions between the locally $SO(2)$ flat domain wall as $\rho\rightarrow+\infty$ and $t\times CP^2$-sliced curved domain walls for $SO(2)$ twist in $SO(2,2)\ltimes \mathbb{R}^{4}$ gauge group. The blue, orange, green, red, and purple curves refer to $g=-0.1, -0.5, -1, -1.7$, and $-4.4$, respectively.}
\label{40_CP2_SO(2)_SO(22)gg_flows}
\end{figure}

\begin{figure}[h!]
  \centering
    \includegraphics[width=\linewidth]{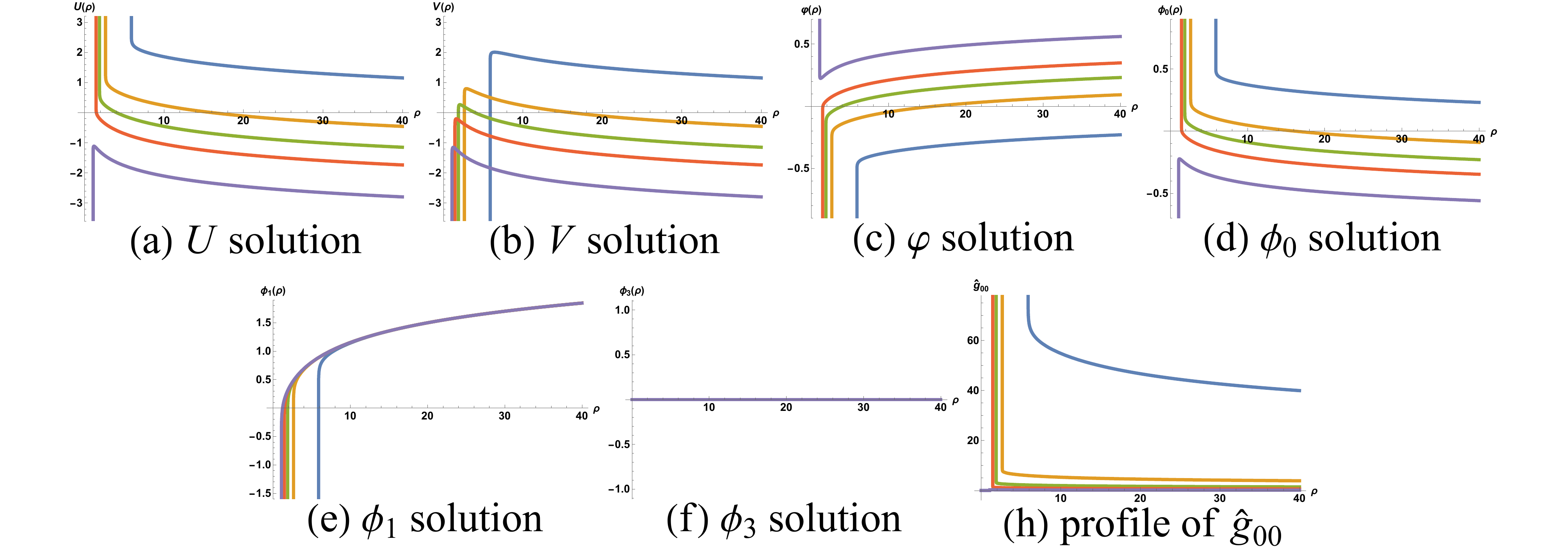}
\caption{Interpolating solutions between the locally $SO(2)$ flat domain wall as $\rho\rightarrow+\infty$ and $t\times CH^2$-sliced curved domain walls for $SO(2)$ twist in $SO(2,2)\ltimes \mathbb{R}^{4}$ gauge group. The blue, orange, green, red, and purple curves refer to $g=-0.1, -0.5, -1, -1.8$, and $-5.2$, respectively.}
\label{40_CH2_SO(2)_SO(22)gg_flows}
\end{figure}

\begin{figure}[h!]
  \centering
    \includegraphics[width=\linewidth]{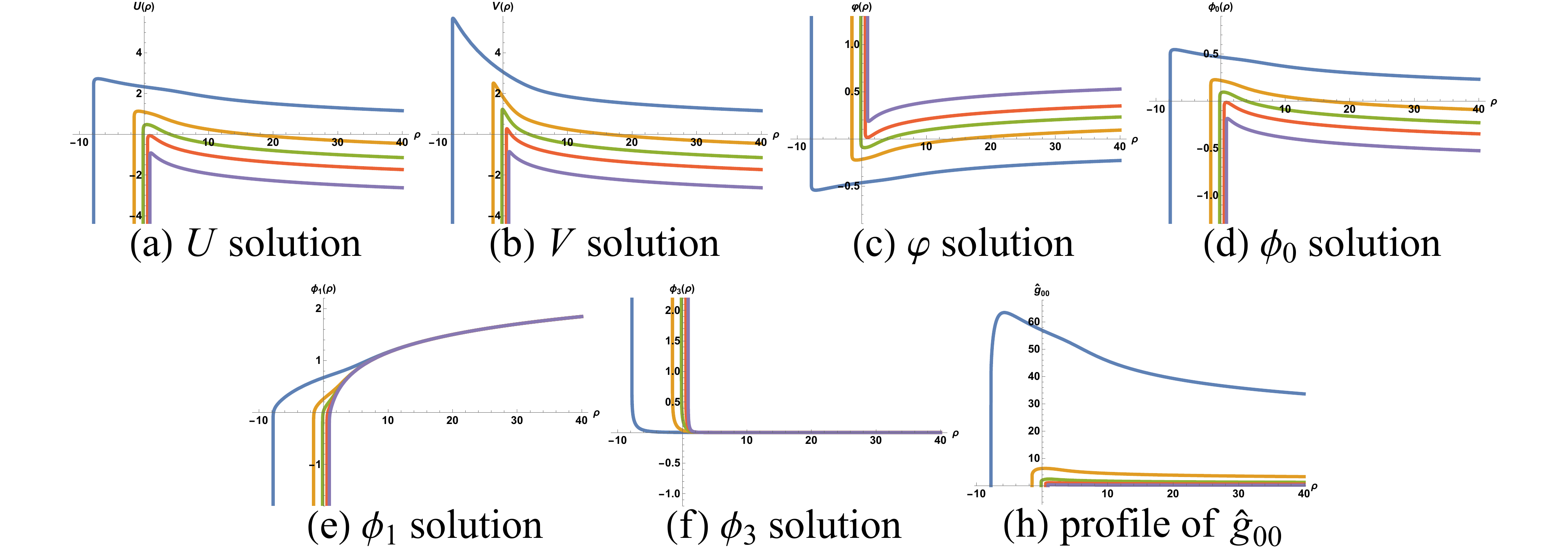}
\caption{Interpolating solutions between the locally $SO(2)$ flat domain wall as $\rho\rightarrow+\infty$ and $t\times CP^2$-sliced curved domain walls for $SO(2)$ twist in $CSO(2,1,1)\ltimes \mathbb{R}^{4}$ gauge group. The blue, orange, green, red, and purple curves refer to $g=-0.1, -0.5, -1, -1.8$, and $-4.4$, respectively.}
\label{40_CP2_SO(2)_CSO(211)gg_flows}
\end{figure}

\clearpage\newpage
\begin{figure}[h!]
  \centering
    \includegraphics[width=\linewidth]{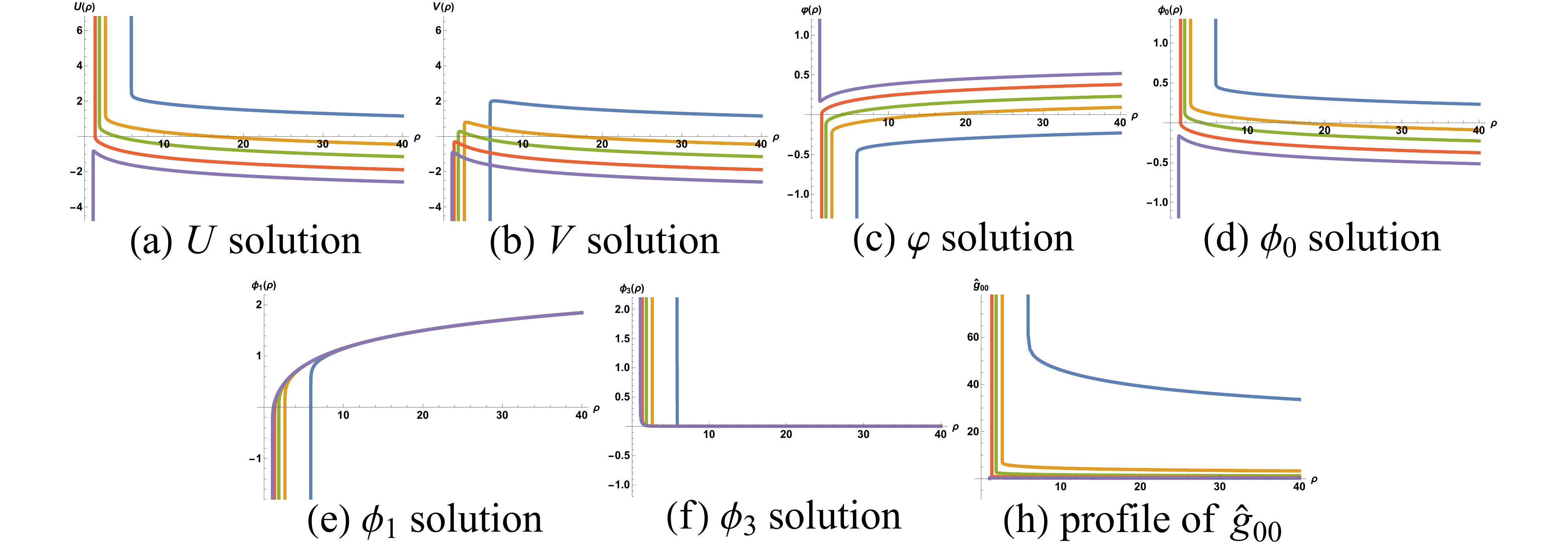}
\caption{Interpolating solutions between the locally $SO(2)$ flat domain wall as $\rho\rightarrow+\infty$ and $t\times CH^2$-sliced curved domain walls for $SO(2)$ twist in $CSO(2,1,1)\ltimes \mathbb{R}^{4}$ gauge group. The blue, orange, green, red, and purple curves refer to $g=-0.1, -0.5, -1, -2.1$, and $-4.2$, respectively.}
\label{40_CH2_SO(2)_CSO(211)gg_flows}
\end{figure}

%%%%%%%%%%%%%%%%%%%%%%%%%%%%%%%%%%%%%%%%%%%%%%%%%%%%%%%%%%%%%%%%%%%%%%%%%%%%%%%%%%%%%%%%%%%%%%%%%%%%%%%%%%%%%%%%%%%%%%%%%%%%%%%%%%%%%%%%%

\end{document}